\documentclass[a4paper,aps, prd, reprint,
  onecolumn, amsfonts, amssymb, amsmath, eqsecnum,
  showkeys, showpacs, superscriptaddress, floatfix, preprintnumbers, nodate]{revtex4-1}

\usepackage{hyperref}
\hypersetup{colorlinks = true,
  linkcolor = blue,
  urlcolor = blue,
  citecolor = blue}
\usepackage{braket}
\usepackage{mathtools} 
\usepackage{mathrsfs}
  \DeclareMathAlphabet{\mathpzc}{OT1}{pzc}{m}{it} 
\usepackage{slashed}

\usepackage{subfig}
\usepackage{array}
 
\newcommand{\D}[2]{\mathrm{d}^{#1}{#2}}

\begin{document}

\title{Perturbative Non-Equilibrium Thermal Field Theory}
\author{Peter Millington}
\email{peter.millington@manchester.ac.uk}
\affiliation{Consortium for Fundamental Physics, School of Mathematics and Statistics,
University of Sheffield, Sheffield S3 7RH, United Kingdom}
\affiliation{Consortium for Fundamental Physics, School of Physics and Astronomy,
University of Manchester, Manchester M13 9PL, United Kingdom}
\author{Apostolos Pilaftsis}
\email{apostolos.pilaftsis@manchester.ac.uk}
\affiliation{Consortium for Fundamental Physics, School of Physics and Astronomy,
University of Manchester, Manchester M13 9PL, United Kingdom}
\preprint{MAN/HEP/2012/04}
\eprint{arXiv: 1211.3152}

\begin{abstract}
  We present a new perturbative formulation of non-equilibrium
  thermal  field theory, based  upon non-homogeneous  free propagators
  and  time-dependent  vertices.    Our  approach  to  non-equilibrium
  dynamics yields time-dependent diagrammatic perturbation series that
  are  free of  pinch singularities,  without  the need  to resort  to
  quasi-particle  approximation or  effective  resummations of  finite
  widths.  In our formalism, the avoidance of pinch singularities is a
  consequence of  the consistent inclusion of  finite-time effects and
  the proper consideration of the time of observation.  After arriving
  at a physically meaningful  definition of particle number densities,
  we   derive  master   time  evolution   equations   for  statistical
  distribution   functions,  which   are  valid   to  all   orders  in
  perturbation  theory.  The resulting  equations do  not rely  upon a
  gradient expansion of Wigner transforms or involve any separation of
  time scales.   To illustrate the  key features of our  formalism, we
  study out-of-equilibrium  decay dynamics of unstable  particles in a
  simple scalar model.  In~particular, we show how finite-time effects
  remove  the pinch  singularities  and lead  to  violation of  energy
  conservation at early times,  giving rise to otherwise kinematically
  forbidden processes.  The non-Markovian nature of the memory effects
  as predicted in our formalism is explicitly demonstrated.
\end{abstract}

\maketitle

\tableofcontents
\flushbottom

\newpage

\section{Introduction}
\label{sec:intro}

As modern particle physics continues to advance at both the energy and
intensity  frontiers,  we are  increasingly  concerned with  transport
phenomena  in  dense  systems  of  ultra-relativistic  particles,  the
so-called   \textit{density  frontier}.   One   such  system   is  the
de\-confined phase of quantum chromodynamics, known as the quark-gluon
plasma~\cite{Blaizot:2001nr}, whose  existence has been  inferred from
the  observation  of  jet  quenching  in  {Pb-Pb}  collisions  by  the
ATLAS~\cite{Aad:2010bu},        CMS~\cite{Chatrchyan:2011sx}       and
ALICE~\cite{Aamodt:2010jd}  experiments at  the  CERN Large
Hadron  Collider (LHC).   In  addition to  laboratory experiments,  an
understanding  of  such ultra-relativistic  many-body  dynamics is  of
interest     in     theoretical     astro-particle     physics     and
cosmology. Predictions about the  evolution of the early Universe rely
upon  our understanding  of  the dynamics  of  states at  the end  and
shortly after the phase of cosmological inflation.

The Wilkinson  Microwave Anisotropy Probe (WMAP)~\cite{Spergel:2003cb,
  Spergel:2006hy}  measured a  baryon-to-photon ratio  at  the present
epoch      of       ${\eta\:      =\:      n_{B}/n_{\gamma}\:      =\:
  6.116^{+0.197}_{-0.249}\times 10^{-10}}$,  where $n_{\gamma}$ is the
number density of photons and ${n_{B}\: =\: n_b\: -\: n_{\bar{b}}}$ is
the difference in number  densities of baryons and anti-baryons.  This
observed  Baryon Asymmetry of  the Universe  (BAU) is  also consistent
with    the     predictions    of    Big     Bang    Nucleo\-synthesis
(BBN)~\cite{Steigman:2005ys}.  The generation  of the BAU requires the
presence of  out-of-equilibrium processes and the  violation of baryon
number ($B$), charge ($C$)  and charge-parity ($CP$), according to the
Sakharov conditions~\cite{Sakharov:1967dj}. One  such set of processes
is    prescribed   by    the   scenarios    of    baryo\-genesis   via
leptogenesis~\cite{Kuzmin:1985mm,  Fukugita:1986hr,  Fukugita:1990gb},
in which  an initial  excess in lepton  number ($L$), provided  by the
decay  of heavy  right-handed Majorana  neutrinos, is  converted  to a
baryon   number  excess   through   the  ${B+L}$-violating   sphaleron
interactions.  The description of  such phenomena require a consistent
approach to the non-equilibrium dynamics of particle number densities.
Further  examples  to which  non-equilibrium  approaches are  relevant
include,    for   instance,   the    phenomena   of    reheating   and
preheating~\cite{Kofman:1994rk,    Boyanovsky:1994me,   Baacke:1996kj,
  Kofman:1997yn}   and   the   generation   of   dark   matter   relic
densities~\cite{Bender:2012gc}.

The classical evolution of particle number densities in the early
Universe is described by Boltzmann transport equations; see for
instance~\cite{Kolb:1979qa, Carena:2002ss, Giudice:2003jh,
  Pilaftsis:2003gt, Buchmuller:2004nz, Pilaftsis:2005rv,
  Davidson:2008bu, Deppisch:2010fr,Blanchet:2011xq}.  A semi-classical
improvement to these equations may be achieved by substituting the
classical Boltzmann distributions with quantum-statistical
Bose--Einstein and Fermi--Dirac distribution functions for bosons and
fermions, respectively.  However, such improved approaches take into
account finite-width and off-shell effects by means of effective
field-theoretic methods.  Hence, a complete and systematic
field-theoretic description of quantum transport phenomena would be
desirable.

The first framework for calculating Ensemble Expectation Values (EEVs)
of field operators  was provided by Matsubara~\cite{Matsubara:1955ws}.
This so-called Imaginary Time  Formalism (ITF) of thermal field theory
is  derived  by interpreting  the  canonical  density  operator as  an
evolution  operator  in negative  imaginary  time.  Real-time  Green's
functions  may  then  be  obtained by  subtle  analytic  continuation.
Nevertheless,  the applicability  of the  ITF remains  limited  to the
description of processes occurring in thermodynamic equilibrium.

The calculation of EEVs of operators in non-static systems is
performed using the so-called Real Time Formalism (RTF); see for
example \cite{Chou:1984es, Landsman:1986uw}.  In particular, for
non-equilibrium systems, one uses the Closed Time Path (CTP)
formalism, or the {\textit{in}-\textit{in}} formalism, due to
Schwinger and Keldysh~\cite{Schwinger:1960qe, Keldysh:1964ud}.  The
correspondence of these results with those obtained by the ITF in the
equilibrium limit are discussed extensively in the
literature~\cite{Kobes:1990kr, Aurenche:1991hi, Xu:1993zh,
  Baier:1993yh, Evans:1994gb, vanEijck:1994rw, Zhou:2001sk}.  A
non-perturbative loopwise expansion of the CTP generating functional
is then provided by the Cornwall--Jackiw--Tomboulis (CJT) effective
action~\cite{Cornwall:1974vz, Carrington:2004sn}, which was
subsequently applied to the CTP formalism by Calzetta and
Hu~\cite{Calzetta:1986ey, Calzetta:1986cq}. The CJT effective action
has been used extensively in applications to $1/N$ expansions far
from equilibrium~\cite{Aarts:2001yn, Aarts:2002dj, Aarts:2003bk,
  Berges:2004pu, Arrizabalaga:2004iw}.

Recently, the computation of the out-of-equilibrium evolution of
particle number densities has received much attention, where several
authors put forward quantum-corrected or quantum Boltzmann
equations~\cite{Berera:1998gx, Boyanovsky:1998pg, Niegawa:1999pn,
  Cassing:1999wx, Dadic:1999bp, Ivanov:1999tj,Buchmuller:2000nd,
  Morawetz:1999bv, Aarts:2001qa, Blaizot:2001nr, Juchem:2003bi,
  Prokopec:2003pj, Prokopec:2004ic, Berges:2004yj,
  Arrizabalaga:2005tf, Berges:2005ai, Lindner:2005kv,
  FillionGourdeau:2006hi, DeSimone:2007rw, Cirigliano:2007hb,
  Garbrecht:2008cb, Garny:2009qn, Cirigliano:2009yt, Beneke:2010dz,
  Anisimov:2010dk, Hamaguchi:2011jy, Fidler:2011yq, Gautier:2012vh,
  Drewes:2012qw}.  Existing approaches generally rely upon the Wigner
transformation and gradient
expansion~\cite{Winter:1986da,Bornath:1996zz} of a system of
Kadanoff--Baym (KB)~\cite{Baym:1961zz, Kadanoff1989} equations,
originally applied in the non-relativistic
regime~\cite{Danielewicz:1982kk,Lipavsky:1986zz,Rammer:1986zz,
  Bornath:1996zz}. Often the truncation of the gradient expansion is
accompanied by quasi-particle ansaetze for the forms of the
propagators.  Similar approaches have recently been applied to the
glasma~\cite{Berges:2012cj}.  Dynamical equations have
also been derived by expansion of the Liouville-von Neumann equation
\cite{Sigl:1992fn,Gagnon:2010kt} and using functional renormalization
techniques~\cite{Gasenzer:2010rq}.

In this article, we present a new approach to non-equilibrium thermal
field theory.  Our approach is based upon a diagrammatic perturbation
series, constructed from non-homogeneous free propagators and
time-dependent vertices, which encode the absolute space-time
dependence of the thermal background.  In particular, we show how the
systematic inclusion of finite-time effects and the proper
consideration of the time of observation render our perturbative
expansion free of pinch singularities, thereby enabling a consistent
treatment of non-equilibrium dynamics.  Unlike other methods, our
approach does not require the use of quasi-particle approximation or
other effective resummations of finite-width effects.

A key element of our formalism has been to define physically
meaningful particle number densities in terms of off-shell Green's
functions.  This definition is unambiguous, as it can be closely
linked to Noether's charge, associated with a partially conserved
current.  Subsequently, we derive master time evolution equations for
the statistical distribution functions related to particle number
densities.  These time evolution equations do not rely on the
truncation of a gradient expansion of the so-called Wigner transforms;
neither do they involve separation of various time scales. Instead,
they are built of non-homogeneous free propagators, with modified
time-dependent Feynman rules, which enable us to analyze the pertinent
kinematics fully.  Our analysis shows that the systematic inclusion of
finite-time effects leads to the microscopic violation of energy
conservation at early times.  Aside from preventing the appearance of
pinch singularities, the effect of a finite time interval of evolution
leads to contributions from processes that would otherwise be
kinematically disallowed on grounds of energy conservation.  Applying
this formalism to a simple scalar model with unstable particles, we
show that these evanescent processes contribute significantly to the
rapidly-oscillating transient evolution of these systems, inducing
late-time memory effects.  We find that the spectral evolution of
two-point correlation functions exhibits an oscillating pattern with
time-varying frequencies.  Such an oscillating pattern signifies a
non-Markovian evolution of memory effects, which is a distinctive
feature governing truly out-of-thermal-equilibrium dynamical
systems. A summary of the main results detailed in this article can be
found in \cite{Millington:2013isa}.

\begin{table}
  \caption{\label{tab:gloss} Glossary for clarifying polysemous notation.}
    \begin{tabular*}{3.4in}[c]{c l}
      \hline\hline
      $T$ & thermodynamic temperature \\
      $t$ & macroscopic time \\
      $\tilde{t}$ & microscopic real time \\
      $\tau$ & microscopic negative imaginary time \\
      $\mathfrak{t}$ & complex time \\
      $\mathrm{T}$($\bar{\mathrm{T}}$) & time-(anti-time-)ordering
      operator \\
      $\mathrm{T}_{\mathcal{C}}$ & path-ordering operator \\
      $Z$ & wavefunction renormalization \\
      $\mathcal{Z}$ & partition function/ generating functional \\
      $f$ & statistical distribution function\\
      $f_{\beta}$ & Boltzmann distribution \\
      $f_{\mathrm{B}}$ & Bose--Einstein distribution \\
      $\tilde{f}$ & ensemble function \\
      $\rho$ & density operator/ density matrix \\
      $n$ & number density \\
      $N$ & total particle number \\
      \hline\hline
    \end{tabular*}
\end{table}

The outline of the paper is as follows. In Section~\ref{sec:canon}, we
review the canonical quantization of a scalar field theory, placing
particular emphasis on the inclusion of a finite time of coincidence
for the three equivalent pictures of quantum mechanics, namely the
Schr\"{o}dinger, Heisenberg and Dirac (interaction) pictures.  In
Section~\ref{sec:CTP}, we introduce the CTP formalism, limiting
ourselves initially to consider its application to quantum field
theory at zero temperature and density. This is followed by a
discussion of the constraints upon the form of the CTP propagator.
With these prerequisites reviewed, we proceed in
Section~\ref{sec:nonhom} to discuss the generalization of the CTP
formalism to finite temperature and density in the presence of both
spatially and temporally inhomogeneous backgrounds.  In the same
section, we derive the most general form of the non-homogeneous free
propagators for a scalar field theory.  The thermodynamic equilibrium
limit is outlined in Section~\ref{sec:eq}.  Subsequently, in
Section~\ref{sec:num}, we define the concept of particle number
density and relate this to a perturbative loopwise expansion of the
resummed CTP propagators.  In Section~\ref{sec:eom}, we derive {\em
  new} master time evolution equations for statistical distribution
functions, which go over to classical Boltzmann equations in the
appro\-priate limits. In Section~\ref{sec:oneloop}, we demonstrate the
absence of pinch singularities in the perturbation series arising in
our approach at the one-loop level.  Section~\ref{sec:toy} studies the
thermalization of unstable particles in a simple scalar model, where
particular emphasis is laid on the early-time behavior and the impact
of the finite-time effects.  Finally, our conclusions and possible
future directions are presented in Section~\ref{sec:con}.

For clarity, a glossary that might  be useful to the reader to clarify
polysemous    notation     is    given    in    Table~\ref{tab:gloss}.
Appendix~\ref{app:rel} provides a  summary of all important propagator
definitions,  along with  their  basic relations  and properties.   In
Appendix~\ref{app:ITF}, we describe the correspondence between the RTF
and ITF  in the  equilibrium limit  at the one-loop  level for  a real
scalar field theory with a cubic self-interaction.  The generalization
of   our  approach   to   complex  scalar   fields   is  outlined   in
Appendix~\ref{app:complex}.  Appendix~\ref{app:dens} contains a series
expansion  of the most  general non-homogeneous  Gaussian-like density
operator.   In Appendix~\ref{app:kb}, we  summarize the  derivation of
the so-called  Kadanoff--Baym equations and  their subsequent gradient
expansion.   Lastly,  in  Appendix~\ref{app:loops},  we  describe  key
technical details  involved in the calculation of  loop integrals with
non-homogeneous free propagators.

\newpage
\section{Canonical Quantization}
\label{sec:canon}

In  this section,  we review  the  basic results  obtained within  the
canonical quantization  formalism for  a massive scalar  field theory.
This discussion  will serve as  a precursor for our  generalization to
finite temperature and density, which follows in subsequent sections.

As a starting point, we consider a simple self-interacting theory of a
real scalar  field~$\Phi(x)$ with mass~$M$, which is  described by the
Lagrangian
\begin{equation}
  \label{eq:scallag}
  \mathcal{L}(x)\ =\
  \tfrac{1}{2}\partial_{\mu}\Phi(x)\partial^{\mu}\Phi(x)
  \: -\: \tfrac{1}{2}M^2\Phi^2(x)\: -\: \tfrac{1}{3!}g\Phi^3(x)
  \: -\: \tfrac{1}{4!}\lambda\Phi^4(x)\; ,
\end{equation}
where $g$ and $\lambda$  are dimensionful and dimensionless couplings,
respectively. Throughout this article, we use the short-hand notation:
$x\:\equiv\: x^{\mu}\: =\: (x^0,\mathbf{x})$, for the four-dimensional
space-time   arguments    of   field   operators,    and   adopt   the
signature~$(+,-,-,-)$      for      the      Minkowski      space-time
metric~$\eta_{\mu\nu}$.

It proves  convenient to start our canonical  quantization approach in
the   Schr\"{o}dinger   picture,   where   the   state   vectors   are
\emph{time dependent}, whilst basis vectors  and operators are, in the
absence  of  external  sources, \emph{time independent}.   Hence,  the
\emph{time-independent}    Schr\"{o}dinger-picture   field   operator,
denoted by  a subscript~$\mathrm{S}$, may  be written in  the familiar
plane-wave decomposition
\begin{equation}
  \Phi_{\mathrm{S}}(\mathbf{x};\tilde{t}_i)\ =\ \int\!\D{}{\Pi_{\mathbf{p}}}\;
  \Big(\,a_{\mathrm{S}}(\mathbf{p};\tilde{t}_i)e^{i\mathbf{p}\cdot\mathbf{x}}
  \: +\: a_{\mathrm{S}}^{\dag}(\mathbf{p};\tilde{t}_i)
  e^{-i\mathbf{p}\cdot\mathbf{x}}\,\Big)\; ,
\end{equation}
where we have introduced the short-hand notation:
\begin{equation}
  \label{eq:lips}
  \!\int\!\D{}{\Pi_{\mathbf{p}}}\ \equiv\ \!\int\!\!
  \frac{\D{3}{\mathbf{p}}}{(2\pi)^3}\,\frac{1}{2E(\mathbf{p})}
  \ =\ \!\int\!\!\frac{\D{4}{p}}{(2\pi)^4}\;
  2\pi\theta(p_0)\delta(p^2-M^2)\; ,
\end{equation}
for  the  Lorentz-invariant  phase space (LIPS).   In~(\ref{eq:lips}),
$E(\mathbf{p})\:  =\: \sqrt{\mathbf{p}^2+M^2}$  is the  energy  of the
single-particle    mode     with    three-momentum~$\mathbf{p}$    and
$\theta(p_0)$ is  the generalized unit step function, defined  by the
Fourier representation
\begin{equation}
  \label{eq:thetastep}
  \theta(p_0)\ \equiv\ i\!\int_{-\infty}^{+\infty}\!\frac{\D{}{\xi}}{2\pi}
  \;\frac{e^{-ip_0\xi}}{\xi\: +\: i\epsilon}\ =\
  \begin{cases}
    1\; ,\qquad &p_0\ >\ 0\\
    \frac{1}{2}\; ,\qquad & p_0\ =\ 0\\
    0\; ,\qquad & p_0\ <\ 0\;,
  \end{cases}
\end{equation}
where $\epsilon=0^+$.  It  is essential to stress here  that we define
the Schr\"{o}dinger-, Heisenberg-  and Interaction (Dirac)-pictures to
be    coincident   at    the   finite    \emph{microscopic}   boundary
time~$\tilde{t}_i$, such that
\begin{equation}
  \Phi_{\mathrm{S}}(\mathbf{x};\tilde{t}_i)\ =\
  \Phi_{\mathrm{H}}(\tilde{t}_i,\mathbf{x};\tilde{t}_i)\ =\
  \Phi_{\mathrm{I}}(\tilde{t}_i,\mathbf{x};\tilde{t}_i)\; ,
\end{equation}
where  implicit  dependence upon  the  boundary time~$\tilde{t}_i$  is
marked by separation from explicit arguments by a semi-colon.

The    \emph{time-independent}    Schr\"{o}dinger-picture    operators
$a_{\mathrm{S}}^{\dag}(\mathbf{p};\tilde{t}_i)$                     and
$a_{\mathrm{S}}(\mathbf{p};\tilde{t}_i)$  are the  usual  creation and
annihilation operators, which  act on the stationary vacuum~$\ket{0}$,
respectively    creating   and    destroying   \emph{time-independent}
single-particle momentum eigenstates.  Their defining properties are:
\begin{subequations}
  \begin{align}
    a_{\mathrm{S}}^{\dag}(\mathbf{p};\tilde{t}_i)\!\ket{0}&\ =\
    \ket{\mathbf{p};\tilde{t}_i}\!\; , \\
    a_{\mathrm{S}}(\mathbf{p};\tilde{t}_i)\!\ket{\mathbf{p}';\tilde{t}_i}&
    \ =\ (2\pi)^32E(\mathbf{p})\delta^{(3)}(\mathbf{p}\: -\: \mathbf{p}')\!
    \ket{0}\!\; , \\
    a_{\mathrm{S}}(\mathbf{p};\tilde{t}_i)\!\ket{0}&\ =\ 0\; .
  \end{align}
\end{subequations}
Note  that  the momentum  eigen\-states~$\ket{\mathbf{p};\tilde{t}_i}$
respect the ortho\-normality condition
\begin{equation}
  \label{eq:orth}
  \braket{\mathbf{p}';\tilde{t}_i|\mathbf{p};\tilde{t}_i}\ =\ 
  (2\pi)^32E(\mathbf{p})\delta^{(3)}(\mathbf{p}\: -\: \mathbf{p}')\; .
\end{equation}

We  then define  the  \emph{time-dependent} interaction-picture  field
operator $\Phi_{\mathrm{I}}(x;\tilde{t}_i)$ via
\begin{equation}
  \label{eq:PhiItoS}
  \Phi_{\mathrm{I}}(x;\tilde{t}_i)\ =\ e^{iH_{\mathrm{S}}^0(x_0\: -\: \tilde{t}_i)}
  \Phi_{\mathrm{S}}(\mathbf{x};\tilde{t}_i)
  e^{-iH_{\mathrm{S}}^0(x_0\: -\: \tilde{t}_i)}\; ,
\end{equation}
where $H_{\mathrm{S}}^0$  is the free part  of the Hamiltonian  in the
Schr\"{o}dinger picture. Using the commutators
\begin{subequations}
  \begin{align}
    \big[\, H_{\mathrm{S}}^0,\:
    a_{\mathrm{S}}(\mathbf{p};\tilde{t}_i)\,\big]&
    \ =\ -E(\mathbf{p})a_{\mathrm{S}}(\mathbf{p};\tilde{t}_i)\; ,\\
    \big[\, H_{\mathrm{S}}^0,\: 
    a^{\dag}_{\mathrm{S}}(\mathbf{p};\tilde{t}_i)\,\big]
    &\ =\ +E(\mathbf{p})a^{\dag}_{\mathrm{S}}(\mathbf{p};\tilde{t}_i)\;,
  \end{align}
\end{subequations}
the interaction-picture field operator may be written
\begin{equation}
  \label{eq:PhiI}
  \Phi_{\mathrm{I}}(x;\tilde{t}_i)\ =\ \int\!\D{}{\Pi_{\mathbf{p}}}\,
  \Big(\,a_{\mathrm{S}}(\mathbf{p};\tilde{t}_i)
  e^{-iE(\mathbf{p})(x_0\: -\: \tilde{t}_i)}e^{i\mathbf{p}\cdot\mathbf{x}}\: +\:
  a_{\mathrm{S}}^{\dag}(\mathbf{p};\tilde{t}_i)
  e^{iE(\mathbf{p})(x_0\: -\: \tilde{t}_i)}
  e^{-i\mathbf{p}\cdot\mathbf{x}}\,\Big)\; ,
\end{equation}
or equivalently, in terms of interaction-picture operators only,
\begin{equation}
  \label{eq:PhiI0}
  \Phi_{\mathrm{I}}(x;\tilde{t}_i)\ =\ \!\int\!\D{}{\Pi_{\mathbf{p}}}\;
  \Big(\,a_{\mathrm{I}}(\mathbf{p},0;\tilde{t}_i)e^{-iE(\mathbf{p})x_0}
  e^{i\mathbf{p}\cdot \mathbf{x}}
  \: +\: a_{\mathrm{I}}^{\dag}(\mathbf{p},0;\tilde{t}_i)e^{iE(\mathbf{p})x_0}
  e^{-i\mathbf{p}\cdot\mathbf{x}}\,\Big)\; .
\end{equation}
Notice    that     in~(\ref{eq:PhiI0})    the    \emph{time-dependent}
interaction-picture     creation    and     annihilation    operators,
$a_{\mathrm{I}}^\dagger     (\mathbf{p},\tilde{t};\tilde{t}_i)$    and
$a_{\mathrm{I}}(\mathbf{p},\tilde{t};\tilde{t}_i)$,  are  evaluated at
the microscopic  time $\tilde{t}\: =\: 0$, after  employing a relation
analogous  to~(\ref{eq:PhiItoS}).  We  may write  the four-dimensional
Fourier transform of the interaction-picture field operator as
\begin{equation}
  \label{eq:fourier}
  \Phi_{\mathrm{I}}(p;\tilde{t}_i)\ =\ \!\int\!\D{4}{x}\; e^{ip\cdot x}\,
  \Phi_{\mathrm{I}}(x;\tilde{t}_i)\ =\ 2\pi\delta(p^2-M^2)
  \Big(\,\theta(p_0)a_{\mathrm{I}}(\mathbf{p},0;\tilde{t}_i)\: +\: 
  \theta(-p_0)a^{\dag}_{\mathrm{I}}(-\mathbf{p},0;\tilde{t}_i)\,\Big)\; .
\end{equation}

In the limit where the  interactions are switched off adiabatically as
$\tilde{t}_i\: \to\: -\infty$, one may define the asymptotic \emph{in}
creation and annihilation operators via
\begin{equation}
  \label{eq:wave}
  a^{(\dag)}_{\mathrm{in}}(\mathbf{p})\ \simeq\ 
  Z^{-1/2}\lim_{\tilde{t}_i\:\to\:-\infty}
  a^{(\dag)}_{\mathrm{I}}(\mathbf{p},0;\tilde{t}_i)\ = \ Z^{-1/2}
  \lim_{\tilde{t}_i\: \to\: -\infty} a^{(\dag)}_{\mathrm{S}}(\mathbf{p};\tilde{t}_i)
  e^{+(-)iE(\mathbf{p})\tilde{t}_i}\; ,
\end{equation}
where  $Z\:  =\:  1\:  +\:  \mathcal{O}(\hbar)$  is  the  wavefunction
renormalization.   Evidently,  keeping track  of  the finite  boundary
time~$\tilde{t}_i$  plays  an  important  role in  ensuring  that  our
forthcoming  generalization  to   perturbative  thermal  field  theory
remains  consistent   with  asymptotic  field  theory   in  the  limit
$\tilde{t}_i\:\to\:   -\infty$.    Hereafter,   we   will   omit   the
subscript~$\mathrm{I}$  on interaction-picture operators  and suppress
the  implicit dependence  on the  boundary time  $\tilde{t}_i$, except
where it is necessary to do otherwise for clarity.

We  start our  quantization procedure  by defining  the  commutator of
interaction-picture fields
\begin{equation}
  \label{eq:caus}
  \big[\,\Phi(x),\: \Phi(y)\,\big]\ \equiv\ i\Delta^0(x,y;M^2)\; ,
\end{equation}
where $i\Delta^0(x,y;M^2)$ is the \emph{free} Pauli-Jordan propagator.
Herein,  we  denote  free  propagators  by  a  superscript  $0$.   The
condition  of micro-causality  requires  that the  interaction-picture
fields  commute for  space-like  separations $(x\:-\:  y)^2\: <\:  0$.
This  restricts  $i\Delta^0(x,y;M^2)$ to  be  invariant under  spatial
translations, having the Poincar\'{e}-invariant form
\begin{equation}
  \label{eq:pjfunc}
  i\Delta^0(x,y;M^2)\ =\ \!\int\!\D{}{\Pi_{\mathbf{p}}}\;
  \Big(\,e^{-iE(\mathbf{p})(x^0\: -\: y^0)}e^{i\mathbf{p}\cdot
    (\mathbf{x}\:-\:\mathbf{y})}\: -\:
    \big(\,x\: \longleftrightarrow\: y\,\big)\,\Big)\; .
\end{equation}
Observe  that $i\Delta^0(x,y;M^2)$  represents the  difference  of two
counter-propagating   packets  of   plane  waves   and   vanishes  for
$(x\:-\:y)^2\: <\: 0$.

It  proves  useful  for  our  forthcoming analysis  to  introduce  the
\emph{double} Fourier transform
\begin{subequations}
  \begin{align}
  \label{eq:pj}
  i\Delta^0(p,p';M^2)&\ =\ \!\iint\!\D{4}{x}\,\D{4}{y}\;
  e^{ip\cdot x}e^{-ip'\cdot y}\,i\Delta^0(x,y;M^2) \\
  \label{eq:pjb}     
  &\ =\ 2\pi\varepsilon(p_0)\delta(p^2-M^2)
  (2\pi)^4\delta^{(4)}(p\:-\: p')\; ,
\end{align}
\end{subequations}
where $\varepsilon(p_0)\: \equiv\:  \theta(p_0)\: -\: \theta(-p_0)$ is
the generalized signum function. Note that we have defined the Fourier
transforms  such  that four-momentum~$p$  flows  \emph{away from}  the
point~$x$ and four-momentum~$p'$ flows \emph{towards} the point~$y$.

From  (\ref{eq:caus})   and  (\ref{eq:pjfunc}),  we   may  derive  the
equal-time commutation relations
\begin{subequations}
  \label{eq:comrel}
  \begin{align}
    i\Delta^0(x,y;M^2)\big|_{x^0\: =\: y^0\: =\: \tilde{t}} &\ =\
    \big[\,\Phi(\tilde{t},\mathbf{x}),\:
    \Phi(\tilde{t},\mathbf{y})\,\big]\ =\ 0\; ,\\
    \label{eq:comrel1}
    \partial_{x_0}i\Delta^0(x,y;M^2)\big|_{x^0\: =\: y^0\: =\: \tilde{t}} &\ =\
    \big[\,\pi(\tilde{t},\mathbf{x}),\: \Phi(\tilde{t},\mathbf{y})
    \,\big]\ =\ -i\delta^{(3)}(\mathbf{x}\: -\: \mathbf{y})\; ,\\
    \partial_{x_0}\partial_{y_0}i\Delta^0(x,y;M^2)
    \big|_{x^0\: =\: y^0\: =\: \tilde{t}}&\ =\ \big[\,\pi(\tilde{t},\mathbf{x})
    ,\: \pi(\tilde{t},\mathbf{y})\,\big]\ =\ 0\; ,
  \end{align}
\end{subequations}
where $\pi(x)\:  =\: \partial_{x_0}\Phi(x)$ is  the conjugate-momentum
operator.  In  order to  satisfy the canonical  quantization relations
(\ref{eq:comrel}),  the  creation   and  annihilation  operators  must
respect the commutation relation
\begin{equation}
  \label{eq:momcomrel}
  \big[\, a(\mathbf{p},\tilde{t}\,),\:
  a^{\dag}(\mathbf{p}',\tilde{t}'\,)\, \big]
  \ =\ (2\pi)^32E(\mathbf{p})\delta^{(3)}(\mathbf{p}\: -\: \mathbf{p}')
  e^{-iE(\mathbf{p})(\tilde{t}\: -\: \tilde{t}')}\; ,
\end{equation}
with  all  other  commutators   vanishing.   Here,  we  emphasize  the
appearance  of an  overall  phase $e^{-iE(\mathbf{p})(\tilde{t}\:  -\:
  \tilde{t}')}$   in~(\ref{eq:momcomrel})   for  $\tilde{t}\:   \neq\:
\tilde{t}'$ [cf.~Section \ref{sec:ctpsd}].

The vacuum  expectation value of the  commutator of Heisenberg-picture
field   operators   may   be   expressed   as   a   superposition   of
interaction-picture    field    commutators    by   means    of    the
K\"{a}ll\'{e}n--Lehmann  spectral  representation \cite{Kallen:1952zz,
  Lehmann:1954xi}:
\begin{equation}
  \label{eq:kl}
  \braket{0|\big[\,\Phi_{\mathrm{H}}(x),\; \Phi_{\mathrm{H}}(y)\,\big]|0}
  \ \equiv\ i\Delta(x,y)\ =\ \!\int_0^{\infty}\!\!\D{}{s}\;\sigma(s)\,
  i\Delta^0(x,y;s)\; ,
\end{equation}
where  $i\Delta^0(x,y;s)$  is  the  free Pauli--Jordan  propagator  in
(\ref{eq:pjfunc}) with $M^2$ replaced by $s$ and $i\Delta(x,y)$ is the
\emph{dressed}  or \emph{resummed}  propagator. The  positive spectral
density~$\sigma(s)$  contains all  information about  the  spectrum of
single- and  multi-particle states produced  by the Heisenberg-picture
field operators~$\Phi_{\mathrm{H}}$.  Note  that for a homogeneous and
stationary  vacuum~$\ket{0}$,   $\sigma(s)$  is  independent   of  the
space-time  coordinates  and  the  resummed  Pauli--Jordan  propagator
maintains its translational  invariance. If $\sigma(s)$ is normalized,
such that
\begin{equation}
  \int_0^{\infty}\!\!\D{}{s}\;\sigma(s)\ =\ 1\; ,
\end{equation}
the equal-time  commutation relations of  Heisenberg-picture operators
maintain  exactly the  form in  (\ref{eq:comrel}). In  this  case, the
spectral  function   cannot  depend  upon  any   fluctuations  in  the
background.  Clearly, for non-trivial `vacua', or thermal backgrounds,
the spectral density becomes a  function also of the coordinates.  The
spectral representation  of the  resummed propagators may  then depend
non-trivially  on space-time  coordinates,  i.e.~$\sigma\: =\:  \sigma
(s;x,y)$;  see for  instance \cite{Aarts:2001qa}.   In this  case, the
convenient factorization of the K\"{a}ll\'{e}n--Lehmann representation
breaks   down.

The retarded and advanced causal propagators are defined as
\begin{equation}
  i\Delta_{\mathrm{R}}(x,y)\ \equiv\ \theta(x_0\: -\: y_0)i\Delta(x,y)\; ,
  \qquad i\Delta_{\mathrm{A}}(x,y)
  \ \equiv\ -\theta(y_0\: -\: x_0)i\Delta(x,y)\; .
\end{equation}
Using  the  Fourier  representation  of  the  unit  step  function  in
(\ref{eq:thetastep}),  we  introduce  a convenient  representation  of
these causal propagators in terms of the convolution
\begin{equation}
  \label{eq:specrep}
  i\Delta_{\mathrm{R}(\mathrm{A})}(p,p')\ =\ i\!\int\!\frac{\D{}{k_0}}{2\pi}\;
  \frac{i\Delta(p_0-k_0,p_0'-k_0;\mathbf{p},\mathbf{p}')}
  {k_0\: +(-)\: i\epsilon}\; .
\end{equation}

The absolutely-ordered Wightman propagators are defined as
\begin{equation}
  i\Delta_>(x,y)\ \equiv\ \braket{\,\Phi(x)\Phi(y)\,}\!\; ,
  \qquad
  i\Delta_<(x,y)\ \equiv\ \braket{\,\Phi(y)\Phi(x)\,}\!\; .
\end{equation}
We  note  that  the  two-point correlation  functions~$\Delta  (x,y)$,
$\Delta_{>,<}(x,y)$   and    $\Delta_{\rm   R,A}(x,y)$   satisfy   the
\emph{causality relation}:
\begin{equation}
  \label{eq:causality}
  \Delta(x,y)\ =\ \Delta_>(x,y)\: -\: \Delta_<(x,y)\ =\ 
  \Delta_{\mathrm{R}}(x,y)\: -\: \Delta_{\mathrm{A}}(x,y)\; .
\end{equation}

Our next step  is to define the non-causal  Hadamard propagator, which
is the vacuum expectation value of the field anti-commutator
\begin{equation}
  \label{eq:had1}
  i\Delta_1(x,y)\ \equiv\ \braket{\,\big\{\,\Phi(x),\: \Phi(y)\,\big\}\,}
  \!\; .
\end{equation}
Correspondingly, the time-ordered  Feynman and anti-time-ordered Dyson
propagators are given by
\begin{equation}
  i\Delta_{\mathrm{F}}(x,y)\ \equiv\
  \braket{\,\mathrm{T}\,\big[\,\Phi(x)\Phi(y)\,\big]\,}
  \!\; ,\qquad i\Delta_{\mathrm{D}}(x,y)\ \equiv\ 
  \braket{\bar{\,\mathrm{T}}\,\big[\,\Phi(x)\Phi(y)\,\big]\,}\!\; ,
\end{equation}
where   $\mathrm{T}$  and   $\bar{\mathrm{T}}$  are   the   time-  and
anti-time-ordering      operators,      respectively.      Explicitly,
$\Delta_{\mathrm{F}}(x,y)$   and  $\Delta_{\mathrm{D}}(x,y)$   may  be
written  in  terms  of  the  absolutely-ordered  Wightman  propagators
$\Delta_>(x,y)$ and $\Delta_<(x,y)$ as
\begin{subequations}
  \begin{align}
    \Delta_{\mathrm{F}}(x,y)\ =\ \theta(x_0\: -\: y_0)\Delta_>(x,y)\: +\:
    \theta(y_0\: -\: x_0)\Delta_<(x,y)\; ,\\
    \Delta_{\mathrm{D}}(x,y)\ =\ \theta(x_0\: -\: y_0)\Delta_<(x,y)\: +\:
    \theta(y_0\: -\: x_0)\Delta_>(x,y)\; .
  \end{align}
\end{subequations}
The propagators  $\Delta_1(x,y)$, $\Delta_{>,<}(x,y)$ and $\Delta_{\rm
  F,D}(x,y)$ obey the \emph{unitarity relations}:
\begin{equation}
  \label{eq:unitarity}
  \Delta_1(x,y)\ =\ \Delta_{\mathrm{F}}(x,y)\: +\: \Delta_{\mathrm{D}}(x,y)
  \ =\ \Delta_>(x,y)\: +\: \Delta_<(x,y)\ =\
  2i\mathrm{Im}\,\Delta_{\mathrm{F}}(x,y)\; .
\end{equation}
Finally, for completeness, we define the principal-part propagator
\begin{equation}
  \label{eq:pprop}
  \Delta_{\mathcal{P}}(x,y)\ =\ \frac{1}{2}\Big(\,\Delta_{\mathrm{R}}(x,y)\: +\:
  \Delta_{\mathrm{A}}(x,y)\,\Big)\ =\ \mathrm{Re}\,\Delta_{\mathrm{F}}(x,y)\; .
\end{equation}
Here, we should bear in mind that
\begin{equation}
  \mathrm{Re}\big(\mathrm{Im}\big)\,\Delta_{\mathrm{F}}(x,y)\ \neq\ 
  \!\iint\!\!\frac{\D{4}{p}}{(2\pi)^4}\;\frac{\D{4}{p'}}{(2\pi)^4}\,
  e^{-ip\cdot x}e^{ip'\cdot y}\,
  \mathrm{Re}\big(\mathrm{Im}\big)\,\Delta_{\mathrm{F}}(p,p')\; ,
\end{equation}
implying that
\begin{equation}
  \Delta_{\mathcal{P}}(p,p')\ \neq\ \mathrm{Re}\,\Delta_{\mathrm{F}}(p,p')\; ,
\end{equation}
unless  $\Delta_{\mathrm{F}}(p,p')\: =\: \Delta_{\mathrm{F}}(-p,-p')$,
which is  not generally true  in non-equilibrium thermal  field theory
[cf.~(\ref{eq:hermitdmF})].

The definitions and  the relations discussed above are  valid for both
free and resummed propagators.  In Appendix~\ref{app:rel}, we list the
properties  of these  propagators in  coordinate, momentum  and Wigner
(see Section~\ref{sec:gradapp})  representations, as well  as a number
of useful identities, which we  will refer to throughout this article.
More  detailed  discussion  related  to these  propagators  and  their
contour-integral  representations may be  found in~\cite{Greiner1996}.
In Appendix~\ref{app:complex},  these considerations and  the analysis
of the following sections are generalized to the complex scalar field.

\bigskip

\section{The CTP Formalism}
\label{sec:CTP}

In this  section, we review the  Closed Time Path  (CTP) formalism, or
the  so-called {\emph{in}-\emph{in}} formalism,  due to  Schwinger and
Keldysh~\cite{Schwinger:1960qe,  Keldysh:1964ud}.  As  an illuminating
exercise,   we  consider  the   CTP  formalism   in  the   context  of
zero-temperature  quantum  field  theory  and  derive  the  associated
$2\times   2$  matrix   propagator,   obeying  basic   field-theoretic
constraints,  such  as $CPT$  invariance,  Hermiticity, causality  and
unitarity.   Finally,  we  discuss  the  properties  of  the  resummed
propagator in the CTP formalism.

In  the calculation  of scattering-matrix  elements, we  are concerned
with  the  transition   amplitude  between  \emph{in}  and  \emph{out}
asymptotic  states, where  single-particle states  are defined  in the
infinitely-distant  past and  future. On  the other  hand,  in quantum
statistical  mechanics,  we  are  interested  in  the  calculation  of
Ensemble  Expectation Values  (EEVs)  of operators  at  a fixed  given
time~$t$.  Specifically, the evaluation  of EEVs of operators requires
a field-theoretic approach that  allows us to determine the transition
amplitude  between  states  evolved  to the  \emph{same  time}.   This
approach is the Schwinger--Keldysh CTP formalism, which we describe in
detail below.

For illustration, let us consider the following observable ${\cal O}$
in the Schr\"{o}dinger picture (suppressing the spatial coordinates
$\mathbf{x}$ and $\mathbf{y}$):
\begin{equation}
  \label{eq:Phixy}
  {\cal O}_{\rm (S)}(\tilde{t}_f;\tilde{t}_i)\ = \
  \!\int\![\D{}{\Phi(\mathbf{z})}]\;\prescript{}{\mathrm{S}}
          {\braket{\Phi(\mathbf{z}),\tilde{t}_f;\tilde{t}_i|
              \Phi_{\mathrm{S}}(\mathbf{x};\tilde{t}_i)
              \Phi_{\mathrm{S}}(\mathbf{y};\tilde{t}_i)
  |\Phi(\mathbf{z}),\tilde{t}_f;\tilde{t}_i}}_{\mathrm{S}}\; ,
\end{equation}
where $[\D{}{\Phi}]$ represents the functional integral over all field
configurations        $\Phi(\mathbf{z})$.         In~(\ref{eq:Phixy}),
$\ket{\Phi(\mathbf{z}),\tilde{t}_f;\tilde{t}_i}_{\mathrm{S}}$   is   a
\emph{time-evolved}   eigenstate    of   the   \emph{time-independent}
Schr\"{o}dinger-picture                 field                 operator
$\Phi_{\mathrm{S}}(\mathbf{x};\tilde{t}_i)$       with      eigenvalue
$\Phi(\mathbf{x})$  at time  $\tilde{t}_f\:=\:\tilde{t}_i$,  where the
implicit  dependence  upon the  boundary  time~$\tilde{t}_i$ has  been
restored.

We  should remark  here that  there are  seven  independent space-time
coordinates involved in  the ob\-servable~(\ref{eq:Phixy}).  These are
the six spatial coordinates, $\mathbf{x}$ and $\mathbf{y}$, {\em plus}
the  microscopic  time~$\tilde{t}_f$.    In  addition,  there  is  one
implicit coordinate: the boundary  time $\tilde{t}_i$. As we will see,
exactly  seven  independent  coordinates  are  required  to  construct
physical observables that are compatible with Heisenberg's uncertainty
principle.  We choose the seven independent coordinates to be
\begin{equation}
  \label{eq:7ind}
  t\ =\ \tilde{t}_f\: -\: \tilde{t}_i\;,\qquad
  \mathbf{X}\ =\ \frac{1}{2}\: \Big(\mathbf{x}\: +\:
  \mathbf{y}\Big)\;,\qquad \mathbf{p}\; ,
\end{equation}
where $t$ and $\mathbf{X}$ are  the macroscopic time and central space
coordinates and $\mathbf{p}$ is  the Fourier-conjugate variable to the
relative                                                        spatial
coordinate~$\mathbf{R}\:=\:\mathbf{x}\:-\:\mathbf{y}$.

In   the   interaction  picture,   the   same  observable~${\cal   O}$
in~(\ref{eq:Phixy}) is given by
\begin{equation}
  \label{eq:intobs}
  {\cal O}_{\rm (I)}(\tilde{t}_f;\tilde{t}_i)\ =\ \!\int\![\D{}{\Phi(\mathbf{z})}]\;
  \prescript{}{\mathrm{I}}{\braket{\Phi(\mathbf{z}),\tilde{t}_f;\tilde{t}_i|
  \Phi_{\mathrm{I}}(\tilde{t}_f,\mathbf{x};\tilde{t}_i)
  \Phi_{\mathrm{I}}(\tilde{t}_f,\mathbf{y};\tilde{t}_i)
  |\Phi(\mathbf{z}),\tilde{t}_f;\tilde{t}_i}}_{\mathrm{I}}\; ;
\end{equation}
and, in the Heisenberg picture, by
\begin{equation}
  \label{eq:heisobs}
  {\cal O}_{\rm (H)}(\tilde{t}_f;\tilde{t}_i)\ =\ \int\![\D{}{\Phi(\mathbf{z})}]\;
  \prescript{}{\mathrm{H}}{\braket{\Phi(\mathbf{z});\tilde{t}_i|
  \Phi_{\mathrm{H}}(\tilde{t}_f,\mathbf{x};\tilde{t}_i)
  \Phi_{\mathrm{H}}(\tilde{t}_f,\mathbf{y};\tilde{t}_i)|
  \Phi(\mathbf{z});\tilde{t}_i}}_{\mathrm{H}}\; .
\end{equation}
Notice  that the  prediction for  the observable~${\cal  O}$  does not
depend   on  which   picture   we  are   using,  i.e.~${\cal   O}_{\rm
  (S)}(\tilde{t}_f;\tilde{t}_i)\:        =\:       {\cal       O}_{\rm
  (I)}(\tilde{t}_f;\tilde{t}_i)\:        =\:       {\cal       O}_{\rm
  (H)}(\tilde{t}_f;\tilde{t}_i)$.  This picture independence of ${\cal
  O}$  is  only  possible   because  the  time-dependent  vectors  and
operators  in ${\cal  O}$  are evaluated  individually at  \emph{equal
  times}.    Otherwise,  any  potential   observable,  built   out  of
time-dependent   vectors   and  operators   that   are  evaluated   at
\emph{different   times},   would   be  \emph{picture dependent}   and
therefore   {\em  unphysical}.   Moreover,   the  prediction   of  the
observable $\mathcal{O}$  should be invariant  under simultaneous time
translations of the boundary  and observation times, $\tilde{t}_i$ and
$\tilde{t}_f$, i.e.
\begin{equation}
  \mathcal{O}(\tilde{t}_f;\tilde{t}_i)\ = \ \mathcal{O}(t;0)
  \ \equiv \ \mathcal{O}(t),
\end{equation}
where $t\:=\:\tilde{t}_f\:-\:\tilde{t}_i$ is  the macroscopic time, as
we  will see  below.   Herein  and throughout  the  remainder of  this
article,  the time arguments  of quantities  that are  invariant under
such simultaneous  translations of the  boundary times are  written in
terms of the macroscopic time $t$ only.

\subsection{The CTP Contour}

In   order   to   evaluate   equal-time  observables   of   the   form
in~(\ref{eq:heisobs}),  we   first  introduce  the   \emph{in}  vacuum
state~$\ket{0_{\mathrm{in}};\tilde{t}_i}$,       which      is      at
time~$\tilde{t}_i$  a time-independent  eigenstate  of the  Heisenberg
field     operator~$\Phi_{\mathrm{H}}(x;\tilde{t}_i)$     with    zero
eigenvalue; see~\cite{Calzetta:1986ey, Calzetta:1986cq}.  We then need
a         means        of         driving         the        amplitude
$\braket{0_{\mathrm{in}};\tilde{t}_i|0_{\mathrm{in}};\tilde{t}_i}$,
which  can be  achieved by  the appropriate  introduction  of external
sources.

\begin{figure}
  \begin{center}
    \includegraphics{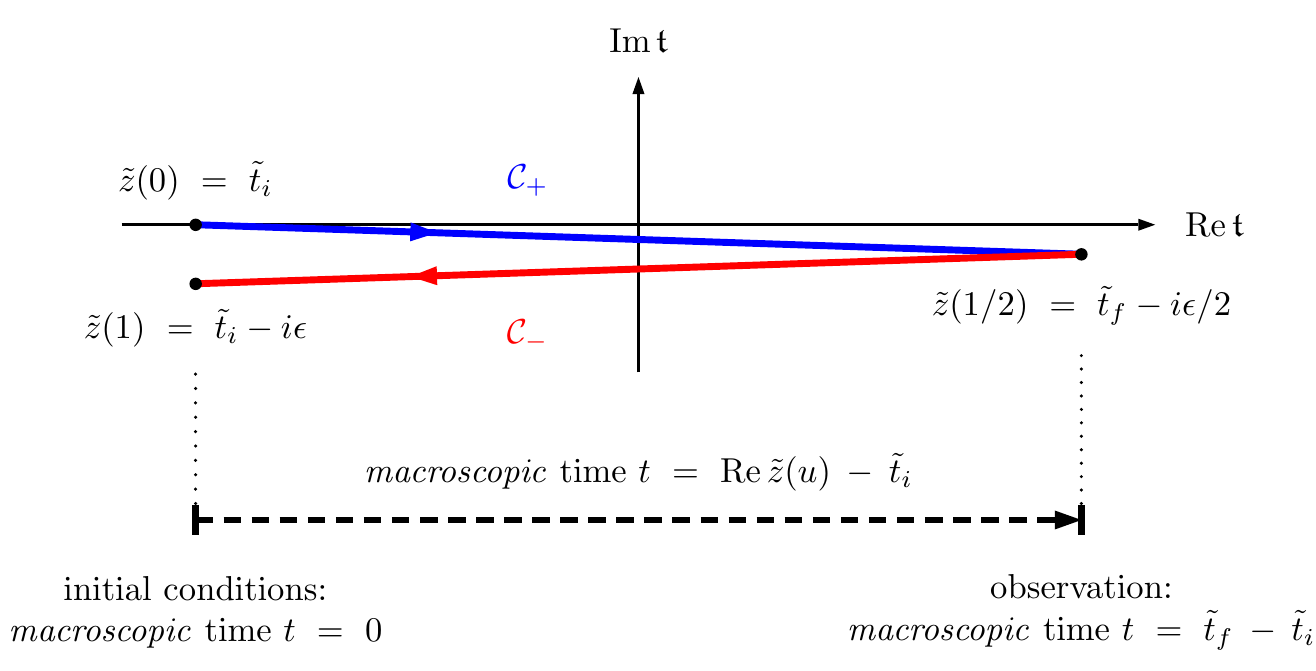}
  \end{center}
  \caption{The closed-time path,
    $\mathcal{C}\ =\ \mathcal{C}_+\:\cup\,\:\mathcal{C}_-$, running
    first along $\mathcal{C}_+$ from $\tilde{t}_i$ to
    $\tilde{t}_f\:-\:i\epsilon/2$ and then returning along
    $\mathcal{C}_-$ from $\tilde{t}_f\:-\:i\epsilon/2$ to
    $\tilde{t}_i\:-\:i\epsilon$.  The relationship between the
    complex \emph{microscopic} time $\tilde{z}(u)$ and the \emph{macroscopic}
    time $t$ is indicated by a dashed black arrow.}
  \label{fig:sk}
\end{figure}

The procedure may  be outlined with the aid  of Figure~\ref{fig:sk} as
follows.     We   imagine    evolving   our    \emph{in}    state   at
time~$\tilde{t}_i$  forwards  in  time  under  the  in\-fluence  of  a
source~$J_+(x)$ to some \emph{out}  state at time~$\tilde{t}_f$ in the
future,  which  will  be  a  superposition over  all  possible  future
states. We  then evolve this  superposition of states  backwards again
under the  in\-fluence of a  second source~$J_-(x)$, returning  to the
same initial time~$\tilde{t}_i$ and the original \emph{in} state.  The
sources~$J_{\pm}(x)$  are  assumed  to  vanish  adiabatically  at  the
boundaries  of  the  interval~$[\tilde{t}_i,\ \tilde{t}_f]$.   We  may
in\-terpret the path of this  evolution as de\-fining a closed contour
~$\mathcal{C}\:  =\: \mathcal{C}_+ \:  \cup\,\: \mathcal{C}_-$  in the
complex-time                plane               ($\mathfrak{t}$-plane,
$\mathfrak{t}\:\in\:\mathbb{C}$),   which   is   the  union   of   two
anti-parallel branches: $\mathcal{C}_+$, running from $\tilde{t}_i$ to
$\tilde{t}_f\:-\:i\epsilon/2$;   and  $\mathcal{C}_-$,   running  from
$\tilde{t}_f\:-\:i\epsilon/2$ back to $\tilde{t}_i\:-\:i\epsilon$.  We
refer   to  $\mathcal{C}_+$  and   $\mathcal{C}_-$  as   the  positive
time-ordered  and negative  anti-time-ordered  branches, respectively.
As  de\-picted  in   Figure~\ref{fig:sk},  the  small  imaginary  part
$\epsilon\:=\:0^+$  has been added  to allow  us to  disting\-uish the
two, essentially co\-incident, branches. We para\-metrize the distance
along  the   contour,  starting   from  $\tilde{t}_i$,  by   the  real
variable~$u\:\in\:[0,\  1]$,  which  increases  mono\-tonically  along
$\mathcal{C}$.   We   may  then  de\-fine   the  contour  by   a  path
$\tilde{z}(u)\:=\:\tilde{t}(u)\:-\:i\tilde{\tau}(u)\:\in\:
\mathfrak{t}$,  where  $\tilde{t}(0)\:=\:\tilde{t}(1)\:=\:\tilde{t}_i$
and   $\tilde{t}(1/2)\:=\:\tilde{t}_f$.    Thus,   the   complex   CTP
contour~$\tilde{z}(u)$ may be written down explicitly as
\begin{equation}
  \tilde{z}(u)\ =\ \theta(\tfrac{1}{2}-u)\, \Big[\,\tilde{t}_i\:+\:
  2u\,\big(\,\tilde{t}_f\:-\:\tilde{t}_i\,\big)\,\Big]\:+\:
  \theta(u-\tfrac{1}{2})\,\Big[\,\tilde{t}_i\:+\:
  2\,\big(\,1\:-\:u\,\big)\,\big(\,\tilde{t}_f\:-\:\tilde{t}_i\,\big)
  \,\Big]\: -\: i\epsilon u\; ,
\end{equation}
with $\theta(0)\:=\:1/2$, as given in~(\ref{eq:thetastep}).

To derive a path-integral representation of the generating functional,
we introduce the
eigenstate~$\ket{\Phi(\mathbf{x}),\tilde{t};\tilde{t}_i}$ of the
Heisenberg field
operator~$\Phi_{\mathrm{H}}(\tilde{t},\mathbf{x};\tilde{t}_i)$,
satisfying the eigenvalue equation
\begin{equation}
  \Phi_{\mathrm{H}}(\tilde{t},\mathbf{x};\tilde{t}_i)\!
  \ket{\Phi(\mathbf{x}),\tilde{t};\tilde{t}_i}\ =\
  \Phi(\mathbf{x})\!
  \ket{\Phi(\mathbf{x}),\tilde{t};\tilde{t}_i}\; .
\end{equation}
The basis vectors $\ket{\Phi(\mathbf{x}),\tilde{t};\tilde{t}_i}$ form
a complete orthonormal basis, respecting the orthonormality condition
\begin{equation}
  \label{eq:eigencomp}
  \int\![\D{}{\Phi(\mathbf{x})}]\;
  \ket{\Phi(\mathbf{x}),\tilde{t};\tilde{t}_i}
            \bra{\Phi(\mathbf{x}),\tilde{t};\tilde{t}_i}\ =\
  \mathbb{I}\; .
\end{equation}
We    may     then    write    the     CTP    generating    functional
$\mathcal{Z}[J_{\pm},t]$ as
\begin{align}
  \label{eq:genfunc1}
  \mathcal{Z}[J_{\pm},t]&\ =\ \!\int\![\D{}{\Phi(\mathbf{x})}]\;
  \prescript{}{J_-}{\braket{0_{\mathrm{in}},\tilde{t}_i;\tilde{t}_i|
      \Phi(\mathbf{x}),\tilde{t}_f;\tilde{t}_i}}
      \braket{\Phi(\mathbf{x}),\tilde{t}_f;\tilde{t}_i|
      0_{\mathrm{in}},\tilde{t}_i;\tilde{t}_i}_{J_+}
  \nonumber\\&\ =\ \!\int\![\D{}{\Phi(\mathbf{x})}]\;
  \braket{0_{\mathrm{in}},\tilde{t}_i;\tilde{t}_i|
      \,\bar{\mathrm{T}}\,\Big(\,
      e^{-i\int_{\tilde{t}_i}^{\tilde{t}_f}\!\D{4}{x}\;J_-(x)\Phi_{\mathrm{H}}(x)}\,\Big)
      |\Phi(\mathbf{x}),\tilde{t}_f;\tilde{t}_i}
  \nonumber\\&\hspace{1.8cm} \times
  \braket{\Phi(\mathbf{x}),\tilde{t}_f;\tilde{t}_i|\,\mathrm{T}\,
      \Big(\,e^{i\int_{\tilde{t}_i}^{\tilde{t}_f}\!\D{4}{x}\;J_+(x)\Phi_{\mathrm{H}}(x)}
      \,\Big)|0_{\mathrm{in}},\tilde{t}_i;\tilde{t}_i}\; ,
\end{align}
where the $x^0$ integra\-tions run from $\tilde{t}_i$ to $\tilde{t}_f$
and the `latest' time (with $u\:=\:1$) appears furthest to the left.

In  order  to preserve  the  correspondence  with  the ordinary  ${\rm
  S}$-matrix       theory      in      the       asymptotic      limit
$\tilde{t}_i\:\to\:-\infty$, we take
\begin{equation}
  \label{eq:symtime}
  \tilde{t}_f\ =\ -\:\tilde{t}_i\;.
\end{equation}
With      this       identification,      the      CTP      generating
functional~$\mathcal{Z}[J_{\pm},t]$  becomes  \emph{manifestly}  $CPT$
invariant.  This  enables one to easily verify  that microscopic $CPT$
invariance continues to hold,  even when time translational invariance
is   broken   by   thermal    backgrounds,   as   we   will   see   in
Section~\ref{sec:nonhom}.  Given that $\tilde{t}_i$ is the microscopic
time at which the three pictures of quantum mechanics coincide and the
interactions  are switched  on,  it is  also  the point  at which  the
boundary conditions may  be specified fully and \emph{instantaneously}
in  terms of  on-shell  free particle  states.   The microscopic  time
$\tilde{t}_i$ is therefore the natural origin for a \emph{macroscopic}
time
\begin{equation}
  t\ =\ \tilde{t}_f\:-\:\tilde{t}_i\ =\ 2\tilde{t}_f\;,
\end{equation}
where the last equality holds for the choice in (\ref{eq:symtime}).
This \emph{macroscopic} time is also the total interval of microscopic
time over which the system has evolved.  This fact is illustrated
graphically in Figure~\ref{fig:sk}.

We    denote    by     $\Phi_{\pm}(x)\:    \equiv\:    \Phi(x^0    \in
\mathcal{C}_{\pm},\mathbf{x})$   fields  with  the   microscopic  time
variable $x^0$ confined  to the positive and negative  branches of the
contour,        respectively.         Following~\cite{Calzetta:1986ey,
  Calzetta:1986cq}, we define the doublets
\begin{equation}
  \Phi^a(x)\ =\ \Big(\Phi_+(x)\,,\ \Phi_-(x)\Big)\;,\qquad
  \Phi_a(x)\ =\ \eta_{ab}\Phi^b(x)\ =\ \Big(
  \Phi_+(x)\,,\ -\Phi_-(x)\Big)\; ,
\end{equation}
where    the     CTP    indices    $a,\    b\:     =\:1,\    2$    and
$\eta_{ab}\:=\:\mathrm{diag}\,(1,\  -1)$ is an  $\mathbb{SO}\,(1,\ 1)$
`metric.'   Inserting   into  (\ref{eq:genfunc1})  complete   sets  of
eigenstates of the Heisenberg field operator at intermediate times, we
may  derive  a  path-integral  representation of  the  CTP  generating
functional:
\begin{equation}
  \label{eq:ZJa}
  \mathcal{Z}[J_a,t]\ =\ \mathcal{N}\!\int [\D{}{\Phi^a}]\;
  \exp\bigg[\,i\,\bigg(S[\Phi^a,t]\:+\:\!\int_{\Omega_t}\!\!\D{4}{x}\;
  J_a(x)\Phi^a(x)\,\bigg)\bigg]\; ,
\end{equation}
where  $\mathcal{N}$  is  some normalization  and 
\begin{equation}
  \label{eq:omegat}
  \Omega_t\ \simeq\ [-\:t/2,\  t/2]\times\mathbb{R}^3
\end{equation}
is the Minkowski space-time volume bounded by the hypersurfaces
$x^0\:=\:\pm t/2$. In~(\ref{eq:ZJa}), the action is
\begin{align}
  \label{eq:SPat}
  S[\Phi^a,t]\ &=\ \!\int_{\Omega_t}\!\!\D{4}{x}\;
  \Big[\,\tfrac{1}{2}\eta_{ab}\partial_{\mu}\Phi^a(x)\partial^{\mu}
  \Phi^b(x)\:-\:\tfrac{1}{2}\big(\,M^2\eta_{ab}\:-\:i\epsilon
  \mathbb{I}_{ab}\,\big)\Phi^a(x)\Phi^b(x)\nonumber\\
  &\qquad -\:\tfrac{1}{3!}g\eta_{abc}\Phi^{a}(x)\Phi^b(x)\Phi^c(x)\: -\: 
  \tfrac{1}{4!}\lambda\eta_{abcd}\Phi^a(x)\Phi^b(x)\Phi^c(x)\Phi^d(x)
  \, \Big]\;,
\end{align}
where  $\eta_{abc\cdots}\:=\:+1$ for $a\:=\:b\:=\:c\:=\:\cdots\:=\:1$,
$\eta_{abc\cdots}\:=\:-1$  for  $a\:=\:b\:=\:c\:=\:\cdots\:=\:2$,  and
$\eta_{abc\cdots}\:=\:0$    otherwise.    In    (\ref{eq:SPat}),   the
$\epsilon\:=\:0^+$  gives  the  usual Feynman  prescription,  ensuring
convergence of the CTP path integral. We note that the damping term is
proportional to  the identity matrix $\mathbb{I}_{ab}$ and  not to the
`metric' $\eta_{ab}$.   This prescription  requires the addition  of a
contour-dependent       damping       term,      proportional       to
$\varepsilon(\tfrac{1}{2}-u)$,  which has  the same  sign on  both the
positive and negative branches of the contour, ${\cal C}_+$ and ${\cal
  C}_-$, respectively.

In    order   to    define   properly    a    path-ordering   operator
$\mathrm{T}_{\mathcal{C}}$,  we introduce  the  contour-dependent step
function
\begin{equation}
  \theta_{\mathcal{C}}(x^0-y^0)\ \equiv\ \theta(u_x-u_y)\; ,
\end{equation}
where $x^0\: =\:  \tilde{z}(u_x)$ and $y^0\: =\: \tilde{z}(u_y)$.  By
       analogy, we introduce a contour-dependent delta function
\begin{equation}
  \delta_{\mathcal{C}}(x^0\:-\:y^0)\ =\ \frac{\delta(u_x\:-\:u_y)}
  {\big|\frac{\D{}{\tilde{z}}}{\D{}{u}}\big|}\ =\
  \frac{\delta(u_x\:-\:u_y)}{2|\tilde{t}_f\:-\:\tilde{t}_i|}
  \ =\ \frac{\delta(u_x\:-\:u_y)}{2t}\; .
\end{equation}
As           a          consequence,           a          path-ordered
propagator~$\Delta_{\mathcal{C}}(x,y)$ may be defined as follows:
\begin{equation}
  i\Delta_{\mathcal{C}}(x,y)\ \equiv\
  \braket{\,\mathrm{T}_{\mathcal{C}}\,\big[\,\Phi(x)\Phi(y)\,\big]\,}\; .
\end{equation}

For  $x^0$  and  $y^0$   on  the  positive  branch~${\cal  C}_+$,  the
path-ordering~$\mathrm{T}_{\mathcal{C}}$ is equivalent to the standard
time-ordering~$\mathrm{T}$  and  we  obtain the  time-ordered  Feynman
propagator~$i\Delta_{\mathrm{F}}(x,y)$.  On the  other hand, for $x^0$
and    $y^0$    on    the    negative   branch~${\cal    C}_-$,    the
path-ordering~$\mathrm{T}_{\mathcal{C}}$      is     equivalent     to
anti-time-ordering~$\bar{\mathrm{T}}$     and     we    obtain     the
anti-time-ordered  Dyson propagator~  $i\Delta_{\mathrm{D}}(x,y)$. For
$x^0$  on ${\cal  C}_+$ and  $y^0$ on  ${\cal C}_-$,  $x^0$  is always
`earlier'   than   $y^0$   and   we  obtain   the   absolutely-ordered
negative-frequency  Wightman propagator~$i\Delta_<(x,y)$.  Conversely,
for $y^0$  on ${\cal C}_+$  and $x^0$ on  ${\cal C}_-$, we  obtain the
positive-frequency   Wightman  propagator~$i\Delta_>(x,y)$.    In  the
$\mathbb{SO}\,(1,\ 1)$  notation, we write  the CTP propagator  as the
$2\times 2$ matrix
\begin{equation}
  \label{eq:lower}
  i\Delta^{ab}(x,y)\ \equiv\ \braket{\,\mathrm{T}_{\mathcal{C}}\,
  \big[\,\Phi^a(x)\Phi^b(y)\,]\,}\ =\ i
  \begin{bmatrix}
    \Delta_{\mathrm{F}}(x,y)   &  \Delta_{<}(x,y)  \\
    \Delta_{>}(x,y)  & \Delta_{\mathrm{D}}(x,y)
  \end{bmatrix}\; .
\end{equation}
In this  notation, the CTP indices  $a,\ b$ are raised  and lowered by
contraction with the `metric'~$\eta_{ab}$, e.g.
\begin{equation}
  \label{eq:raise}
  i\Delta_{ab}(x,y)\ =\ \eta_{ac}\,i\Delta^{cd}(x,y)\,\eta_{db} \ =\ i
  \begin{bmatrix}
    \Delta_{\mathrm{F}}(x,y)  &  -\Delta_{<}(x,y) \\
    -\Delta_{>}(x,y)  & \Delta_{\mathrm{D}}(x,y)
  \end{bmatrix}\; .
\end{equation}
Notice  the  difference  in  sign  of  the  off-diagonal  elements  in
(\ref{eq:raise})  compared   with  (\ref{eq:lower}).   An  alternative
definition~$i\widetilde{\Delta}^{ab}(x,y)$ of  the CTP propagator uses
the so-called Keldysh  basis~\cite{vanEijck:1994rw} and is obtained by
means of an orthogonal transformation:
\begin{equation}
  \label{eq:physrep}
  \widetilde{\Delta}^{ab}(x,y)\ \equiv\ O^{a}_{\    c}\,O^{b}_{\    d}
  \,\Delta^{cd}(x,y)\ =\
  \begin{bmatrix}
    0 & \Delta_{\mathrm{A}}(x,y) \\
    \Delta_{\mathrm{R}}(x,y) & \Delta_1(x,y)
  \end{bmatrix}\!\; ,\quad
  O^{ab}\ =\ \frac{1}{\sqrt{2}}\begin{bmatrix}
    1 &  1 \\ 1 & -1
  \end{bmatrix}\!\; .
\end{equation}

In  Section~\ref{sec:nonhom},  we  will  generalize these  results  to
macroscopic ensembles by incorporating  background effects in terms of
physical    sources.    In   this    case,   the    surface   integral
$\oint_{\partial\Omega_t}\D{}{s_{\mu}}\;\Phi^a(x)\partial^{\mu}\Phi^b(x)$
may     not     in      general     vanish     on     the     boundary
hypersurface~$\partial\Omega_t$ of the volume~$\Omega_t$.  However, by
requiring the `$+$'- and `$-$'-type fields to satisfy
\begin{equation}
  \label{eq:surf}
  \Phi_+(x)\partial^{\mu}\Phi_+(x)\big|_{x^{\mu}\:\in\:
  \partial\Omega_t}\ =\ \Phi_-(x)\partial^{\mu}\Phi_-(x)\big|_{x^{\mu}\:\in\:
  \partial\Omega_t}\; ,
\end{equation}
we  can ensure  that surface  terms  cancel between  the positive  and
negative  branches, ${\cal  C}_+$ and  ${\cal C}_-$,  respectively. In
this case, the free part of the action may be rewritten as
\begin{equation}
  \label{eq:freeS}
  S^0[\Phi^a,t]\ =\ \iint_{\Omega_t}\!\!\D{4}{x}\,\D{4}{y}\;\tfrac{1}{2}
  \Phi^a(x)\Delta_{ab}^{0,\,-1}(x,y)\Phi^b(y)\; ,
\end{equation}
where
\begin{equation}
  \label{eq:freeinv}
  \Delta_{ab}^{0,\,-1}(x,y)\ =\ \delta^{(4)}(x\:-\:y)\Big[\,
  -\big(\Box_x^2\:+\: M^2\big)\eta_{ab}\:+\:i\epsilon\mathbb{I}_{ab}\,
  \Big]
\end{equation}
is      the       free      inverse      CTP       propagator      and
$\Box_x^2\:\equiv\:\frac{\partial}{\partial
  x^{\mu}}\,\frac{\partial}{\partial  x_{\mu}}$  is the  d'Alembertian
operator.   Note that the  variational principle  remains well-defined
irrespective of (\ref{eq:surf}), since  we are always free to choose
the variation  of the field~$\delta\Phi^a(x)$ to  vanish for $x^{\mu}$
on $\partial \Omega_t$.

We  may complete  the square  in the  exponent of  the  CTP generating
functional~$\mathcal{Z}[J_a,t]$   in  (\ref{eq:ZJa})  by   making  the
following shift in the field:
\begin{equation}
  \Phi^a(x)\ \equiv \ {\Phi'}^a(x)\:-
  \:\!\int_{\Omega_t}\!\!\D{4}{y}\;\Delta^{0,\, ab}(x,y)J_{b}(y)\; ,
\end{equation}
where $i\Delta^{0,\, ab}(x,y)$ is the free CTP propagator. We may then
re\-write $\mathcal{Z}[J_a,t]$ in the form
\begin{equation}
  \mathcal{Z}[J_a,t]\ =\ \mathcal{Z}^0[0,t]
  \exp\bigg[\,i\!\int_{\Omega_t}\!\!\D{4}{x}\;
    \mathcal{L}^{\mathrm{int}}\bigg(\frac{1}{i}\frac{\delta}
    {\delta J_a}\bigg)\bigg]
  \exp\bigg[-\frac{i}{2}\!\iint_{\Omega_t}\!
    \!\D{4}{x}\,\D{4}{y}\;J_a(x)\Delta^{0,\,ab}(x,y)J_b(y)\,\bigg]\; ,
\end{equation}
where  $\mathcal{L}^{\mathrm{int}}$  is the  interaction  part of  the
Lagrangian and $\mathcal{Z}^0[0,t]$ is the free part of the generating
functional
\begin{equation}
  \mathcal{Z}^0[0,t]\ =\ \mathcal{N}\!
  \int[\D{}{\Phi^a}]\;e^{iS^0[\Phi^a,\,t]}\; ,
\end{equation}
with the free action $S^0[\Phi^a,t]$ given by (\ref{eq:freeS}). We may
then       express       the       resummed       CTP       propagator
$i\Delta^{ab}(x,y,\tilde{t}_f;\tilde{t}_i)$ as follows:
\begin{equation}
  \label{eq:funcdif}
  i\Delta^{ab}(x,y,\tilde{t}_f;\tilde{t}_i)\ =\ 
  \frac{1}{\mathcal{Z}[0,t]}\,\frac{1}{i}\,
  \frac{\delta}{\delta J_a(x)}\,\frac{1}{i}\,
  \frac{\delta}{\delta J_b(y)}\,\mathcal{Z}[J_a,t]\bigg|_{J_{a}\:=\:0}\; ,
\end{equation}
where  $\mathcal{Z}[0,t]$  is   the  generating  functional  with  the
external sources $J_a$ set to zero. The functional derivatives satisfy
\begin{equation}
  \frac{\delta}{\delta J_a(x)}
  \int_{\Omega_t}\!\!\D{4}{y}\;J^b(y)\ =\ \eta^{ab}\delta^{(4)}(x\:-\:y)\; ,
\end{equation}
with   $x^{\mu},\    y^{\mu}\:\in\:\Omega_t$.    We   will    see   in
Section~\ref{sec:nonhomprop}   that   the   resummed  CTP   propagator
$\Delta^{ab}(x,y,\tilde{t}_f;\tilde{t}_i)$  is  not  in  general  time
translationally invariant.

In the absence of  interactions, eigen\-states of the free Hamiltonian
will propagate uninterrupted from times infinitely distant in the past
to times  infinitely far  in the  future. As such,  we may  extend the
limits of integration  in the free part of the  action to positive and
negative infinity, since
\begin{equation}
  \big(\Box_x^2\:+\:M^2\big)\Phi^a(x)\big|_{x^0\: \notin\: [-t/2,\ t/2]}
  \ =\ 0\; ,
\end{equation}
i.e.~the     sources~$J_{a}(x)$      vanish     for     $x^0\:\notin\:
[-t/2,\ t/2]$. The free  CTP propagator $\Delta^{0,\,ab}(x,y)$ is then
obtained  by  inverting  (\ref{eq:freeinv})  subject  to  the  inverse
relation
\begin{equation}
  \label{eq:invdef}
  \int\!\D{4}{z}\;\Delta^{0,\,-1}_{ab}(x,z)\Delta^{0,\,bc}(z,y)\ =\
  \eta_{a}^{\ c}\delta^{(4)}(x\: -\: y)\; ,
\end{equation}
where the domain of integration over $z^0$ is extended to infinity. We
expect    to    recover    the    familiar    propagators    of    the
{\emph{in}-\emph{out}}  formalism of  asymptotic  field theory,  which
occur in ${\rm S}$-matrix elements  and in the reduction formalism due
to   Lehmann,  Symanzik  and   Zimmermann~\cite{Lehmann:1954rq}.   The
propagators    will    also    satisfy   unitarity    cutting    rules
\cite{'tHooft:1973pz,    Kobes:1985kc,   Kobes:1986za,   Veltman1994},
thereby   maintaining   perturbative    unitarity   of   the   theory.
Specifically,     the     free     Feynman     (Dyson)     propagators
$i\Delta^0_{\mathrm{F}(\mathrm{D})}(x,y)$  satisfy  the  inhomogeneous
Klein--Gordon equation
\begin{equation}
  \label{eq:inhomkg}
  -\big(\Box_x^2\: +\: M^2\big)i\Delta^0_{\mathrm{F}(\mathrm{D})}(x,y)
  \ =\ (-)i\delta^{(4)}(x\:-\:y)\; ;
\end{equation}
and the free  Wightman propagators $i\Delta^0_{\gtrless}(x,y)$ satisfy
the homogeneous equation
\begin{equation}
  \label{eq:homkg}
  -\big(\Box_x^2\: +\: M^2\big)i\Delta^0_{>(<)}(x,y)\ =\ 0\; .
\end{equation}

In the  double momentum  representation, the free  part of  the action
(\ref{eq:freeS}) may be written as
\begin{equation}
  S^0[\Phi^a]\ =\ \!\iint\!\!
  \frac{\D{4}{p}}{\big(2\pi\big)^4}\,
  \frac{\D{4}{p'}}{\big(2\pi\big)^4}\;\frac{1}{2}\Phi^a(p)
  \Delta_{ab}^{0,\,-1}(p,p')\Phi^b(-p')\; ,
\end{equation}
where
\begin{equation}
  \label{eq:inv}
  \Delta_{ab}^{0,\,-1}(p,p')\ =\ 
  \begin{bmatrix}
    p^2\:-\:m^2\:+\: i\epsilon & 0 \\
    0 & -\big(p^2\: -\: m^2\: -\: i\epsilon\big)
  \end{bmatrix}
  \!(2\pi)^4\delta^{(4)}(p\:-\:p')
\end{equation}
is  the  double  momentum  representation  of  the  free  inverse  CTP
propagator, satisfying the inverse relation
\begin{equation}
  \label{eq:dminv}
  \int\!\!\frac{\D{4}{q}}{(2\pi)^4}\;\Delta_{ab}^{0,\,-1}(p,q)
  \Delta^{0,\,bc}(q,p')\ =\ \eta_a^{\ c}(2\pi)^4\delta^{(4)}(p\:-\:p')\; .
\end{equation}
Given  that the  free  inverse  CTP propagator  is  proportional to  a
four-dimensional delta function of the  two momenta, it may be written
more conveniently in the \emph{single} Fourier representation as
\begin{equation}
  \Delta_{ab}^{0,\,-1}(p)\ =\
  \begin{bmatrix}
    p^2\:-\:m^2\:+\:i\epsilon  & 0  \\
    0 & -\big(p^2\:-\:m^2\:-\:i\epsilon\big)
  \end{bmatrix}\; .
\end{equation}
Hence, for \emph{translationally invariant} backgrounds, we may recast
(\ref{eq:inv}) in the form
\begin{equation}
  \Delta_{ab}^{0,\,-1}(p)\Delta^{0,\, bc}(p)\ =\ \eta_{a}^{\ c}\; ,
\end{equation}
where
\begin{equation}
  \Delta^{0,\,ab}(p)\ \equiv \ \!\int\!\!
  \frac{\D{4}{p'}}{(2\pi)^4}\;\Delta^{0,\, ab}(p,p')\; 
\end{equation}
is the single-momentum representation of the free CTP propagator.

\subsection{The Free CTP Propagator}
\label{sec:CTPprop}

We  proceed now  to make  the following  ansatz for  the  most general
\emph{translationally  invariant}  form of  the  free CTP  propagator,
without evaluating the correlation functions directly:
\begin{equation}
\label{eq:ansatzzero}
\Delta^{0,\,ab}(p)\ =\ 
\begin{bmatrix}
  \big(p^2\,-\,M^2\,+\,i\epsilon\big)^{-1}\:
  +\:\tilde{c}_1(p)\delta(p^2\,-\,M^2)  &
  \tilde{c}_3(p)\delta(p^2\,-\,M^2) \\
  \tilde{c}_2(p)\delta(p^2\,-\,M^2) &
  -\big(p^2\,-\,M^2\,-\,i\epsilon\big)^{-1}\:
  +\:\tilde{c}_4(p)\delta(p^2\,-\,M^2)\end{bmatrix}\!\; .
\end{equation}
The
$\tilde{c}_i(p)\:\equiv\:\theta(p_0)c_i(p)\:+\:\theta(-p_0)c'_i(p)$
are as  yet undetermined analytic functions  of the four-momentum~$p$,
which  may in  general  be  complex.  The  diagonal  elements are  the
Fourier  transforms  of  the  most general  translationally  invariant
solutions        to       the        inhomogeneous       Klein--Gordon
equation~(\ref{eq:inhomkg}), whilst the  off-diagonal elements are the
most  general translationally invariant  solutions to  the homogeneous
Klein--Gordon equation~(\ref{eq:homkg}).

The remaining freedom in  the matrix elements of $\Delta^{0,\, ab}(p)$
is determined by the following field-theoretic requirements:

\medskip
\noindent
(i)  {\bf  \emph{CPT}  Invariance.}\null~Since  the action  should  be
invariant  under $CPT$, the  real scalar  field~$\Phi$ should  be even
under   $CPT$.    From  the   definitions   of   the  propagators   in
(\ref{eq:pdefs}),    we     obtain    the    $CPT$     relations    in
(\ref{eq:hermit}).  Consequently,  the momentum representation of
these relations in (\ref{eq:hermitdm}) imply that
\begin{equation}
  \label{eq:CPTreq}
  \tilde{c}_{1(4)}(p)\ =\ \tilde{c}_{1(4)}(-p)\; , \qquad
  \tilde{c}_2(p)\ =\ \tilde{c}_3(-p)\; .
\end{equation}

\medskip
\noindent
(ii)  {\bf   Hermiticity.}\null~The  Hermiticity  properties   of  the
correlation  functions defined  in (\ref{eq:pdefs})  give rise  to the
Hermiticity  relations outlined  in~(\ref{eq:hermitdm}).   These imply
that
\begin{equation}
  \tilde{c}_4(p)\ =\ -\tilde{c}^*_1(p)\; , \qquad
  \tilde{c}_2(p)\ =\ -\tilde{c}^*_3(-p)\; .
\end{equation}
In    conjunction   with    (\ref{eq:CPTreq}),   we    conclude   that
$\tilde{c}_2(p)$ and $\tilde{c}_3(p)$  must be purely imaginary-valued
functions of the four-momentum $p$.

\medskip
\noindent
(iii)   {\bf   Causality.}\null~The   free  Pauli--Jordan   propagator
$\Delta^0(x,y)$  is proportional  only to  the real  part of  the free
Feynman      propagator      $\mathrm{Re}\,\Delta^0_{\mathrm{F}}(x,y)$
[cf.~(\ref{eq:PJrealF})].   The addition  of  an even-parity  on-shell
dispersive  part  to  the   Fourier  transform  of  the  free  Feynman
propagator  $\Delta^0_{\mathrm{F}}(p)$  will  contribute to  the  free
Pauli--Jordan propagator  terms that are  non-vanishing for space-like
separations $(x\:-\:y)^2\: <\:  0$, thus violating the micro-causality
condition outlined  in Section~\ref{sec:canon}.  It  follows then that
$\tilde{c}_1(p)$ and $\tilde{c}_4(p)$ are also purely imaginary-valued
functions.   We shall  therefore replace  the $\tilde{c}_i(p)$  by the
real-valued         functions         $\tilde{f}_i(p)$         through
$\tilde{c}_i(p)\:\equiv\:-2\pi i\tilde{f}_i(p)$,  where the minus sign
and factor  of $2\pi$ have  been included for later  convenience.  The
explicit form of the  free Pauli--Jordan propagator in (\ref{eq:pjb}),
along with  the causality relation (\ref{eq:causality}),  give rise to
the constraint
\begin{equation}
  \tilde{f}_2(p)\: -\: \tilde{f}_3(p)\ =\ \varepsilon(p_0)\; .
\end{equation}

\medskip
\noindent
(iv)  {\bf   Unitarity.}\null~Finally,  the  unitarity   relations  in
(\ref{eq:unitarity}) require that
\begin{equation}
  \tilde{f}_2(p)\:+\:\tilde{f}_3(p)
  \ = \ 1\:+\:\tilde{f}_1(p)\:+\:\tilde{f}_4(p)\; .
\end{equation}

\medskip

Solving   the    system   of   the   above    four   constraints   for
$\tilde{f}_{1,2,3,4}(p)$,  we arrive at  the following  expression for
the most general translationally invariant free CTP propagator:
\begin{equation}
  \label{eq:singctp}
  \Delta^{0,\,ab}(p) \ = \
  \begin{bmatrix}
    \big(p^2\:-\:M^2\:+\:i\epsilon\big)^{-1} &
    -2\pi i\, \theta(-p_0)\delta(p^2\:-\:M^2) \\ 
    -2\pi i\, \theta(p_0)\delta(p^2\: -\: M^2) &
    -\big(p^2\: -\: M^2\: -\: i\epsilon\big)^{-1}
  \end{bmatrix}\:-\:2\pi i\tilde{f}(p)\delta(p^2\:-\:M^2)\!
  \begin{bmatrix}
    1 & 1 \\
    1 & 1 \\
  \end{bmatrix}\!\; .
\end{equation}
All  elements of  $\Delta^{0,\,ab}(p)$ contain  terms dependent  upon the
same function
\begin{equation}
  \label{eq:ftilde}
  \tilde{f}(p)\ \equiv\ \tilde{f}_1(p)\
  = \ \theta(p_0)f(p)\:+\:\theta(-p_0)f(-p)\; .
\end{equation}
These terms correspond to the vacuum expectation of the normal-ordered
product of fields $\braket{\mathbf{:}\Phi(x)\Phi(y)\mathbf{:}}$, which
vanishes  for  the  trivial  vacuum  $\ket{0}$.   Therefore,  we  must
conclude  that $\tilde{f}(p)$  also  vanishes in  this  case. We  then
obtain      the       free      \emph{vacuum}      CTP      propagator
$\widehat{\Delta}^{0,\,ab}(p)$, which contains  the set of propagators
familiar  from   the  unitarity  cutting  rules   of  absorptive  part
theory~\cite{'tHooft:1973pz, Veltman1994}:
\begin{equation}
  \label{eq:zerotemp}
  \widehat{\Delta}^{0,\, ab}(p)\  =\
  \begin{bmatrix}
    \big(p^2\:-\:M^2\:+\:i\epsilon\big)^{-1} &
    -2\pi i\, \theta(-p_0)\delta(p^2\:-\:M^2) \\
    -2\pi i\, \theta(p_0)\delta(p^2\: -\: M^2) &
    -\big(p^2\:-\:M^2\:-\:i\epsilon\big)^{-1}
  \end{bmatrix}\; .
\end{equation}

We may similarly arrive  at (\ref{eq:singctp}) by considering the free
CTP      propagator       in      the      Keldysh      representation
$\widetilde{\Delta}^{0,\,ab}(p)$    from    (\ref{eq:physrep}).    The
constraints  outlined above  allow  us  to add  to  the free  Hadamard
propagator $\Delta_1^0(p)$  any purely-imaginary even  function of $p$
proportional to $\delta(p^2\:-\: M^2)$, that is
\begin{equation}
  \widetilde{\Delta}^{0,\,ab}(p)\ =\
  \begin{bmatrix}
    0 & \big[(p_0\:-\:i\epsilon)^2\:-\:\mathbf{p}^2\:-\:M^2\big]^{-1} \\
    \big[(p_0\:+\:i\epsilon)^2\:-\:\mathbf{p}^2\:-\:M^2\big]^{-1} &
    -2\pi i\, \delta(p^2\:- \: M^2)
  \end{bmatrix}\:-\: 2\pi i\, 2\tilde{f}(p)\delta(p^2\:-\:M^2)\!
  \begin{bmatrix}
    0 & 0 \\
    0 & 1
  \end{bmatrix}\; .
\end{equation}
We  note that  there  is no  such freedom  to  add terms  to the  free
retarded  and  advanced  propagators,  $\Delta^0_{\mathrm{R}}(p)$  and
$\Delta^0_{\mathrm{A}}(p)$,   which   is    a   consequence   of   the
micro-causality  constraints on  the  form of  the free  Pauli--Jordan
propagator $\Delta^0(p)$. Employing the fact that
\begin{equation}
  \mathbf{O}^{\mathsf{T}}
  \begin{bmatrix}
    0 & 0 \\
    0 & 1
  \end{bmatrix}
  \mathbf{O}\   =\  \frac{1}{2}
  \begin{bmatrix}
    1 & 1 \\
    1 & 1 \end{bmatrix}\!\;, \qquad \mathbf{O}\ =\ \{O^{a}_{\ b}\}=\begin{bmatrix} 1 & -1 \\ 1 & 1 \end{bmatrix}\;,
\end{equation}
we  recover~(\ref{eq:singctp}),  which  serves as  a  self-consistency
check for the correctness of our ansatz for the free CTP propagator.

\subsection{The Resummed CTP Propagator}

In order to obtain the  \emph{resummed} CTP propagator, we must invert
the  inverse   resummed  CTP  propagator  on   the  restricted  domain
$[-t/2,\ t/2]$ subject to the inverse relation
\begin{equation}
  \label{eq:resinvdef}
  \int_{\Omega_t}\!\!\D{4}{z}\;
  \Delta^{-1}_{ab}(x,z,\tilde{t}_f;\tilde{t}_i)
  \Delta^{bc}(z,y,\tilde{t}_f;\tilde{t}_i)
  \ =\ \eta_{a}^{\ c}\delta^{(4)}(x\:-\: y)\;,
\end{equation}
for   $x^{\mu},\   y^{\mu}\:\in\:   \Omega_t$.    We  shall   see   in
Section~\ref{sec:ctpsd}  that  this  restriction  of the  time  domain
implies that  a closed analytic  form for the resummed  CTP propagator
$\Delta^{ab}(x,y,\tilde{t}_f;\tilde{t}_i)$ is in general not possible,
for systems  out of  thermal equilibrium (see  also our  discussion in
Section~\ref{sec:eq}).

The     double    momentum     representation    of     the    inverse
relation~(\ref{eq:resinvdef}) takes on the form
\begin{equation}
  \label{eq:invdmom}
  \iint\!\!\frac{\D{4}{q}}{\big(2\pi\big)^4}\,
  \frac{\D{4}{q'}}{\big(2\pi\big)^4}\;
  \Delta_{ab}^{-1}(p,q,\tilde{t}_f;\tilde{t}_i)(2\pi)^4
  \delta^{(4)}_{t}(q\:-\:q')\Delta^{bc}(q',p',\tilde{t}_f;\tilde{t}_i)
  \ =\ \eta_{a}^{\ c}(2\pi)^4\delta^{(4)}_t(p\:-\:p')\;,
\end{equation}
where we have defined
\begin{equation}
  \label{eq:deltat4}
  \delta^{(4)}_t(p\:-\:p')\ \equiv \ \delta_t(p_0\:-\: p_0')
  \delta^{(3)}(\mathbf{p}\:-\:\mathbf{p}')
  \ =\ \frac{1}{(2\pi)^4}\!\iint_{\Omega_t}\!\D{4}{x}\,\D{4}{y}\;
  e^{ip\cdot x}e^{-ip'\cdot y}\,\delta^{(4)}(x\:-\:y)\;.
\end{equation}
The  restriction of  the domain  of time  integration has  led  to the
introduction of the analytic weight function
\begin{equation}
  \label{eq:deltat}
  \delta_t(p_0\:-\:p_0')\ =\ \frac{t}{2\pi}\,\mathrm{sinc}
  \Big[\Big(\frac{p_0\:-\:p_0'}{2}\Big)t\Big]\ =\
  \frac{1}{\pi}\,\frac{\sin\!\big[\big(p_0-p_0'\big)\frac{t}{2}\big]}{p_0-p_0'}\; ,
\end{equation}
which has  replaced the ordinary energy-conserving  delta function. As
expected, we have
\begin{equation}
  \label{eq:deltatlim}
  \lim_{t\:\to\:\infty}\delta_t(p_0\:-\:p_0')\ =\ \delta(p_0\:-\:p_0')\; ,
\end{equation}
so that the standard description of asymptotic quantum field theory is
recovered     in    the     limit     $\tilde{t}_i\:=\:-\:t/2    \:\to\:
-\:\infty$.  Moreover,  the  weight  function $\delta_t$  satisfies  the
convolution
\begin{equation}
  \label{eq:dciden}
  \int_{-\infty}^{+\infty}\!\D{}{q_0}\;\delta_t(p_0\:-\:q_0)
  \delta_t(q_0\:-\:p_0')\ =\ \delta_t(p_0\:-\:p_0')\; .
\end{equation}

The  emergence of  the  function~$\delta_t$ is  a  consequence of  our
requirement  that   the  time   evolution  and  the   mapping  between
quantum-mechanical pictures (see Section~\ref{sec:canon}) are governed
by the standard interaction-picture evolution operator
\begin{equation}
  \label{eq:evoop}
  U(\tilde{t}_f,\tilde{t}_i)\ =\
  \mathrm{T}\exp\bigg(\,
  -i\!\int_{\tilde{t}_i}^{\tilde{t}_f}\!\!\D{}{\tilde{t}}\,
  H^{\mathrm{int}}(\tilde{t};\tilde{t}_i\,)\bigg)\;.
\end{equation}
This evolution is defined for times greater than the boundary
time~$\tilde{t}_i$, at which point the three pictures are
coincident. We stress here that the weight function $\delta_t$ in
(\ref{eq:deltat}) is neither a prescription, nor is it an \emph{a priori}
regularization of the Dirac delta function.

\begin{figure}
\begin{center}
\includegraphics[scale=0.48]{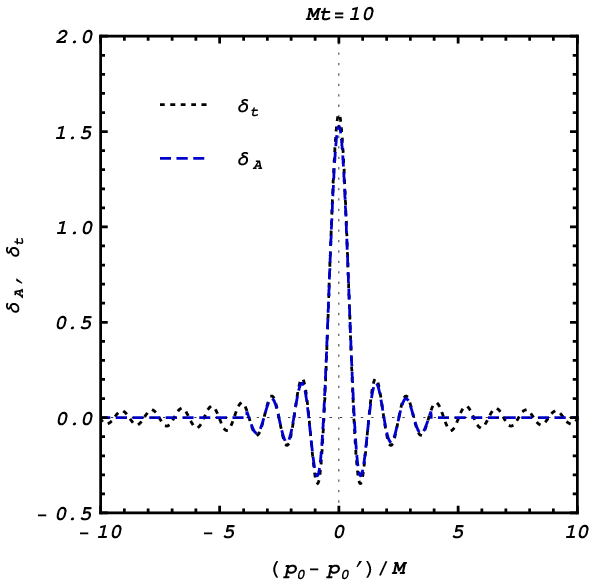}
\end{center}
\caption{Comparison  of  $\delta_t(p_0\:-\:p_0')$  (black dotted)  and
  $\delta_A(p_0\:-\:p_0')$ (blue  dashed).  The arbitrary  mass $M$ is
  included so that axes are dimensionless.}
\label{fig:adiabatic}
\end{figure}

As we will see later, the oscillatory behavior of the sinc function in
$\delta_t$ is fundamentally important to the dynamical behavior of the
system.  Let  us therefore convince ourselves  that these oscillations
persist, even  if we  smear the switching  on of the  interactions, or
equivalently,  if  we  impose   an  adiabatic  switching  off  of  the
interaction  Hamiltonian for  microscopic times  outside  the interval
$[-\:t/2,\  t/2]$.   To this  end,  we  introduce  to the  interaction
Hamiltonian $H^{\rm int}$ in~(\ref{eq:evoop}) the Gaussian function
\begin{equation}
  A_t(\tilde{t})\ =\ \exp\!\bigg(\!-\frac{\tilde{t}^2}{2t^2}\bigg)\;,
\end{equation}
such that the evolution operator takes the form
\begin{equation}
  U(\tilde{t}_f,\tilde{t}_i)\ =\ 
  \mathrm{T}\exp\bigg(\,-i\!\int_{\tilde{t}_i}^{\tilde{t}_f}\!\!
  \D{}{\tilde{t}}\; A_t(\tilde{t})
  H^{\mathrm{int}}(\tilde{t};\tilde{t}_i)\bigg)\; .
\end{equation}
Clearly, for $\tilde{t}\:\gg\: t$, $A_t(\tilde{t})\: \to\: 0$, whereas
for $\tilde{t}\:\ll\: t$, $A_t(\tilde{t})\:  \to\: 1$.  To account for
the  effect  of  $A_{t}(\tilde{t})$   in  the  action,  the  following
replacement needs to be made:
\begin{align}
  \label{eq:deltaA}
  & \delta_t(p_0-p_0')\ \to\ \delta_{A}(p_0\:-\:p_0')\ \equiv\
  \frac{1}{2\pi}\!\int_{-t/2}^{+t/2}\!\D{}{\tilde{t}}\;
  e^{-i(p_0\:-\:p_0')\tilde{t}}\,A_{t}(\tilde{t})\nonumber \\
  & =\ \frac{t}{2\sqrt{2\pi}}\,e^{-\frac{1}{2}(p_0\:-\:p_0')^2t^2}
  \bigg[\,\mathrm{Erf}\bigg(\frac{1\,-\,2i(p_0\,-\,p_0')t}{2\sqrt{2}}
  \bigg)\,+\,
  \mathrm{Erf}\bigg(\frac{1\,+\,2i(p_0\,-\,p_0')t}{2\sqrt{2}}
  \bigg)\bigg]\;.
\end{align}
Due to the error  functions of complex arguments in~(\ref{eq:deltaA}),
the oscillatory behavior remains.  The analytic behavior of both the
functions  $\delta_t(p_0\:-\:p_0')$  and  $\delta_A(p_0\:-\:p_0')$  is
shown in  Figure~\ref{fig:adiabatic} in which  we see that  the smooth
smearing by  the Gaussian function $A_t(\tilde{t})$  has little effect
on the central region of the sinc function, as one would expect.

\bigskip

\section{Non-Homogeneous Backgrounds}
\label{sec:nonhom}

Until now, we  have considered the vacuum to be  an `empty' state with
all quantum  numbers zero.  In  this section, we replace  that `empty'
vacuum state with some macroscopic background, which may in general be
in\-homogeneous  and  in\-coherent.    This  non-trivial  `vacuum'  is
described by  the density operator~$\rho$.  Following  a derivation of
the CTP Schwinger--Dyson equation, we  show that it is not possible to
obtain a closed analytic form  for the resummed CTP propagators in the
presence  of time-dependent backgrounds.   Finally, we  generalize the
discussions  in  Section~\ref{sec:CTPprop}  to obtain  non-homogeneous
free  propagators  in  which  space-time translational  invariance  is
explicitly broken.

The  density operator  $\rho$  is necessarily  Hermitian  and, for  an
isolated system,  evolves in the interaction picture  according to the
von Neumann or quantum Liouville equation
\begin{equation}
  \label{eq:qLiou}
  \frac{\D{}{\rho(\tilde{t};\tilde{t}_i)}}
       {\D{}{(\tilde{t}-\tilde{t}_i)}}
  \ =\ -i\,\big[H^{\mathrm{int}}(\tilde{t};\tilde{t}_i),\
    \rho(\tilde{t};\tilde{t}_i)\,\big]\;,
\end{equation}
where  $H^{\mathrm{int}}(\tilde{t};\tilde{t}_i)$  is  the  interaction
part  of  the  Hamiltonian   in  the  interaction  picture,  which  is
time-dependent.  Notice that the  time derivative appearing on the LHS
of  (\ref{eq:qLiou})   is  taken   with  respect  to   the  \emph{time
  translationally   invariant}  quantity  $\tilde{t}\:-\:\tilde{t}_i$.
Developing the usual Neumann series, we find that
\begin{equation}
  \rho(\tilde{t};\tilde{t}_i)\ =\
  U(\tilde{t},\tilde{t}_i)\,\rho(\tilde{t}_i;\tilde{t}_i)\,
  U^{-1}(\tilde{t},\tilde{t}_i)\; ,
\end{equation}
where $U$ is the evolution operator in (\ref{eq:evoop}). Hence, in the
absence of external  sources and given the unitarity  of the evolution
operator, the partition function $\mathcal{Z}\: =\: \mathrm{Tr}\,\rho$
is time independent.  On the other  hand, the partition function of an
open  or closed  subsystem is  in  general time dependent  due to  the
presence of external sources.

We  are interested in  evaluating time-dependent  Ensemble Expectation
Values   (EEVs)  of  field   operators  $\braket{\bullet}_t$   at  the
\emph{macro}scopic     time~$t$,    which    corresponds     to    the
\emph{micro}scopic      time~$\tilde{t}_f\:=\:t/2$,      where     the
\emph{bra}-\emph{ket} now denotes the weighted expectation
\begin{equation}
  \label{eq:thermbra}
  \braket{\bullet}_t\ =\ \frac{\mathrm{Tr}\,
  \big(\,\rho(\tilde{t}_f;\tilde{t}_i)\,\bullet\,\big)}
  {\mathrm{Tr}\,\rho(\tilde{t}_f;\tilde{t}_i)}\; .
\end{equation}
In this case, EEVs of two-point products of field operators begin with
a total of  nine independent coordinates: the microscopic  time of the
density operator  and the two  four-dimensional space-time coordinates
of   the  field  operators.    As  discussed   in  the   beginning  of
Section~\ref{sec:CTP} [cf.~(\ref{eq:7ind})], this number is reduced to
the  required seven  coordinates, i.e.~one  temporal and  six spatial,
after  setting all microscopic  times equal  to $\tilde{t}_f\:=\:t/2$.
Hence,  physical  observables  in  the interaction  picture  are,  for
instance, of the form
\begin{equation}
  \braket{\Phi(\tilde{t}_f,\mathbf{x};\tilde{t}_i)
    \Phi(\tilde{t}_f,\mathbf{y};\tilde{t}_i)}_{t}\ =\ 
  \frac{\mathrm{Tr}\,\big(\,\rho(\tilde{t}_f;\tilde{t}_i)
  \Phi(\tilde{t}_f,\mathbf{x};\tilde{t}_i)
  \Phi(\tilde{t}_f,\mathbf{y};\tilde{t}_i)\,\big)}
  {\mathrm{Tr}\,\rho(\tilde{t}_f;\tilde{t}_i)}\; .
\end{equation}

In the presence of a non-trivial background, the $\emph{out}$ state of
Section~\ref{sec:CTP} is  replaced by  the density operator  $\rho$ at
the  time  of observation  $\tilde{t}_f\:=\:  t/2$. Consequently,  the
starting point for the CTP generating functional of EEVs is
\begin{equation}
  \label{eq:genout}
  \mathcal{Z}[\rho,J_{\pm},t]\ =\ \mathrm{Tr}\,\Big[\Big(
  \bar{\mathrm{T}}e^{-i\!\int_{\Omega_t}\!\D{4}{x}\,J_-(x)\Phi_{\mathrm{H}}(x)}\Big)\,
  \rho_{\mathrm{H}}(\tilde{t}_f;\tilde{t}_i)\,
  \Big(\mathrm{T}e^{i\!\int_{\Omega_t}\!\D{4}{x}\,J_+(x)\Phi_{\mathrm{H}}(x)}\Big)
  \Big]\; .
\end{equation}
Within the  generating functional $\mathcal{Z}$  in (\ref{eq:genout}),
the  Heisenberg-picture density  operator~$\rho_{\rm H}$  has explicit
time dependence,  as it is built  out of state vectors  that depend on
time due  to the presence  of the external sources~$J_{\pm}$.   In the
absence of such  sources, however, the state vectors  do not evolve in
time,   so   $\rho_{\rm   H}$   and  the   partition   function~${\cal
  Z}[\rho,J_{\pm}=0,t]\:=\:\mathrm{Tr}\,\rho$      in~(\ref{eq:genout})
become time independent quantities.

It  is  important to  emphasize  that  the  explicit microscopic  time
$\tilde{t}_f\:=\:t/2$       of        the       density       operator
$\rho_{\mathrm{H}}(\tilde{t}_f;\tilde{t}_i)$  appearing   in  the  CTP
generating   functional  (\ref{eq:genout})   is   the  \emph{time   of
  observation}. This is in contrast to existing interpretations of the
CTP  formalism, see  for instance  \cite{Berges:2004yj}, in  which the
density operator  replaces the \emph{in} state and  is therefore fixed
at  the  initial  time  ${\tilde{t}_i\:=\:-t/2}$,  encoding  only  the
boundary  conditions.  As we  shall see  in Section~\ref{sec:oneloop},
this new interpretation of the  CTP formalism will lead to the absence
of pinch singularities in the resulting perturbation series.

\subsection{The Schwinger--Dyson Equation in the CTP Formalism}
\label{sec:ctpsd}

In order to generate a perturbation series of correlation functions in
the  presence  of  non-homogeneous  backgrounds, we  must  derive  the
Schwinger--Dyson equation in the CTP formalism. Of particular interest
is the explicit form of the Feynman--Dyson series for the expansion of
the resummed CTP propagator. We  will show that, in the time-dependent
case,  this series  does not  collapse to  the resummation  known from
zero-temperature field  theory. In particular,  we find that  a closed
analytic form  for the  resummed CTP propagator  is not  attainable in
general.

We proceed  by inserting into the  generating functional $\mathcal{Z}$
in (\ref{eq:genout})  complete sets  of eigenstates of  the Heisenberg
field   operator  $\Phi_{\mathrm{H}}$   at   intermediate  times   via
(\ref{eq:eigencomp}).   In  this  way,  we  obtain  the  path-integral
representation
\begin{equation}
  \mathcal{Z}[\rho,J_a,t]\ =\ \!\int\![\D{}{\Phi^a(\mathbf{x})}]\;
            \braket{\Phi_-(\mathbf{x}),\tilde{t}_f;\tilde{t}_i|\,
            \rho_{\mathrm{H}}(\tilde{t}_f;\tilde{t}_i)\,|
            \Phi_+(\mathbf{x}),\tilde{t}_f;\tilde{t}_i}
 \exp\!\bigg[i\bigg(\!S[\Phi^a,t]\:+\:\!
  \int_{\Omega_t}\!\!\D{4}{x}\;J_a(x)\Phi^a(x)\bigg)\bigg]\; .
\end{equation}
As before, we may extend the  limits of integration to infinity in the
free part  of the action and  the $J$-dependent term, due  to the fact
that   the  external   sources  vanish   outside  the   time  interval
$[-t/2,\ t/2]$. \emph{It is only in the interaction part of the action
  that the finite domain of integration must remain.}

Following     \cite{Calzetta:1986cq},    we    write     the    kernel
$\braket{\Phi_-(\mathbf{x}),\tilde{t}_f;\tilde{t}_i|
  \,\rho_{\mathrm{H}}(\tilde{t}_f;\tilde{t}_i)\,
  |\Phi_+(\mathbf{x}),\tilde{t}_f;\tilde{t}_i}$  as an
infinite series of poly-local sources:
\begin{equation}
  \braket{\Phi_-(\mathbf{x}),\tilde{t}_f;\tilde{t}_i|
    \,\rho_{\mathrm{H}}(\tilde{t}_f;\tilde{t}_i)\,|
    \Phi_+(\mathbf{x}),\tilde{t}_f;\tilde{t}_i}
  \ =\ \exp\big(iK[\Phi^a,t\,]\big)\; ,
\end{equation}
where
\begin{equation}
  \label{eq:Kt}
  K[\Phi^a,t\,]\ =\ K\:+\: \!\int_{\Omega_t}\!\!\D{4}{x}\;
  K_a(x,\tilde{t}_f;\tilde{t}_i)\Phi^a(x)\:+\:
  \frac{1}{2}\!\iint_{\Omega_t}\!\!\D{4}{x}\,
  \D{4}{x'}\;K_{ab}(x,x',\tilde{t}_f;\tilde{t}_i)
  \Phi^a(x)\Phi^b(x')\:+\:\cdots\; 
\end{equation}
is a time translationally invariant quantity and only depends on
$t\:=\:\tilde{t}_f\:-\:\tilde{t}_i$.  The poly-local sources
$K_{ab\cdots}$ encode the state of the system at the microscopic time
of observation~$\tilde{t}_f\:=\:t/2$, i.e.~the time at which the EEV
is evaluated, according to Figure~\ref{fig:sk}.  It follows that these
sources must contribute only for
$x_0\:=\:x_0'\:=\:\cdots\:=\:\tilde{t}_f$ and therefore be
proportional to delta functions of the form
$\delta(x_0\:-\:\tilde{t}_f)\delta(x_0'\:-\:\tilde{t}_f)\cdots$.  For
instance, the bi-local source $K_{ab}$ in the double momentum
representation must have the form
\begin{equation}
  \label{eq:kmom}
  K_{ab}(x,x',\tilde{t}_f;\tilde{t}_i)\ =\ \!\iint\!\!
  \frac{\D{4}{p}}{(2\pi)^4}\,\frac{\D{4}{p'}}{(2\pi)^4}\;
  e^{-ip\cdot x}\,e^{ip'\cdot x'}\,
  e^{i(p_0\:-\:p_0')\tilde{t}_f}K_{ab}(\mathbf{p},\mathbf{p}',t)\; ,
\end{equation}
so that  the $p_0$  and $p_0'$ integrations  yield the  required delta
functions. Here,  it is understood that the  bi-local $K_{ab}$ sources
occurring on  the LHS and  RHS of~(\ref{eq:kmom}) are  distinguished by
the    form     of    their    arguments.      We    emphasize    that
$K_{ab}(x,x',\tilde{t}_f;\tilde{t}_i)$  is not a  time translationally
invariant quantity  due to the explicit  dependence upon $\tilde{t}_f$
on      the       RHS      of~(\ref{eq:kmom}).       In      contrast,
$K_{ab}(\mathbf{p},\mathbf{p}',t)$   \emph{is}   time  translationally
invariant.

Notice that we could extend the limits of integration to infinity for
the time integrals in the expansion of the kernel given
in~(\ref{eq:Kt}) also.  Nevertheless, for the following derivation,
all space-time integrals are taken to run over the
hypervolume~$\Omega_t$ in (\ref{eq:omegat}) for consistency.  We
should reiterate here that the limits of time integration can be
extended to an infinite domain $\Omega_\infty$ in all but the
interaction part of the action. We will also suppress the
time dependencies of the action $S$ and sources $K_{ab\cdots}$ for
notational convenience.

We  now absorb  the  constant $K$  in~(\ref{eq:Kt})  into the  overall
normalization  of the  CTP generating  functional ${\cal  Z}$  and the
local   source   $K_a$   into   a   redefinition   of   the   external
source~$J_a$. Then, ${\cal Z}$ may be written down as
\begin{align}
  \label{eq:chgen}
  \mathcal{Z}[J_a,K_{ab},\cdots]\ &=\ \!\int\![\D{}{\Phi^a(x)}]\;
  \exp\!\bigg[i\bigg(\!S[\Phi^a]\:+\:\!\int\!\D{4}{x}\;J_a(x)\Phi^a(x)
  \:+\: \frac{1}{2}\!\iint\!\D{4}{x}\,\D{4}{y}\;
  K_{ab}(x,x')\Phi^a(x)\Phi^b(x')\nonumber\\
  &\qquad  +\: \frac{1}{6}\!\iiint\!\D{4}{x}\,\D{4}{x'}\,\D{4}{x''}\;
  K_{abc}(x,x',x'')\Phi^a(x)\Phi^b(x')\Phi^c(x'')\:+\:\cdots\bigg)\bigg]
  \; .
\end{align}

The     Cornwall--Jackiw--Tomboulis     (CJT)     effective     action
\cite{Cornwall:1974vz} is given by the following Legendre transform:
\begin{align}
  \Gamma[\widehat{\Phi}^a,\mathcal{G}^{ab},\mathcal{G}^{abc},\cdots]\ &=\ 
  \mathcal{W}[J_a,K_{ab},K_{abc}]\:-\:\!\int\!\D{4}{x}\;
  J_a(x)\widehat{\Phi}^a(x)\nonumber\\&\qquad
  -\: \frac{1}{2}\!\iint\!\D{4}{x}\,\D{4}{x'}\;K_{ab}(x,x')\,
  \Big(\,\widehat{\Phi}^a(x)\widehat{\Phi}^b(x')\:+\:i\hbar
  \mathcal{G}^{ab}(x,x')\,\Big)\nonumber\\&
  \qquad -\: \frac{1}{6}\!\iiint\!\D{4}{x}\,\D{4}{x'}\,\D{4}{x''}\;
  K_{abc}(x,x',x'')\,\Big(\,\widehat{\Phi}^a(x)\widehat{\Phi}^b(x')
  \widehat{\Phi}^c(x'')\nonumber\\&
  \qquad +\: 3i\hbar\mathcal{G}^{(ab}(x,x')\widehat{\Phi}^{c)}(x'')\:-\:
  \hbar^2\mathcal{G}^{abc}(x,x',x'')\,\Big)\:+\:\cdots\; ,
\end{align}
where      $\mathcal{W}[J_a,K_{ab},K_{abc},\cdots]\:=\:     -i\hbar\ln
\mathcal{Z}[J_a,K_{ab},K_{abc},\cdots]$  is the  generating functional
of connected ensemble Green's  functions. We obtain an infinite system
of equations:
\begin{subequations}
  \label{eq:sys1}
  \begin{align}
    \label{eq:varphi}
    \widehat{\Phi}^a(x)&\ =\ \frac{\delta\mathcal{W}}{\delta
    J_a(x)}\ =\ \braket{\Phi^a(x)}\;,\\
    \label{eq:twocon}
    i\hbar\mathcal{G}^{ab}(x,x')&\ =\ 2\frac{\delta\mathcal{W}}
    {\delta K_{ab}(x,x')}\:-\:\widehat{\Phi}^a(x)\widehat{\Phi}^b(x')\nonumber
    \\ &\ =\ -i\hbar\frac{\delta^2 \mathcal{W}}
    {\delta J_a(x)\delta J_b(x')}\ =\ \braket{\,
    \mathrm{T}_{\mathcal{C}}\,\big[\Phi^a(x)\Phi^b(x')\,\big]}
    \:-\: \braket{\Phi^a(x)}\!\braket{\Phi^b(x')}\;,\\
    \label{eq:Gabc}
    -\hbar^2\mathcal{G}^{abc}(x,x',x'')&\ =\ 6\frac{\delta \mathcal{W}}
    {\delta K_{abc}(x,x',x'')}\:-\:3i\hbar
    \mathcal{G}^{(ab}(x,x')\widehat{\Phi}^{c)}(x'')
    \:-\:\widehat{\Phi}^a(x)\widehat{\Phi}^b(x')\widehat{\Phi}^c(x'')\nonumber\\&
    \ =\ -\hbar^2\frac{\delta^3 \mathcal{W}}
    {\delta J_a(x)\delta J_b(x')\delta J_c(x'')}\; ,
\end{align}
\end{subequations}
and
\begin{subequations}
  \label{eq:sys2}
  \begin{align}
  \frac{\delta \Gamma}{\delta
  \widehat{\Phi}^a(x)}\ &=\ -J_a(x)\:-\:\int\!\D{4}{x'}\;K_{ab}(x,x')
  \widehat{\Phi}^b(x')\nonumber \\
  &\qquad -\: \frac{1}{2}\!\iint\!\D{4}{x'}\,\D{4}{x''}\;K_{abc}(x,x',x'')\,
  \Big(\,\widehat{\Phi}^b(x')\widehat{\Phi}^c(x'')\:+\:
  i\hbar\mathcal{G}^{bc}(x',x'')\,\Big)\:-\:\cdots\;,\\
  \label{eq:dGab}
  \frac{\delta \Gamma}{\delta \mathcal{G}^{ab}(x,x')}\ & =\ 
  -\frac{i\hbar}{2}K_{ab}(x,x')\:-\:\frac{i\hbar}{2}\!\int\!
  \D{4}{x''}\;K_{abc}(x,x',x'')\widehat{\Phi}^c(x'')\:-\:\cdots\; ,\\
  \frac{\delta \Gamma}{\delta \mathcal{G}^{abc}(x,x',x'')}\ &
   = \ \frac{\hbar^2}{6}K_{abc}(x,x',x'')\:+\:\cdots\; ,
\end{align}
\end{subequations}
where  the parentheses~$(abc)$  on the  RHS of  (\ref{eq:Gabc}) denote
cyclic permutation with respect to the indices $a$, $b$, $c$.

The above infinite system of equations~(\ref{eq:sys1}) and
(\ref{eq:sys2}) may be simplified by assuming that the density
operator $\rho$ is Gaussian, as we will do later in
Section~\ref{sec:nonhomprop}. In this case, the tri-local and higher
kernels ($K_{abc}$, $K_{abcd}$, $\cdots$) can be set to zero
in~(\ref{eq:sys2}), neglecting contributions from thermally-corrected
vertices [see (\ref{eq:ntherm})], which would otherwise be present for
non-Gaussian density operators.  Within the Gaussian approximation,
the three- and higher-point connected Green's functions
($\mathcal{G}^{abc}$, $\mathcal{G}^{abcd}$, $\cdots$) may be
eliminated as dynamical variables by performing a second Legendre
transform
\begin{equation}
  \Gamma[\widehat{\Phi}^a,\mathcal{G}^{ab}]\ \equiv\ 
  \Gamma[\widehat{\Phi}^a,\mathcal{G}^{ab},\widetilde{\mathcal{G}}^{abc},
  \cdots]\; ,
\end{equation}
where    the    $\widetilde{\mathcal{G}}$'s    are   functionals    of
$\widehat{\Phi}^a$ and $\mathcal{G}^{ab}$ given by
\begin{equation}
  \frac{\delta \Gamma}{\delta \mathcal{G}^{abc\cdots}}
  [\widehat{\Phi}^a,\mathcal{G}^{ab},\widetilde{\mathcal{G}}^{abc},\cdots]
  \ =\ 0\; .
\end{equation}

The  effective action is  evaluated by  expanding around  the constant
background                                                        field
{$\Phi_0^a(x)\ =\ \Phi^a(x)\:-\:\hbar^{1/2}\phi^a(x)$}, defined at the
saddle point
\begin{equation}
  \frac{\delta S[\Phi^a]}{\delta \Phi^a(x)}\Bigg|_{\Phi\:=\:\Phi_0}\:+\:
  J_a(x)\:+\:\!\int\!\D{4}{x'}\;K_{ab}(x,x')\Phi_0^b(x')\ =\ 0\; .
\end{equation}
The  result of  this  expansion is  well known  \cite{Cornwall:1974vz,
  Carrington:2004sn} and, truncating to order $\hbar^2$, we obtain the
two-particle-irreducible (2PI) CJT effective action
\begin{equation}
  \label{eq:CJTaction}
  \Gamma[\widehat{\Phi}^a,\mathcal{G}^{ab}]\ =\ 
  S[\widehat{\Phi}^a]\:+\:\frac{i\hbar}{2}\,\mathrm{Tr}_{x}\Big[\,
  \mathrm{Ln}_x\,\mathrm{Det}_{ab}\,\mathcal{G}_{ab}^{-1}\:+\:
  \Big(\,G_{ab}^{-1}\:-\:K_{ab}\,\Big)*\mathcal{G}^{ab}\:-\:\eta_a^{\;a}\,\Big]\:+\:
  \hbar^2\Gamma_2[\widehat{\Phi}^a,\mathcal{G}^{ab}]\;, 
\end{equation}
where a subscript~$x$ and the $*$'s indicate that the trace, logarithm
and  products  should be  understood  as  functional operations.   The
operator $G^{-1}_{ab}$ is defined by
\begin{equation}
  G_{ab}^{-1}(\widehat{\Phi}^a;x,x')\ =\ \frac{\delta^2 S[\widehat{\Phi}^a]}
  {\delta \widehat{\Phi}^a(x)\delta \widehat{\Phi}^b(x')}
  \: +\: K_{ab}(x,x')
  \ =\ \Delta_{ab}^{0,\,-1}(x,x')\: +\: 
  \frac{\delta^2S^{\mathrm{int}}[\widehat{\Phi}^a]}{\delta\widehat{\Phi}^a(x)
  \delta\widehat{\Phi}^b(x')}\:+\:K_{ab}(x,x')\; ,
\end{equation}
where $\Delta_{ab}^{0,\,-1}(x,x')$ is  the free inverse CTP propagator
in (\ref{eq:freeinv})  and $S^{\mathrm{int}}[\widehat{\Phi}^a]$ is the
interaction  part of  the  action.  Obviously,  all Green's  functions
depend  upon the  state  of the  system  at the  macroscopic time  $t$
through  the   bi-local  source  $K_{ab}$.   For   the  Lagrangian  in
(\ref{eq:scallag}), we have
\begin{equation}
  G_{ab}^{-1}(\widehat{\Phi}^a;x,x')\ =\ \delta^{(4)}(x\:-\:x')
  \Big[\,-\big(\Box^2_x\:+\:M^2\big)\eta_{ab}\: +\:i\epsilon\mathbb{I}_{ab}
  \:-\: g\eta_{abc}\widehat{\Phi}^c(x)\:-\:\frac{1}{2}
  \lambda\eta_{abcd}\widehat{\Phi}^c(x) \widehat{\Phi}^d(x)\Big]\:+\: K_{ab}(x,x')\;,
\end{equation}
where       $\eta_{abc\cdots}\:=\:+1$       for      all       indices
$a\:=\:b\:=\:\cdots\:=\:1$, $\eta_{abc\cdots}\:=\:-1$  for all indices
$a\:=\:b\:=\:\cdots\:=\:2$ and $\eta_{abc\cdots}\:=\:0$ otherwise.

The  overall normalization ($\eta_{a}^{\;a}$)  of (\ref{eq:CJTaction})
has been  chosen so that  when ${K_{ab}(x,x')\:=\:0}$, we  may recover
the  conventional effective  action~\cite{Jackiw:1974cv}  by making  a
further  Legendre  transform  to  eliminate  $\mathcal{G}^{ab}$  as  a
dynamical variable:
\begin{equation}
  \label{eq:conveact}
  \Gamma[\widehat{\Phi}^a]\ \equiv\ \Gamma[\widehat{\Phi}^a,
  \widetilde{\mathcal{G}}^{ab}]\ =\ S[\widehat{\Phi}^a]\:+\:
  \frac{i\hbar}{2}\mathrm{Tr}_{x}\,\mathrm{Ln}_x\,\mathrm{Det}_{ab}
  \,G_{ab}^{-1}\:+\:\mathcal{O}(\hbar^2)\;.
\end{equation}
Here, $\mathcal{G}^{-1}_{ab}$  has been replaced  by $G^{-1}_{ab}$ and
$\widetilde{\mathcal{G}}^{ab}$ is a functional of $\widehat{\Phi}^a$.

In      the       CJT      effective      action~(\ref{eq:CJTaction}),
$\Gamma_2[\widehat{\Phi}^a,\mathcal{G}^{ab}]$  is the  sum of  all 2PI
vacuum graphs:
\vspace{-1em}
\begin{equation}
  \Gamma_2[\widehat{\Phi}^a,\mathcal{G}^{ab}]\ =\ 
  -i\sum_{a,\,b\:=\:1,\,2}\Bigg[\frac{1}{8}\,\parbox[][][c]{2.2cm}{
  \includegraphics{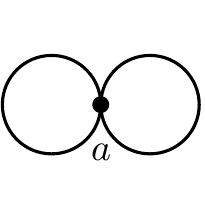}}\delta_{ab}\,\:+\:
  \frac{1}{12}\parbox[][][c]{1.9cm}{\includegraphics{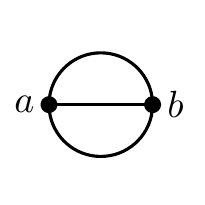}}\,
  \Bigg]\; ,
\vspace{-1em}
\end{equation}
where  combinatorial  factors  have  been written  explicitly  and  we
associate with each $n$-point vertex a factor of
\begin{equation}
  iS^{(n)}_{a}(\widehat{\Phi}^a;x)\ = \ i\frac{\delta^n S[\widehat{\Phi}^a]}
  {\delta \big(\widehat{\Phi}^a(x)\big)^n}\; ,
\end{equation}
and          each          line          a          factor          of
$i\mathcal{G}^{ab}(\widehat{\Phi}^a;x,y)$.  The three-  and four-point
vertices are
\begin{equation}  
  iS_a^{(3)}(\widehat{\Phi}^a;x)\ =\ -ig\eta_{aaa}\:-\:
  i\lambda\eta_{aaaa}\widehat{\Phi}^a(x)\;, \qquad
  iS_a^{(4)}(\widehat{\Phi}^a;x)\ =\ -i\lambda\eta_{aaaa}\; .
\end{equation}

Upon    functional    differentiation    of    the    CJT    effective
action~(\ref{eq:CJTaction})  with respect  to $\mathcal{G}^{ab}(x,y)$,
we obtain by virtue of (\ref{eq:dGab}) the Schwinger--Dyson equation
\begin{equation}
  \label{eq:consd}
  \mathcal{G}^{-1}_{ab}(\widehat{\Phi}^a;x,y)\ =\ G_{ab}^{-1}(\widehat{\Phi}^a;x,y)
  \:+\:\Pi_{ab}(\widehat{\Phi}^a,\mathcal{G}^{ab};x,y)\; ,
\end{equation}
where
\begin{equation}
  \Pi_{ab}(\widehat{\Phi}^a,\mathcal{G}^{ab};x,y)\ =\ -2i\hbar\frac{\delta
  \Gamma_2[\widehat{\Phi}^a,\mathcal{G}^{ab}]}{\delta\mathcal{G}^{ab}(x,y)}\ =\ 
  -i\hbar\Bigg[\:\parbox[][][t]{3.2cm}{
  \includegraphics{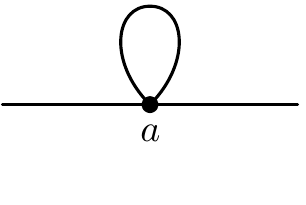}}\delta_{ab}\:+\:
  \,\parbox[][][c]{3.2cm}{\includegraphics{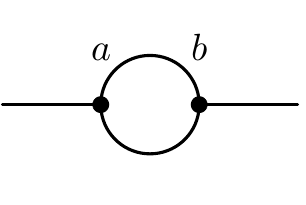}}\Bigg]
\end{equation}
is the one-loop truncated  CTP self-energy.  A combinatorial factor of
$\frac{1}{2}$ has been absorbed into the diagrammatics.

Suppressing  the $\widehat{\Phi}^a$  and  $\mathcal{G}^{ab}$ arguments
for notational convenience, the CTP self-energy $\Pi_{ab}(x,y)$ may be
written in matrix form as
\begin{equation}
  \Pi_{ab}(x,y)\ =\ 
  \begin{bmatrix}
    \Pi(x,y) & -\Pi_<(x,y) \\
    -\Pi_>(x,y) & -\Pi^{*}(x,y)
  \end{bmatrix}\!\; ,
\end{equation}
where $\Pi(x,y)$ and $-\Pi^*(x,y)$ are the time- and anti-time-ordered
self-energies; and $\Pi_>(x,y)$ and $\Pi_<(x,y)$ are the positive- and
negative-frequency absolutely-ordered  self-energies, respectively. In
analogy    to     the    propagator    definitions     discussed    in
Section~\ref{sec:canon} and Appendix~\ref{app:rel}, we also define the
self-energy functions
\begin{subequations}
\begin{align}
  \label{eq:hadself}
  \Pi_1(x,y)&\ =\ \Pi_>(x,y)\:+\:\Pi_<(x,y)\ =\
  \Pi(x,y)\:-\: \Pi^*(x,y)\ =\ 2i\,\mathrm{Im}\,\Pi(x,y)\; ,\\
  \label{eq:pself}
  \Pi_{\mathcal{P}}(x,y)&\ =\ \frac{1}{2}
  \Big(\,\Pi_{\mathrm{R}}(x,y)\:+\:\Pi_{\mathrm{A}}(x,y)\,\Big)
  \ = \ \mathrm{Re}\,\Pi(x,y)\; ,\\
  \label{eq:BreitWco}
  2iM\Gamma(x,y)&\ =\ \Pi_>(x,y)\: -\: \Pi_<(x,y)\ =\ 
  \Pi_{\mathrm{R}}(x,y)\: -\: \Pi_{\mathrm{A}}(x,y)\ =\
  2i\,\mathrm{Im}\,\Pi_{\mathrm{R}}(x,y)\; ,
\end{align}
\end{subequations}
which   satisfy   relations    analogous   to   those   described   in
Appendix~\ref{app:rel}.    $\Gamma(x,y)$  in   (\ref{eq:BreitWco})  is
related  to  the usual  Breit--Wigner  width  in  the equilibrium  and
zero-temperature      limits.      The      Keldysh     representation
[see~(\ref{eq:physrep})]   $\widetilde{\Pi}_{ab}(x,y)$   of  the   CTP
self-energy reads:
\begin{equation}
  \widetilde{\Pi}_{ab}(x,y)\ = \
  \begin{bmatrix}
    \Pi_1(x,y) & \Pi_{\mathrm{R}}(x,y) \\
    \Pi_{\mathrm{A}}(x,y) & 0
  \end{bmatrix}\!\; .
\end{equation}

In  the  limit  $\widehat{\Phi}^a(x)\:\to\: 0$,  the  Schwinger--Dyson
equation (\ref{eq:consd}) reduces to
\begin{equation}
  \label{eq:sd}
  \Delta^{-1}_{ab}(x,y,\tilde{t}_f;\tilde{t}_i)\ =\ \Delta_{ab}^{0,\,-1}(x,y)
  \:+\:K_{ab}(x,y,\tilde{t}_f;\tilde{t}_i)\:
  +\:\Pi_{ab}(x,y,\tilde{t}_f;\tilde{t}_i)\; ,
\end{equation}
in  which $\Delta^{-1}_{ab}(x,y,\tilde{t}_f;\tilde{t}_i) \:  \equiv \:
\mathcal{G}_{ab}^{-1}(\widehat{\Phi}^a=0;x,y,\tilde{t}_f;\tilde{t}_i)$
and  $\Delta_{ab}^{0,\,-1}(x,y)$ is  the free  inverse  CTP propagator
defined in  (\ref{eq:freeinv}).  We have  re-introduced the dependence
upon $\tilde{t}_f$  and $\tilde{t}_i$ for clarity. Notice  that due to
the   explicit  $\tilde{t}_f$ dependence   of   the  bi-local   source
$K_{ab}(x,y,\tilde{t}_f;\tilde{t}_i)$ in  (\ref{eq:kmom}), the inverse
resummed                         CTP                        propagator
${\Delta^{-1}_{ab}(x,y,\tilde{t}_f;\tilde{t}_i)}$    and    the    CTP
self-energy  ${\Pi_{ab}(x,y,\tilde{t}_f;\tilde{t}_i)}$  are  not  time
translationally invariant quantities.

In   order   to   develop   a   self-consistent   inversion   of   the
Schwinger--Dyson  equation  in   (\ref{eq:sd}),  the  bi-local  source
$K_{ab}(x,y,\tilde{t}_f;\tilde{t}_i)$  is  absorbed  into  an  inverse
non-homogeneous CTP propagator
\begin{equation}
  \label{eq:treesd}
  D^{0,\,-1}_{ab}(x,y,\tilde{t}_f;\tilde{t}_i)\ = \
  \Delta_{ab}^{0,\,-1}(x,y)\:+\:K_{ab}(x,y,\tilde{t}_f;\tilde{t}_i),
\end{equation}
whose  inverse,  to  leading  order  in  $K_{ab}$,  is  the  free  CTP
propagator $\Delta^{0,\,ab}(x,y,\tilde{t}_f;\tilde{t}_i)$, i.e.
\begin{equation}
  D^{0,\,ab}(x,y,\tilde{t}_f;\tilde{t}_i)\ \equiv\ 
  \Delta^{0,\,ab}(x,y,\tilde{t}_f;\tilde{t}_i)\:+\:\mathcal{O}(K^2)\;,
\end{equation}
as   we   will  illustrate   in   Sections  \ref{sec:nonhomprop}   and
\ref{sec:eq}. The contribution of  the bi-local source $K_{ab}$ is now
absorbed        into       the        free        CTP       propagator
$\Delta^{0,\,ab}(x,y,\tilde{t}_f;\tilde{t}_i)$,       whose       time
translational invariance is broken  as a result.  The Schwinger--Dyson
equation  (\ref{eq:sd}) may  then be  written in  the  double momentum
representation as
\begin{equation}
  \label{eq:sddmom}
  \Delta^{-1}_{ab}(p,p',\tilde{t}_f;\tilde{t}_i)\ =\ 
  \Delta_{ab}^{0,\,-1}(p,p')\: +\: \Pi_{ab}(p,p',\tilde{t}_f;\tilde{t}_i)\; .
\end{equation}

Since the stationary vacuum~$\ket{0}$ has been replaced by the density
operator~$\rho$ at the microscopic time $\tilde{t}_f\:=\:t/2$, we must
consider the following field-particle duality relation in the Wick
contraction of interaction-picture fields:
\begin{align}
  \braket{0|\Phi(x;\tilde{t}_i)
    a^{\dag}(\mathbf{k},\tilde{t}_f;\tilde{t}_i)|0}&\ =\ e^{-ik\cdot x}
  e^{iE(\mathbf{k})\tilde{t}_f}\;.
\end{align}
Here, the extra  phase~$e^{iE(\mathbf{k})\tilde{t}_f}$ arises from the
fact   that   the  creation   and   annihilation   operators  of   the
interaction-picture field~$\Phi (x;\tilde{t}_i)$  are evaluated at the
microscopic    time~$\tilde{t}\:=\:0$    [cf.~(\ref{eq:fourier})   and
  (\ref{eq:momcomrel})],                  whereas                  the
operator~$a^{\dag}(\mathbf{k},\tilde{t}_f;\tilde{t}_i)$,      resulting
from  the  expansion  of  the  density operator  $\rho$  (see  Section
\ref{sec:nonhomprop}),    is     evaluated    at    the    microscopic
time~$\tilde{t}_f\:=\:t/2$.  Analytically  continuing this extra phase
to  off-shell energies  and  in consistency  with (\ref{eq:kmom}),  we
associate      with     each      external      vertex     of      the
self-energy~$\Pi_{ab}(p,p',\tilde{t}_f;\tilde{t}_i)$                 in
(\ref{eq:sddmom}) a phase:
\begin{equation}
e^{ip_0 \tilde{t}_f}\,,
\end{equation}
where $p_0$ is the energy flowing \emph{into} the vertex. This amounts
to the absorption of an overall phase
\begin{equation}
e^{i(p_0\:-\:p_0')\tilde{t}_f}
\end{equation}
into            the            definition            of            the
self-energy~$\Pi_{ab}(p,p',\tilde{t}_f;\tilde{t}_i)$.

Convoluting  from left  and right  on both  sides of~(\ref{eq:sddmom})
first with the weight function $(2\pi)^4\delta_t^{(4)}(p\:-\:p')$ from
(\ref{eq:deltat})              and              then              with
$\Delta^{0,\,ab}(p,p',\tilde{t}_f;\tilde{t}_i)$                     and
$\Delta^{ab}(p,p',\tilde{t}_f;\tilde{t}_i)$,  respectively,  we obtain
the Feynman--Dyson series
\begin{align}
  \Delta^{ab}(p,p',\tilde{t}_f;\tilde{t}_i)\ &=\
  \Delta^{0,\,ab}(p,p',\tilde{t}_f;\tilde{t}_i)\:-\:\!\idotsint\!\!
  \frac{\D{4}{q}}{(2\pi)^4}\,\frac{\D{4}{q'}}{(2\pi)^4}\,
  \frac{\D{4}{q''}}{(2\pi)^4}\,\frac{\D{4}{q'''}}{(2\pi)^4}
  \nonumber\\ &\qquad \times \: \Delta^{0,\,ac}(p,q,\tilde{t}_f;\tilde{t}_i) 
  (2\pi)^4\delta_t^{(4)}(q\:-\:q')
  \Pi_{cd}(q',q'',\tilde{t}_f;\tilde{t}_i)
  (2\pi)^4\delta_t^{(4)}(q''\:-\:q''')
  \Delta^{db}(q''',p',\tilde{t}_f;\tilde{t}_i)\;,
\end{align}
where  $\delta_t^{(4)}(p\:-\:p')$  is  defined in  (\ref{eq:deltat4}).
Because of the  form of $\delta_t(p_0\:-\:p_0')$ in (\ref{eq:deltat}),
we see that this series does  not collapse to an algebraic equation of
resummation, as known from  zero-temperature field theory.  As we will
see  in  Section~\ref{sec:gradapp}, one  cannot  write  down a  closed
analytic     form     for      the     resummed     CTP     propagator
$\Delta^{ab}(p,p',\tilde{t}_f;\tilde{t}_i)$,     except     in     the
thermodynamic equilibrium limit, see Section~\ref{sec:eq}.

Given that $\delta_t$  satisfies the convolution in (\ref{eq:dciden}),
the weight functions may be absorbed into the external vertices of the
self-energy~$\Pi_{ab}(p,p',\tilde{t}_f;\tilde{t}_i)$;               see
Section~\ref{sec:toy}.  The Feynman--Dyson  series may then be written
in the more concise form
\begin{equation}
  \label{eq:fd}
  \Delta^{ab}(p,p',\tilde{t}_f;\tilde{t}_i)\ =\
  \Delta^{0,\,ab}(p,p',\tilde{t}_f;\tilde{t}_i)\:-\:\!\iint\!\!
  \frac{\D{4}{q}}{(2\pi)^4}\,\frac{\D{4}{q'}}{(2\pi)^4}\;
  \Delta^{0,\,ac}(p,q,\tilde{t}_f;\tilde{t}_i)
  \Pi_{cd}(q,q',\tilde{t}_f;\tilde{t}_i)
  \Delta^{db}(q',p',\tilde{t}_f;\tilde{t}_i)\;.
\end{equation}
Note that for finite $t$, $\delta_t(p_0\:-\:p_0')$ is analytic for all
$p_0$, including $p_0\: =\: p_0'$.  As we shall see in
Section~\ref{sec:oneloop}, the systematic incorporation of these
finite-time effects ensures that the perturbation expansion is free of
pinch singularities.

\subsection{Applicability of the Gradient Expansion}
\label{sec:gradapp}

Here,   we  will   look   more  closely   at   the  inverse   relation
(\ref{eq:resinvdef}) that  determines the resummed  CTP propagator. We
will show  that a full matrix  inversion may only be  performed in the
thermo\-dynamic   equilibrium  limit.    Hence,  the   application  of
truncated  gradient  expansions  and  the use  of  partially  resummed
quasi-particle  propagators,  particularly  for  early  times,  become
questionable in out-of-equilibrium systems.

We define the relative and central coordinates
\begin{equation}
  \label{eq:relcen}
  R_{xy}^{\mu}\ =\ x^{\mu}\:-\: y^{\mu}\; , \qquad
  X_{xy}^{\mu}\ =\ \frac{x^{\mu}\:+\:y^{\mu}}{2}\; ,
\end{equation}
such that
\begin{equation}
  x^{\mu}\ =\ X_{xy}^{\mu}\:+\:\frac{1}{2}R_{xy}^{\mu}\; ,\qquad
  y^{\mu}\ =\ X^{\mu}_{xy}\:-\:\frac{1}{2}R_{xy}^{\mu}\; .
\end{equation}
We  then introduce  the Wigner  transform  (see \cite{Winter:1986da}),
namely   the  Fourier   transform   with  respect   to  the   relative
coordinate~$R^\mu_{xy}$ only.   Explicitly, the Wigner  transform of a
function~$F(R,X)$ is
\begin{equation}
  F(p,X)\ =\ \!\int\!\D{4}{R}\;e^{i p\cdot R}\,F(R,X)\; .
\end{equation}

The  resummed CTP propagator  $\Delta^{ab}(x,y)$ respects  the inverse
relation   in    (\ref{eq:resinvdef}).    Here,   we    suppress   the
$\tilde{t}_{f}$ and  $\tilde{t}_i$ dependence of the  propagators for
notational  convenience.   Inserting  into  (\ref{eq:resinvdef}),  the
Wigner   transforms  of   the  resummed   and  inverse   resummed  CTP
propagators, the inverse relation takes the form
\begin{equation}
  \int_{\Omega_t}\!\D{4}{z}\iint\!\!\frac{\D{4}{p}}{(2\pi)^4}\,
  \frac{\D{4}{p'}}{(2\pi)^4}\,e^{-ip\cdot R_{xz}}\,e^{-ip'\cdot R_{zy}}\,
  \Delta_{ab}^{-1}(p,X_{xz})\Delta^{bc}(p',X_{zy})\ =\ \iint\!\!
  \frac{\D{4}{p}}{(2\pi)^4}\,\frac{\D{4}{p'}}{(2\pi)^4}\,
  e^{-ip\cdot x}\,e^{i p'\cdot y}\,\eta_{a}^{\;c}(2\pi)^4
  \delta^{(4)}_t(p\: -\: p')\; ,
\end{equation}
where the $z^0$ domain of integration is restricted to be in the range
$[-t/2,\ t/2]$.

In the case where deviations from homogeneity are small, i.e.~when the
characteristic scale  of macroscopic  variations in the  background is
large  in  comparison  to  that  of  the  microscopic  single-particle
excitations,  we  may perform  a  gradient  expansion  of the  inverse
relation  in terms  of the  soft derivative~$\partial_{X_{xy}}^{\mu}\:
\equiv\:        \partial/\partial        X_{xy,\,\mu}$.        Writing
$X_{xz}\:=\:X_{xy}\:+\:R_{zy}/2$  and $X_{zy}\:=\:X_{xy}\:-\:R_{xz}/2$
and after integrating by parts, we obtain
\begin{align}
  \label{eq:winv}
  &\iint\!\!\frac{\D{4}{p}}{(2\pi)^4}\,\frac{\D{4}{p'}}{(2\pi)^4}
  \,e^{-ip\cdot   x}\,e^{i   p'\cdot   y}\,(2\pi)^4\delta_t^{(4)}(p-p')
  \Big\{\Delta_{ab}^{-1}(p,X)\exp\!\Big[-\frac{i}{2}
  \big(\overleftarrow{\partial_{p}}\cdot\overrightarrow{\partial_X}
  \:-\:\overleftarrow{\partial_X}\cdot\overrightarrow{\partial_{p'}}
  \big)\Big]\Delta^{bc}(p',X)\Big\}\nonumber \\&\qquad \qquad
  =\ \!\iint\!\!\frac{\D{4}{p}}{(2\pi)^4}\,
  \frac{\D{4}{p'}}{(2\pi)^4}\,e^{-ip\cdot x}\,e^{ip'\cdot y}\,
  \eta_{a}^{\;c}(2\pi)^4\delta^{(4)}_t(p\:-\:p')\; ,
\end{align}
where $X\:\equiv\: X_{xy}$ and the derivatives act only within the curly brackets.

We now define the central and relative momenta
\begin{equation}
  q^{\mu}\ =\ \frac{p^{\mu}\:+\:p'^{\mu}}{2}\;, \qquad
  Q^{\mu}\ =\ p^{\mu}\:-\:p'^{\mu}\; ,
\end{equation}
which  are  the  Fourier   conjugates  to  the  relative  and  central
coordinates, $R^\mu$ and $X^\mu$, respectively. It follows that
\begin{equation}
  p^{\mu}\ =\ q^{\mu}\: +\: \frac{1}{2}Q^{\mu}\; ,\qquad
  p'^{\mu}\ =\ q^{\mu}\: -\: \frac{1}{2}Q^{\mu}\; .
\end{equation}
We may then Fourier transform (\ref{eq:winv}) with respect to $R_{xy}$
to obtain
\begin{align}
  \label{eq:GradI}
  &\int\!\!\frac{\D{4}{Q}}{(2\pi)^4}\;e^{-iQ\cdot X}\,(2\pi)^4
  \delta_t^{(4)}(Q)\exp\!
  \Big[\,-i\,\Big(\,\diamondsuit_{q,X}^-\:+\:2\diamondsuit^+_{Q,X}\,\Big)\Big]
  \{\Delta_{ab}^{-1}(q+\tfrac{Q}{2},X)\}\{\Delta^{bc}(q-\tfrac{Q}{2},X)\}\nonumber\\& \qquad \qquad
  =\ \int\!\!\frac{\D{4}{Q}}{(2\pi)^4}\,e^{-iQ\cdot X}\,\eta_{a}^{\;c}
  (2\pi)^4\delta_t^{(4)}(Q)\; ,
\end{align}
where, following~\cite{Cassing:1999wx, Prokopec:2003pj,
  Prokopec:2004ic}, we have introduced the diamond operator
\begin{equation}
  \label{eq:diam}
  \diamondsuit_{p,X}^{\pm}\{A\}\{B\}\ =\ \frac{1}{2}\{A,\ B\}^{\pm}_{p,X}
\end{equation}
and  $\{A,\ B\}^{\pm}_{p,X}$  denote the  symmetric  and anti-symmetric
Poisson brackets
\begin{equation}
  \label{eq:pois}
  \{A,\ B\}^{\pm}_{p,X}\ \equiv\ \frac{\partial A}{\partial p^{\mu}}
  \,\frac{\partial B}{\partial X_{\mu}}\:\pm\:
  \frac{\partial A}{\partial X^{\mu}}\,
  \frac{\partial B}{\partial p_{\mu}}\; .
\end{equation}
For $t>0$, we may perform the integral on the RHS of~(\ref{eq:GradI}),
yielding
\begin{equation}
  \label{eq:gradinv}
  \int\!\!\frac{\D{4}{Q}}{(2\pi)^4}\;e^{-iQ\cdot X}\,(2\pi)^4
  \delta_t^{(4)}(Q)\exp\!
  \Big[-i\,\Big(\,\diamondsuit_{q,X}^-\:+\:2\diamondsuit^+_{Q,X}\,\Big)\Big]
  \{\Delta_{ab}^{-1}(q+\tfrac{Q}{2},X)\}\{\Delta^{bc}(q-\tfrac{Q}{2},X)\}\ =\
  \eta_a^{\;c}\theta(t-2|X_0|)\; ,
\end{equation}

In   the   above   expressions,   of   particular   concern   is   the
$\diamondsuit^{+}_{Q,X}$ operator, where the relative momentum $Q$ and
the central coordinate $X$ are  conjugate to one another. Thus, if the
derivatives  with respect to  $X$ are  assumed to  be small,  then the
derivatives  with respect to  $Q$ must  be large.   In this  case, all
orders  of  the  gradient  expansion  may be  significant,  so  it  is
inappropriate  to truncate  to a  given order  in the  soft derivative
$\partial_X^{\mu}$.

As  $t\:\to\:\infty$,  we have  the  transition $\delta^{(4)}_t  (Q)\:
\to\: \delta^{(4)} (Q)$ and (\ref{eq:gradinv}) reduces to
\begin{equation}
  \label{eq:gradexp}
  e^{-i\diamondsuit^-_{q,X}}\{\Delta^{-1}_{ab}(q,X)\}\{\Delta^{bc}(q,X)\}
  \ =\ \eta_{a}^{\;c}\; .
\end{equation}
Even for these late times, we can perform the matrix inversion exactly
only if  we truncate  the gradient expansion  in~(\ref{eq:gradexp}) to
\emph{zeroth} order.   However, such  a truncation appears  valid only
for   time-independent  {\em   and}  spatially   homogeneous  systems.
Employing  a  suitable  quasi-particle  approximation  to  the  Wigner
representation     of      the     propagators,     it      can     be
shown~\cite{Bornath:1996zz}  that this inversion  may be  performed at
first   order   in  the   gradient   expansion.   However,   off-shell
contributions are not fully accounted for in such an approximation.

In conclusion, a closed analytic  form for the resummed CTP propagator
may only be obtained in the time-independent thermodynamic equilibrium
limit.  The  truncation of the  gradient expansion may  be justifiable
only to the late-time evolution  of systems very close to equilibrium,
even  for  spatially   homogeneous  thermal  backgrounds.   A  similar
conclusion is drawn from different arguments in~\cite{Berges:2005md}.

\subsection{Non-Homogeneous Free Propagators}
\label{sec:nonhomprop}

Unlike the  resummed CTP  propagator, the free  CTP propagator  can be
derived  analytically, even in  the presence  of a  time-dependent and
spatially   inhomogeneous   background.    The  non-homogeneous   free
propagator  will account  explicitly for  the violation  of space-time
translational invariance. Our derivation  relies on the algebra of the
canonical  quantization  commutators   of  creation  and  annihilation
operators described in Section~\ref{sec:canon}.  Subsequently, we make
connection of our results with the path-integral representation of the
CTP generating functional in~(\ref{eq:chgen}). Finally, we introduce a
diagrammatic   representation  for   the   non-homogeneous  free   CTP
propagator.

We note that the derived propagators are `free' in the sense that
their spectral structure is that of single-particle states,
corresponding to the free part of the action (see
Section~\ref{sec:CTP}).  Their statistical structure, on the other
hand, will turn out to contain a summation over contributions from all
possible multi-particle states.  The time-dependent statistical
distribution function appearing in these propagators is therefore a
statistically-dressed object.  This subtle point is significant for
the consistent definition of the number density in
Section~\ref{sec:num}, the derivation of the master time evolution
equations in Section~\ref{sec:eom} and the absence of pinch singularities, described in Section~\ref{sec:oneloop}.

The starting point of our canonical derivation is the explicit form of
the density operator~$\rho$.  We  relax any assumptions about the form
of the density operator and take  it to be in general non-diagonal but
Hermitian  within the  general  Fock  space.  We  may  write the  most
general interaction-picture  density operator at  the microscopic time
$\tilde{t}_f\:=\:t/2$ as
\begin{align}
  \label{eq:gendens}
  \rho(\tilde{t}_f;\tilde{t}_i)\ &=\
  C\exp\!\bigg[-\!\int\!\D{}{\Pi_{\mathbf{k}_1}}
  \;W_{10}(\mathbf{k}_1:0)a^{\dag}(\mathbf{k}_1,\tilde{t}_f)\:-\:
  \!\int\!\D{}{\Pi_{\mathbf{k}_1'}}\;W_{01}(0:\mathbf{k}_1')
  a(\mathbf{k}_1',\tilde{t}_f)\nonumber \\&
  \qquad -\: \!\iint\!\D{}{\Pi_{\mathbf{k}_1}}\,\D{}{\Pi_{\mathbf{k}_1'}}\;
  W_{11}(\mathbf{k}_1:\mathbf{k}_1')a^{\dag}(\mathbf{k}_1,\tilde{t}_f)
  a(\mathbf{k}_1',\tilde{t}_f)\:-\:\cdots\nonumber\\&
  \qquad -\: \!\frac{1}{n!}\frac{1}{m!}\!\idotsint\!\Bigg(\prod_{i=1}^n
  \D{}{\Pi_{\mathbf{k}_i}}\bigg)\bigg(\prod_{j=1}^m\D{}{\Pi_{\mathbf{k}_j'}}\bigg)
  \;W_{nm}(\{\mathbf{k}_i\}:\{\mathbf{k}_j'\})
  \prod_{i=1}^na^{\dag}(\mathbf{k}_i,\tilde{t}_f)
  \prod_{j=1}^ma(\mathbf{k}_j',\tilde{t}_f)\bigg]\; ,
\end{align}
where the constant $C$ can be set to unity without loss of generality.
The                       complex-valued                       weights
$W_{nm}(\{\mathbf{k}\}_n;\{\mathbf{k}'\}_m)$  depend on  the  state of
the system  at time~$\tilde{t}_f\:=\:t/2$ and  satisfy the Hermiticity
constraint:
\begin{equation}
  W_{nm}(\{\mathbf{k}\}_n:\{\mathbf{k}'\}_m)\ =\ 
  W_{mn}^*(\{\mathbf{k}'\}_m:\{\mathbf{k}\}_n)\; .
\end{equation}
The density  operator $\rho$ may be  written in the  basis of momentum
eigenstates by multiplying  the exponential form in (\ref{eq:gendens})
by  the  completeness  relation  of   the  basis  of  Fock  states  at
time~$\tilde{t}_f\:=\:t/2$:
\begin{equation}
  \mathbf{I}\ =\ \ket{0}\!\bra{0}\:+\:\sum_{\ell\:=\:1}^{\infty}
  \frac{1}{\ell!}\bigg(\prod_{k=1}^{\ell}\int\!\D{}{\Pi_{\mathbf{p}_k}}\bigg)
  \!\ket{\{\mathbf{p}\}_{\ell},\tilde{t}_f}\!
  \bra{\{\mathbf{p}\}_{\ell},\tilde{t}_f}\; ,
\end{equation}
where $\ket{\{\mathbf{p}\}_{\ell},\tilde{t}_f}$ is the multi-mode Fock
state
$\ket{\mathbf{p}_1,\tilde{t}_f}\:\otimes\:\ket{\mathbf{p}_2,\tilde{t}_f}
\:\otimes\:      \cdots\:\otimes\:\ket{\mathbf{p}_{\ell},\tilde{t}_f}$.
This    usually   gives    an   intractable    infinite    series   of
$n$-to-$m$-particle     correlations.       Taking     all     weights
$W_{nm}(\{\mathbf{k}\}_n:\{\mathbf{k}\}_m)$    to     be    zero    if
$n\:+\:m\:>\:2$, i.e.~taking  a Gaussian-like density  operator, it is
still   possible   to   generate  all   possible   $n$-to-$m$-particle
correlations.   In Appendix~\ref{app:dens}, we  give the  expansion of
the  general  Gaussian-like density  operator,  where only  sufficient
terms are included to help us visualize its analytic form.

We may account for our ignorance of the series expansion of the
density operator by defining the following bilinear EEVs of
interaction-picture creation and annihilation operators as
\begin{subequations}
  \label{eq:fsgs}
\begin{align}
  \label{eq:fs}
  \braket{a^{\dag}(\mathbf{p}',\tilde{t}_f;\tilde{t}_i)
  a(\mathbf{p},\tilde{t}_f;\tilde{t}_i)}_t&\ =\
  2\,\mathscr{E}(\mathbf{p},\mathbf{p}')f^0(\mathbf{p},\mathbf{p}',t)\;,
  \\ \label{eq:gs}
  \braket{a(\mathbf{p}',\tilde{t}_f;\tilde{t}_i)
  a(\mathbf{p},\tilde{t}_f;\tilde{t}_i)}_t&\ =\ 
  2\,\mathscr{E}(\mathbf{p},\mathbf{p}')g^0(\mathbf{p},\mathbf{p}',t)\;,
\end{align}
\end{subequations}
consistent with the commutation relations in (\ref{eq:momcomrel}).
The energy factor $2\mathscr{E}(\mathbf{p},\mathbf{p}')$, having
dimensions $\mathrm{E}^1$, arises from the fact that the `number
operator'
$a^{\dag}(\mathbf{p},\tilde{t};\tilde{t}_i)a(\mathbf{p},\tilde{t};
\tilde{t}_i)$ of quantum field theory has dimensions
$\mathrm{E}^{-2}$, i.e.~it does \emph{not} have the dimensions of a
number.  Bearing in mind that the density operator is constructed from
on-shell Fock states, a natural ansatz for this energy factor is
\begin{equation}
  \label{eq:efact}
  \mathscr{E}(\mathbf{p},\mathbf{p}')
  \ =\ \sqrt{E(\mathbf{p})E(\mathbf{p}')}\; .
\end{equation}
The complex-valued distributions $f^0$ and $g^0$ have dimensions
$\mathrm{E}^{-3}$ and satisfy the identities:
\begin{subequations}
  \label{eq:restrict}
  \begin{align}
    \label{eq:restrictf}
    f^0(\mathbf{p},\mathbf{p}',t)&\ =\ f^{0*}(\mathbf{p}',\mathbf{p},t)\;,\\
    g^0(\mathbf{p},\mathbf{p}',t)&\ =\ g^0(\mathbf{p}',\mathbf{p},t)\;.
  \end{align}
\end{subequations}  
We refer to $f$ and $g$ as \emph{statistical distribution functions}.
In particular, we interpret the Wigner transform
\begin{equation}
  \label{eq:fnumd}
  n^0(\mathbf{q},\mathbf{X},t)\ =\
  \int\!\frac{\D{3}{\mathbf{Q}}}{(2\pi)^3}\;e^{i\mathbf{Q}\cdot\mathbf{X}}
  f^0(\mathbf{q}+\tfrac{\mathbf{Q}}{2},
  \mathbf{q}-\tfrac{\mathbf{Q}}{2},t)
\end{equation}
as the number density of \emph{spectrally-free} particles at
macroscopic time~$t$ in the phase-space hypervolume between
$\mathbf{q}$ and $\mathbf{q}\:+\:\D{}{\mathbf{q}}$ and $\mathbf{X}$
and $\mathbf{X}\:+\:\D{}{\mathbf{X}}$.  Notice that
$n^0(\mathbf{q},\mathbf{X},t)$ is real thanks to the Hermiticity
constraint~(\ref{eq:restrictf}). Hereafter, except where it is
necessary to make the distinction, we will omit the superscript $0$ on
the spectrally-free statistical distribution functions for notational
convenience.

The         EEV        of         the         two-point        product
$\braket{a(\mathbf{p},\tilde{t}_f;\tilde{t}_i)
  a^{\dag}(\mathbf{p}',\tilde{t}_f;\tilde{t}_i)}_{t}$ follows from the
definition  (\ref{eq:fs}) and  the canonical  commutation  relation in
(\ref{eq:momcomrel}), giving
\begin{equation}
  \label{eq:fs2}
  \braket{a(\mathbf{p},\tilde{t}_f;\tilde{t}_i)
    a^{\dag}(\mathbf{p}',\tilde{t}_f;\tilde{t}_i)}_{t}\ = \
  (2\pi)^32E(\mathbf{p})\delta^{(3)}(\mathbf{p}\:-\:\mathbf{p}')\:+\:
  2\,\mathscr{E}(\mathbf{p},\mathbf{p}')f(\mathbf{p},\mathbf{p}',t)\;.
\end{equation}
Hermitian conjugation of (\ref{eq:gs}) yields
\begin{equation}
  \label{eq:gs2}
  \braket{a^{\dag}(\mathbf{p},\tilde{t}_f;\tilde{t}_i)
    a^{\dag}(\mathbf{p}',\tilde{t}_f;\tilde{t}_i)}_t\ =\
  2\,\mathscr{E}(\mathbf{p},\mathbf{p}')g^*(\mathbf{p},\mathbf{p}',t)\;.
\end{equation}
Note  that  (\ref{eq:fsgs}), (\ref{eq:fs2})  and
(\ref{eq:gs2}) are consistent with the canonical quantization rules in
(\ref{eq:caus}) and (\ref{eq:comrel}).

When the linear  terms in the exponent of  the density operator $\rho$
in (\ref{eq:gendens}) are non-zero, we may consider the EEVs of single
creation or annihilation operators
\begin{equation}
  \label{eq:onepointdis}
  \braket{a(\mathbf{p},\tilde{t}_f;\tilde{t}_i)}_{t}\ =\
  \sqrt{2E(\mathbf{p})}w(\mathbf{p},t)\; .
\end{equation}
In this case, we may define the connected distribution functions
\begin{subequations}
  \label{eq:condis}
  \begin{align}
    f_{\mathrm{con}}(\mathbf{p},\mathbf{p}',t)&\ \equiv\ 
    f(\mathbf{p},\mathbf{p}',t)\:-\:w(\mathbf{p},t)w(\mathbf{p}',t)\;,
    \\ g_{\mathrm{con}}(\mathbf{p},\mathbf{p}',t)&\ \equiv\ 
    g(\mathbf{p},\mathbf{p}',t)\:-\:w(\mathbf{p},t)w(\mathbf{p}',t)\;,
  \end{align}
\end{subequations}
which obey the same symmetry properties given in (\ref{eq:restrict}).

We are now in a position to derive the most general form of the double
momentum  representation of the  non-homogeneous free  CTP propagator,
satisfying  the inverse  relation~(\ref{eq:dminv}).  Proceeding as  in
Section~\ref{sec:CTPprop}, we  make the following ansatz  for the most
general solution of the  Klein--Gordon equation in the double momentum
representation:
\begin{multline}
   \label{eq:ansatz}
   \Delta^{0,\,ab}(p,p',\tilde{t}_f;\tilde{t}_i)\ =\
   \begin{bmatrix}
     \big(p^2\:-\:M^2\:+\:i\epsilon\big)^{-1}  &
     -i2\pi\theta(-p_0)\delta(p^2\:-\:M^2)  \\
     -i2\pi\theta(p_0)\delta(p^2\:-\:M^2) &
     -\big(p^2\:-\:M^2\:-\:i\epsilon\big)^{-1}
   \end{bmatrix}\!(2\pi)^4\delta^{(4)}(p\:-\:p') \\
   -i2\pi|2p_0|^{1/2}\delta(p^2\:-\:M^2)\tilde{f}(p,p',t)
   e^{i(p_0\:-\:p_0')\tilde{t}_f}2\pi|2p_0'|^{1/2}\delta(p'^2\:-\:M^2)
   \!\begin{bmatrix}
     1 & 1 \\
     1 & 1
   \end{bmatrix}\!\;,
\end{multline}
which we confirm by evaluating the EEVs directly, using the algebra of
(\ref{eq:fsgs}).

In   (\ref{eq:ansatz})  the   phase factor $e^{i(p_0\:-\:p_0')\tilde{t}_f}$
arises  from the  fact that  the creation  and  annihilation operators
appearing  in the  Fourier transform  of the  field operator  given in
(\ref{eq:fourier})  are evaluated at  the time  $\tilde{t}\:=\:0$. The
density  operator,  on  the  other  hand, is  evaluated  at  the  time
$\tilde{t}_f$.  As  a consequence,  in the evaluation  of the  EEV, we
have, for instance,
\begin{equation}
  \braket{a^{\dag}(\mathbf{p}',0;\tilde{t}_i)
  a(\mathbf{p},0;\tilde{t}_i)}_t\ =\
  2\,\mathscr{E}(\mathbf{p},\mathbf{p}')f(\mathbf{p},\mathbf{p}',t)
  e^{i[E(\mathbf{p})\:-\:E(\mathbf{p}')]\tilde{t}_f}\;,
\end{equation}
which directly results from (\ref{eq:fs}).

The form of the function $\tilde{f}(p,p',t)$ is
\begin{equation}
  \label{eq:dftilde}
  \tilde{f}(p,p',t)\ =\ \theta(p_0)\theta(p_0')f(\mathbf{p},\mathbf{p}',t)
  \:+\:\theta(-p_0)\theta(-p_0')f^*(-\mathbf{p},-\mathbf{p}',t)
  \:+\:\theta(p_0)\theta(-p_0')g(\mathbf{p},-\mathbf{p}',t)
  \:+\:\theta(-p_0)\theta(p_0')g^*(-\mathbf{p},\mathbf{p}',t)\;.
\end{equation}
The     function      $\tilde{f}$     satisfies     the     relations:
$\tilde{f}(p,p',t)\:=\:\tilde{f}(-p',-p,t)\:=\:\tilde{f}^*(-p,-p',t)$,
consistent  with  the  properties  in  (\ref{eq:hermitdm}).   It  also
contains  all information  about  the  state of  the  ensemble at  the
macroscopic time~$t$.  For this reason, we refer to $\tilde{f}$ as the
\emph{ensemble function}.

In  the  double momentum  representation,  the retarded  and  advanced
propagators are
\begin{equation}
  \label{eq:causgen}
  \Delta_{\mathrm{R}(\mathrm{A})}^0(p,p')\ =\
  \frac{1}{(p_0\:+(-)\:i\epsilon)^2\:-\:\mathbf{p}^2\:-\:M^2}
  (2\pi)^4\delta^{(4)}(p\:-\:p')\; .
\end{equation}
The                                      Pauli--Jordan~$\Delta^0(p,p')$,
Hadamard~$\Delta^0_1(p,p',\tilde{t}_f;\tilde{t}_i)$                 and
principal-part~$\Delta^0_{\mathcal{P}}(p,p')$ propagators become
\begin{subequations}
  \label{eq:pjh}
  \begin{align}
    \label{eq:pjvary}
    \Delta^0(p,p')\ =\ &-i2\pi\varepsilon(p_0)\delta(p^2\:-\:M^2)
    (2\pi)^4\delta^{(4)}(p\:-\:p')\;,\\
     \label{eq:had0}
    \Delta_1^0(p,p',\tilde{t}_f;\tilde{t}_i)\ =\
    &-i2\pi\delta(p^2\:-\:M^2)(2\pi)^4
    \delta^{(4)}(p\:-\:p')\nonumber\\
    &-\: i2\pi|2p_0|^{1/2}\delta(p^2\:-\:M^2)2\tilde{f}(p,p',t)
    e^{i(p_0\:-\:p_0')\tilde{t}_f}2\pi|2p_0'|^{1/2}\delta(p'^2\:-\:M^2)\;,\\
    \Delta^0_{\mathcal{P}}(p,p')\ =\ &\mathcal{P}\frac{1}{p^2\:-\:M^2}
    (2\pi)^4\delta^{(4)}(p\:-\:p')\; .
  \end{align}
\end{subequations}
Thus,  at  the tree  level,  only  the  Hadamard correlation  function
$\Delta^0_1 (p,p',\tilde{t}_f;\tilde{t}_i)$  depends explicitly on the
background  and macroscopic  time~$t$, through  the  ensemble function
$\tilde{f} (p,p',t)$  in~(\ref{eq:dftilde}). This is  a consequence of
the  causality of the  theory, as  we would  expect from  the spectral
decomposition   (\ref{eq:specrep})  of   the  retarded   and  advanced
propagators  $\Delta^0_{\mathrm{R}(\mathrm{A})}(x,y)$ in terms  of the
canonical  commutation  relation   (\ref{eq:caus}).  Notice  that  the
complex   phase  factor   $e^{i(p_0\:-\:p_0')\tilde{t}_f}$   has  only
appeared in the Hadamard propagator~(\ref{eq:had0}) and so it does not
spoil causality.  Beyond the  tree level, the background contributions
are expected to  modify the structure of the  Pauli--Jordan and causal
propagators,     according     to     our    discussion     of     the
K\"{a}ll\'{e}n--Lehmann        spectral        representation       in
Section~\ref{sec:canon}.  The full  complement of non-homogeneous free
propagators is listed in Table \ref{tab:props}.

\begin{table}
\caption{\label{tab:props} The  full  complement of non-homogeneous
  free propagators, where $\tilde{f}(p,p',t)$ is the ensemble function
  defined in (\ref{eq:dftilde}) and $\mathcal{P}$ denotes the 
  Cauchy principal value.}
\renewcommand{\arraystretch}{2}
\begin{ruledtabular}
\begin{tabular}{ c l }
{\bfseries Propagator} & {\bfseries Double Momentum Representation} \\
\hline
Feynman (Dyson) & $\begin{array}{l}
  \vspace{-1.5em} \\ 
  i\Delta^0_{\mathrm{F}(\mathrm{D})}(p,p',\tilde{t}_f;\tilde{t}_i)
  \ =\ \displaystyle\frac{(-)i}{p^2\:-\:M^2\:+(-)\:i\epsilon}
  \,(2\pi)^4\delta^{(4)}(p\:-\:p')\\\hspace{11em}
  +\:2\pi|2p_0|^{1/2}\delta(p^2\:-\:M^2)\tilde{f}(p,p',t)
  e^{i(p_0\:-\:p_0')\tilde{t}_f}2\pi|2p_0'|^{1/2}\delta(p'^2\:-\:M^2)
\end{array}$ \\
$+$($-$)ve-freq. Wightman & $\begin{array}{l}
  i\Delta_{>(<)}^0(p,p',\tilde{t}_f;\tilde{t}_i)\ = \ 2\pi\theta(+(-)p_0)\delta(p^2\:-\:M^2)
  (2\pi)^4\delta^{(4)}(p\:-\:p')\\\hspace{11em}
  +\:2\pi|2p_0|^{1/2}\delta(p^2\:-\:M^2)\tilde{f}(p,p',t)
  e^{i(p_0\:-\:p_0')\tilde{t}_f}2\pi|2p_0'|^{1/2}\delta(p'^2\:-\:M^2)
\end{array}$ \\
Retarded (Advanced) & $\begin{array}{l}
  \vspace{-2em} \\
  i\Delta_{\mathrm{R}(\mathrm{A})}^0(p,p')\ =\ \displaystyle
  \frac{i}{(p_0\:+(-)\:i\epsilon)^2\:-\:\mathbf{p}^2\:-\:M^2}
  (2\pi)^4\delta^{(4)}(p\:-\:p')
\end{array}$ \\
Pauli--Jordan & $\begin{array}{l}
  \vspace{0.3em}
  i\Delta^0(p,p')\ =\ 2\pi\varepsilon(p_0)\delta(p^2\:-\:M^2)
  (2\pi)^4\delta^{(4)}(p\:-\:p')
\end{array}$ \\
Hadamard & $\begin{array}{l} 
  i\Delta_1^0(p,p',\tilde{t}_f;\tilde{t}_i)\ =\ 2\pi\delta(p^2\:-\:M^2)
  (2\pi)^4\delta^{(4)}(p\:-\:p') \\
  \hspace{11em} +\:2\pi|2p_0|^{1/2}\delta(p^2\:-\:M^2)2\tilde{f}(p,p',t)
  e^{i(p_0\:-\:p_0')\tilde{t}_f}2\pi|2p_0'|^{1/2}\delta(p'^2\:-\:M^2)
\end{array}$ \\
Principal-part & $\begin{array}{l}
  \vspace{-1.6em} \\ \vspace{0.4em}
  \displaystyle
  i\Delta^0_{\mathcal{P}}(p,p')\ =\ \mathcal{P}\frac{i}{p^2\:-\:M^2}
  (2\pi)^4\delta^{(4)}(p\:-\:p')
\end{array}$ \\
\end{tabular}
\end{ruledtabular}
\renewcommand{\arraystretch}{1}
\end{table}

For the most general non-Gaussian density operator, we must account
for all $n$-linear EEVs of creation and annihilation operators. We
will then obtain $n$-point thermally-corrected vertex functions, given by
\begin{align}
  \label{eq:ntherm}
  \Gamma_{n}(p_1,p_2,\cdots,p_n,\tilde{t}_f;\tilde{t}_i)\ &\equiv
  \ \braket{\Phi(p_1;\tilde{t}_i)\Phi(p_2;\tilde{t}_i)
    \cdots\Phi(p_n;\tilde{t}_i)}_t\nonumber\\ &=\ 
  \bigg(\prod_{i\,=\,1}^{n}2\pi|2p_i^0|^{1/2}
  \delta(p_i^2-M^2)e^{ip^0_i\tilde{t}_f}\bigg)
  \tilde{f}_n(p_1,p_2,\cdots,p_n,t)\;,
\end{align}
where the interaction-picture field operator $\Phi(p;\tilde{t}_i)$ is
defined in (\ref{eq:fourier}). The $n$-point ensemble
function\linebreak $\tilde{f}_n(p_1,p_2,\cdots,p_n,t)$ generalizes
(\ref{eq:dftilde}).  In the remainder of this article, we will work
only with Gaussian density operators, as discussed in
Section~\ref{sec:ctpsd}, for which all but
$W_{11}(\mathbf{k}:\mathbf{k}')$ in $(\ref{eq:gendens})$ are zero.

It  would  be interesting  to  establish  a  connection between  these
canonically derived non-homogeneous free propagators and those derived
by the  path-integral representation of the  CTP generating functional
$\mathcal{Z}$. This will be  achieved through the bi-local source
$K_{ab}$   via    the   tree-level   Schwinger--Dyson    equation   in
(\ref{eq:treesd}).  The role of the bi-local source $K_{ab}$ will
be illustrated further,  when discussing the thermodynamic equilibrium
limit in Section~\ref{sec:eq}.

We proceed by replacing the  exponent of the CTP generating functional
$\mathcal{Z}$   in    (\ref{eq:chgen})   by   its    double   momentum
representation.  Subsequently,  we may  complete  the  square in  this
exponent by making the following shift in the field:
\begin{equation}
  \label{eq:shift}
  \Phi^a(p)\ =\ {\Phi'}^{a}(p)\:-\:
  \widehat{\Delta}^{0,\,a}_{\ \ \ b}(p)J^b(-p)\; ,
\end{equation}
where $\widehat{\Delta}^{0,\,ab}(p)$ is the free vacuum CTP propagator
in (\ref{eq:zerotemp}) in which the ensemble function~$\tilde{f}$ of
(\ref{eq:ansatz}) is set to zero.  Notice that the normal-ordered
contribution does not appear in~$\widehat{\Delta}^{0,\,ab}(p)$, as it
is sourced from the bi-local term~$K_{ab}$.  Upon substitution of
(\ref{eq:shift}) into the momentum representation of (\ref{eq:chgen}),
the CTP generating functional $\mathcal{Z}$ takes on the form
\begin{align}
  \label{eq:bilocgen}
  \mathcal{Z}[J,K,t]\ &=\ \mathcal{Z}^0[0,K,t]\exp\bigg\{iS^{\mathrm{int}}
  \bigg[\frac{1}{i}\frac{\delta}{\delta J_a},t\bigg]\bigg\}
  \!\int\![\D{}{{\Phi'}^a}]\;\exp\bigg\{-\,\frac{i}{2}\!\iint\!\!
  \frac{\D{4}{p}}{(2\pi)^4}\,\frac{\D{4}{p}'}{(2\pi)^4}\nonumber\\
  &\qquad \Big[\,J^a(p)\widehat{\Delta}_{\ \ a}^{0,\ \ c}(p)K_{cb}(p,p',t)
  {\Phi'}^{b}(-p')\:+\:{\Phi'}^a(p)K_{ac}(p,p',t)
  \widehat{\Delta}^{0,\,c}_{\ \ \ b}(p')J^{b}(-p')\nonumber\\
  &\qquad+\:J^a(p)\,\Big(\,\widehat{\Delta}^0_{ab}(p)(2\pi)^4\delta^{(4)}(p\,-\,p')
  \:-\:\widehat{\Delta}_{\ \ a}^{0,\ \; c}(p)K_{cd}(p,p',t)
  \widehat{\Delta}^{0,\,d}_{\ \ \ b}(p')\,\Big)\,J^b(-p')\,\Big]\bigg\}\;.
\end{align}
For the  Lagrangian in (\ref{eq:scallag}),  the cubic self-interaction
part ($-g\Phi^3$) of the action may be written down explicitly as
\begin{equation}
  \frac{-i g}{3!}\!\iiint\!\!\frac{\D{4}{p_1}}{(2\pi)^4}\,
  \frac{\D{4}{p_2}}{(2\pi)^4}\,\frac{\D{4}{p_3}}{(2\pi)^4}
  \, \eta_{abc}(2\pi)^4\delta^{(4)}_t(p_1\:+\:p_2\:+\:p_3)
  \frac{1}{i}\frac{\delta}{\delta J_a(p_1)}
  \frac{1}{i}\frac{\delta}{\delta J_b(p_2)}
  \frac{1}{i}\frac{\delta}{\delta J_c(p_3)}\ \subset \ 
  iS^{\mathrm{int}}\bigg[\frac{1}{i}\frac{\delta}{\delta J_a},t\bigg]\;.
\end{equation}
Hence, in  the three-point  vertex, the usual  energy-conserving delta
function    has   been    replaced   by    $\delta_t$,    defined   in
(\ref{eq:deltat}),  as  a  result   of  the  systematic  inclusion  of
finite-time effects.  This  time-dependent modification of the Feynman
rules is  fundamental to our perturbative  approach to non-equilibrium
thermal   field   theory   and    will   be   discussed   further   in
Section~\ref{sec:toy} in the context of a simple scalar model.

\begin{figure}
\begin{center}
\includegraphics{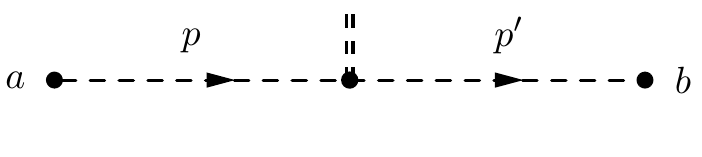}
\vspace{-2em}
\end{center}
\caption{The     Feynman--diagrammatic    interpretation     of    the
  non-homogeneous            free            CTP            propagator
  $i\Delta^{0,\,ab}(p,p',\tilde{t}_f;\tilde{t}_i)$ for the real scalar
  $\Phi$, where the double line represents momentum-violating coupling
  to  the  thermal  background  through the  bi-local  source  $K_{ab}
  (p,p',\tilde{t}_f;\tilde{t}_i)$.}
\label{fig:scalprop}
\end{figure}

In  (\ref{eq:bilocgen}), the  remaining terms  linear in  the external
source $J$ yield contributions  to the free propagator proportional to
$K^2$ upon double functional  differentiation with respect to~$J$.  As
we  shall  see in  Section~\ref{sec:eq},  these  contributions may  be
neglected.     Employing~(\ref{eq:funcdif}),   we   find    that   the
non-homogeneous      free       CTP      propagator      $\Delta^{0,\,
  ab}(p,p',\tilde{t}_f;\tilde{t}_i)$ may be  expressed in terms of the
free  vacuum CTP  propagator $\widehat{\Delta}^{0,\,  ab}(p)$  and the
bi-local source $K_{ab}(p,p',\tilde{t}_f;\tilde{t}_i)$ as follows:
\begin{equation}
  \label{eq:Dcjt}
  i\Delta^{0,\,ab}(p,p',\tilde{t}_f;\tilde{t}_i)\ =\
  i\widehat{\Delta}^{0,\,ab}(p)
  (2\pi)^4\delta^{(4)}(p\:-\:p')\:+\:
  i\widehat{\Delta}^{0,\,ac}(p)iK_{cd}(p,p',\tilde{t}_f;\tilde{t}_i)
  i\widehat{\Delta}^{0,\,db}(p')\; ,
\end{equation}
where
\begin{equation}
  K_{ab}(p,p',\tilde{t}_f;\tilde{t}_i)\ =\ e^{i(p_0\:-\:p_0')\tilde{t}_f}\,
  K_{ab}(\mathbf{p},\mathbf{p}',t)\; .
\end{equation}
The       form       of       the      free       CTP       propagator
$\Delta^{0,\,ab}(p,p',\tilde{t}_f;\tilde{t}_i)$  in (\ref{eq:Dcjt}) is
consistent  with  a  perturbative  inversion of  (\ref{eq:treesd})  to
leading order in the bi-local  source $K_{ab}$.  It is also consistent
with  the  canonically  derived   form  of  the  non-homogeneous  free
propagators in~(\ref{eq:ansatz}).

The  result in  (\ref{eq:Dcjt}) may  be  interpreted diagrammatically,
where      the      non-homogeneous      free      CTP      propagator
$i\Delta^{0,\,ab}(p,p',\tilde{t}_f;\tilde{t}_i)$    for    the    real
scalar~$\Phi$  is associated  with  the Feynman  diagram displayed  in
Figure~\ref{fig:scalprop}.          The         bi-local        source
$K_{ab}(p,p',\tilde{t}_f;\tilde{t}_i)$    plays   the   role    of   a
three-momentum-violating vertex  that gives  rise to the  violation of
translational invariance,  thus encoding the  spatial inhomogeneity of
the background.

\section{The Thermodynamic Equilibrium Limit}
\label{sec:eq}

In  this section,  we  derive the  analytical  forms of  the free  and
resummed  CTP propagators  in the  limit of  thermal  equilibrium. The
results  of  this  section   are  of  particular  importance  for  the
discussion  of pinch  singularities in  Section~\ref{sec:oneloop}.  We
also  show  the  connection  between  the  equilibrium  Bose--Einstein
distribution function  and the bi-local source  $K_{ab}$ introduced in
Section~\ref{sec:nonhomprop}.

In the  limit of thermal  equilibrium, the density  operator~$\rho$ is
diagonal in particle number,  so all amplitudes except $W_{11}$ vanish
in~(\ref{eq:gendens}).    In   this   limit,   the   general   density
operator~$\rho$, given explicitly in~(\ref{eq:densop}), reduces to the
series
\begin{align}
  \label{eq:rhoeq}
  \rho\ &=\ \ket{0}\!\bra{0}\:+\:
  \!\iint\!\D{}{\Pi_{\mathbf{k}_1}}\,\D{}{\Pi_{\mathbf{k}_1'}}\;
  \Big(\,(2\pi)^32E(\mathbf{k}_1)\delta^{(3)}(\mathbf{k}_1\:-\:\mathbf{k}_1')
  \:-\:W_{11}(\mathbf{k}_1:\mathbf{k}_1')\nonumber\\&\qquad \qquad
  +\:\frac{1}{2}\!\int\!\D{}{\Pi_{\mathbf{q}_1}}\;
  W_{11}(\mathbf{k}_1:\mathbf{q}_1)W_{11}(\mathbf{q}_1:\mathbf{k}_1')
  \:+\:\cdots\,\Big)\!\ket{\mathbf{k}_1}\!\bra{\mathbf{k}_1'}\nonumber\\&
  \qquad +\: \frac{1}{2}\!\idotsint\!\D{}{\Pi_{\mathbf{k}_1}}\,
  \D{}{\Pi_{\mathbf{k}_2}}\,\D{}{\Pi_{\mathbf{k}_1'}}\,\D{}{\Pi_{\mathbf{k}_2'}}\;
  \Big(\,(2\pi)^32E(\mathbf{k}_1)\delta^{(3)}(\mathbf{k}_1\:-\:\mathbf{k}_1')
  \:-\:W_{11}(\mathbf{k}_1:\mathbf{k}_1')\nonumber\\
  &\qquad \qquad +\:\frac{1}{2}\!\int\!
  \D{}{\Pi_{\mathbf{q}_1}}\;W_{11}(\mathbf{k}_1:\mathbf{q}_1)
  W_{11}(\mathbf{q}_1:\mathbf{k}_1')\:+\:\cdots\,\Big)
  \Big(\,(2\pi)^32E(\mathbf{k}_2)\delta^{(3)}(\mathbf{k}_2\:-\:\mathbf{k}_2')
  \nonumber\\&\qquad \qquad
  -\:W_{11}(\mathbf{k}_2:\mathbf{k}_2')\:+\:\frac{1}{2}\!\int\!
  \D{}{\Pi_{\mathbf{q}_2}}\;W_{11}(\mathbf{k}_2:\mathbf{q}_2)
  W_{11}(\mathbf{q}_2:\mathbf{k}_2')\:+\:\cdots\,\Big)
  \!\ket{\mathbf{k}_1,\mathbf{k}_2}\!\bra{\mathbf{k}_1',\mathbf{k}_2'}
  \:+\:\cdots\;,
\end{align}
where  the time  arguments in  the  multi-mode Fock  states have  been
omitted  for notational  convenience.  In  the equilibrium  limit, the
statistical  distribution  function  $g(\mathbf{p},\mathbf{p}',t)$  is
trivially   zero.   Instead,   $f(\mathbf{p},\mathbf{p}',t)$ is
calculated from~(\ref{eq:fsgs}) and takes the form of the series
\begin{align}
  \label{eq:feq}
  2\mathcal{E}(\mathbf{p},\mathbf{p}')f(\mathbf{p},\mathbf{p}',t)
  \ &=\ (2\pi)^32E(\mathbf{p})\delta^{(3)}(\mathbf{p}\:-\:\mathbf{p}')
  \:-\:W_{11}(\mathbf{p}:\mathbf{p}')\:+\:\frac{1}{2}\!
  \int\!\D{}{\Pi_{\mathbf{q}}}\;W_{11}(\mathbf{p}:\mathbf{q})
  W_{11}(\mathbf{q}:\mathbf{p}')\:+\:\cdots\nonumber\\
  &\qquad +\: \!\int\!\D{}{\Pi_{\mathbf{q}}}\;
  \Big(\,(2\pi)^32E(\mathbf{p})\delta^{(3)}(\mathbf{p}\:-\:\mathbf{q})
  \:-\:W_{11}(\mathbf{p}:\mathbf{q})\:+\:\cdots\,\Big)\nonumber\\
  &\qquad \quad \quad \times \: \Big(\,(2\pi)^32E(\mathbf{q})
  \delta^{(3)}(\mathbf{q}\:-\:\mathbf{p}')\:-\:
  W_{11}(\mathbf{q}:\mathbf{p}')\:+\:\cdots\,\Big)
  \:+\:\cdots\;,
\end{align}
where  disconnected parts  have been  canceled order-by-order  in the
expansion     by    the    normalization     $\mathrm{Tr}\,\rho$    in
(\ref{eq:thermbra}).                  The                  factor
$\mathcal{E}(\mathbf{p},\mathbf{p}')$ is defined in (\ref{eq:efact}).

The equilibrium density operator $\rho_{\rm eq}$ must also be diagonal
in the momenta and is thus obtained by making the replacement:
\begin{equation}
  \label{eq:wlim}
  W_{11}(\mathbf{k}:\mathbf{k}')\ \to\ \beta
  E(\mathbf{k})(2\pi)^32E(\mathbf{k})\delta^{(3)}(\mathbf{k}\:-\:\mathbf{k}')
\end{equation}
in~(\ref{eq:rhoeq}),    where   $\beta\:=\:1/T$    is    the   inverse
thermodynamic temperature. In detail, we find
\begin{equation}
  \rho_{\rm eq}\ =\ \ket{0}\!\bra{0}\:+\:
  \sum_{n=1}^{\infty}\frac{1}{n!}\!\idotsint\!\prod_{i=1}^n
  \Big(\,\D{}{\Pi_{\mathbf{k}_i}}\,f_{\beta}(\mathbf{k}_i)\,\Big)
  \bigotimes_{i=1}^n\ket{\mathbf{k}_i}\bigotimes_{i=1}^n
  \bra{\mathbf{k}_i}\;,
\end{equation}
where the amplitudes are the Boltzmann distributions
$f_{\beta}(\mathbf{k})\:=\:e^{-\beta E(\mathbf{k})}$
[cf.~(\ref{eq:sums})].  This last expression of~$\rho_{\rm eq}$ can be
shown to be fully equivalent to the Gaussian form
\begin{equation}
  \label{eq:expform}
  \rho_{\rm  eq}\  =\  \exp\!\bigg(-\beta\!\int\!\D{}{\Pi_{\mathbf{k}}}\,
  E(\mathbf{k})a^{\dag}(\mathbf{k})a(\mathbf{k})\bigg)\;,
\end{equation}
which corresponds to the standard Boltzmann density operator
\begin{equation}
  \label{eq:eqdens}
  \rho_{\rm eq}\ =\ e^{-\beta H^0}\; ,
\end{equation}
where $H^0$  is the free part of  the interaction-picture Hamiltonian.
Note that our convention for the normalization of the density operator
$\rho$,  including $\rho_{\rm eq}$,  is chosen  so that  the canonical
partition    function   $\mathcal{Z}(\beta)\:=\:\mathrm{Tr}\,e^{-\beta
  H^0}$  appears   explicitly  in  the   definition  of  the   EEV  in
(\ref{eq:thermbra}).

We  may  now  substitute  the limit~(\ref{eq:wlim})  into  the  series
expansion    of   the    statistical    distribution   function    $f$
in~(\ref{eq:feq}).  Using the identities of summation
\begin{equation}
  \label{eq:sums}
  \sum_{n=0}^{\infty}\frac{(-x)^n}{n!}\ =\ e^{-x}\;, \qquad
  \sum_{n=1}^{\infty}x^n\ =\ \frac{x}{1-x}\; ,
\end{equation}
we find the following correspondence in the equilibrium limit:
\begin{subequations}
  \label{eq:eqlim}
  \begin{align}
    f(\mathbf{p},\mathbf{p}',t)&\ \underset{\mathrm{eq.}}
    {\longrightarrow}\ f_{\mathrm{eq}}(\mathbf{p},\mathbf{p}')\ = \ 
    (2\pi)^3\delta^{(3)}(\mathbf{p}\:-\:\mathbf{p}')
    f_{\mathrm{B}}\big(E(\mathbf{p})\big)\;,\\
    g(\mathbf{p},\mathbf{p}',t)&\ \underset{\mathrm{eq.}}
    {\longrightarrow}\ g_{\mathrm{eq}}(\mathbf{p},\mathbf{p}')\ =\ 0\; ,
  \end{align}
\end{subequations}
where
\begin{equation}
  \label{eq:fB}
f_{\mathrm{B}}(p_0)\ =\ \frac{1}{e^{\beta p_0}\: -\: 1}
\end{equation}
is   the  Bose--Einstein   distribution  function.    The  equilibrium
statistical distribution functions  in (\ref{eq:eqlim}) depend only on
the  magnitude of  the  three-momentum $\mathbf{p}$  via the  on-shell
energy $E(\mathbf{p})$.  This is a consequence of the homogeneity
and  isotropy  implied by  thermodynamic  equilibrium.  Moreover,  the
multiplying  factor $\delta^{(3)}(\mathbf{p}\:-\:\mathbf{p}')$  on the
RHS   of~(\ref{eq:eqlim})  necessarily   restores   translational  and
rotational invariance.

It is well-known that the pinch singularities present in perturbative
expansions cancel in the equilibrium limit~\cite{Landsman:1986uw} (see
Section~\ref{sec:oneloop}) and we can safely take the limit
$t\:\to\:\infty$ throughout the CTP generating functional, as we
should expect for a system with static macroscopic properties.
Working then in the single momentum representation, we obtain from
(\ref{eq:singctp}) and (\ref{eq:ftilde}) the free equilibrium CTP
propagators
\begin{subequations}
  \label{eq:equilprop}
  \begin{align}
    \label{eq:eqF}
    i\Delta^0_{\mathrm{F}}(p)&\ =\ \Big(i\Delta_{\rm D}^0(p)\Big)^*\ =\
    i\Big(p^2\:-\:M^2\:+\:i\varepsilon\Big)^{-1}\:+\:
    2\pi f_{\mathrm{B}}(|p_0|)\delta(p^2\:-\:M^2)\;,\\[2mm]
    \label{eq:equilpropg}
    i\Delta^0_>(p)&\ =\ 2\pi\Big(\theta(p_0)\:+\:f_{\mathrm{B}}(|p_0|)\Big)
    \delta(p^2\:-\:M^2)\ \equiv\
    2\pi\varepsilon(p_0)\Big(1\:+\:f_{\mathrm{B}}(p_0)\Big)
    \delta(p^2\:-\:M^2)\;,\\
    \label{eq:equilpropl}
    i\Delta^0_<(p)&\ =\ 2\pi\Big(\theta(-p_0)\:+\:f_{\mathrm{B}}(|p_0|)\Big)
    \delta(p^2-M^2)\ \equiv\ 2\pi\varepsilon(p_0)f_{\mathrm{B}}(p_0)
    \delta(p^2\:-\:M^2)\;.
  \end{align}
\end{subequations}
The form  of the Wightman propagators  written in terms  of the signum
function $\varepsilon  (p_0)$ prove very useful in  the calculation of
loop diagrams, as detailed in Appendix~\ref{app:ITF}.

Returning  to  the  free  CTP  generating  functional  ${\cal  Z}$  in
(\ref{eq:bilocgen}),  it  follows  from  the  results  above  that  in
equilibrium the bi-local source~$K_{ab}(p,p',\tilde{t}_f;\tilde{t}_i)$
must  be proportional  to  a four-dimensional  delta  function of  the
momenta, i.e.
\begin{equation}
  K_{ab}(p,p',\tilde{t}_f;\tilde{t}_i)\ \underset{\mathrm{eq.}}
  {\longrightarrow}\ K_{ab}^{\mathrm{eq}}(p,p')\ =\
  (2\pi)^4\delta^{(4)}(p\:-\: p')K^{\mathrm{eq}}_{ab}(p)\;.
\end{equation}
In addition, $K^{\mathrm{eq}}_{ab}(p)$ must satisfy
\begin{equation}
  \label{eq:kabrels}
  \widehat{\Delta}^{0,\,ac}(p)K^{\mathrm{eq}}_{cd}(p)\widehat{\Delta}^{0,\,db}(p)
  \ =\ 2\pi i\,\delta(p^2\:-\;M^2)f_{\mathrm{B}}\big(E(\mathbf{p})\big)
  \!\begin{bmatrix}
    1 & 1 \\
    1 & 1
  \end{bmatrix}\!\;.
\end{equation}
Solving the  resulting system of  equations, keeping terms  to leading
order in $\epsilon$,  and noting that $K^{\mathrm{eq}}_{ab}(p)$ should
be written in terms of the three-momentum only, we find
\begin{equation}
  \label{eq:kab}
  K^{\mathrm{eq}}_{ab}(p)\ =\ 2i\epsilon f_{\mathrm{B}}\big(E(\mathbf{p})\big)
  \!\begin{bmatrix}
    1 & -1 \\
    -1 & 1
  \end{bmatrix}\!\;.
\end{equation}
By virtue of the limit representation of the delta function
\begin{equation}
  \label{eq:limdel}
  \delta(x)\ =\ \lim_{\epsilon\:\to\:0}\,\frac{1}{\pi}\,
  \frac{\epsilon}{x^2\:+\:\epsilon^2}\;,
\end{equation}
we can verify that we do indeed recover~(\ref{eq:kabrels}) and the
correct free CTP propagator by means of~(\ref{eq:Dcjt}).  We also
confirm that the terms linear in $J$ remaining in (\ref{eq:bilocgen})
may safely be ignored, since they yield contributions to the free
propagator proportional to $K^2\:\sim\: \epsilon^2$ upon double
functional differentiation with respect to $J$.

Alternatively,    interpreting   the   Boltzmann    density   operator
in~(\ref{eq:eqdens})  as an evolution  operator in  negative imaginary
time and  using the cyclicity of the  trace in the EEV,  we derive the
Kubo--Martin--Schwinger    (KMS)    relation    (see   for    instance
\cite{LeBellac2000})
\begin{equation}
  \label{eq:KMSgl}
  \Delta_{>}(x^0-y^0,\mathbf{x}-\mathbf{y})\ =\
  \Delta_{<}(x^0-y^0+i\beta,\mathbf{x}-\mathbf{y})\; .
\end{equation}
In the momentum representation, the KMS relation reads:
\begin{equation}
  \label{eq:KMS}
  \Delta_>(p)\ =\ e^{\beta p_0}\Delta_<(p)\; ,
\end{equation}
which   offers    the   final   constraint    on   $\tilde{f}(p)$   in
(\ref{eq:ftilde}):
\begin{equation}
  \label{eq:fconst}
  \tilde{f}(p)\ =\
  \theta(p_0)f_{\mathrm{B}}(p_0)+\theta(-p_0)f_{\mathrm{B}}(-p_0)\ =\
  f_{\mathrm{B}}(|p_0|)\; .
\end{equation}
Furthermore,    the     KMS    relation    also     leads    to    the
fluctuation-dissipation theorem
\begin{equation}
  \label{eq:flucdis}
  \Delta_1(p)\ =\ \Big(\,1\:+\:2f_{\mathrm{B}}(p_0)\Big)\, \Delta(p)\; ,
\end{equation}
relating the causality and unitarity relations in (\ref{eq:causality})
and (\ref{eq:unitarity}).  Subsequently, by means of the KMS relation,
we   may   write   all   propagators   in  terms   of   the   retarded
propagator~$\Delta_{\mathrm{R}}(p)$:
\begin{subequations}
  \label{eq:retident}
  \begin{align}
    \mathrm{Re}\,\Delta_{\mathrm{F}}(p)&\ =\
    \mathrm{Re}\,\Delta_{\mathrm{R}}(p)\;,\\
    \mathrm{Im}\,\Delta_{\mathrm{F}}(p)&\ =\ \varepsilon(p_0)
    \Big(1\:+\:2f_{\mathrm{B}}(|p_0|)\Big)
    \mathrm{Im}\,\Delta_{\mathrm{R}}(p)\;,\\
    \Delta_>(p)&\ =\ 2i\varepsilon(p_0)
    \Big(\theta(p_0)\:+\:f_{\mathrm{B}}(|p_0|)\Big)\mathrm{Im}\,
    \Delta_{\mathrm{R}}(p)\;,\\
    \Delta_<(p)&\ =\ 2i\varepsilon(p_0)
    \Big(\theta(-p_0)\:+\:f_{\mathrm{B}}(|p_0|)\Big)\mathrm{Im}\,
    \Delta_{\mathrm{R}}(p)\;.
\end{align}
\end{subequations}

In the homogeneous equilibrium  limit of the Schwinger--Dyson equation
in (\ref{eq:sd}), the inverse resummed CTP propagator is given by
\begin{equation}
  \label{eq:homsd}
  \Delta^{-1}_{ab}(p)\ =\ \Delta_{ab}^{0,\,-1}(p)\:+\:
  \Pi_{ab}(p)\;.
\end{equation}
In  the  absence of  self-energy  effects,  the  free equilibrium  CTP
propagator  is   obtained  by   inverting  the  equilbrium   limit  of
(\ref{eq:treesd}):
\begin{equation}
\label{eq:eqtreesd}
D^{0,\,-1}_{ab}(p)\ =\ \Delta^{0,\,-1}_{ab}(p)\:+\:K_{ab}^{\mathrm{eq}}(p)\;.
\end{equation}
Knowing   that   $K^{\mathrm{eq}}_{ab}   \:  \sim\:   \epsilon$   from
(\ref{eq:kab}),  the  inversion  of  (\ref{eq:eqtreesd}) can  be  done
perturbatively to leading order in $K^{\mathrm{eq}}_{ab}(p)$, in which
case  the  expression~(\ref{eq:Dcjt})  gets  reproduced.   Beyond  the
tree-level,  however,  the   contribution  from  the  bi-local  source
$K^{\mathrm{eq}}_{ab}(p)$  may be  neglected next  to  the self-energy
term  $\Pi_{ab}(p)$  and  the   inverse  resummed  CTP  propagator  is
explicitly given by
\begin{equation}
  \label{eq:resinv}
  \Delta_{ab}^{-1}(p)\ =\ 
  \begin{bmatrix}
    p^2\:-\:M^2\:+\:\Pi(p) & -\Pi_<(p) \\
    -\Pi_>(p) & -p^2\:+\:M^2\:-\:\Pi^{*}(p)
  \end{bmatrix}\!\;.
\end{equation}
In  this  equilibrium  limit, (\ref{eq:resinv})  may  be  inverted
exactly, yielding the {\em equilibrium} resummed CTP propagator
\begin{equation}
  \label{eq:resmatprop}
  \Delta^{ab}(p)\ =\ 
  \Big[\,\Big(\,p^2\:-\:M^2\:+\:\mathrm{Re}\,\Pi_{\mathrm{R}}(p)\,\Big)^2\:+\:
  \Big(\,\mathrm{Im}\,\Pi_{\mathrm{R}}(p)\,\Big)^2\,\Big]^{-1}
  \begin{bmatrix}
    p^2\:-\:M^2\:+\:\Pi^{*}(p) & -\Pi_<(p) \\
    -\Pi_>(p) & -p^2\:+\:M^2\:-\:\Pi(p)
  \end{bmatrix}\!\;.
\end{equation}
The  results  obtained above  in  (\ref{eq:resmatprop})  may only  be
compared    with    existing    resummations,   see    for    instance
\cite{Altherr:1994jc,   Garbrecht:2011xw},    in   the   thermodynamic
equilibrium  limit, as we  have discussed  in Sections~\ref{sec:ctpsd}
and \ref{sec:gradapp}.

In this single momentum representation, the self-energies satisfy the
unitarity and causality relations
\begin{subequations}
  \label{eq:piuc}
  \begin{align}
    \Pi_1(p)\ =\ \Pi_>(p)\:+\:\Pi_<(p) &\ =\ \Pi(p)\:-\:\Pi^{*}(p)
    \ =\ 2i\mathrm{Im}\,\Pi(p)\;,\\
    \label{eq:BreitW}
    2 i M\Gamma(p)\ =\ \Pi_>(p)\:-\:\Pi_<(p)
    &\ =\ \Pi_{\mathrm{R}}(p)\:-\:\Pi_{\mathrm{A}}(p)
    \ =\ 2i\mathrm{Im}\,\Pi_{\mathrm{R}}(p)\; ,
  \end{align}
\end{subequations}
where $\Gamma(p)$ is the  Breit--Wigner width, relating the absorptive
part  of the  retarded self-energy  $\Pi_{\mathrm{R}}(p)$  to physical
reaction rates~\cite{Kobes:1990ua,  vanEijck:1992mq}.  Notice that the
KMS  relation  (\ref{eq:KMSgl}) leads  also  to  the detailed  balance
condition
\begin{equation}
  \label{eq:db}
  \Pi_>(p)\ =\ e^{\beta p_0}\, \Pi_<(p)\; .
\end{equation}
Given the  relations in (\ref{eq:piuc}),  we find, in  compliance with
(\ref{eq:retident}), an analogous set of relations for the elements of
the CTP self-energy:
\begin{subequations}
  \label{eq:pirels}
  \begin{align}
    \mathrm{Re}\,\Pi(p)&\ =\ \mathrm{Re}\,\Pi_{\mathrm{R}}(p)\;,\\
    \mathrm{Im}\,\Pi(p)&\ =\ \varepsilon(p_0)
    \Big(1\:+\:2f_{\mathrm{B}}(|p_0|)\Big)\mathrm{Im}\,
    \Pi_{\mathrm{R}}(p)\;,\\
    \label{eq:pirelsg}
    \Pi_>(p)&\ =\ 2i\varepsilon(p_0)
    \Big(\theta(p_0)\:+\:f_{\mathrm{B}}(|p_0|)\Big)
    \mathrm{Im}\,\Pi_{\mathrm{R}}(p)\;,\\
    \Pi_<(p)&\ =\ 2i\varepsilon(p_0)
    \Big(\theta(-p_0)\:+\:f_{\mathrm{B}}(|p_0|)\Big)
    \mathrm{Im}\,\Pi_{\mathrm{R}}(p)\;.
\end{align}
\end{subequations}

Ignoring the dispersive parts of the self-energy, we expect to recover
the free  CTP propagators given  in (\ref{eq:equilprop}) in  the limit
$\mathrm{Im}\,\Pi(p)\:\to\:  \epsilon\:=\:0^+$.   This  limit  is
equivalent to
\begin{equation}
  \label{eq:epsR}
  \mathrm{Im}\,\Pi_{\mathrm{R}}(p)\ \to\ \epsilon_{\mathrm{R}}\ \equiv\ 
  \varepsilon(p_0)\,\epsilon\; .
\end{equation}
Expressing    the    equilibrium    resummed   CTP    propagator    in
(\ref{eq:resmatprop}) in terms  of the retarded absorptive self-energy
$\mathrm{Im}\,\Pi_{\mathrm{R}}(p)$, we can  convince ourselves that we
do     indeed      reproduce     the     free      equilibrium     CTP
propagators~(\ref{eq:equilprop}) in the limit~(\ref{eq:epsR}).

In  Appendix~\ref{app:ITF},  we  discuss  the  correspondence  of  the
results  of this  section  with the  Imaginary  Time Formalsim  (ITF),
clarifying the analytic  continuation of the imaginary-time propagator
and self-energy.

\section{The Particle Number Density}
\label{sec:num}

It is important to establish a direct connection between off-shell
Green's functions and physical observables.  Such observables include
the particle number density for which various interpretations have
been reported in the literature~\cite{Aarts:2001qa, Juchem:2003bi,
  Prokopec:2003pj, Prokopec:2004ic, Berges:2004yj, Lindner:2005kv,
  DeSimone:2007rw, Cirigliano:2007hb, Garny:2009qn, Cirigliano:2009yt,
  Beneke:2010dz, Anisimov:2010dk}.  In this section, we derive a
physically meaningful definition of the particle number density in
terms of the resummed CTP propagators.

In order to count off-shell contributions systematically, we suggest
to `measure' the particle number density in terms of charges, rather
than by quanta of energy.  The latter approach would necessitate the
use of a quasi-particle approximation to identify `single-particle'
energies, which we do not follow here.  Instead, we analytically
continue the real scalar field to a pair of complex scalar fields
($\Phi,\ \Phi^{\dag}$). We may then introduce the Noether charge
$\mathcal{Q}(x_0;\tilde{t}_i)$ of the global $U(1)$ symmetry for the
Heisenberg-picture field $\Phi_{\mathrm{H}}(x;\tilde{t}_i)$ operator:
\begin{equation}
  \label{eq:Q}
  \mathcal{Q}(x_0;\tilde{t}_i)\ =\ i\!\int\!\D{3}{\mathbf{x}}\;
  \Big(\,\Phi_{\mathrm{H}}^{\dag}(x;\tilde{t}_i)\pi_{\mathrm{H}}^{\dag}(x;\tilde{t}_i)
  \:-\:\pi_{\mathrm{H}}(x;\tilde{t}_i)\Phi_{\mathrm{H}}(x;\tilde{t}_i)\,\Big)\;.
\end{equation}
Here, $\pi_{\mathrm{H}}(x;\tilde{t}_i)\:=\:\partial_{x_0}
\Phi_{\mathrm{H}}^{\dag}(x;\tilde{t}_i)$ is the conjugate momentum
operator and we include all time dependencies explicitly for
clarity. In the absence of derivative interactions, the Noether charge
depends only on the quadratic form of the kinetic term in the
Lagrangian. Hence, this analytic continuation may be employed even for
real scalar theories with interaction terms that break the
$\mathbb{Z}_2$ symmetry. Up to the infinite $T\:=\:0$ vacuum
contribution, the EEV of the operator $\mathcal{Q}(x_0;\tilde{t}_i)$
in (\ref{eq:Q}) is zero on analytically continuing back to the
original real scalar field $\Phi$, since the identical particle and
anti-particle contributions cancel.  Therefore, we need to devise a
method by which to separate the particle from the anti-particle
degrees of freedom in the EEV of (\ref{eq:Q}).

We note that the Noether charge of the local $U(1)$ symmetry of the
complex scalar theory is gauge-dependent and therefore unphysical.
The physical conserved matter charge remains that of the global $U(1)$
symmetry and is recovered in the temporal gauge $A^0\:=\:0$. In this
case, the conserved charge in (\ref{eq:Q}) would be written in terms
of the fields and their time derivatives and not the conjugate momenta
of the full Lagrangian.

For the general case of a spatially and temporally inhomogeneous
background, we need to generalize the Noether charge operator
$\mathcal{Q}(x_0;\tilde{t}_i)$ by writing it in terms of a charge
density operator $\mathcal{Q}(\mathbf{p},\mathbf{X},X_0;\tilde{t}_i)$
as
\begin{equation}
  \mathcal{Q}(X_0;\tilde{t}_i)\ =\ \!\int\D{3}{\mathbf{X}}\int\!\!
  \frac{\D{3}{\mathbf{p}}}{(2\pi)^3}\;
  \mathcal{Q}(\mathbf{p},\mathbf{X},X_0;\tilde{t}_i)\;.
\end{equation}
In  the above,  the three-momentum  $\mathbf{p}$ is  conjugate  to the
spatial    part     of    the    relative     space-time    coordinate
$R^{\mu}\:=\:x^{\mu}\:-\:y^{\mu}$  and $X^{\mu}\:=\:(X^0,\mathbf{X})\:
=\:  (x^\mu\:  +\:  y^\mu)/2$  is the  central  space-time  coordinate
[cf.~(\ref{eq:7ind})].  To this end, we proceed by inserting into
(\ref{eq:Q}) unity in the following form:
\begin{equation}
1\  = \  \!\int\!\D{4}{y}\;\delta^{(4)}(x\:-\:y)
 \  = \ \!\int\!\D{4}{y}\int\!\!\frac{\D{3}{\mathbf{p}}}{(2\pi)^3}\;
  e^{-i\mathbf{p}\cdot(\mathbf{x}\:-\:\mathbf{y})}\delta(x_0\:-\:y_0)\;.
\end{equation}
Observe  that   in  what  follows,   $x^0\:=\:y^0\:=\:X^0$  thanks  to
$\delta(x_0\:-\:y_0)$.   Subsequently, symmetrizing  the  integrand in
$x$ and $y$, we may write the charge operator as
\begin{equation}
  \label{eq:Qsym}
  \mathcal{Q}(X_0;\tilde{t}_i)\ =\ \frac{i}{2}
  \!\int\!\D{3}{\mathbf{x}}\int\!\D{4}{y}
  \!\int\!\!\frac{\D{3}{\mathbf{p}}}{(2\pi)^3}\;
  e^{-i\mathbf{p}\cdot(\mathbf{x}\:-\:\mathbf{y})}\delta(x_0\:-\:y_0)\Big(\,\Phi_{\mathrm{H}}^{\dag}(y;\tilde{t}_i)
  \pi_{\mathrm{H}}^{\dag}(x;\tilde{t}_i)
  \:-\:\pi_{\mathrm{H}}(y;\tilde{t}_i)\Phi_{\mathrm{H}}(x;\tilde{t}_i)\:+\:
  (x\:\leftrightarrow\:y)\,\Big)\;.
\end{equation}
In  terms  of  the  central  and  relative  coordinates,  $X^\mu$  and
$R^{\mu}$,         the         charge         density         operator
$\mathcal{Q}(\mathbf{p},\mathbf{X},X_0;\tilde{t}_i)$       may      be
appropriately identified from (\ref{eq:Qsym}) as
\begin{equation}
  \label{eq:Qdef}
  \mathcal{Q}(\mathbf{p},\mathbf{X},X_0;\tilde{t}_i)\ =\
  \frac{i}{2}\!\int\!\D{4}{R}\;e^{-i\mathbf{p}\cdot\mathbf{R}}\,\delta(R_0)\Big(\,
  \Phi_{\mathrm{H}}^{\dag}(X-\tfrac{R}{2};\tilde{t}_i)
  \pi_{\mathrm{H}}^{\dag}(X+\tfrac{R}{2};\tilde{t}_i)\:-\:
  \pi_{\mathrm{H}}(X-\tfrac{R}{2};\tilde{t}_i)
  \Phi_{\mathrm{H}}(X+\tfrac{R}{2};\tilde{t}_i)\:+\:
  (R\:\leftrightarrow\:-R)\,\Big)\;.
\end{equation}
Substituting for  the definitions of  the conjugate-momentum operators
$\pi_{\mathrm{H}}$  and   $\pi_{\mathrm{H}}^{\dag}$,  we  may  rewrite
$\mathcal{Q}(\mathbf{p},\mathbf{X},X_0;\tilde{t}_i)$  in the following
form:
\begin{equation}
  \label{eq:Qdef2}
  \mathcal{Q}(\mathbf{p},\mathbf{X},X_0;\tilde{t}_i)\ =\ 
  i\!\int\!\D{4}{R}\;e^{-i\mathbf{p}\cdot\mathbf{R}}\,\delta(R_0)\partial_{R_0}\Big(
  \Phi_{\mathrm{H}}^{\dag}(X-\tfrac{R}{2};\tilde{t}_i)
  \Phi_{\mathrm{H}}(X+\tfrac{R}{2};\tilde{t}_i)\:-\:
  \Phi_{\mathrm{H}}^{\dag}(X+\tfrac{R}{2};\tilde{t}_i)
  \Phi_{\mathrm{H}}(X-\tfrac{R}{2};\tilde{t}_i)\Big)\;.
\end{equation}

The EEV of $\mathcal{Q}(\mathbf{p},\mathbf{X},X_0;\tilde{t}_i)$ at the
macroscopic time  $t$ is  then obtained by  taking the trace  with the
density  operator~$\rho$  as   given  in  (\ref{eq:thermbra})  in  the
equal-time   limit    $X_0\:=\:\tilde{t}_f$.    We   have    seen   in
Section~\ref{sec:CTP} that the equal-time limit is necessary to ensure
that the observable charge density is picture independent and that the
number   of   independent  coordinates   is   reduced   to  seven   as
required. Thus, we have
\begin{equation}
  \label{eq:qbar}
  \braket{\mathcal{Q}(\mathbf{p},\mathbf{X},\tilde{t}_f;\tilde{t}_i)}_t
  \ =\ \lim_{X_0\:\to\:\tilde{t}_f}\,i\!\int\!\D{4}{R}\;
  e^{-i\mathbf{p}\cdot\mathbf{R}}\,\delta(R_0)\partial_{R_0}
  \Big(\,i\Delta_<(R,X,\tilde{t}_f;\tilde{t}_i)\:-\:
  i\Delta_<(-R,X,\tilde{t}_f;\tilde{t}_i)\,\Big)\;,
\end{equation}
where we use the notation
\begin{equation}
  i\Delta_<(R,X,\tilde{t}_f;\tilde{t}_i)\ =\ 
  \braket{\Phi_{\mathrm{H}}^{\dag}(X-\tfrac{R}{2};\tilde{t}_i)
  \Phi_{\mathrm{H}}(X+\tfrac{R}{2};\tilde{t}_i)}_t
\end{equation}
for the resummed CTP Wightman propagator.

Let        us       comment        on       the        two       terms
$i\Delta_<(R,X,\tilde{t}_f;\tilde{t}_i)$                            and
$i\Delta_<(-R,X,\tilde{t}_f;\tilde{t}_i)$  that occur  on  the RHS  of
(\ref{eq:qbar}).               The              first             term
$i\Delta_<(R,X,\tilde{t}_f;\tilde{t}_i)$   comprises   \emph{ensemble}
positive-frequency  particle  modes  and \emph{ensemble  plus  vacuum}
negative-frequency    anti-particle    modes.     The    second    one
$i\Delta_<(-R,X,\tilde{t}_f;\tilde{t}_i)$   comprises   \emph{ensemble
  plus    vacuum}   positive-frequency    anti-particle    modes   and
\emph{ensemble}  negative-frequency  particle  modes.  Hence,  we  may
extract  the number  density of  particles by  taking the  sum  of the
positive-frequency                  contribution                  from
$i\Delta_<(R,X,\tilde{t}_f;\tilde{t}_i)$  and  the  negative-frequency
contribution from $i\Delta_<(-R,X,\tilde{t}_f;\tilde{t}_i)$.

We   separate   the   positive-   and  negative-frequency   parts   of
(\ref{eq:qbar})   by  decomposing   the   equal-time  delta   function
$\delta(R_0)$ via the limit representation
\begin{equation}
  \delta(R_0)\ =\ \frac{i}{2\pi}
  \bigg(\frac{1}{R_0\:+\:i\epsilon}\:-\:
  \frac{1}{R_0\:-\:i\epsilon}\bigg)\;,
\end{equation}
with $\epsilon  = 0^+$.  Thus,  a physically meaningful  definition of
the number density  of particles at the macroscopic  time $t$ is given
by
\begin{equation}
  n(\mathbf{p},\mathbf{X},\tilde{t}_f;\tilde{t}_i)\ =\
  -\lim_{X_0\:\to\: \tilde{t}_f}\,\!
  \int\!\D{3}{\mathbf{R}}\;e^{-i\mathbf{p}\cdot\mathbf{R}}
  \int\!\frac{\D{}{R_0}}{2\pi}
  \bigg(\frac{1}{R_0\:+\:i\epsilon}\,\partial_{R_0}
  i\Delta_<(R,X,\tilde{t}_f;\tilde{t}_i)\:+\:
  \frac{1}{R_0\:-\:i\epsilon}\,\partial_{R_0}
  i\Delta_<(-R,X,\tilde{t}_f;\tilde{t}_i)\bigg)\;.
\end{equation}
Using  time   translational  invariance  of  the   CTP  contour,  this
observable  may be  recast in  terms of  the Wigner  transform  of the
Wightman propagators as
\begin{equation}
  \label{eq:numdensfin}
  n(\mathbf{p},\mathbf{X},t)\ \equiv\ n(\mathbf{p},\mathbf{X},t;0)
  \ =\ \lim_{X_0\:\to\:t}
  \,\!\int\!\frac{\D{}{p_0}}{2\pi}\;p_0\Big(\,\theta(p_0)
  i\Delta_<(p,X,t;0)\:-\:\theta(-p_0)i\Delta_<(-p,X,t;0)\,\Big)\;.
\end{equation}
Note     that     the     number     density     of     anti-particles
$n^{C}(\mathbf{p},\mathbf{X},t)$  is obtained  by  $C$-conjugating the
two negative-frequency  Wightman propagators in~(\ref{eq:numdensfin}).
Useful   relations    between   correlation   functions    and   their
$C$-conjugated  counterparts  are  given  in  Appendices~\ref{app:rel}
and~\ref{app:complex}.

We       reiterate      from      (\ref{eq:fnumd})       that      the
$n(\mathbf{p},\mathbf{X},t)\,\D{3}{\mathbf{p}}\,\D{3}{\mathbf{X}}$   is
interpreted as the number of  particles at macroscopic time $t$ in the
volume      of      phase space      between     $\mathbf{p}$      and
$\mathbf{p}\:+\:\D{}{\mathbf{p}}$      and      $\mathbf{X}$      and
$\mathbf{X}\:+\:\D{}{\mathbf{X}}$.   The  particle  number  per  unit
volume is obtained by integrating over all momentum modes, i.e.
\begin{equation}
  \label{eq:numvol}
  n(\mathbf{X},t)\ =\ \!\int\!\!\frac{\D{3}{\mathbf{p}}}{(2\pi)^3}
  \;n(\mathbf{p},\mathbf{X},t)\;,
\end{equation}
and finally the total particle  number, by integrating over all space,
i.e.
\begin{equation}
  N(t)\ =\ \!\int\!\D{3}{\mathbf{X}}\;n(\mathbf{X},t)\;.
\end{equation}

By  inserting the  inverse  Wigner transform
\begin{equation}
  i\Delta_<(p,X,t;0)\ =\ \!\int\!\!\frac{\D{4}{P}}{(2\pi)^4}\;
  e^{-iP\cdot X}\,
   i\Delta_<(p+\tfrac{P}{2},p-\tfrac{P}{2},t;0)\;,
\end{equation}
into  (\ref{eq:numdensfin}),  the  particle  number  per  unit  volume
$n(\mathbf{X},t)$  may be expressed  in terms  of the  double momentum
representation  of  the  Wightman propagators  via  (\ref{eq:numvol}).
After  making  the   coordinate  transformation  $p\:\to\:-\:p$  in  the
negative-frequency contribution, we then obtain
\begin{equation}
  \label{eq:numdensdoub}
  n(\mathbf{X},t)\ =\ \lim_{X_0\:\to\: t}\,2\!\iint\!\!
  \frac{\D{4}{p}}{(2\pi)^4}\,\frac{\D{4}{P}}{(2\pi)^4}\;
  e^{-iP\cdot X}\,\theta(p_0)p_0
  i\Delta_<(p+\tfrac{P}{2},p-\tfrac{P}{2},t;0)\; ,
\end{equation}
The particle number per unit volume $n(\mathbf{X},t)$ in
(\ref{eq:numdensdoub}) is related to the statistical distribution
function $f$, through
\begin{equation}
  \label{eq:nf}
  n(\mathbf{X},t)\ =\ \!\int\!\!\frac{\D{3}{\mathbf{p}}}{(2\pi)^3}
  \int\!\!\frac{\D{3}{\mathbf{P}}}{(2\pi)^3}\;
  e^{i\mathbf{P}\cdot\mathbf{X}}\,
  f(\mathbf{p}+\tfrac{\mathbf{P}}{2},\mathbf{p}-\tfrac{\mathbf{P}}{2},t)\;,
\end{equation}
cf.~(\ref{eq:fnumd}). Here, we must emphasize that (\ref{eq:nf}) is
understood in the Heisenberg picture. Working instead in the
interaction picture, (\ref{eq:numdensdoub}) and (\ref{eq:nf}) define
the statistical distribution function at a given order in perturbation
theory. Explicitly, the $n$-loop statistical distribution function is
defined in terms of the $n$-loop negative-frequency Wightman
propagator via
\begin{equation}
  \label{eq:statpert}
  f^{(n)}(\mathbf{p}+\tfrac{\mathbf{P}}{2},\mathbf{p}-\tfrac{\mathbf{P}}{2},t)
  \ =\ \lim_{X_0\:\to\: t}\,2\!\iint\!\!
  \frac{\D{}{p_0}}{2\pi}\,\frac{\D{}{P_0}}{2\pi}\;
  e^{-iP_0X_0}\,\theta(p_0)p_0
  i\Delta^{(n)}_<(p+\tfrac{P}{2},p-\tfrac{P}{2},t;0)\;.
\end{equation}

It is instructive to check that our definitions for the number density
lead to the expected results for the free and quasi-particle
equilibrium cases.  Substituting the free equilibrium Wightman
propagator (\ref{eq:equilpropl}) of the real scalar field into
(\ref{eq:numdensfin}), we obtain
\begin{equation}
  n^{0}_{\mathrm{eq}}(\mathbf{p},\mathbf{X},t)\ =\
  f_{\mathrm{B}}\big(E(\mathbf{p})\big)\;,
\end{equation}
exactly as we would expect for the number density of spectrally-free
particles in thermodynamic equilibrium.  Inserting instead the
resummed equilibrium Wightman propagator given in
(\ref{eq:resmatprop}) in the narrow width limit, we get
\begin{equation}
  n_{\mathrm{qp}}(\mathbf{p},\mathbf{X},t)\ =\
  f_{\mathrm{B}}\big(\mathcal{E}(\mathbf{p})\big)\;,
\end{equation}
where $\mathcal{E}(\mathbf{p})$ is the solution to the gap equation,
\begin{equation}
  \mathcal{E}^2(\mathbf{p})\ =\ \mathbf{p}^2\:+\:M^2\:-\:
  \mathrm{Re}\,\Pi_{\mathrm{R}}(\mathcal{E}(\mathbf{p})+i\epsilon,
  \mathbf{p})\;,
\end{equation}
and  $n_{\mathrm{qp}}(\mathbf{p},\mathbf{X},t)$  then  represents  the
number density of quasi-particles.

\section{Master Time Evolution Equations for Particle Number Densities}
\label{sec:eom}

Having established a direct relationship between the non-homogeneous
CTP propagators and the particle number density in
Section~\ref{sec:num}, we are now in a position to derive in this
section master time evolution equations for the particle number
density~$n({\bf X}, t)$ and the statistical distribution function
$f(\mathbf{p},\mathbf{p}',t)$.  This is achieved in analogy to the
derivation of the well-known Kadanoff--Baym
equations~\cite{Baym:1961zz, Kadanoff1989} by partially inverting the
CTP Schwinger--Dyson equation obtained in Section~\ref{sec:ctpsd}.
Our approach, however, differs significantly from other methods in that we do
not rely on a truncation of a gradient expansion of the resulting
expressions.  More details of the gradient expansion can be found in
Section~\ref{sec:gradapp} and Appendix~\ref{app:kb}.  In the next
section, we will employ a loopwise truncation of the time evolution
equations in terms of non-homogeneous free CTP propagators.  As we
will see, these dynamical equations are nonetheless resummed \emph{to
  all orders} in a gradient expansion.

We  begin our  derivation of  the  time evolution  equations with  the
double  momentum representation  of the  Schwinger--Dyson  equation in
(\ref{eq:sddmom}).  We  convolute (\ref{eq:sddmom}) consecutively from
the            right             with            the            weight
function \linebreak $(2\pi)^4\delta_t^{(4)}(q_1\:-\:q_2)$                 defined
in~(\ref{eq:deltat4})    and    then    with    the    resummed    CTP
propagator~$\Delta^{ab}(q_2,p_2,\tilde{t}_f;\tilde{t}_i)$.   By making
use  of   the  inverse  relation~(\ref{eq:invdmom}),   we  obtain  the
following expression:
\begin{align}
  \label{eq:kbdmom}
  &\iint\!\!\frac{\D{4}{q_1}}{(2\pi)^4}\,\frac{\D{4}{q_2}}{(2\pi)^4}
  \;\Delta_{ac}^{0,\,-1}(p_1,q_1)(2\pi)^4\delta_t^{(4)}(q_1\:-\:q_2)
  \Delta^{c}_{\;b}(q_2,p_2,\tilde{t}_f;\tilde{t}_i)\nonumber\\
  &\qquad =\ \eta_{ab}
  (2\pi)^4\delta^{(4)}_t(p_1\:-\:p_2)\:-\:\!\iint\!\!
  \frac{\D{4}{q_1}}{(2\pi)^4}\,\frac{\D{4}{q_2}}{(2\pi)^4}\;
  \Pi_{ac}(p_1,q_1,\tilde{t}_f;\tilde{t}_i)(2\pi)^4\delta^{(4)}_t(q_1\:-\:q_2)
  \Delta_{\;b}^c(q_2,p_2,\tilde{t}_f;\tilde{t}_i)\;,
\end{align}
where the  contribution of the  bi-local source $K_{ab}$  is neglected
next to the self-energies.  It is  essential to remark that the LHS of
(\ref{eq:kbdmom}) has the following coordinate-space representation:
\begin{equation}
  \label{eq:LHSxy}
  \int_{\Omega_t}\!\!\D{4}{z}\;\Delta^{0,\,-1}_{ab}(x,z)
  \Delta^{b}_{\; c}(z,y,\tilde{t}_f;\tilde{t}_i)\;.
\end{equation}
Substituting     for    the     free     inverse    CTP     propagator
$\Delta^{0,\,-1}_{ab}(x,y)$ given  in (\ref{eq:freeinv}), we  may then
confirm  via  (\ref{eq:LHSxy}) that  evaluating  the  $q_1$ and  $q_2$
integrals on the LHS of (\ref{eq:kbdmom}) yields
\begin{equation}
  \iint\!\!\frac{\D{4}{q_1}}{(2\pi)^4}\,\frac{\D{4}{q_2}}{(2\pi)^4}
  \;\Delta_{ac}^{0,\,-1}(p_1,q_1)(2\pi)^4\delta_t^{(4)}(q_1\:-\:q_2)
  \Delta^{c}_{\;b}(q_2,p_2,\tilde{t}_f;\tilde{t}_i)\ = \
  \Big(\,p_1^2\:-\:M^2\,\Big)\Delta_{ab}(p_1,p_2,\tilde{t}_f;\tilde{t}_i)\;.
\end{equation}
Recalling               that              the              self-energy
$\Pi_{ab}(p,p',\tilde{t}_f;\tilde{t}_i)$                       contains
$\delta^{(4)}_t(p\:-\:p')$ functions  in the vertices,  we may perform
the $q_1$  integral on the RHS  of (\ref{eq:kbdmom}) by  making use of
(\ref{eq:dciden}).   Consequently,  (\ref{eq:kbdmom})  may be  written
down in the following concise form:
\begin{align}
  \label{eq:kbsimple}
  \Big(\,p_1^2\:-\:M^2\,\Big)\Delta_{ab}(p_1,p_2,\tilde{t}_f;\tilde{t}_i)
  \ &=\ \eta_{ab}(2\pi)^4\delta_t^{(4)}(p_1\:-\:p_2)\:-\:\!\int\!\!\frac{\D{4}{q}}{(2\pi)^4}\;
  \Pi_{ac}(p_1,q,\tilde{t}_f;\tilde{t}_i)
  \Delta^{c}_{\;b}(q,p_2,\tilde{t}_f;\tilde{t}_i)\;.
\end{align}

At this point, it is essential to remark that, in any loopwise
truncation of (\ref{eq:kbsimple}), the external propagator
$\Delta_{ab}(p,p',\tilde{t}_f;\tilde{t}_i)$, appearing on both the
left- and right-hand sides, must be evaluated at the same order. This
constraint ensures that the delta function on the RHS of
(\ref{eq:kbsimple}) is present order-by-order in such an expansion and
that the convolution in (\ref{eq:kbdmom}) remains self-consistent.

With  the   definition  of  the   particle  number  per   unit  volume
$n(\mathbf{X},t)$  from  (\ref{eq:numvol})  in  mind,  we  equate  the
element  $(a,b)\:=\:(1,2)$  of  each  side of  (\ref{eq:kbsimple})  to
extract    the    interacting    Klein--Gordon   equation    of    the
negative-frequency resummed Wightman propagator:
\begin{equation}
  \label{eq:doubKG}
  \Big(\,p_1^2\:-\:M^2\,\Big)\Delta_{<}(p_1,p_2,\tilde{t}_f;\tilde{t}_i)
  \ =\ -\!\int\!\!\frac{\D{4}{q}}{(2\pi)^4}\;
  \Big(\,\Pi(p_1,q,\tilde{t}_f;\tilde{t}_i)
  \Delta_{<}(q,p_2,\tilde{t}_f;\tilde{t}_i)
  \:-\:\Pi_<(p_1,q,\tilde{t}_f;\tilde{t}_i)
  \Delta_{\mathrm{D}}(q,p_2,\tilde{t}_f;\tilde{t}_i)\,\Big)\;.
\end{equation}
Using                the                decomposition               of
$\Delta_{\mathrm{D}}(p,p',\tilde{t}_f;\tilde{t}_i)$                from
(\ref{eq:decomF})      and     an      analogous      identity     for
$\Pi(p,p',\tilde{t}_f;\tilde{t}_i)$,    we   may   rewrite
(\ref{eq:doubKG}) as
\begin{align}
  \label{eq:doubKGexp}
  &\Big(\,p_1^2\:-\:M^2\,\Big)\Delta_{<}(p_1,p_2,\tilde{t}_f;\tilde{t}_i)
  \:+\:\!\int\!\!\frac{\D{4}{q}}{(2\pi)^4}\;
  \Pi_{\mathcal{P}}(p_1,q,\tilde{t}_f;\tilde{t}_i)
  \Delta_<(q,p_2,\tilde{t}_f;\tilde{t}_i)\nonumber\\
  &\qquad \qquad=\ -\frac{1}{2}\!\int\!\!
  \frac{\D{4}{q}}{(2\pi)^4}\;
  \Big[\,\Pi_>(p_1,q,\tilde{t}_f;\tilde{t}_i)
  \Delta_<(q,p_2,\tilde{t}_f;\tilde{t}_i)\:-\:\Pi_<(p_1,q,\tilde{t}_f;\tilde{t}_i)
  \,\Big(\,\Delta_>(q,p_2,\tilde{t}_f;\tilde{t}_i)
  \:-\:2\Delta_{\mathcal{P}}(q,p_2,\tilde{t}_f;\tilde{t}_i)\,\Big)\,\Big]\;,
\end{align}
where the subscript $\mathcal{P}$ denotes principal-part evaluation of
the functions given in (\ref{eq:pprop}) and (\ref{eq:pself}).

Introducing the central  and relative momenta, $p\:=\:(p_1\:+\:p_2)/2$
and  $P\:=\:p_1\:-\:p_2$, respectively, we  write (\ref{eq:doubKGexp})
in the following form:
\begin{equation}
  \label{eq:dmomBE}
  \Big[\,\big(\,p_0\:+\:\tfrac{P_0}{2}\big)^2\:-\:E^2(\mathbf{p}\:
  +\:\tfrac{\mathbf{P}}{2})\,\Big]
  \Delta_<(p+\tfrac{P}{2},p-\tfrac{P}{2},\tilde{t}_f;\tilde{t}_i)\:+\:
  \mathscr{F}(p+\tfrac{P}{2},p-\tfrac{P}{2},\tilde{t}_f;\tilde{t}_i)\ =\
  \mathscr{C}(p+\tfrac{P}{2},p-\tfrac{P}{2},\tilde{t}_f;\tilde{t}_i)\; ,
\end{equation}
where we have defined
\begin{subequations}
\begin{align}
  \label{eq:fdef}
  \mathscr{F}(p+\tfrac{P}{2},p-\tfrac{P}{2},\tilde{t}_f;\tilde{t}_i)\ &\equiv\ 
  -\!\int\!\!\frac{\D{4}{q}}{(2\pi)^4}\;
  i\Pi_{\mathcal{P}}(p+\tfrac{P}{2},q,\tilde{t}_f;\tilde{t}_i)\:
  i\Delta_<(q,p-\tfrac{P}{2},\tilde{t}_f;\tilde{t}_i),\\
  \label{eq:cdef}
  \mathscr{C}(p+\tfrac{P}{2},p-\tfrac{P}{2},\tilde{t}_f;\tilde{t}_i)\ &\equiv\
  \frac{1}{2}\!\int\!\!\frac{\D{4}{q}}{(2\pi)^4}\;
  \Big[\,i\Pi_>(p+\tfrac{P}{2},q,\tilde{t}_f;\tilde{t}_i)\:
  i\Delta_<(q,p-\tfrac{P}{2},\tilde{t}_f;\tilde{t}_i)
  \nonumber\\&\qquad
  -\: i\Pi_<(p+\tfrac{P}{2},q,\tilde{t}_f;\tilde{t}_i)\,
  \Big(\,i\Delta_>(q,p-\tfrac{P}{2},\tilde{t}_f;\tilde{t}_i)
  \:-\:2i\Delta_{\mathcal{P}}(q,p-\tfrac{P}{2},\tilde{t}_f;\tilde{t}_i)\,\Big)\,\Big]\;.
\end{align}
\end{subequations}

With the  aim of  finding a master time  evolution equation for  the particle
number density~$n({\bf X},t)$  in~(\ref{eq:numvol}), we integrate both
sides of (\ref{eq:dmomBE}) with the measure
\begin{equation}
  \iint\!\!\frac{\D{4}{p}}{(2\pi)^4}\frac{\D{4}{P}}{(2\pi)^4}
  \;e^{-iP\cdot X}\,\theta(p_0)\;.
\end{equation}
Explicitly, this gives
\begin{align}
  \label{eq:kbinted}
  &\iint\!\!\frac{\D{4}{p}}{(2\pi)^4}\,\frac{\D{4}{P}}{(2\pi)^4}\;
  e^{-iP\cdot X}\,\theta(p_0)\Big[\,\big(p_0\:+\:\tfrac{P_0}{2}\big)^2\:-\:
  E^2(\mathbf{p}+\tfrac{\mathbf{P}}{2})\,\Big]
  \Delta_<(p+\tfrac{P}{2},p-\tfrac{P}{2},\tilde{t}_f;\tilde{t}_i)\nonumber\\&\qquad +\: \!\iint\!\!
  \frac{\D{4}{p}}{(2\pi)^4}\,\frac{\D{4}{P}}{(2\pi)^4}\;
  e^{-iP\cdot X}\,\theta(p_0)\,
  \mathscr{F}(p+\tfrac{P}{2},p-\tfrac{P}{2},\tilde{t}_f;\tilde{t}_i)\ = \ \!\iint\!\!
  \frac{\D{4}{p}}{(2\pi)^4}\,\frac{\D{4}{P}}{(2\pi)^4}\;
  e^{-iP\cdot X}\,\theta(p_0)\,
  \mathscr{C}(p+\tfrac{P}{2},p-\tfrac{P}{2},\tilde{t}_f;\tilde{t}_i)\;.
\end{align}
Adding  to  (\ref{eq:kbinted})  the  complex  conjugate  of  the  same
expression   with  $P\:\to\:   -P$   and  using   the  identities   in
(\ref{eq:hermitdm}), we may extract  the terms proportional to $p\cdot
P$ on the LHS of (\ref{eq:kbinted}). In~this~way, we obtain
\begin{align}
  \label{eq:kbpdotP}
  &2\!\iint\!\!\frac{\D{4}{p}}{(2\pi)^4}\,\frac{\D{4}{P}}{(2\pi)^4}
  \;e^{-iP\cdot X}\,\theta(p_0)\,p\cdot
  P\,\Delta_<(p+\tfrac{P}{2},p-\tfrac{P}{2},\tilde{t}_f;\tilde{t}_i)\nonumber\\&\quad
  +\: \!\iint\!\!\frac{\D{4}{p}}{(2\pi)^4}\,\frac{\D{4}{P}}{(2\pi)^4}
  \;e^{-iP\cdot X}\,\theta(p_0)\Big(\,
  \mathscr{F}(p+\tfrac{P}{2},p-\tfrac{P}{2},\tilde{t}_f;\tilde{t}_i)\:+\:
  \mathscr{F}^*(p-\tfrac{P}{2},p+\tfrac{P}{2},\tilde{t}_f;\tilde{t}_i)\,\Big)\nonumber\\&
  \quad \quad =\ \!\iint\!\!\frac{\D{4}{p}}{(2\pi)^4}\,
  \frac{\D{4}{P}}{(2\pi)^4}\,e^{-iP\cdot X}\,\theta(p_0)\Big(\,
  \mathscr{C}(p+\tfrac{P}{2},p-\tfrac{P}{2},\tilde{t}_f;\tilde{t}_i)\:+\:
  \mathscr{C}^*(p-\tfrac{P}{2},p+\tfrac{P}{2},\tilde{t}_f;\tilde{t}_i)\,\Big)\;.
\end{align}
The first term on the LHS of (\ref{eq:kbpdotP}) may be rewritten as
\begin{equation}
  \label{eq:drift}
  2\!\iint\!\!\frac{\D{4}{p}}{(2\pi)^4}\,\frac{\D{4}{P}}{(2\pi)^4}
  \;\theta(p_0)\Big(\,ip_0\partial_{X_0}\:-\:
  \mathbf{p}\cdot\mathbf{P}\,\Big)
  e^{-iP\cdot X}\,\Delta_<(p+\tfrac{P}{2},p-\tfrac{P}{2},\tilde{t}_f;\tilde{t}_i)\;.
\end{equation}
Using time translational invariance of  the CTP contour and taking the
limit  $X_0\:\to\:  t$  in  (\ref{eq:drift}), we  recognize  that  the
derivative term  with respect to~$t$ is  precisely the time derivative
of    the    particle     number    density    $n(\mathbf{X},t)$    in
(\ref{eq:numvol}).     Hence,    from   (\ref{eq:kbpdotP})    and
(\ref{eq:drift}), we arrive at  our master time evolution equation for
$n(\mathbf{X},t)$:
\begin{align}
  \label{eq:kbn}
  &\partial_{t}n(\mathbf{X},t)\:-\:2\!\iint\!\!\frac{\D{4}{p}}{(2\pi)^4}
  \,\frac{\D{4}{P}}{(2\pi)^4}\;e^{-iP\cdot X}\,\mathbf{p}\cdot\mathbf{P}\,
  \theta(p_0)\Delta_<(p+\tfrac{P}{2},p-\tfrac{P}{2},t;0)\nonumber\\&
  \qquad +\:\!\iint\!\!\frac{\D{4}{p}}{(2\pi)^4}\,
  \frac{\D{4}{P}}{(2\pi)^4}\;e^{-iP\cdot X}\,\theta(p_0)
  \Big(\,\mathscr{F}(p+\tfrac{P}{2},p-\tfrac{P}{2},t;0)\:+\:
  \mathscr{F}^{*}(p-\tfrac{P}{2},p+\tfrac{P}{2},t;0)\,\Big)\nonumber\\&
  \qquad \qquad =\ \!\iint\!\!\frac{\D{4}{p}}{(2\pi)^4}\,
  \frac{\D{4}{P}}{(2\pi)^4}\;e^{-iP\cdot X}\theta(p_0)
  \Big(\,\mathscr{C}(p+\tfrac{P}{2},p-\tfrac{P}{2},t;0)\:+\:
  \mathscr{C}^{*}(p-\tfrac{P}{2},p+\tfrac{P}{2},t;0)\,\Big)\;,
\end{align}
with $X_0\:=\:t$, where  $\mathscr{F}$ and $\mathscr{C}$ are defined
in (\ref{eq:fdef}) and (\ref{eq:cdef}).  Comparing with the full
non-truncated   form  of  the   Kadanoff--Baym  kinetic   equation  in
(\ref{eq:kinfull}),   the  series  of   nested  Poisson   brackets  in
(\ref{eq:kinfull})  has  been   replaced  by  a single  convolution
integral over the central momentum $P$ in (\ref{eq:kbn}).

In addition  to the  master time evolution  equation for  the particle
number    density   $n(\mathbf{X},t)$   in~(\ref{eq:kbn}),    we   may
respectively  find  a  time  evolution equation  for  the  statistical
distribution            function            $f(\mathbf{p}            +
\tfrac{\mathbf{P}}{2},\mathbf{p}-\tfrac{\mathbf{P}}{2},t)$. Specifically,
given  the  relation   (\ref{eq:nf})  and  the  differential  equation
(\ref{eq:kbn}), the  following time evolution equation  may be derived
for the \emph{resummed} statistical distribution function $f$:
\begin{align}
  \label{eq:evo}
  &\partial_{t}
  f(\mathbf{p}+\tfrac{\mathbf{P}}{2},\mathbf{p}-\tfrac{\mathbf{P}}{2},t)\:-\:
  2\!\iint\!\frac{\D{}{p_0}}{2\pi}\,\frac{\D{}{P_0}}{2\pi}\;e^{-iP_0t}
  \,\mathbf{p}\cdot\mathbf{P}\,\theta(p_0)
  \Delta_<(p+\tfrac{P}{2},p-\tfrac{P}{2},t;0)\nonumber\\&
  \qquad +\: \!\iint\!\frac{\D{}{p_0}}{2\pi}\,
  \frac{\D{}{P_0}}{2\pi}\;e^{-iP_0t}\,\theta(p_0)
  \Big(\,\mathscr{F}(p+\tfrac{P}{2},p-\tfrac{P}{2},t;0)\:+\:
  \mathscr{F}^{*}(p-\tfrac{P}{2},p+\tfrac{P}{2},t;0)\,\Big)\nonumber\\&
  \qquad \qquad =\ \!\iint\!\frac{\D{}{p_0}}{2\pi}\,
  \frac{\D{}{P_0}}{2\pi}\;e^{-iP_0t}\,\theta(p_0)
  \Big(\,\mathscr{C}(p+\tfrac{P}{2},p-\tfrac{P}{2},t;0)\:+\:
  \mathscr{C}^{*}(p-\tfrac{P}{2},p+\tfrac{P}{2},t;0)\,\Big)\;.
\end{align}
It  is  important  to  stress  here  that  (\ref{eq:evo})  provides  a
self-consistent  time evolution  equation for~$f$  valid  \emph{to all
  orders} in perturbation theory  and to \emph{all orders} in gradient
expansion.
 
We  note  that a  physical  interpretation  may  be attributed  to  the
different  terms  contributing  to~(\ref{eq:evo}).  Specifically,  all
terms on  the LHS of (\ref{eq:evo})  may be associated  with the total
derivative in the phase space $(\mathbf{X},\  \mathbf{p})$, appearing
in  the  classical  Boltzmann  transport
equation~\cite{KolbTurner}:
\begin{equation}
  D_t\ =\ \partial_t\:+\:\mathbf{v}\cdot\mathbf{\nabla}_{\mathbf{X}}\:+\:
  {\bf  F}\cdot\mathbf{\nabla}_{\mathbf{p}}\; ,
\end{equation}  
where  $\mathbf{v}$ is  the average  non-relativistic velocity  of the
particle  distribution and  ${\bf  F}$  is the  force  acting on  this
distribution.  In  particular, the $\mathscr{F}$  terms on the  LHS of
(\ref{eq:evo}) are the \emph{force}  terms, generated by the potential
due  to  the   dispersive  part  of  the  self-energy.    On  the  RHS
of~(\ref{eq:evo}),    the    $\mathscr{C}$    terms   represent    the
\emph{collision} terms.

It  would be interesting  to discuss  the spatially  homogeneous limit
of~(\ref{eq:evo})  at late  times. In  this case,  energy conservation
holds to a good  approximation, so the time-dependent weight functions
$\delta_t$ may be replaced by the standard Dirac delta functions, even
in  the   vertices  of  the  self-energies  contained   in  the  force
$\mathscr{F}$  and collision  $\mathscr{C}$  terms of  (\ref{eq:evo}).
Moreover,  as a consequence  of the  assumed spatial  homogeneity, all
propagators and self-energies, which  in general depend on two momenta
$p_1$  and $p_2$,  will now  be proportional  to  the four-dimensional
delta  function~$(2\pi)^4\delta^{(4)}  (p_1\:-\:p_2)$.  Likewise,  the
statistical distribution function~$f$ takes on the form
\begin{equation}
  f(\mathbf{p}+\tfrac{\mathbf{P}}{2},
  \mathbf{p}-\tfrac{\mathbf{P}}{2},t)\ =\ 
  (2\pi)^3\delta^{(3)}(\mathbf{P})f(|\mathbf{p}|,t)\;.
\end{equation}
Because of the above simplifications,  one can then work in the single
momentum  representation,  by   integrating  over  the  three-momentum
$\mathbf{P}$.   Thus,  we   find  the  following  time  evolution
equation for $f(|\mathbf{p}|,t)$:
\begin{equation}
  \label{eq:boltz}
  \partial_{t}f(|\mathbf{p}|,t)\ =\ \!\int\!
  \frac{\D{}{p_0}}{2\pi}\;\theta(p_0)
  \Big(\,i\Pi_>(p,t)i\Delta_<(p,t)\:-\:i\Pi_<(p,t)i\Delta_>(p,t)
  \,\Big)\;,
\end{equation}
where the purely-imaginary force term $\mathscr{F}$ and off-shell
effects from $\Delta_{\mathcal{P}}$ have vanished for $P^{\mu}\: =\:
(0,\mathbf{0})$.  This result corresponds to the semi-classical
Boltzmann transport equation or, equivalently, to the zeroth-order
truncation of the gradient-expanded Kadanoff--Baym kinetic equation in
(\ref{eq:kinetic}) with $X_0\:=\:t$.

It is a formidable task to provide an evaluation of~(\ref{eq:evo}) to
all orders in perturbation theory. For this reason, let us now
consider the perturbative loopwise truncation of
(\ref{eq:evo}). Notice however that this truncation will remain valid
to all orders in a gradient expansion.

As identified immediately below (\ref{eq:kbsimple}), in any
perturbative loopwise truncation, the external propagators must be
evaluated at the same order. At lowest order, we insert the free
propagators of Section~\ref{sec:nonhomprop} for the external legs in
(\ref{eq:evo}). By (\ref{eq:statpert}), the LHS of (\ref{eq:evo}) must
depend on the tree-level statistical distribution function $f^0$,
where we have written the superscript $0$ explicitly for clarity. For
the homogeneous limit in (\ref{eq:boltz}), we would then obtain
\begin{equation}
  \partial_{t}f^0(|\mathbf{p}|,t)\ =\ \!\int\!
  \frac{\D{}{p_0}}{2\pi}\;\theta(p_0)
  \Big(\,i\Pi_>(p,t)i\Delta^0_<(p,t)\:-\:i\Pi_<(p,t)i\Delta^0_>(p,t)
  \,\Big)\;.
\end{equation}
Hereafter, the choice of order of the external legs
$i\Delta_{\gtrless,\,\mathcal{P}}$ in (\ref{eq:evo}) is referred to as
\emph{spectral} truncation.

Notice that the external self-energies $i\Pi_{\gtrless,\,\mathcal{P}}$
may be truncated \emph{independently} at a \emph{different} order to
the external legs $i\Delta_{\gtrless,\,\mathcal{P}}$.  Inserting a
given loop order of external self-energy in (\ref{eq:evo}), we
restrict the set of processes contributing to the statistical
evolution. The choice of external self-energy is therefore referred to
as \emph{statistical} truncation. As was the case for the external
legs, the set of external self-energies must be truncated to the same
order amongst themselves. This ensures that the master time evolution
equations vanish in the equilibrium limit by virtue of the KMS
relation (\ref{eq:KMS}) and detailed balance condition (\ref{eq:db}).

The origin of these two independent perturbative truncations can be
understood by recalling that, in the interaction picture, the relevant
objects of quantum statistical mechanics are EEVs of operators of the
form
\begin{equation}
  \label{eq:eevexp}
  \mathrm{Tr}\,\rho(\tilde{t}_f;\tilde{t}_i)
  \mathcal{O}(\tilde{t}_f;\tilde{t}_i)\;.
\end{equation}
In (\ref{eq:eevexp}), there are two distinct objects: the density
operator $\rho$, describing the background ensemble, and the operator
$\mathcal{O}$, corresponding to our chosen observable. The time
evolution of $\rho$ is determined by the quantum Liouville
equation (\ref{eq:qLiou}) and $\mathcal{O}$, by the
interaction-picture analogue of the Heisenberg equation of motion. In
the context of the master time evolution equations, the perturbative
truncation of the former corresponds to the statistical truncation,
restricting the set of processes driving the background evolution. The
truncation of the latter corresponds to the spectral truncation,
determining what we have chosen to observe through
$\mathcal{O}$. Thus, with the external insertion of free propagators,
(\ref{eq:evo}) describes the statistical evolution of the number
density of \emph{spectrally-free} particles, due to a given set of
processes.

Inserting instead one-loop propagators in the external legs, the LHS
of (\ref{eq:evo}) depends on the one-loop statistical distribution
function $f^{(1)}$ by (\ref{eq:statpert}). In the homogeneous limit,
we then obtain
\begin{equation}
  \label{eq:Boltz2}
  \partial_{t}f^{(1)}(|\mathbf{p}|,t)\ =\ \!\int\!
  \frac{\D{}{p_0}}{2\pi}\;\theta(p_0)
  \Big(\,i\Pi_>(p,t)i\Delta^{(1)}_<(p,t)\:-\:i\Pi_<(p,t)i\Delta^{(1)}_>(p,t)
  \,\Big)\;.
\end{equation}
The evolution equation now describes the statistical evolution of
one-loop \emph{spectrally-dressed} particles. Again, we are free to
insert any order of self-energy.

In summary, there are two independent perturbative loopwise
truncations of (\ref{eq:evo}): (i) the spectral truncation of the
external leg determines what we have chosen to count as a ``particle''
and (ii) the statistical truncation of the external self-energy
restricts the set of processes contributing to the statistical
evolution. In the next section, we describe this perturbative loopwise
expansion explicitly at the one-loop spectral and $n$-loop statistical
level, with reference to the absence of pinch singularities and the
approach to equilibrium at late times.

\section{Perturbative One-Loop Spectral Expansion without Pinch Singularities}
\label{sec:oneloop}

Pinch singularities, or so-called secular terms, normally spoil the
perturbative expansion of non-equilibrium Green's
functions~\cite{Weldon:1991ek,Altherr:1994fx, Altherr:1994jc,Dadic:1998yd}. These mathematical
pathologies arise from ill-defined products of delta functions with
identical arguments.  In this section, we demonstrate explicitly, in
contrast to \cite{Greiner:1998ri, Berges:2004yj}, that such pinch
singularities do not occur in our perturbative approach.

The absence of pinch singularities is ensured by (a) the systematic
inclusion of finite-time effects, as shown in Figure~\ref{fig:sk}, and
(b) the proper consideration of the dependence upon the time of
observation $t$. The finite-time effects in (a) result in finite upper
and lower bounds on interaction-dependent time integrals, leading to
the microscopic violation of energy conservation. We emphasize that
these finite-time effects are not included \emph{a priori} and that
the removal of pinch singularities is achieved without any \emph{ad
  hoc} regularization or obscure resummation. In addition, we note
that our treatment differs from the semi-infinite time domains
employed in the existing literature \cite{Bedaque:1994di,
  Dadic:1999bp}. As a result of (b), the statistical distribution
functions appearing in the non-homogeneous free propagators are
evaluated at the macroscopic time $t$, as noted in
Section~\ref{sec:nonhomprop}. In this case, the role of the
Feynman--Dyson series is to dress the \emph{spectral} structure of the
propagators only. The evolution of the statistical distribution
functions is determined by the master time evolution equations,
derived in Section~\ref{sec:eom}.

For early times, the microscopic violation of energy conservation
prevents the appearance of pinch singularities. On the other hand, at
infinitely-late times, the system thermalizes and the time-dependent
statistical distribution functions are replaced by the equilibrium
Bose--Einstein distributions. In this equilibrium limit, pinch
singularities are known to cancel by virtue of the KMS relation
(\ref{eq:KMSgl})~\cite{Landsman:1986uw}. In this section, we
demonstrate the absence of pinch singularities explicitly at the
one-loop level. In addition, we illustrate how the perturbative
loopwise truncation, used in our approach, successfully captures the
dynamics on all time scales. To this end, we calculate the one-loop
non-equilibrium CTP propagator in the late-time limit.

Proceeding perturbatively, we truncate the Feynman--Dyson series in
(\ref{eq:fd}) to leading order in the couplings and set the bi-local
source $K_{ab}$ to zero. This corresponds to keeping free CTP
propagators $\Delta^{0,\,ab}(p,p',\tilde{t}_f;\tilde{t}_i)$ and
one-loop CTP self-energies
$\Pi^{(1)}_{ab}(p,p',\tilde{t}_f;\tilde{t}_i)$, containing free
propagators, on the RHS of (\ref{eq:fd}).  We then have the
one-loop-inserted CTP propagator
\begin{equation}
  \label{eq:oneins}
  \Delta^{(1),\,ab}(p,p',\tilde{t}_f;\tilde{t}_i)\ =\
  \Delta^{0,\,ab}(p,p',\tilde{t}_f;\tilde{t}_i)\:-\:
  \!\iint\!\!\frac{\D{4}{q}}{(2\pi)^4}\,
  \frac{\D{4}{q'}}{(2\pi)^4}\;
  \Delta^{0,\,ac}(p,q,\tilde{t}_f;\tilde{t}_i)
  \Pi^{(1)}_{cd}(q,q',\tilde{t}_f;\tilde{t}_i)
  \Delta^{0,\,db}(q',p',\tilde{t}_f;\tilde{t}_i)\;.
\end{equation}
It  will prove convenient  to work  in a  mixed CTP--Keldysh  basis by
inserting  the transformation  outlined in  (\ref{eq:physrep}) between
the external  legs and self-energies on the  RHS of (\ref{eq:oneins}).
In this mixed CTP--Keldysh basis, we have
\begin{equation}
  \Delta^{(1),\,ab}(p,p',\tilde{t}_f;\tilde{t}_i)\ =\
  \Delta^{0,\, ab}(p,p',\tilde{t}_f;\tilde{t}_i)\:-\:
  \frac{1}{2}\!\iint\!\!\frac{\D{4}{q}}{(2\pi)^4}\,
  \frac{\D{4}{q'}}{(2\pi)^4}\;
  \Delta_{\mathrm{ret}}^{0,\,ac}(p,q,\tilde{t}_f;\tilde{t}_i)
  {\widetilde{\Pi}}^{(1)}_{cd}(q,q',\tilde{t}_f;\tilde{t}_i)
  \Delta_{\mathrm{adv}}^{0,\,db}(q',p',\tilde{t}_f;\tilde{t}_i)\;,
\end{equation}
where,  making  use  of  the  relations  in  (\ref{eq:propidens}),
\begin{align}
  \Delta_{\mathrm{ret}}^{ac}(p,q,\tilde{t}_f;\tilde{t}_i)\ &=\
  \begin{bmatrix}
    \Delta_{\mathrm{F}}(p,q,\tilde{t}_f;\tilde{t}_i) &
    \Delta_<(p,q,\tilde{t}_f;\tilde{t}_i) \\
    \Delta_>(p,q,\tilde{t}_f;\tilde{t}_i) &
    -\Delta^*_{\mathrm{F}}(p,q,\tilde{t}_f;\tilde{t}_i)
  \end{bmatrix}\!
  \begin{bmatrix}
    1 & 1 \\ 
    -1 & 1
  \end{bmatrix}\nonumber\\
   &=\
  \begin{bmatrix}
    \Delta_{\mathrm{R}}(p,q,\tilde{t}_f;\tilde{t_i}) & \
    \Delta_{\mathrm{R}}(p,q,\tilde{t}_f;\tilde{t}_i)
    \:+\:2\Delta_{<}(p,q,\tilde{t}_f;\tilde{t}_i) \ \\
    \Delta_{\mathrm{R}}(p,q,\tilde{t}_f;\tilde{t}_i) & \
    -\Delta_{\mathrm{R}}(p,q,\tilde{t}_f;\tilde{t}_i)
    \:+\:2\Delta_>(p,q,\tilde{t}_f;\tilde{t}_i) \
  \end{bmatrix}\!\;,
\end{align}
and
\begin{align}
  \Delta_{\mathrm{adv}}^{ac}(p,q,\tilde{t}_f;\tilde{t}_i)\ &=\
  \begin{bmatrix}
    1 & -1 \\
    1 & 1
  \end{bmatrix}\!
  \begin{bmatrix}
    \Delta_{\mathrm{F}}(p,q,\tilde{t}_f;\tilde{t}_i) &
    \Delta_<(p,q,\tilde{t}_f;\tilde{t}_i) \\ 
    \Delta_>(p,q,\tilde{t}_f;\tilde{t}_i) &
    -\Delta^*_{\mathrm{F}}(p,q,\tilde{t}_f;\tilde{t}_i)
  \end{bmatrix}\nonumber\\
   &=\
  \begin{bmatrix}
    \Delta_{\mathrm{A}}(p,q,\tilde{t}_f;\tilde{t}_i) & \
    \Delta_{\mathrm{A}}(p,q,\tilde{t}_f;\tilde{t}_i) \ \\
    \Delta_{\mathrm{A}}(p,q,\tilde{t}_f;\tilde{t}_i)
    \:+\:2\Delta_>(p,q,\tilde{t}_f;\tilde{t}_i) &
    \ -\Delta_{\mathrm{A}}(p,q,\tilde{t}_f;\tilde{t}_i)
    \:+\:2\Delta_<(p,q,\tilde{t}_f;\tilde{t}_i) \
  \end{bmatrix}\!\;.
\end{align}

In the  same mixed  CTP--Keldysh basis, the  one-loop-inserted Feynman
$\Delta_{\mathrm{F}}^{(1)}(p,p',\tilde{t}_f;\tilde{t}_i)$           and
                                             Wightman
$\Delta_{\gtrless}^{(1)}(p,p',\tilde{t}_f;\tilde{t}_i)$  propagators  may  be
written in the following forms:
\begin{subequations}
\begin{align}
  \label{eq:F1}
  &\Delta^{(1)}_{\mathrm{F}}(p,p',\tilde{t}_f;\tilde{t}_i)\ =\ \Delta^0_{\mathrm{F}}(p,p',\tilde{t}_f;\tilde{t}_i)
  \:-\:\!\iint\!\!\frac{\D{4}{q}}{(2\pi)^4}\,\frac{\D{4}{q'}}
  {(2\pi)^4}\;\Big(\,\Delta_{\mathrm{R}}^0(p,q)\Pi^{(1)}(q,q',\tilde{t}_f;\tilde{t}_i)
  \Delta_{\mathrm{A}}^0(q',p')\nonumber\\
  &\qquad +\: \Delta_{\mathrm{R}}^0(p,q){\Pi}_{\mathrm{R}}^{(1)}(q,q',\tilde{t}_f;\tilde{t}_i)
  \Delta_{>}^0(q',p',\tilde{t}_f;\tilde{t}_i)\:+\:\Delta_{<}^0(p,q,\tilde{t}_f;\tilde{t}_i)
  {\Pi}_{\mathrm{A}}^{(1)}(q,q',\tilde{t}_f;\tilde{t}_i)
  \Delta_{\mathrm{A}}^0(q',p')\,\Big)\;,\\
  \label{eq:gr1}
  &\Delta^{(1)}_{\gtrless}(p,p',\tilde{t}_f;\tilde{t}_i)\ =\ \Delta^0_{\gtrless}(p,p',\tilde{t}_f;\tilde{t}_i)
  \:-\:\!\iint\!\!\frac{\D{4}{q}}{(2\pi)^4}\,\frac{\D{4}{q'}}
  {(2\pi)^4}\;\Big(\,\Delta_{\mathrm{R}}^0(p,q)\Pi^{(1)}_{\gtrless}(q,q',\tilde{t}_f;\tilde{t}_i)
  \Delta_{\mathrm{A}}^0(q',p')\nonumber\\
  &\qquad +\: \Delta_{\mathrm{R}}^0(p,q){\Pi}_{\mathrm{R}}^{(1)}(q,q',\tilde{t}_f;\tilde{t}_i)
  \Delta_{\gtrless}^0(q',p',\tilde{t}_f;\tilde{t}_i)\:+\:\Delta_{\gtrless}^0(p,q,\tilde{t}_f;\tilde{t}_i)
  {\Pi}_{\mathrm{A}}^{(1)}(q,q',\tilde{t}_f;\tilde{t}_i)\Delta_{\mathrm{A}}^0(q',p')\,\Big)\;,
\end{align}
\end{subequations}
where we have used the identities
\begin{subequations}
  \begin{align}
    \Pi_1 (p,p',\tilde{t}_f;\tilde{t}_i)\:+\:
    \Pi_{\mathrm{R}}(p,p',\tilde{t}_f;\tilde{t}_i)\:+\:
    \Pi_{\mathrm{A}}(p,p',\tilde{t}_f;\tilde{t}_i)
    &\ =\ 2\Pi (p,p',\tilde{t}_f;\tilde{t}_i)\;,\\
    \Pi_1(p,p',\tilde{t}_f;\tilde{t}_i)\:\pm\:
    \Pi_{\mathrm{R}}(p,p',\tilde{t}_f;\tilde{t}_i)\:\mp\:
    \Pi_{\mathrm{A}}(p,p',\tilde{t}_f;\tilde{t}_i)
    &\ =\ 2\Pi_{\gtrless}(p,p',\tilde{t}_f;\tilde{t}_i)\;,
  \end{align}
\end{subequations}
which  can  be  derived   from  relations  between  the  self-energies
analogous to those in (\ref{eq:propidens}).

At late times or, equivalently, infinitesimal departures from
equilibrium, the free CTP propagator is spatially homogeneous and may
be written as
\begin{equation}
  \label{eq:homlim}
  \Delta^{0,\,ab}(p,p',\tilde{t}_f;\tilde{t}_i)\
  \simeq\ \Delta^{0,\,ab}(p,t)(2\pi)^4\delta^{(4)}(p-p')\;.
\end{equation}
In addition, the one-loop self-energy may
be written as
\begin{equation}
\label{eq:Pipinch}
 \Pi^{(1)}_{ab}(p,p',\tilde{t}_f;\tilde{t}_i)\ \simeq\ 
 \Pi^{(1)}_{ab}(p,t)(2\pi)^4\delta_t^{(4)}(p-p')e^{i(p_0-p_0')\tilde{t}_f}\;.
\end{equation}
We note that the one-loop CTP self-energy
$i\Pi_{ab}^{(1)}(p,p',\tilde{t}_f;\tilde{t}_i)$ is manifestly free of
pinch singularities for all times (see Appendices~\ref{app:ITF} and
\ref{app:loops}).  Using (\ref{eq:homlim}) and (\ref{eq:Pipinch}), the
one-loop expansions in (\ref{eq:F1}) and (\ref{eq:gr1}) become
\begin{subequations}
\begin{align}
  \label{eq:F2}
  &\Delta^{(1)}_{\mathrm{F}}(p,p',\tilde{t}_f;\tilde{t}_i)\ =\
  e^{i(p_0-p_0')\tilde{t}_f}\bigg[\Delta^0_{\mathrm{F}}(p,t)(2\pi)^4\delta^{(4)}(p-p')
  \:-\:\Big(\,\Delta_{\mathrm{R}}^0(p)\Pi^{(1)}(p,t)
  \Delta_{\mathrm{A}}^0(p')\nonumber\\
  &\qquad +\: \Delta_{\mathrm{R}}^0(p){\Pi}_{\mathrm{R}}^{(1)}(p,t)
  \Delta_{>}^0(p',t)\:+\:\Delta_{<}^0(p,t)
  {\Pi}_{\mathrm{A}}^{(1)}(p,t)
  \Delta_{\mathrm{A}}^0(p')\,\Big)(2\pi)^4\delta^{(4)}_t(p-p')\bigg]\;,\\
  \label{eq:gr2}
  &\Delta^{(1)}_{\gtrless}(p,p',\tilde{t}_f;\tilde{t}_i)\ =\
  e^{i(p_0-p_0')\tilde{t}_f}\bigg[\Delta^0_{\gtrless}(p,t)(2\pi)^4\delta^{(4)}(p-p')
  \:-\:\Big(\,\Delta_{\mathrm{R}}^0(p)\Pi^{(1)}_{\gtrless}(p,t)
  \Delta_{\mathrm{A}}^0(p')\nonumber\\
  &\qquad +\: \Delta_{\mathrm{R}}^0(p){\Pi}_{\mathrm{R}}^{(1)}(p,t)
  \Delta_{\gtrless}^0(p',t)\:+\:\Delta_{\gtrless}^0(p,t)
  {\Pi}_{\mathrm{A}}^{(1)}(p,t)\Delta_{\mathrm{A}}^0(p')\,\Big)(2\pi)^4\delta^{(4)}(p-p')\bigg]\;,
\end{align}
\end{subequations}
We note the appearance of the overall phase
$e^{i(p_0-p_0')\tilde{t}_f}$ in (\ref{eq:F2}) and (\ref{eq:gr2}). This
free phase ensures that the inverse Fourier transform of
$\Delta^{(1),\,ab}(p,p',\tilde{t}_f;\tilde{t}_i)$ with respect to
$p_0$ and $p_0'$ depends only on the macroscopic time $t\:=\:\tilde{t}_f\:-\:\tilde{t}_i$ in the
equal-time limit, i.e.
\begin{equation}
  \Delta^{(1),\,ab}(\mathbf{x},\mathbf{y},t)\ \equiv\
  \Delta^{(1),\,ab}(x,y,\tilde{t}_f;\tilde{t}_i)\big|_{x_0\:=\:y_0=\:\tilde{t}_f}\ =\
  \lim_{x_0,\,y_0\,\to\,\tilde{t}_f}
  \iint\!\frac{\D{4}{p}}{(2\pi)^4}\frac{\D{4}{p'}}{(2\pi)^4}\;
  e^{-ip\cdot x}e^{ip'\cdot
    y}\Delta^{(1),\,ab}(p,p',\tilde{t}_f;\tilde{t}_i)\;.
\end{equation}

Expanding around the equilibrium Bose--Einstein distribution
$f_{\mathrm{B}}$, we write the spectrally-free statistical
distribution function
\begin{equation}
  \label{eq:deltaf}
  f^0(|\mathbf{p}|,t)\ =\ f_{\mathrm{B}}\big(E(\mathbf{p})\big)
  \:+\:\delta\!f^0(|\mathbf{p}|,t)\;.
\end{equation}
Using the expansion in (\ref{eq:deltaf}), the single-momentum
representation of the spatially-homogeneous free CTP propagator
$i\Delta^{0,\,ab}(p,t)$ in (\ref{eq:homlim}) may be decomposed as
\begin{equation}
  \label{eq:propexp}
  i\Delta^{0,\,ab}(p,t)\ =\ i\Delta^{0,\,ab}_{\,\mathrm{eq}}(p)\:+
  \:2\pi\delta(p^2-M^2)\,\delta\!f^0(|\mathbf{p}|,t)\!
  \begin{bmatrix}
    1 & 1 \\ 1 & 1
  \end{bmatrix}\!\;,
\end{equation}
where $i\Delta^{0,\,ab}_{\,\mathrm{eq}}(p)$ is the free equilibrium
CTP propagator discussed in Section~\ref{sec:eq}. Analogously, we may
introduce the following decomposition of the homogeneous one-loop
self-energy $\Pi^{(1)}_{ab}(p,t)$:
\begin{equation}
  \label{eq:selfexp}
  i\Pi^{(1)}_{ab}(p,t)\ =\
  i\Pi^{(1)}_{\mathrm{eq},\,ab}(p)\:+\:i\delta\Pi^{(1)}_{ab}(p,t)\;,
\end{equation}
where $\delta\Pi^{(1)}_{ab}(p,t)$ contains terms of order $\delta\! f^0$
and higher. Substituting the expansions (\ref{eq:propexp}) and
(\ref{eq:selfexp}) into (\ref{eq:F2}), we obtain, for the one-loop
Feynman propagator,
\begin{align}
  \label{eq:Feynexp}
  &i\Delta_{\mathrm{F}}^{(1)}(p,p',\tilde{t}_f;\tilde{t}_i)\ =\
  \Big(i\Delta^{(1)}_{\mathrm{F},\,\mathrm{eq}}(p)\:+\:
  2\pi\delta(p^2-M^2)\delta\!f^0(|\mathbf{p}|,t)\Big)(2\pi)^4\delta^{(4)}(p-p')
  \nonumber\\&\qquad +\
  e^{i(p_0-p_0')\tilde{t}_f}\Big(i\Delta_{\mathrm{R}}^0(p)i\Pi_{\mathrm{R}}^{(1)}(p,t)
  2\pi\delta(p'^2-M^2)\delta\!f^0(|\mathbf{p}'|,t)\:+\:
  2\pi\delta(p^2-M^2)\delta\!f^0(|\mathbf{p}|,t)
  i\Pi_{\mathrm{A}}^{(1)}(p,t)i\Delta_{\mathrm{A}}^0(p')\nonumber\\&\qquad \qquad
  +\:i\Delta_{\mathrm{R}}^0(p)i\delta\Pi^{(1)}(p,t)i\Delta_{\mathrm{A}}^0(p')\:
  +\:i\Delta_{\mathrm{R}}^0(p)i\delta\Pi^{(1)}_{\mathrm{R}}(p,t)i\Delta^0_{>,\,\mathrm{eq}}(p')\:
  +\:i\Delta_{<,\,\mathrm{eq}}^0(p)i\delta\Pi_{\mathrm{A}}^{(1)}(p,t)i\Delta_{\mathrm{A}}^0(p')\Big)
  \nonumber\\&\qquad \qquad\times\:(2\pi)^4\delta_t^{(4)}(p-p')\;.
\end{align}
Potential pinch singularities arise only from the $\delta\!f^0$ and
$\delta\Pi$ dependent terms in (\ref{eq:Feynexp}), since pinch
singularities cancel in the equilibrium contribution
$i\Delta^{(1)}_{\mathrm{F},\,\mathrm{eq}}(p)$, as we will see below.

Let us first consider the terms
\begin{align}
  \label{eq:danger}
  &i\Delta_{\mathrm{R}}^0(p)i\Pi^{(1)}_{\mathrm{R}}(p,t)(2\pi)^4\delta_t^{(4)}(p-p')
  2\pi\delta\!f^0(|\mathbf{p}'|,t)\delta(p'^2-M^2)\nonumber\\
  &\qquad +\:2\pi\delta\!f^0(|\mathbf{p}|,t)\delta(p^2-M^2)i\Pi^{(1)}_{\mathrm{A}}(p,t)
  (2\pi)^4\delta_t^{(4)}(p-p')i\Delta^0_{\mathrm{A}}(p')\;.
\end{align}
The real and imaginary parts of the free retarded propagator
$\Delta_{\mathrm{R}}^0(p)$ are given by
\begin{equation}
  \mathrm{Re}\,\Delta_{\mathrm{R}}^0(p)\ =\ \mathcal{P}
  \frac{1}{p^2\:-\:M^2}\;,
\end{equation}
where $\mathcal{P}$ denotes the principal value integral, and
\begin{equation}
  \label{eq:imdeltar}
  \mathrm{Im}\,\Delta_{\mathrm{R}}^0(p)\ =\ -\pi\varepsilon(p_0)
  \delta(p^2\:-\:M^2)\;.
\end{equation}
By considering the limit representation of the Cauchy principal value
\begin{equation}
  \mathcal{P}\frac{1}{x}\ =\ \lim_{\epsilon\:\to\: 0}\,
  \frac{x}{x^2\:+\:\epsilon^2}\;
\end{equation}
and   the   limit   representation    of   the   delta   function   in
(\ref{eq:limdel}), we may then show that the product
\begin{equation}
  \label{eq:remtermb}
  \mathrm{Re}\,\Delta_{\mathrm{R}}^0(p)\,\mathrm{Im}\,\Delta_{\mathrm{R}}^0(p)
  \ =\ \frac{\pi}{2}\varepsilon(p_0)\delta'(p^2\:-\:M^2)\;,
\end{equation}
where $\delta'(x)$ is the derivative of the delta function, satisfying
\begin{equation}
  \int_{-\infty}^{+\infty}\!\!\D{}{x}\; \delta'(x)y(x)\ =\ -y'(0)\; ,
\end{equation}
provided the function $y(x)$ is analytically well-behaved at
$x\:=\:0$. Hence, for late times, in which only the potential pinching
regime $p_0\:=\:p_0'$ survives, the terms in (\ref{eq:danger}) yield
\begin{align}
  \label{eq:danger2}
  2\pi\bigg(\mathrm{Re}\,\Pi^{(1)}_{\mathrm{R}}(p,t)\delta'(p^2-M^2)\:-\:
  \frac{t}{E(\mathbf{p})}\,\varepsilon(p_0)
  \mathrm{Im}\,\Pi^{(1)}_{\mathrm{R}}(p,t)\delta(p^2-M^2)\bigg)
  \delta\!f^0(|\mathbf{p}|,t)(2\pi)^4\delta^{(4)}(p-p')\;,
\end{align}
where we have used the fact that
\begin{equation}
  2\pi\delta_{t}(p_0-p_0')\big|_{p_0\:=\:p_0'}\ =\ t\;,
\end{equation}
by l'H\^{o}pital's rule.

Proceeding similarly for the $\delta\Pi$ dependent terms in
(\ref{eq:Feynexp}), we obtain the contribution
\begin{align}
  \label{eq:selfdang}
  &\bigg[-\big(\hat{\Delta}^0(p)\big)^2i\delta\Pi^{(1)}(p,t)\:+\:
    2\pi f_{\mathrm{B}}(|p_0|)\delta'(p^2-M^2)\delta\hat{\Pi}^{*(1)}(p,t)\:+\:
    \pi\bigg(\frac{t}{E(\mathbf{p})}\,\delta(p^2-M^2)\:-\:i\delta'(p^2-M^2)\bigg)
    \nonumber\\&\qquad\times\:\bigg(\mathrm{Im}\,\Pi^{(1)}(p,t)\:-\:
    \varepsilon(p_0)\big(1\:+\:2f_{\mathrm{B}}(|p_0|)\big)
    \mathrm{Im}\,\Pi^{(1)}_{\mathrm{R}}(p,t)\bigg)\bigg](2\pi)^4\delta^{(4)}(p-p')\;.
\end{align}

Putting everything back together, we obtain the one-loop Feynman
propagator
\begin{align}
  \label{eq:Feynfin}
  &i\Delta_{\mathrm{F}}^{(1)}(p,t)\ =\
  \frac{i\big(p^2-M^2+i\epsilon-\Pi^{(1)}(p,t)\big)}{\big(p^2-M^2+i\epsilon\big)^2}\nonumber\\&\qquad+\:
  2\pi\delta(p^2-M^2)\bigg[f^0(|\mathbf{p}|,t)\:-\:
    \frac{t}{2E(\mathbf{p})}\bigg(\varepsilon(p_0)\big(1\:+\:2f^0(|\mathbf{p}|,t)\big)
    \mathrm{Im}\,\Pi^{(1)}_{\mathrm{R}}(p,t)\:-\:\mathrm{Im}\,\Pi^{(1)}(p,t)\bigg)\bigg]\nonumber\\&\qquad+\:2\pi\delta'(p^2-M^2)
  \bigg[f^0(|\mathbf{p}|,t)\hat{\Pi}^{*(1)}(p,t)\:+\:\frac{i}{2}\Big(\varepsilon(p_0)\big(1\:+\:2f^0(|\mathbf{p}|,t)\big)
    \mathrm{Im}\,\Pi^{(1)}_{\mathrm{R}}(p,t)\:-\:\mathrm{Im}\,\Pi^{(1)}(p,t)\bigg)\bigg]\;,
\end{align}
where 
\begin{equation}
  \widehat{\Pi}(p)\ =\ \mathrm{Re}\,\Pi_{\mathrm{R}}(p)\:+\:
  i\varepsilon(p_0)\mathrm{Im}\,\Pi_{\mathrm{R}}(p)
\end{equation}
is a self-energy-like function, bearing by itself no direct physical
meaning. In the thermodynamic equilibrium limit, the
fluctuation-dissipation theorem in (\ref{eq:flucdis}) is restored and
the term linear in $t$ in (\ref{eq:Feynfin}) vanishes by virtue of
(\ref{eq:pirels}). We then obtain the equilibrium one-loop Feynman
propagator, which is free of pinch singularities. Notice that in the
zero-temperature limit, only the first term in (\ref{eq:Feynfin})
survives, as we would expect.

For the one-loop Wightman propagators, we obtain
\begin{align}
  \label{eq:Wightfin}
  &i\Delta_{\gtrless}^{(1)}(p,t)\ =\
  \frac{-\:i\Pi_{\gtrless}^{(1)}(p,t)}{\big(p^2-M^2+i\epsilon\big)^2}\nonumber\\&\ +\:
  2\pi\delta(p^2-M^2)\bigg[\Big(\theta(\pm p_0)\:+\:f^0(|\mathbf{p}|,t)\Big)\:-\:
    \frac{t}{2E(\mathbf{p})}\bigg(2\varepsilon(p_0)\big(\theta(\pm p_0)\:
    +\:f^0(|\mathbf{p}|,t)\big)
    \mathrm{Im}\,\Pi^{(1)}_{\mathrm{R}}(p,t)\:
    +\:i\Pi_{\gtrless}^{(1)}(p,t)\bigg)\bigg]\nonumber\\&\ 
  +\:2\pi\delta'(p^2-M^2)
  \bigg[\Big(\theta(\pm p_0)\:+\:f^0(|\mathbf{p}|,t)\Big)\hat{\Pi}^{*(1)}(p,t)\:
    +\:\frac{i}{2}\bigg(2\varepsilon(p_0)\big(\theta(\pm p_0)\:
    +\:f^0(|\mathbf{p}|,t)\big)
    \mathrm{Im}\,\Pi^{(1)}_{\mathrm{R}}(p,t)\:+\:i\Pi_{\gtrless}^{(1)}(p,t)\bigg)\bigg]\;
\end{align}
in which the potential pinch singularity again cancels in the
equilibrium limit by virtue of (\ref{eq:pirels}).  The one-loop
propagators in (\ref{eq:Feynfin}) and (\ref{eq:Wightfin}) are
consistent with the properties and relations in
Appendix~\ref{app:rel}.

From (\ref{eq:danger2}), we see that these potential pinch
singularities are controlled by $t\delta\!f^0(|\mathbf{p}|,t)$. In
addition, the potential pinch singularities are proportional to the
Breit--Wigner width
$\Gamma^{(1)}(p,t)\:=\:\mathrm{Im}\,\Pi_{\mathrm{R}}^{(1)}(p,t)/M$,
illustrating that the origin of these dangerous terms lies in the
resummation of absorptive effects. This is consistent with conclusions
in existing approaches that pinch singularities arise as a result of
Fermi's Golden Rule~\cite{Greiner:1998ri}.

In order to show that pinch singularities do not appear in this
one-loop spectral expansion for late, but nonetheless finite times, we
must show that $f^0$ approaches equilibrium more rapidly than a power
law in $t$. Moreover, in order to demonstrate that we capture the
late-time dynamics correctly, the approach to equilibrium of $f^0$
must be such that the appearance of terms linear in $t$ does not lead
to time-sensitivity in the perturbative truncation. For instance, one
might be concerned that if
$\delta\!f^0(|\mathbf{p}|,t)\:\sim\:e^{-\Gamma(p)t}$, the $n$-th order
truncation would contain terms controlled by $t^ne^{-\Gamma(p)t}$,
delaying the thermalization of the system to later and later times.
In order to show that this is not the case, we will now consider the
master time evolution equation of the one-loop spectrally-dressed
statistical distribution function $f^{(1)}$ for late times.

By inserting the one-loop negative frequency Wightman propagator from
(\ref{eq:Wightfin}) into (\ref{eq:statpert}), we obtain the one-loop
spectrally-dressed statistical distribution $f^{(1)}$:
\begin{align}
  \label{eq:statf1a}
  &f^{(1)}(|\mathbf{p}|,t)\ =\ f^0(|\mathbf{p}|,t)
  \bigg(1-\frac{1}{2E(\mathbf{p})}\,\frac{\partial}{\partial
    p_0}\,\frac{p_0}{E(\mathbf{p})}
  \,\mathrm{Re}\,\Pi^{(1)}_{\mathrm{R}}(p,t)
  \:-\:\frac{t}{E(\mathbf{p})}\,\mathrm{Im}\,\Pi^{(1)}_{\mathrm{R}}(p,t)\bigg)\bigg|_{p_0\:=\:E(\mathbf{p})}
  \nonumber\\&\qquad
  -\:\frac{t}{2E(\mathbf{p})}\,i\Pi_{<}^{(1)}(p,t)\bigg|_{p_0\:=\:E(\mathbf{p})}
  \:+\:\frac{1}{4E(\mathbf{p})}\,\frac{\partial}{\partial
    p_0}\,\frac{p_0}{E(\mathbf{p})}\,\Pi_<^{(1)}(p,t)
  \:-\:\!\int\;\frac{\D{}{p_0}}{2\pi}\;2\theta(p_0)p_0\,
  \frac{i\Pi^{(1)}_{<}(p,t)}{\big(p^2-M^2+i\epsilon\big)^2}\;,
\end{align}
where we reiterate that the one-loop self-energy $\Pi^{(1)}$ contains
free propagators. Using the fact that
\begin{equation}
  \frac{1}{\big(p^2-M^2+i\epsilon\big)^2}\ =\
  \mathcal{P}\frac{1}{\big(p^2-M^2\big)^2}
  \:+\:i\pi\,\delta'(p^2-M^2)\;,
\end{equation}
the imaginary parts of the last two terms on the RHS of
(\ref{eq:statf1a}) cancel and we obtain
\begin{align}
  \label{eq:statf1b}
  f^{(1)}(|\mathbf{p}|,t)\ &=\ f^0(|\mathbf{p}|,t)
  \bigg(1-\frac{1}{2E(\mathbf{p})}\,\frac{\partial}{\partial p_0}\,\frac{p_0}{E(\mathbf{p})}
    \,\mathrm{Re}\,\Pi^{(1)}_{\mathrm{R}}(p,t)
    \:-\:\frac{t}{E(\mathbf{p})}\,\mathrm{Im}\,\Pi^{(1)}_{\mathrm{R}}(p,t)\bigg)\bigg|_{p_0\:=\:E(\mathbf{p})}
    \nonumber\\&\qquad -\:\frac{t}{2E(\mathbf{p})}\,i\Pi_{<}^{(1)}(p,t)\bigg|_{p_0\:=\:E(\mathbf{p})}
    \:-\:\!\int_{0}^{+\infty}\!\frac{\D{}{p_0^2}}{2\pi}\;\theta(p_0)\mathcal{P}
    \frac{i\Pi^{(1)}_{<}(p,t)}{\big(p^2-M^2\big)^2}\;.
\end{align}
In (\ref{eq:statf1b}), the final term on the RHS counts all off-shell
contributions with $p_0\:\neq\:E(\mathbf{p})\:>\:0$. Notice that the
terms linear in $t$ in $f^{(1)}$ cancel in equilibrium by virtue of
(\ref{eq:pirels}).

We truncate the Markovian approximation of the master time evolution
equation in (\ref{eq:Boltz2}) spectrally and statistically at the one-
and $n$-loop levels, respectively. Hereafter,
we neglect the off-shell and dispersive contributions in (\ref{eq:statf1b}). With this simplification, we have
\begin{align}
  \label{eq:f1evo}
  \frac{\D{}{f^{(1)}(|\mathbf{p}|,t)}}{\D{}{t}}\ \simeq\ &
  -\:\frac{M}{E(\mathbf{p})}\,\Gamma_>^{(n)}(p,t)f^{(1)}(|\mathbf{p}|,t)\:
  +\:\frac{M}{E(\mathbf{p})}\,\Gamma_<^{(n)}(p,t)\Big(1+f^{(1)}(|\mathbf{p}|,t)\Big)\;.
\end{align}
where the partial widths
{$\Gamma_{\gtrless}^{(n)}(p,t)\:=\:-i\Pi^{(n)}_{\gtrless}(p,t)/2M$}
relate to the absorptive part of the respective $n$-loop
self-energies. In (\ref{eq:f1evo}), the four-momentum $p$ is
understood to be on-shell with $p_0\:=\:E(\mathbf{p})$. Substituting
(\ref{eq:statf1b}) into (\ref{eq:f1evo}) and approximating the partial
widths by their equilibrium values, we obtain the following evolution
equation for $f^0(|\mathbf{p}|,t)$:
\begin{align}
  \frac{\D{}{f^{0}(|\mathbf{p}|,t)}}{\D{}{t}}\ &\simeq \
  -\:\frac{M}{E(\mathbf{p})}\Big(\Gamma^{(n)}_{\mathrm{eq}}(p)
  f^0(|\mathbf{p}|,t)\:-\:\Gamma^{(n)}_{<,\,\mathrm{eq}}(p)\Big)\,\frac{1}{1-\frac{M}{E(\mathbf{p})}
  t\Gamma^{(1)}_{\mathrm{eq}}(p)}\nonumber\\&\qquad \qquad +\:
  \frac{M}{E(\mathbf{p})}\,\Big(\Gamma^{(1)}_{\mathrm{eq}}(p)
  f^0(|\mathbf{p}|,t)\:-\:\Gamma^{(1)}_{<,\,\mathrm{eq}}(p)\Big)\,\frac{1+\frac{M}{E(\mathbf{p})}
  t\Gamma^{(n)}_{\mathrm{eq}}(p)}{1-\frac{M}{E(\mathbf{p})}
  t\Gamma^{(1)}_{\mathrm{eq}}(p)}\;. 
\end{align}
Using the expansion in (\ref{eq:deltaf}), the terms proportional to
$f_{\mathrm{B}}\big(E(\mathbf{p})\big)$ cancel by virtue of the
identities in (\ref{eq:pirels}) and we find that the deviation from
equilibrium is
\begin{equation}
  \label{eq:deltafsol}
  \delta\!f^0(|\mathbf{p}|,t)\ =\
  \frac{e^{-\frac{M}{E(\mathbf{p})}\Gamma^{(n)}_{\mathrm{eq}}(p)t}}{1-\frac{M}{E(\mathbf{p})}
  \Gamma^{(1)}_{\mathrm{eq}}(p)t}\,\delta\!f^0(|\mathbf{p}|,t_0)\;,
\end{equation}
valid for 
\begin{equation}
  t\ >\ t_0\ \gg\ \frac{1}{\frac{M}{E(\mathbf{p})}\Gamma^{(1)}_{\mathrm{eq}}(p)}\;.
\end{equation}
Note that the factor
$(1-\frac{M}{E(\mathbf{p})}\Gamma_{\mathrm{eq}}^{(1)}(p)t)^{-1}$ in
(\ref{eq:deltafsol}) originates from threshold effects, see
e.g. \cite{Joichi:1997xn}, and becomes singular at
$t\:=\:(\frac{M}{E(\mathbf{p})}\Gamma_{\mathrm{eq}}^{(1)}(p))^{-1}$. However,
this singularity is cancelled in the one-loop spectrally-corrected
distribution function $f^{(1)}$. Returning then to (\ref{eq:danger2}),
we see that for times $t\:>\:t_0$, the potential pinch singularity
goes like
\begin{equation}
  -2\pi\frac{t}{E(\mathbf{p})}\,\varepsilon(p_0)
  \mathrm{Im}\,\Pi^{(1)}_{\mathrm{R}}(p,t)\delta(p^2-M^2)
  \delta\!f^0(|\mathbf{p}|,t)\ \longrightarrow
  \ 2\pi\varepsilon(p_0)\delta(p^2-M^2)\,\frac{\Gamma^{(1)}(p,t)}{\Gamma^{(1)}_{\mathrm{eq}}(p)}
  \,e^{-\frac{M}{E(\mathbf{p})}\Gamma^{(n)}_{\mathrm{eq}}(p)t}\delta\!f^0(|\mathbf{p}|,t_0)\;.
\end{equation}
We conclude therefore that the terms linear in $t$ appearing in
(\ref{eq:Feynfin}) and (\ref{eq:Wightfin}) do not lead to
time-sensitivity in the perturbative loopwise truncation of the master
time evolution equations and that the late-time dynamics is correctly
captured.

In summary, for early times, the analytic $t$-dependent vertices lead
to microscopic violation of energy conservation.  This energy
non-conservation regularizes potential pinch singularities.  For
intermediate times, the free time-dependent statistical distribution
functions evolve towards equilibrium. For times
$t\:\gtrsim\:1/\Gamma$, the approach to equilibrium occurs faster than
energy conservation is restored and such that the perturbative
truncation does not induce time-sensitivity.  In the limit $t\:\to\:
\infty$, the time-dependent statistical distribution functions
appearing in the non-homogeneous free propagators are replaced by
their equilibrium forms via the correspondence in (\ref{eq:eqlim}).
We then obtain the well-known equilibrium thermal field theory in
which energy conservation is fully restored and pinch singularities
cancel exactly by virtue of the KMS relation.  In conclusion, we have
demonstrated explicitly at the one-loop level how pinch singularities
do not arise in our perturbative approach.

\section{Thermalization in a Scalar Model}
\label{sec:toy}

We now apply  the formalism developed in the  proceeding sections to a
simple scalar model. In  particular, we introduce the modified Feynman
rules that result from the systematic inclusion of finite-time effects
and  the   violation  of  both  energy   conservation  and  space-time
translational invariance. As a  playground for studying the kinematics
in  the  early-time  energy-non-conserving  regime, we  calculate  the
time-dependent thermal width  of a heavy scalar $\Phi$.   We show that
processes,   which  would   normally   be  kinematically   disallowed,
contribute  significantly to  the  prompt \emph{shock  regime} of  the
initial evolution.  We also  show that the subsequent dynamics exhibit
non-Markovian  behavior, acquiring  oscillations  with time-dependent
frequencies.   This  evolution   signifies  the  occurrence  of  memory
effects,  as   are  expected  in   truly  out-of-equilibrium  systems.
Finally, we  look in  more detail at  the time evolution  equations of
this simple model  and demonstrate the importance of  the violation of
energy conservation to the statistical dynamics.

We consider  a simple  scalar theory, which  comprises one  heavy real
scalar  field $\Phi$  and  one  light pair  of  complex scalar  fields
$(\chi^{\dag}$, $\chi)$, described by the Lagrangian
\begin{equation}
  \label{eq:model}
  \mathcal{L}(x)\ =\ \tfrac{1}{2}\partial_{\mu}\Phi(x)\partial^{\mu}
  \Phi(x)\:-\:\tfrac{1}{2}M^2\Phi^2(x)\:+\:\partial_{\mu}
  \chi^{\dag}(x)\partial^{\mu}\chi(x)\:-\:m^2\chi^{\dag}(x)\chi(x)
  \: -\: g\Phi(x)\chi^{\dag}(x)\chi(x)\:-\:
  \tfrac{1}{4}\lambda\big[\chi^{\dag}(x)\chi(x)\big]^2\;,
\end{equation}
where  $M\:\gg\:  m$.   Appendix~\ref{app:complex} describes  the
generalization of our approach to the complex scalar field $\chi$.

We  formulate  a perturbative  approach based  upon  the following
\emph{modified} Feynman rules:

\begin{itemize}

\item sum over all topologically distinct diagrams at a given order in
  perturbation theory.

\item assign to each $\Phi$-propagator line a factor of
\begin{equation*}
\parbox[][35pt][t]{110pt}{\centering
  \includegraphics{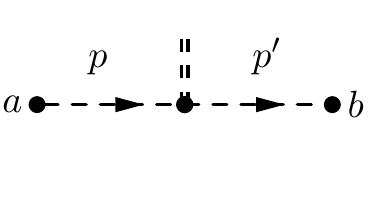}}\ =\
  i\Delta^{0,\,ab}_{\Phi}(p,p',\tilde{t}_f;\tilde{t}_i)\;.
\end{equation*}

\item assign to each $\chi$-propagator line a factor of
\begin{equation*}
\parbox[][35pt][t]{110pt}{\centering
  \includegraphics{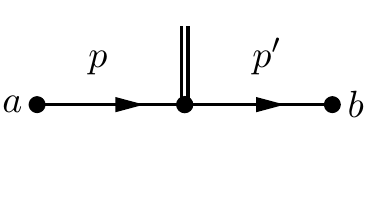}}\ =\
  i\Delta^{0,\,ab}_{\chi}(p,p',\tilde{t}_f;\tilde{t}_i)\;.
\end{equation*}

\item assign to each three-point $\Phi$ vertex a factor of
\begin{equation*}
  \parbox[][110pt][c]{110pt}{\centering
  \includegraphics{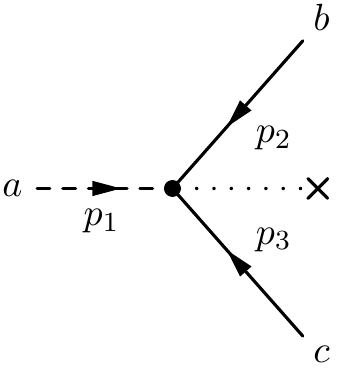}}\ =\ -ig\eta_{abc}\,(2\pi)^4
  \delta_t\big(\textstyle \sum_{i=1}^3p_{0,i}\big)\;
  \delta^{(3)}\big(\textstyle \sum_{i=1}^3\mathbf{p}_i\big)\;,
\end{equation*}
where $\delta_t$  is defined in (\ref{eq:deltat}) and the
pre-factors $\eta_{abc\dots}$ are given after~(\ref{eq:SPat}).

\item assign to each four-point $\chi$ vertex a factor of
\begin{equation*}
  \parbox[][100pt][c]{110pt}{\centering
  \includegraphics{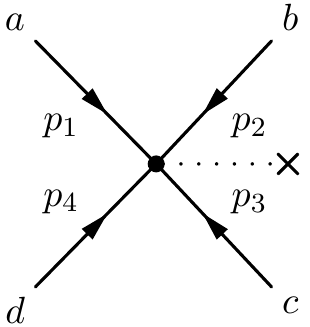}}\ =\ -i\lambda\eta_{abcd}\,(2\pi)^4
  \delta_t\big(\textstyle \sum_{i=1}^4 p_{0,i}\big)\;
  \delta^{(3)}\big(\textstyle\sum_{i=1}^4\mathbf{p}_i\big)\;.
\end{equation*}

\item  associate  with  each external vertex a phase
\begin{equation*}
e^{ip_0\tilde{t}_f}\;,
\end{equation*}
where $p_0$ is the energy flowing \emph{into} the vertex.

\item contract all internal CTP indices.

\item integrate with the measure
\begin{equation*}
  \int\!\!\frac{\D{4}{p}}{(2\pi)^4}\;
\end{equation*}
over  the four-momentum associated  with each contracted pair  of CTP
indices.

\item consider the combinatorial symmetry factors, where appropriate.
\end{itemize}

Notice that  there are a number  of modifications with  respect to the
standard Feynman rules.  In particular, the familiar energy-conserving
delta function has been replaced  by $\delta_t$ in the vertices.  This
is  indicated diagrammatically  by  the dotted  line  terminated in  a
cross, representing the violation of energy conservation. This loss of
energy  conservation  is  a  consequence of  Heisenberg's  uncertainty
principle, due to the finite macroscopic time of observation $t$, over
which  the interactions  have  been switched  on.  The  time-dependent
vertices vanish  in the limit $t\:\to\:  0$, as we  should expect. The
loss of space-time translational invariance leads to a doubling of the
number  of  integrations  with  respect to  the  zero-temperature  and
equilibrium cases.

\begin{figure}
\begin{center}
\parbox[][3.5cm][c]{10cm}{\centering \includegraphics{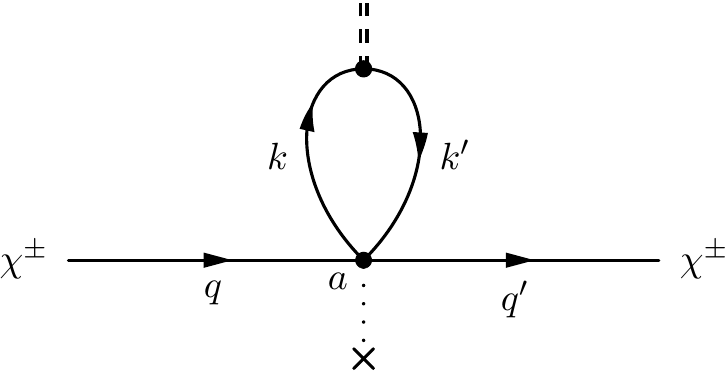}}
\end{center}
\caption{One-loop          local          $\chi$          self-energy:
  $i\Pi^{\mathrm{loc}(1)}_{\chi,\,ab}(q,q',\tilde{t}_f;\tilde{t}_i)\:\propto\:\eta_{ab}$.}
\label{fig:loc}
\end{figure}

\begin{figure}
\begin{center}
\subfloat[$i\Pi^{(1)}_{\Phi,\,ab}(q,q',\tilde{t}_f;\tilde{t}_i)$]{\parbox[][5.5cm][c]{7cm}{\centering
  \includegraphics{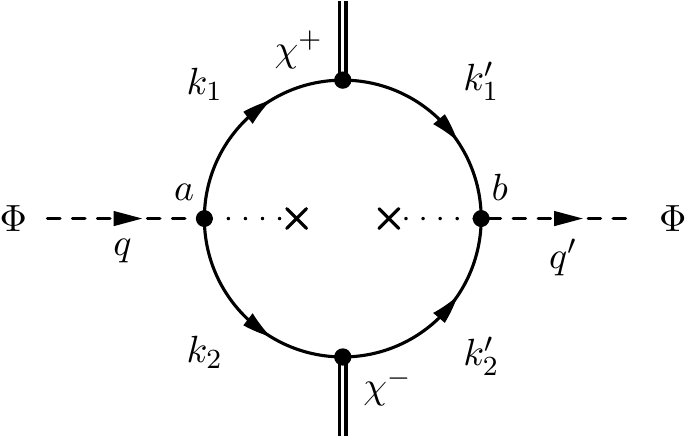}}}                                  \quad
\subfloat[$i\Pi^{(1)}_{\chi,\,ab}(q,q',\tilde{t}_f;\tilde{t}_i)$]{\parbox[][5.5cm][c]{7cm}{\centering
  \includegraphics{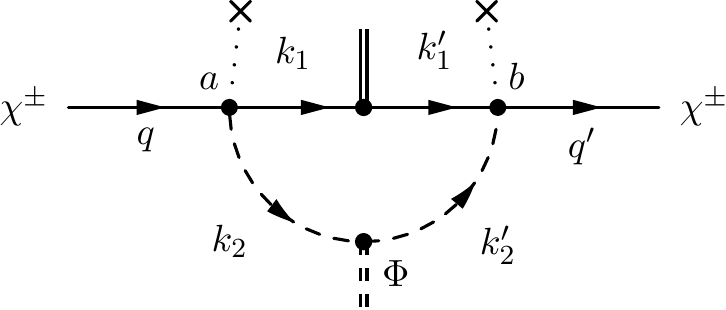}}}
\end{center}
\caption{Non-local one-loop $\Phi$ (a) and $\chi$ (b) self-energies.}
\label{fig:selfs}
\end{figure}

At  the one-loop  level, we  have  three diagrams.   The local  $\chi$
self-energy shown in Figure~\ref{fig:loc}:
\begin{equation}
  i\Pi^{\mathrm{loc}(1)}_{\chi,\,ab}(q,q',\tilde{t}_f;\tilde{t}_i)\ =\
  \frac{-i\lambda}{2!}(2\pi \mu)^{2\epsilon}e^{i(q_0\:-\:q_0')\tilde{t}_f}\!\iint\!\!
  \frac{\D{d}{k}}{(2\pi)^d}\,\frac{\D{4}{k'}}{(2\pi)^4}\;
  (2\pi)^4\delta_t^{(4)}(q\:-\:q'\:-\:k\:+\:k')
  \eta_{abcd}i\Delta^{0,\,cd}_{\chi}(k,k',\tilde{t}_f;\tilde{t}_i)\;,
\end{equation}
and the two non-local diagrams for the $\Phi$ and $\chi$ self-energies
shown in Figure~\ref{fig:selfs}:
\begin{subequations}
\begin{align}
  \label{eq:phiself}
  &i\Pi_{\Phi,\,ab}^{(1)}(q,q',\tilde{t}_f;\tilde{t}_i)\ =\ \frac{(-ig)^2}{2!}
  (2\pi\mu)^{2\epsilon}e^{i(q_0\:-\:q_0')\tilde{t}_f}\!\idotsint\!\!
  \frac{\D{d}{k_1}}{(2\pi)^d}\,\frac{\D{4}{k_1'}}{(2\pi)^4}\,
  \frac{\D{4}{k_2}}{(2\pi)^4}\,\frac{\D{4}{k_2'}}{(2\pi)^4}\nonumber\\
  &\qquad \times\: (2\pi)^4\delta^{(4)}_t(q\:-\:k_{1}\:-\:k_{2})
  (2\pi)^4\delta^{(4)}_t(q'\:-\:k_{1}'\:-\:k_{2}')
  \eta_{acd}i\Delta_{\chi}^{0,\,ce}(k_1,k_1',\tilde{t}_f;\tilde{t}_i)
  i\Delta_{\chi}^{C,\,0,\,df}(k_2,k_2',\tilde{t}_f;\tilde{t}_i)\eta_{efb}\;,\\
  \label{eq:chiself}
  &i\Pi_{\chi,\,ab}^{(1)}(q,q',\tilde{t}_f;\tilde{t}_i)\ =\ \frac{(-ig)^2}{2!}(2\pi\mu)^{2\epsilon}e^{i(q_0\:-\:q_0')\tilde{t}_f}
  \!\idotsint\!\!\frac{\D{d}{k_1}}{(2\pi)^d}\,\frac{\D{4}{k_1'}}{(2\pi)^4}\,
  \frac{\D{4}{k_2}}{(2\pi)^4}\,\frac{\D{4}{k_2'}}{(2\pi)^4}\nonumber\\
  &\qquad \times\: (2\pi)^4\delta^{(4)}_t(q\:-\:k_{1}\:-\:k_{2})
  (2\pi)^4\delta^{(4)}_t(q'\:-\:k_{1}'\:-\:k_{2}')
  \eta_{acd}i\Delta^{0,\,ce}_{\chi}(k_1,k_1',\tilde{t}_f;\tilde{t}_i)
  i\Delta_{\Phi}^{0,\,df}(k_2,k_2',\tilde{t}_f;\tilde{t}_i)\eta_{efb}\;.
\end{align}
\end{subequations}
A  detailed description of  the techniques  required to  perform these
loop integrals is provided in Appendix~\ref{app:loops}.

\subsection{Time-Dependent Width}
\label{sec:width}

In this section, we study the time-dependent width of the heavy scalar
$\Phi$.   In particular,  we investigate  the spectral  evolution that
results from  the restoration of energy  conservation, without solving
explicitly  the  system of  evolution  equations  for the  statistical
dynamics.

We  consider the  following  situation. We  prepare  two isolated  but
coincident  subsystems~$\mathscr{S}_{\Phi}$  and~$\mathscr{S}_{\chi}$,
both  separately   in  thermodynamic  equilibrium  and   at  the  same
temperature  $T$, with  all  interactions turned  off.  The  subsystem
$\mathscr{S}_{\Phi}$  contains  only  the  real  scalar  field~$\Phi$,
whilst $\mathscr{S}_{\chi}$  contains only the  complex scalar $\chi$.
At macroscopic time  $t\:=\:0$, we turn on the  interactions and allow
the                                                              system
$\mathscr{S}\:=\:\mathscr{S}_{\Phi}\:\cup\:\mathscr{S}_{\chi}$       to
re-thermalize.  For  our numerical analysis, we  take for definiteness
the thermodynamic temperature to be $T\:=\:10\ \mathrm{GeV}$, the mass
of the  heavy $\Phi$-scalar $M\:=\:1\  \mathrm{GeV}$, the mass  of the
complex $\chi$-scalar $m\:=\:0.01\  \mathrm{GeV}$, and their trilinear
coupling $g\:=\:0.1\ \mathrm{GeV}$.

The free propagators of the fields $\Phi$ and $\chi$ at time $t\:=\:0$
are   the   equilibrium   propagators  in   (\ref{eq:equilprop})   and
(\ref{eq:eqchi}),  containing   the  Bose--Einstein  distributions  at
temperature $T$.  We take the chemical potential of the complex scalar
to be  vanishingly small in  comparison to the temperature, i.e.~$\mu/T\:\ll\:
1$,        such        that       $f_{\chi}(|\mathbf{p}|,0)\:        =
\:f^{C}_{\chi}(|\mathbf{p}|,0)\:                                      =
\:f_{\mathrm{B}}\big(E_{\chi}(\mathbf{p})\big)$.   Without solving the
system of  time evolution equations,  the form of the  the statistical
distribution functions of the $\Phi$ and $\chi$ scalars is unknown for
$t\:\neq\:  0$.   We  assume  that   the  heat  bath  of  $\chi$'s  is
sufficiently large so as to  remain unperturbed by the addition of the
real scalar $\Phi$.  Specifically,  we may consider the number density
of  $\chi$'s  to remain  unchanged  and  the  free equilibrium  $\chi$
propagators in (\ref{eq:eqchi}) to persist for all times.

By the optical  theorem, the width $\Gamma_\Phi$ of  the scalar $\Phi$
is  defined in terms  of the  absorptive part  of the  retarded $\Phi$
self-energy $\mathrm{Im}\,\Pi_{\Phi,\,\mathrm{R}}$ via
\begin{equation}
  \Gamma_{\Phi}(q_1,q_2,\tilde{t}_f;\tilde{t}_i)\ =\ \frac{1}{M}\mathrm{Im}\,
  \Pi_{\Phi,\,\mathrm{R}}(q_1,q_2,\tilde{t}_f;\tilde{t}_i)\;,
\end{equation}
where
\begin{equation}
  \mathrm{Im}\,\Pi_{\Phi,\,\mathrm{R}}(q_1,q_2,\tilde{t}_f;\tilde{t}_i)
  \ =\ \frac{1}{2i}
  \Big(\,\Pi_{\Phi,\,>}(q_1,q_2,\tilde{t}_f;\tilde{t}_i)
  \:-\:\Pi_{\Phi,\,<}(q_1,q_2,\tilde{t}_f;\tilde{t}_i)\,\Big)\;.
\end{equation}
At       the       one-loop       level,       the       self-energies
$\Pi^{(1)}_{\Phi,\,\gtrless}(q_1,q_2,\tilde{t}_f;\tilde{t}_i)$      are
given by (\ref{eq:phiself}).

Employing  the relative and  central momenta,  $Q\:=\:q_1\:-\:q_2$ and
$q\:=\:(q_1\:+\:q_2)/2$,     and      using     the     results     of
Appendix~{\ref{sec:eqself}, the Laplace  transform with respect to the
  macroscopic time $t$ of the one-loop $\Phi$ width is
\begin{align}
  \label{eq:widthlap}
  &\Gamma^{(1)}_{\Phi}(q+Q/2,q-Q/2,s)\ =\ (2\pi)^4\frac{1}{\pi}\,
  \frac{s}{Q_0^2\:+\:4s^2}e^{iQ_0 \tilde{t}_f}\delta^{(3)}(\mathbf{Q})\nonumber\\
  & \hspace{3cm}\times \ \frac{g^2}{32\pi^2M}\sum_{\{\alpha\}}
  \int\!\D{3}{\mathbf{k}}\;\frac{1}{\pi}\,\frac{\alpha_1\alpha_2}
  {E_1E_2}\,
  \frac{1\:+\:f_{\mathrm{B}}\big(\alpha_1E_1\big)\:+\:
  f_{\mathrm{B}}\big(\alpha_2E_2\big)}
  {\big(q_0^2\:-\:\alpha_1 E_1\:-\:\alpha_2
  E_2\big)^2\:+\:s^2}\;,
\end{align}
where we  use the short-hand  notation $\{\alpha\}$ for  the summation
over $\alpha_1,\ \alpha_2\:=\:\pm1$.  For the sake of generality,
we  distinguish  the  $\chi^{+}$  and  $\chi^{-}$  decay  products  by
assigning them different masses $m_1$ and $m_2$, respectively, so that
\begin{subequations}
  \begin{align}
    E_1\ &\equiv \ E_1(\mathbf{k})\ =\ \sqrt{|\mathbf{k}|^2\:+\:m_1^2}\;,\\
    E_2\ &\equiv \ E_2(\mathbf{q}-\mathbf{k})\ =\ \sqrt{|\mathbf{k}|^2\:-
    \:2|\mathbf{k}||\mathbf{q}|\cos\theta\:+
    \:|\mathbf{q}|^2\:+\:m_2^2}\;.
  \end{align}
\end{subequations}
In our numerical analysis, however, we take $m_1\:=\:m_2\:=\:m$.

Performing  the  inverse  Wigner  transform  with respect  to  $Q$  of
(\ref{eq:widthlap}) in the  equal-time limit $X_0\:=\:\tilde{t}_f$ and
subsequently  the inverse Laplace  transform with  respect to  $s$, we
obtain the time-dependent $\Phi$ width
\begin{equation}
  \label{eq:phiwidth}
  \Gamma^{(1)}_{\Phi}(q,t)\ =\ \frac{g^2}{64\pi^2M}
  \sum_{\{\alpha\}}\int\!\D{3}{\mathbf{k}}\;
  \frac{\alpha_1\alpha_2}{E_1E_2}\,\frac{t}{\pi}\,
  \mathrm{sinc}\big[\big(q_0\:-\:\alpha_1E_1\:-\:\alpha_2E_2\big)t\big]
  \Big(\,1\:+\:f_{\mathrm{B}}\big(\alpha_1E_1\big)\:+\:
  f_{\mathrm{B}}\big(\alpha_2E_2\big)\,\Big)\;.
\end{equation}
In  the  limit $t\:\to\:\infty$,  the  sinc  function  on the  RHS  of
(\ref{eq:phiwidth})   yields  the  standard   energy-conserving  delta
function:
\begin{equation}
  \label{eq:tinf}
  \lim_{t\:\to\:\infty}\,\frac{t}{\pi}\,\mathrm{sinc}\big[
  \big(q_0\:-\:\alpha_1E_{1}\:-\:\alpha_2
  E_{2}\big)t\big]\ =\
  \delta\big(q_0\:-\:\alpha_1E_{1}\:-\:
  \alpha_2E_{2}\big)\; ,
\end{equation}
thereby recovering the known equilibrium result.

\begin{figure}
\begin{center}
\includegraphics[scale=0.48]{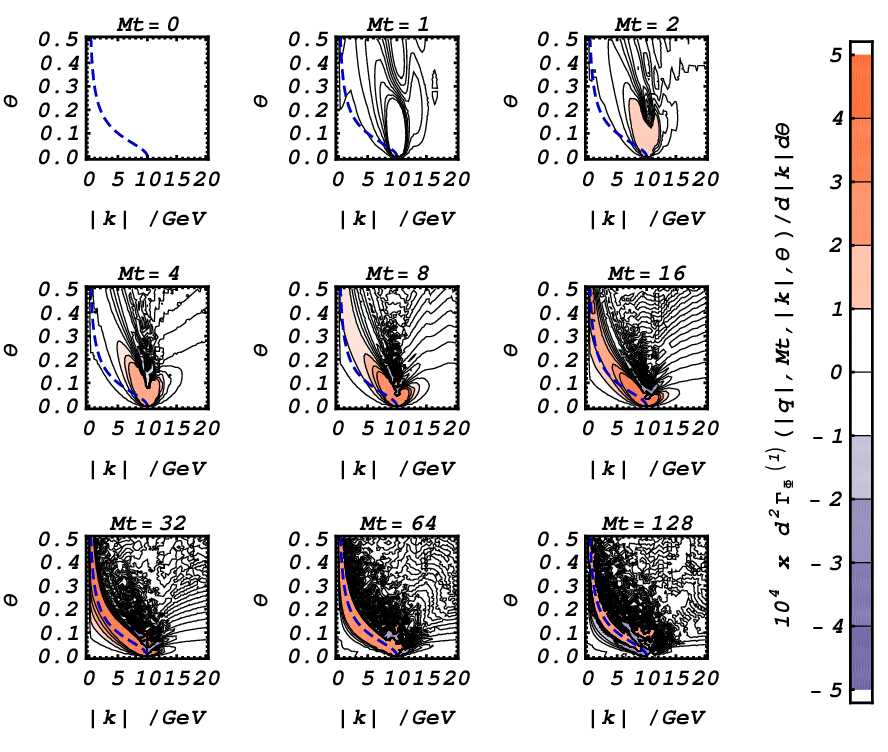}
\end{center}
\caption{Contour   plots  of   $|\mathbf{k}|$   versus  $\theta$   for
  $\D{2}{\Gamma_{\Phi}^{(1)}}$  for  discrete  values  of  $Mt$,  with
  $q^2\:=\:M^2$  and  $|\mathbf{q}|\:=\:10\, \mathrm{GeV}$.   The
  region  of phase space  permitted in  the $t\:\to\:\infty$  limit is
  shown by the  blue dashed line, corresponding to  the delta function
  in (\ref{eq:tinf}).}
\label{fig:kcont}
\end{figure}

Figure~\ref{fig:kcont}  contains  a series  of  contour  plots of  the
differential   one-loop   $\Phi$  width   $\D{2}{\Gamma^{(1)}_{\Phi}}$
evaluated  over  the dominant  region  of the  $|\mathbf{k}|$-$\theta$
phase space, where  the four-momentum $q^\mu$ of the  $\Phi$ scalar is
on-shell,  i.e.~$q^2=M^2$. For  late  times, the  integrand is  highly
oscillatory and  we expect the dominant  peak of the  sinc function to
approach   the  region  of   phase  space   permitted  in   the  limit
$t\:\to\:\infty$ [cf.~(\ref{eq:tinf})], for which
\begin{equation}
  |\mathbf{k}|\ =\ \frac{M^2|\mathbf{q}|\cos\theta\:+\:
  \sqrt{\big(|\mathbf{q}|^2\:+\:M^2\big)\big[M^4\:-\:4m^2\big(M^2\:+\:
  |\mathbf{q}|^2\sin^2\!\theta\big)\big]}}
  {2\big(M^2\:+\:|\mathbf{q}|^2\sin^2\!\theta\big)}\;.
\end{equation}
For early times,  the admissible phase space is  greatly expanded. For
later times, the  frequency of oscillation increases and  the width of
the central peak of the sinc  function narrows, lying along a curve in
the phase space.  To proceed further, we need to  develop a method for
dealing  with   the  kinematics  in   the  absence  of   exact  energy
conservation.

\subsection{Generalized Two-Body Decay Kinematics}
\label{sec:kin}

At  zero temperature  and density,  it  is convenient  to analyze  the
two-body decay  kinematics by performing  a Lorentz boost to  the rest
frame of  the decaying particle.  However, at finite  temperature, the
dependence on  the thermodynamic temperature  of the heat  bath breaks
the Lorentz covariance  of the integral. As such,  we cannot eliminate
dependence upon the the three-momentum of the decaying particle by any
such  Lorentz boost;  the dependence  will reappear  in  the `boosted'
temperature:
\begin{equation}
  T'\ =\ \gamma\, T\;,
\end{equation}
where
\begin{equation}
 \gamma\ =\ \Bigg(1\:+\:\frac{|\mathbf{q}|^2}{M^2}\Bigg)^{1/2}
\end{equation}
is  the  usual  Lorentz  boost  factor  for  the  heavy  scalar  field
$\Phi$.  As a result, we  are compelled to analyze the kinematics
of the  two-body decay in  the rest frame  of the heat bath,  which we
define to be the frame in which the EEV of the three-momentum operator
$\braket{\widehat{\bf P}}$ is minimized.   For an isotropic heat bath,
this is the  frame in which $\braket{\widehat{\bf P}}\:=\:\mathbf{0}$,
that  is the  \emph{comoving} frame.  In  this section,  we look  more
closely at the kinematics in the absence of energy conservation.

For this purpose, let us introduce the variable
\begin{equation}
  \label{eq:eact}
  u\ \equiv\ \big(q_0\:-\:\alpha_1 E_1\:-\:\alpha_2 E_2\big)t\;,
\end{equation}
which may  be interpreted in  terms of energy \emph{borrowed  from} or
\emph{lent to} the heat bath. We shall hereafter refer to the variable
$u$ as the  \emph{evanescent action} of the process,  since it has the
correct dimensions  and quantifies the  extended kinematically allowed
phase-space  configurations.  We   also  define  the  \emph{evanescent
  energy}
\begin{equation}
  q_u(t)\ \equiv\ q_0\:-\:\frac{u}{t}\;,
\end{equation}
which satisfies
\begin{equation}
  \lim_{u/t\:\to\:0}\,q_u(t)\ =\ q_0\;.
\end{equation}
With this substitution, we obtain the kinematic constraint
\begin{equation}
  \label{eq:kincon}
  q_u(t)\:-\:\alpha_1 E_1\:-\:\alpha_2 E_2\ =\ 0\;.
\end{equation}
Since $u$ can take large  positive values, $q_u(t)$ is not necessarily
restricted to  positive values for  early times, even when  $q_0\: =\:
\sqrt{|\mathbf{q}|^2\:+\:M^2}\:>\:0$   is  on-shell.   Processes  with
$u/t\:\neq\:0$ are referred to as \emph{evanescent}.

In order to  make the coordinate transformation $|\mathbf{k}|\:\to\:u$
in  (\ref{eq:phiwidth}),  we  must  solve  (\ref{eq:kincon})  for  the
magnitude  of  the  three-momentum  $|\mathbf{k}(t)|$,  which  becomes
implicitly time dependent. Specifically, we find
\begin{align}
  \label{eq:koft}
  |\mathbf{k}^{(b)}(t)|\ &=\ \frac{1}{2\big(q_u^2(t)\:-\:|\mathbf{q}|^2
  \cos^2\!\theta\big)}
  \Big[\,\big(q_u^2(t)\:-\:|\mathbf{q}|^2
  \:+\:m_1^2\:-\:m_2^2\big)|\mathbf{q}|\cos\theta\nonumber\\
  &\qquad + \: b\,  q_u(t)
  \sqrt{\lambda\big(q_u^2(t)\:-\:|\mathbf{q}|^2,m_1^2,m_2^2\big)\:-\:
  4m_1^2|\mathbf{q}|^2\sin^2\!\theta}\;\Big]\;,
\end{align}
with $b\:=\:\pm 1$ and 
\begin{equation}
  \lambda(x,y,z)\ =\ (x-y-z)^2\:-\:4yz\;.
\end{equation}
After a little algebra, we obtain the energies
\begin{align}
  E_1^{(b)}(t)\ & =\ \alpha_1\frac{1}{2\big(q_u^2(t)\:-\:|\mathbf{q}|^2\cos^2
  \!\theta\big)}\Big[\,\big(q_u^2(t)\:-\:
  |\mathbf{q}|^2\:+\:m_1^2\:-\:m_2^2\big)q_u(t)\nonumber\\
  &\qquad +\: b\, |\mathbf{q}|\cos\theta\sqrt{\lambda\big(q_u^2(t)\:-\:
  |\mathbf{q}|^2,m_1^2,m_2^2\big)\:-\:4m_1^2|\mathbf{q}|^2
  \sin^2\!\theta}\;\Big]\;,\\
  E_2^{(b)}(t)\ &=\ \alpha_2\frac{1}{2\big(q_u^2(t)\:-\:|\mathbf{q}|^2
  \cos^2\!\theta\big)}\Big[\,\big(q_u^2(t)\:-\:|\mathbf{q}|^2
  \cos2\theta\:-\:m_1^2\:+\:m_2^2\big)q_u(t)\nonumber\\&\qquad -\: b\, 
  |\mathbf{q}|\cos\theta\sqrt{\lambda\big(q_u^2(t)\:-\:
  |\mathbf{q}|^2,m_1^2,m_2^2\big)\:-\:4m_1^2
  |\mathbf{q}|^2\sin^2\!\theta}\;\Big]\;,
\end{align}
where the  overall factors of $\alpha_1$ and  $\alpha_2$ are necessary
to satisfy the initial constraint (\ref{eq:kincon}).  It is clear
that these results collapse  to the kinematics of equilibrium
field theory in the limit $u/t\:\to\: 0$.

Keeping $m_1$ and $m_2$ distinct, the $\Phi$ width for $t\:>\:0$ is given by
\begin{align}
  \label{eq:phiwidthu}
  \Gamma^{(1)}_{\Phi}(q,t)\ &=\ \frac{g^2}{64\pi^2M}
  \sum_{\{\alpha\},\,b\:=\:\pm1}\int_0^{\pi}\!\D{}{\theta}
  \int_{u_-(t)}^{u_+(t)}\!\!\D{}{u}\,\,\mathrm{sinc}(u)\,
  \frac{\sin\theta}{\big(q_u^2(t)\:-\:|\mathbf{q}|^2
  \cos^2\!\theta\big)^2}\nonumber\\&\qquad \times\:
  \Big[\,\lambda\big(q_u^2(t)\:-\:|\mathbf{q}|^2,m_1^2,m_2^2\big)
  \:-\:4m_1^2|\mathbf{q}|^2\sin^2\!\theta\,\Big]^{-1/2}\nonumber\\&\qquad \times\: 
  \Big[\,\big(q_u^2(t)\:-\:|\mathbf{q}|^2\:+\:m_1^2
  \:-\:m_2^2\big)|\mathbf{q}|\cos\theta\:+\: b\,q_u(t)
  \sqrt{\lambda\big(q_u^2(t)\:-\:|\mathbf{q}|^2,m_1^2,m_2^2\big)\:-\:
  4m_1^2|\mathbf{q}|^2\sin^2\!\theta}\,\Big]^2\nonumber\\&\qquad \times\: 
  \alpha_1\alpha_2\Big(\,1\:+\:f_{\mathrm{B}}\big(\alpha_1E^{(b)}_{1}(t)\big)\:+\:
  f_{\mathrm{B}}\big(\alpha_2E^{(b)}_{2}(t)\big)\,\Big)\;.
\end{align}
The reality of the  loop momentum $|\mathbf{k}(t)|$ in (\ref{eq:koft})
requires that the discriminant
\begin{equation}
  \lambda\big(q_u^2(t)\:-\:|\mathbf{q}|^2,m_1^2,m_2^2\big)
  \:-\:4m_1^2|\mathbf{q}|^2\sin^2\!\theta\ \geq\ 0\;,
\end{equation}
which   is   now   a   \emph{time-dependent}   kinematic   constraint.
Furthermore,    we    require    $|\mathbf{k}^{(b)}(t)|\:\geq\:    0$,
$E^{(b)}_1(t)\:\geq\:   m_1$  and  $E^{(b)}_2(t)\:\geq\:   m_2$.   For
$t\:>\:0$,  the  limits  of  integration  $u_{\pm}(t)$  are  given  in
Table~\ref{tab:lims}, where we have defined
\begin{equation}
  \label{eq:w0}
  \omega_0(q,\alpha_1,\alpha_2)\ \equiv\ q_0\:-\:\big(\alpha_1m_1
  \:+\:\alpha_2m_2\big)\Bigg(1\:+\:
  \dfrac{|\mathbf{q}|^2}{\big(m_1\:+\:m_2\big)^2}\Bigg)^{\!\! 1/2}\;,
\end{equation}
which is the angular  frequency of the sine-integral-like oscillations
of the integral.  For  $t\:=\:0$, $u_{\pm}(t)\:=\:0$ and the domain of
integration over  $u$ collapses to zero. Given  the analytic behavior
of the integrand  in the limit $t\:\to\:0$, the  integral vanishes, as
we expect.

\begin{table}
\renewcommand{\arraystretch}{1.5}
\caption{\label{tab:lims} Limits of integration of the evanescent
  action $u$ for each of the four processes contributing to the
  non-equilibrium thermal $\Phi$ width of our specific model for
  $t\:>\:0$, where the angular frequency
  $\omega_0(q,\alpha_1,\alpha_2)$ is defined in (\ref{eq:w0}).}
\begin{tabular*}{3.4in}[c]{ c  c  c  c }
\hline\hline
& $\alpha_1\ =\ \alpha_2\ =\ +1$ & $\alpha_1\ =\ -\alpha_2$ &
$\alpha_1\ =\ \alpha_2\ =\ -1$ \\
\hline
$u_+(t)$ & $\omega_0(q,\alpha_1,\alpha_2)t$ & 
$\omega_0(q,\alpha_1,\alpha_2)t$ & $+\infty$ \\ 
$u_-(t)$ & $-\infty$ & $0$ & $\omega_0(q,\alpha_1,\alpha_2)t$ \\
\hline\hline
\end{tabular*}
\renewcommand{\arraystretch}{1}
\end{table}

\begin{figure}
\begin{center}
\includegraphics[scale=0.48]{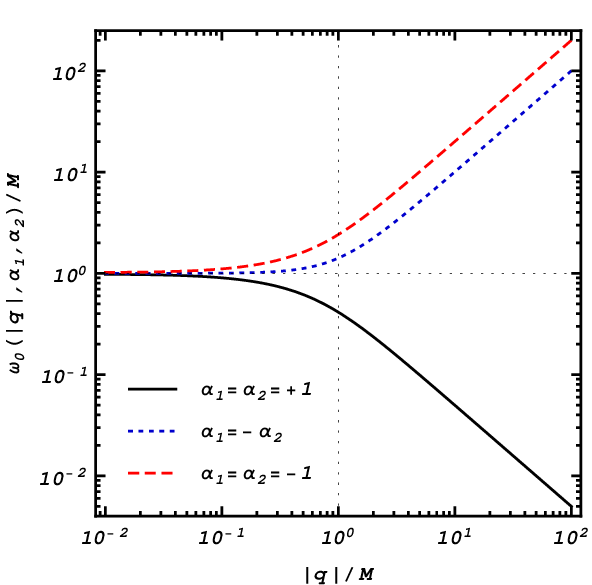}
\end{center}
\caption{The dependence of
  $\omega_0(|\mathbf{q}|,\alpha_1,\alpha_2)/M$ on $|\mathbf{q}|/M$ for
  processes related to $1\:\to\: 2$ decay (solid black), Landau
  damping (blue dotted) and $3\:\to\: 0$ total annihilation (red
  dashed), where $q^2\:=\:M^2$ and
  $m_1\:=\:m_2\:=\:m\:=0.01\,\mathrm{GeV}$.  The latter two of these
  processes become highly oscillatory for large $|\mathbf{q}|/M$.}
\label{fig:sincosc}
\end{figure}

For  on-shell decay  modes with  $q^2\:=\:M^2$, the  angular frequency
$\omega_0(q,1,1)$ in (\ref{eq:w0}) becomes
\begin{equation}
  \label{eq:freq}
  \omega_0(|\mathbf{q}|,1,1)\ \equiv\ 
  \omega_0(q,1,1)\big|_{q^2\:=\:M^2}\ =\
  \sqrt{|\mathbf{q}|^2\:+\:M^2}\:-\:\sqrt{|\mathbf{q}|^2
  \:+\:\big(m_1\:+\:m_2\big)^2}\;.
\end{equation}
Thus, the  evolution of the  phase space for on-threshold  decays with
$M^2\:=\:(m_1\:+\:m_2)^2$ is  \emph{critically damped}.  We  note that
in the large momentum limit $|\mathbf{q}|\:\gg\:M$,
\begin{equation}
  \label{eq:qcritic}
  \omega_0(|\mathbf{q}|,1,1)\big|_{|\mathbf{q}|\:\gg\:M}\
  \simeq\ \frac{M^2\:-\:\big(m_1\:+\:m_2\big)^2}{2|\mathbf{q}|}\;,
\end{equation}
such that the evolution of  the phase space for high-momentum modes is
similarly      damped.       The      momentum      dependence      of
$\omega_0(|\mathbf{q}|,\alpha_1,\alpha_2)$       is      shown      in
Figure~\ref{fig:sincosc}.

\begin{figure}
\begin{center}
\subfloat[$1\rightarrow  2$ decay]{\label{fig:kinsa}{\centering
    \includegraphics{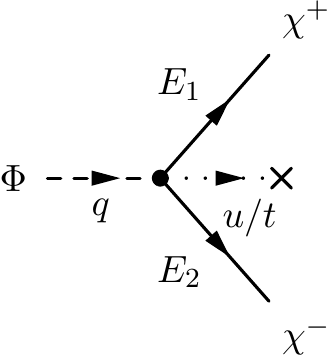}}\quad}  \subfloat[$2\rightarrow  1$
  Landau                         damping]{\label{fig:kinsb}{\centering
    \includegraphics{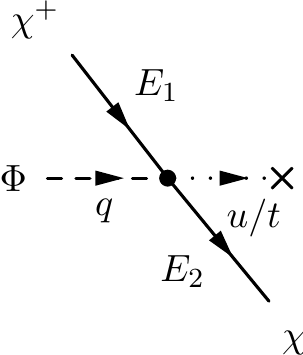}   \quad \includegraphics{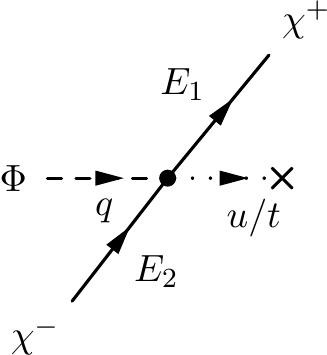}}}
\subfloat[$3\rightarrow                    0$                    total
  annihilation]{\label{fig:kinsc}{\centering                     \qquad
    \includegraphics{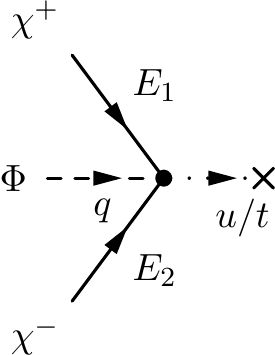} \qquad}}
\end{center}
\caption{Processes  contributing to  the time-dependent  $\Phi$ width:
  (a) the familiar $1\:\rightarrow\:2$ decay; (b) $2\:\rightarrow\: 1$
  Landau damping; and (c) $3\: \rightarrow\: 0$ total annihilation.}
\label{fig:kins}
\end{figure}

\begin{figure}
\begin{center}
\includegraphics[scale=0.48]{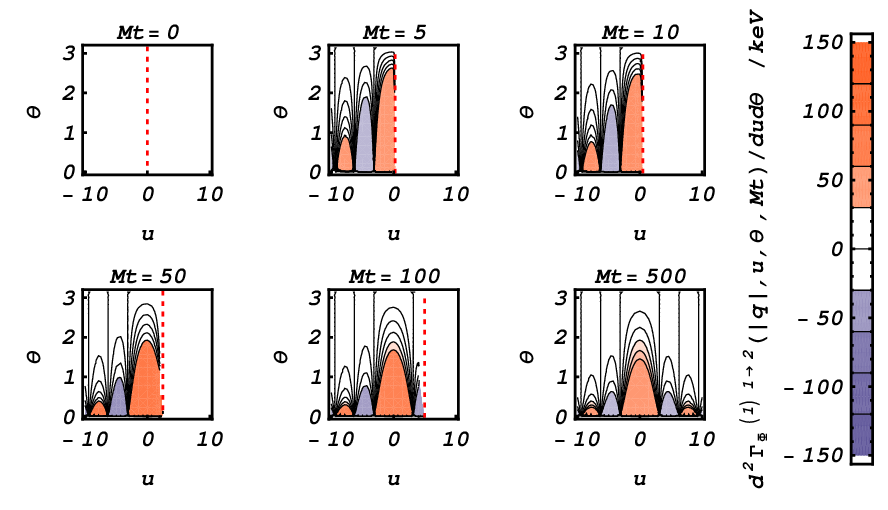}
\end{center}
\caption{Contour    plots   of   $u$    versus   $\theta$    for   the
  $1\:\rightarrow\:        2$       decay        contribution       to
  $\D{2}{\Gamma_{\Phi}^{(1)}}$  for  discrete  values  of  $Mt$,  with
  $q^2\:=\:M^2$  and $|\mathbf{q}|\:=\:10\  \mathrm{GeV}$.   The solid
  excluded regions to the right of the red dotted line lie exterior to
  the limits of integration over $u$.}
\label{fig:decayc}
\end{figure}
 
The summation over $\alpha_1$ and $\alpha_2$ yields four distinct
contributions to the decay width \cite{Berges:2004pu,
  Berges:2004yj}. For $\alpha_1\:=\:\alpha_2\:=\:+1$, we obtain the
contribution from the familiar $1\:\rightarrow\: 2$ decay process
presented in Figure~\ref{fig:kinsa}.  For $\alpha_1\:=\:-\alpha_2\:=\:
\pm 1$, we obtain the two $2\:\rightarrow\: 1$ \emph{Landau--damping}
contributions displayed in Figure~\ref{fig:kinsb}. For
$\alpha_1\:=\:\alpha_2\:=\:-1$, we obtain the $3\:\rightarrow\: 0$
\emph{total annihilation} process shown in Figure~\ref{fig:kinsc}.  In
the latter, the decay `products' appear in the \emph{initial state}
along with the decaying particle.  For late times, the Landau--damping
and total annihilation processes are kinematically disallowed, as we
would expect.  They are only permitted in the \emph{evanescent regime}
at early times.

\begin{figure}
\begin{center}
\includegraphics[scale=0.48]{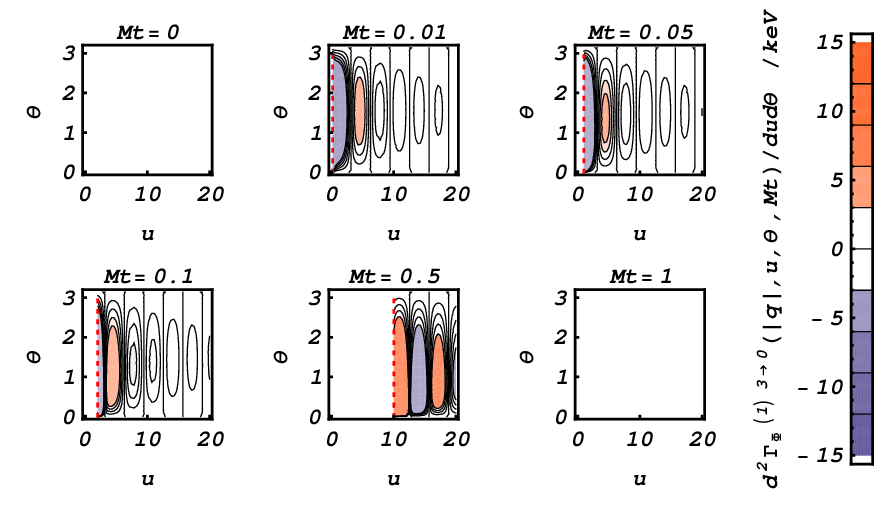}
\end{center}
\caption{Contour    plots   of   $u$    versus   $\theta$    for   the
  $3\:\rightarrow\:    0$   total    annihilation    contribution   to
  $\D{2}{\Gamma_{\Phi}^{(1)}}$  for  discrete  values of  $Mt$,  where
  $q^2\:=\:M^2$ and  $|\mathbf{q}|\:=\:10\ \mathrm{GeV}$.  The regions
  to the  left of the  red dotted line  are exterior to the  domain of
  integration over $u$.}
\label{fig:annic}
\end{figure}

\begin{figure}
\begin{center}
\subfloat[$\alpha_1\ = \ -\alpha_2\ = \ +1$]
{\includegraphics[scale=0.48]{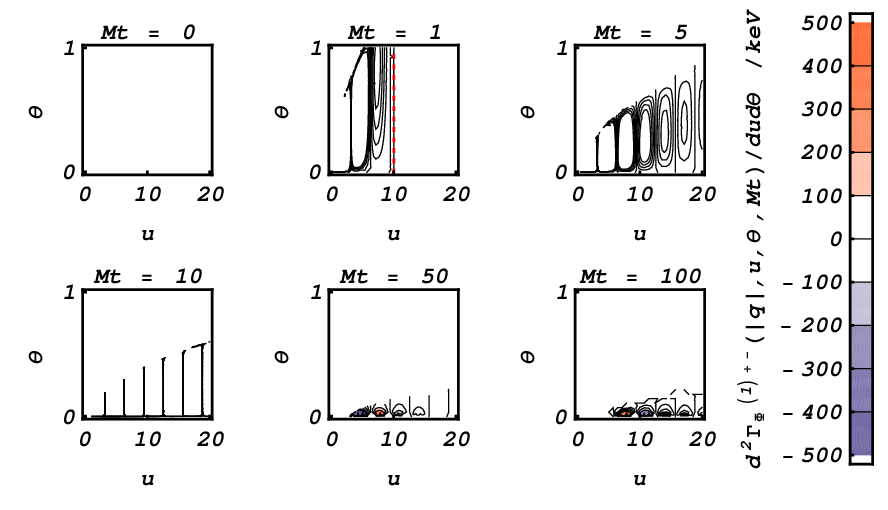}}\\
\subfloat[$\alpha_1\ =\ -\alpha_2\ =\ -1$]{\includegraphics[scale=0.48]{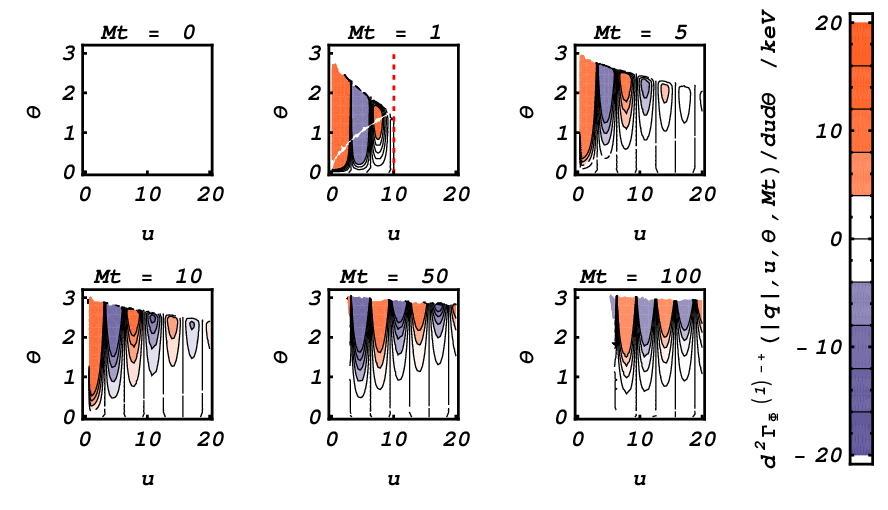}}
\end{center}
\caption{Contour plots of $u$ versus $\theta$ for the two
  Landau--damping contributions (a) ${\alpha_1\: =\: -\alpha_2\: =\: +
    1}$ and (b) $\alpha_1\:=\:-\alpha_2\:=\:-1$ to
  $\D{2}{\Gamma_{\Phi}^{(1)}}$ for discrete values of $Mt$, assuming
  $q^2\:=\:M^2$ and ${|\mathbf{q}|\:=\:10\ \mathrm{GeV}}$.  The
  regions to the right of the red dotted line are exterior to the
  domain of integration over $u$.  The contour plots differ between
  the two contributions due to the asymmetry of the integrands in the
  $\Phi$ three-momentum.}
\label{fig:landc}
\end{figure}

\begin{figure}
\begin{center}
\includegraphics[scale=0.48]{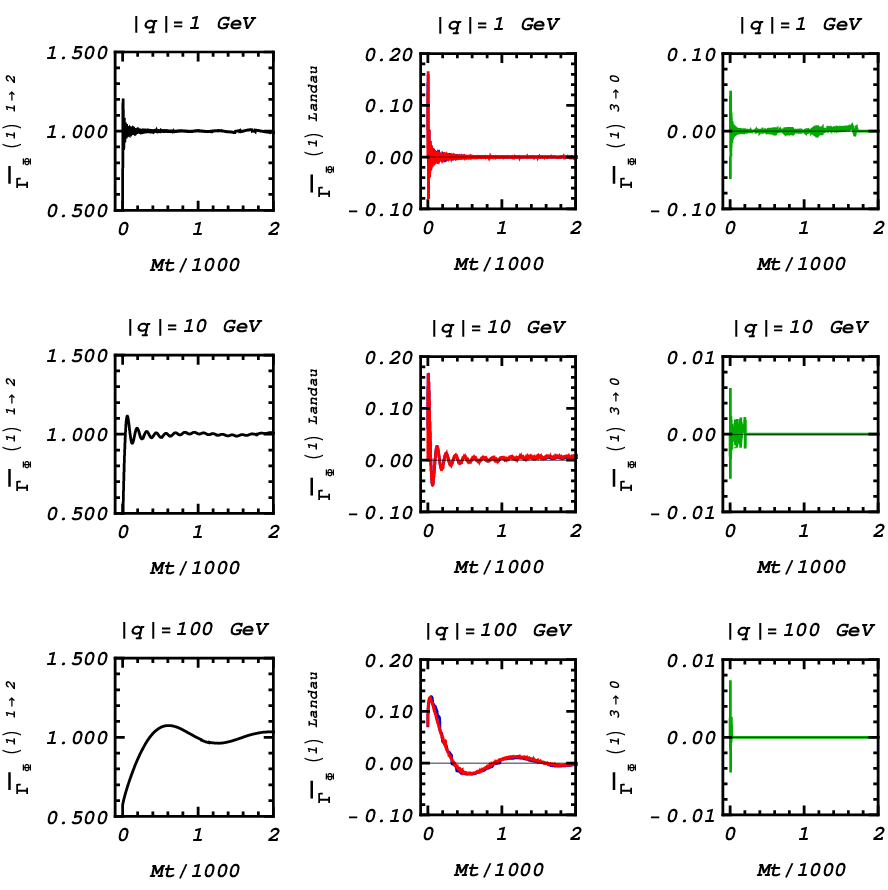}
\end{center}
\caption{The    four    separate    contributions   to    the    ratio
  $\bar{\Gamma}_{\Phi}^{(1)}$  in (\ref{eq:gambar})  versus  $Mt$, for
  on-shell    decays    with    $|\mathbf{q}|\:=\:1\    \mathrm{GeV}$,
  $10\ \mathrm{GeV}$ and $100\ \mathrm{GeV}$.  The two Landau--damping
  contributions are identical up to numerical errors.}
\label{fig:seps}
\end{figure}

Figures \ref{fig:decayc}--\ref{fig:landc} contain contour plots of $u$
versus   $\theta$  for   the  four   contributions  to   the  on-shell
differential  width   $\D{2}{\Gamma_{\Phi}^{(1)}}$.   The  equilibrium
kinematics are  obtained in  the large-time limit  $t\:\to\:\infty$ as
follows.  As can be seen from the $(u,\theta)$ contour plots in Figure
\ref{fig:decayc}, for  the $1\:\to\: 2$  decay process, the  limits of
integration grow to encompass the full range of the sinc function.  At
the  same time,  the  $u$ dependence  of  the phase-space  pre-factors
vanishes,  since  $q_u(t)\:\to \:q_0$  as  $t\:\to\:\infty$.  For  the
$3\:\to\:  0$ total  annihilation process,  the domain  of integration
vanishes  in the  large-time limit  $t\:\to\:\infty$.  Given  that the
integrand  is finite  in the  same limit,  the  contribution therefore
vanishes as  expected, which is confirmed by  our $(u,\theta)$ contour
plots  in   Figure  \ref{fig:annic}.   For   the  two  Landau--damping
contributions,  the  large-time  behavior  becomes more  subtle.   As
$t\:\to\:\infty$, the  domain of integration  covers approximately all
positive $u$.   However, the kinematically allowed  phase space cannot
be  attained with  any values  of $u$  and $\theta$,  so that  the two
Landau--damping  contributions also  vanish in  the  large-time limit.
This  behavior is  reflected  in our  $(u,\theta)$  contour plots  of
Figure   \ref{fig:landc}.    Thus,  only   the   usual  $1\:\to\:   2$
energy-conserving decay remains for late times.

In order to reduce the statistical error in our Monte Carlo
integration over the $(u,\theta)$ phase space, we use a Gaussian
sampling bias to ensure that the majority of sampling points fall over
the dominant region of the sinc function of $u$ in
(\ref{eq:phiwidthu}). We define the weight function
\begin{equation}
  \label{eq:gaussw}
  \varpi(u)\ =\ \frac{\D{}{u}}{\D{}{r}}\ \equiv\
  \exp\bigg(-\frac{(u\:-\:u_0)^2}{2\sigma_u^2}\bigg)
\end{equation}
where
\begin{equation}
  r(u)\ \equiv\ \frac{\mathrm{Erf}\Big(\frac{1}{\sqrt{2}}\,
  \frac{u\:-\:u_0}{\sigma_u}\Big)\:-\:\mathrm{Erf}\Big(
  \frac{1}{\sqrt{2}}\,\frac{u_-\:-\:u_0}{\sigma_u}\Big)}
  {\mathrm{Erf}\Big(\frac{1}{\sqrt{2}}\,\frac{u_+\:-\:u_0}{\sigma_u}
  \Big)\:-\:\mathrm{Erf}\Big(\frac{1}{\sqrt{2}}\,
  \frac{u_-\:-\:u_0}{\sigma_u}\Big)}\ \in\ [0,\ 1]\;.
\end{equation}
After   performing  the   change  of   the  variable   $u\:\to\:r$  in
(\ref{eq:phiwidthu}) for $u_0\:=\:0$, the limits of integration become
time-independent and  are identical for  the decay, total-annihilation
and  Landau--damping contributions.   The dependence  upon  the limits
$u_+(t)$ and $u_-(t)$ appears instead within the transformed integrand
of (\ref{eq:phiwidthu}) in the new variables $(r,\theta)$.

\begin{figure}
\begin{center}
\includegraphics[scale=0.48]{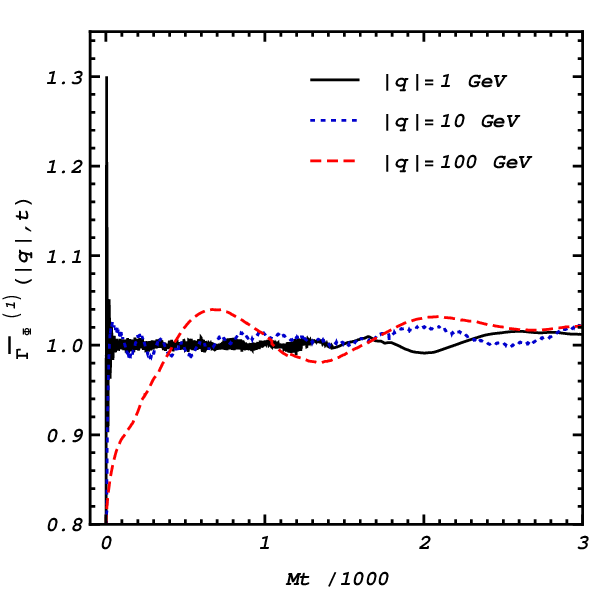}
\end{center}
\caption{The  ratio  $\bar{\Gamma}_{\Phi}^{(1)}$ in  (\ref{eq:gambar})
  versus       $Mt$,        for       on-shell       decays       with
  $|\mathbf{q}|\:=\:1\ \mathrm{GeV}$ (solid black), $10\ \mathrm{GeV}$
  (blue dotted) and $100\ \mathrm{GeV}$ (red dashed).}
\label{fig:twidth}
\end{figure}

The width of the dominant region of the $u$ phase-space is taken to be
the distance between the central  maximum and the maximum at which the
amplitude of the  sinc function has fallen to  $0.1\%$ with respect to
the central maximum.  Hence, we require for the  distant maximum to be
at  the location  where $\mathrm{sinc}\,u_n\:\sim  \:0.001$.  The
extrema $u_n$ of the sinc function satisfy the transcendental equation
\begin{equation}
  u_n\ =\ \tan u_n\;,
\end{equation}
whose  solutions for  $n\:\geq\: 1$ may be expressed as
\cite{Burniston1973}
\begin{equation}
  \label{eq:un}
  u_n\ =\ \pm n\pi\exp\bigg\{\frac{1}{\pi}
  \!\int_0^1\!\D{}{\xi}\;\frac{1}{\xi}\,\mathrm{arg}\bigg[
  \bigg(1\:+\:\frac{1}{2}\xi\ln\frac{1\:-\;\xi}{1\:+\:\xi}\:\pm\:
  \frac{1}{2}\pi  i  \xi\bigg)^{\!\!  2}\!\:+\:n^2\pi^2\xi^2\,\bigg]
  \bigg\}\;.
\end{equation}
The   required   extremum   is   then  given   by   $n\:=\:318$,   the
159\textsuperscript{th}   maximum   of   the   sinc   function,   with
$u_{318}\:=\:1000$.  The Gaussian  weight function $\varpi(u)$ in
(\ref{eq:gaussw})   is   then   taken    to   have   a   variance   of
$\sigma_u^2\:=\:(u_{318}/2)^2$,  such  that  $95\%$  of  the  sampling
points fall within this dominant region.

Our  interest  is  in  the  deviation of  the  one-loop  $\Phi$  width
$\Gamma_{\Phi}^{(1)}$  from  the  known  equilibrium  result.   It  is
therefore       convenient       to       define       the       ratio
$\bar{\Gamma}_{\Phi}^{(1)}(|\mathbf{q}|,t)$ of   the  time-dependent
width to its late-time equilibrium value for $q^2\:=\:M^2$:
\begin{equation}
  \label{eq:gambar}
  \bar{\Gamma}_{\Phi}^{(1)}(|\mathbf{q}|,t)\ =\
  \frac{\Gamma_{\Phi}^{(1)}(|\mathbf{q}|,t)}
  {\Gamma_{\Phi}^{(1)}(|\mathbf{q}|,t\:\to\:\infty)}
  \ =\ \frac{\Gamma_{\Phi}^{(1)1\rightarrow 2}(|\mathbf{q}|,t)\:+\:
  2\Gamma_{\Phi}^{(1)\mathrm{Landau}}(|\mathbf{q}|,t)\:+\:
  \Gamma_{\Phi}^{(1)3\rightarrow 0}(|\mathbf{q}|,t)}
  {\Gamma_{\Phi}^{(1)}(|\mathbf{q}|,t\to \infty)}\;.
\end{equation}
In  Figure~\ref{fig:seps}, we plot  separately the  four contributions
from Figure~\ref{fig:kins} to  this ratio as a function  of $Mt$ for a
series of  discrete momenta. The  evanescent Landau--damping processes
yield a prompt  contribution, which can be as  high as $10$--$20\%$ at
early  times.  The evanescent  total annihilation  process contributes
similarly at the level of about $5\%$.

In    Figure~\ref{fig:twidth},    we     plot    the    total    ratio
$\bar{\Gamma}_{\Phi}^{(1)}(|\mathbf{q}|,t)$  as  a  function of  $Mt$.
The   total  $\Phi$  width   $\Gamma^{(1)}_{\Phi}(|\mathbf{q}|,t)$  is
vanishing  for ${Mt\:=\:0}$,  as we  would expect.   This  is followed
promptly  by a  sharp rise,  which is  particularly pronounced  in the
infra-red  modes,  resulting  from  the  sudden switching  on  of  the
interactions.  This  so-called \emph{shock regime} is  followed by the
superposition   of  transient   oscillations   of  angular   frequency
$\omega_0$   with   short   time   scales   and   \emph{non-Markovian}
oscillations of longer time  scales.  The latter of these oscillations
exhibit  time-dependent  frequencies,  the  origin of  which  will  be
discussed in the next section.

\subsection{Non-Markovian Oscillations}
\label{sec:nonm}

As we  have seen in  Figures \ref{fig:seps} and  \ref{fig:twidth}, the
time-dependent  $\Phi$  width   contains  a  superposition  of  damped
oscillatory  contributions.  The  longer lived  of these  oscillations
exhibit  time-dependent frequencies. These  non-Markovian oscillations
are  illustrated more  clearly in  Figure~\ref{fig:nonmwidth}  for the
$|\mathbf{q}|\:=\:1\  \mathrm{GeV}$  mode  in  which  a  moving  time
average  is carried  out to  eliminate the  higher-frequency Markovian
oscillations.   In  this section,  we  describe  the  origin of  the
time-dependent  oscillations and show  that they  are not  a numerical
artefact,  but  are  instead  an  intrinsic feature  inherent  to  the
dynamics  of  truly  out-of-equilibrium  systems.  To this end, we
consider the high-temperature  limit $T\:\gg\:M$ of the time-dependent
$\Phi$ width.

\begin{figure}
\begin{center}
\includegraphics[scale=0.48]{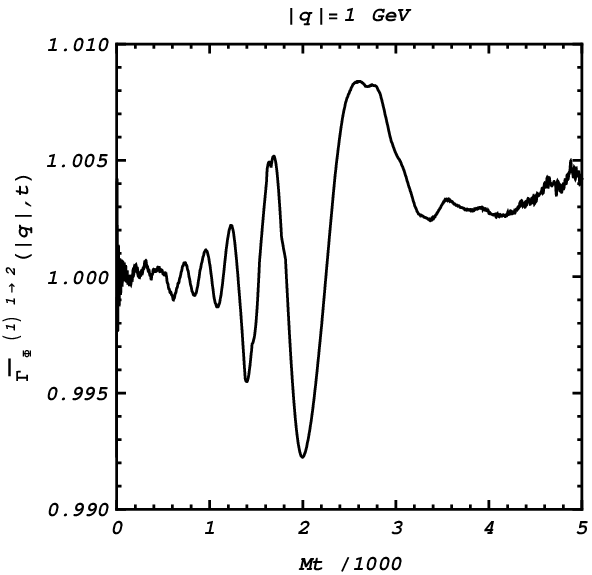}
\end{center}
\caption{Non-Markovian   oscillations  of   the  $1\:\to   \:2$  decay
  contribution    to   the   ratio    $\bar{\Gamma}_{\Phi}^{(1)}$   in
  (\ref{eq:gambar}) against $Mt$, by  performing a moving average over
  bins of $150$ in $Mt$.}
\label{fig:nonmwidth}
\end{figure}

In   this  high-temperature   limit  $T\:\gg\:M$,   the  Bose--Einstein
distribution may be approximated as follows:
\begin{equation}
  \label{eq:fbhight}
  f_{\mathrm{B}}(E)\ \approx\ \frac{T}{E}\;,
\end{equation}
Returning  to  the $\Phi$  width  $\Gamma^{(1)}_{\Phi}$  in
(\ref{eq:phiwidth})  and substituting  for (\ref{eq:fbhight}),  we may
perform  the angular  integration by  making the  following  change of
variables:
\begin{equation}
  \cos\theta\ =\ \frac{|\mathbf{k}|^2\:+\:|\mathbf{q}|^2\:+\:m_2^2}
  {2|\mathbf{k}||\mathbf{q}|}\,(1\:-\:x^2)\;.
\end{equation}
Then,  the high-temperature  limit $T\:\gg\:M$  of  the time-dependent
width of the heavy scalar~$\Phi$ becomes
\begin{align}
  &\Gamma^{(1),\,T\:\gg\:M}_{\Phi}(q,t)\ = \ \frac{g^2}{32\pi^2 M}\sum_{\{\alpha\}}
  \alpha_1\alpha_{\theta}\!\int_0^{\infty}\!\!\D{}{|\mathbf{k}|}\;
  \frac{|\mathbf{k}|}{|\mathbf{q}|}\,\frac{1}{E_1}\nonumber\\&\quad \quad \times\
  \Bigg[\Bigg(1\:+\:\frac{T}{\alpha_1E_1}\:+\:\frac{T}{q_0\:-\:\alpha_1
  E_1}\Bigg)\mathrm{Si}\big[\big(q_0\:-\:\alpha_1 E_1\:-\:\alpha_2
  E_2\big)t\big]\nonumber\\&\quad \quad -\ \frac{T}{q_0\:-\:\alpha_1E_1}
  \Bigg(\sin\!\big[\big(q_0\:-\:\alpha_1E_1\big)t\big]
  \mathrm{Ci}\big(\alpha_2E_2t\big)\:-\:\cos\!\big[\big(q_0\:-\:\alpha_1
  E_1\big)t\big]\mathrm{Si}\big(\alpha_2 E_2t\big)\Bigg)\Bigg]\;,
\end{align}
where  $\mathrm{Si}(x)$  and  $\mathrm{Ci}(x)$  are  respectively  the  sine
integral  and  cosine  integral  functions.  We  have  introduced  the
short-hand   notations:   $\{\alpha\}$    for   the   summation   over
$\alpha_1,\ \alpha_2,\ \alpha_{\theta}\:=\:\pm 1$; and
\begin{align}
  E_1&\ =\ \sqrt{|\mathbf{k}|^2\:+\:m_1^2}\:,\\
  E_2&\ =\ \sqrt{|\mathbf{k}|^2\:-\:2\alpha_{\theta}|\mathbf{k}|
  |\mathbf{q}|\:+\:|\mathbf{q}|^2\:+\:m_2^2}\;.
\end{align}

\begin{figure}
\begin{center}
\includegraphics[scale=0.48]{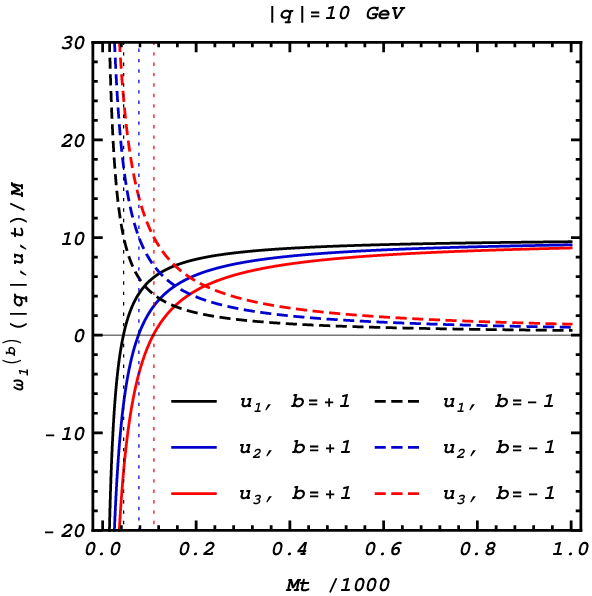}
\end{center}
\caption{$\omega^{(b)}_1(q,u,t)/M$ versus $Mt$ for the first three
  extrema of $\mathrm{sinc}(u)$: $u_1\:=\:4.49$ (left-most black
  lines), $u_2\:=\:7.73$ (central blue lines) and $u_3\:=\:10.90$
  (right-most red lines), obtained by (\ref{eq:un}), for
  $q^2\:=\:M^2$, $|\mathbf{q}|\:=\:10\ \mathrm{GeV}$ and
  $\alpha_{\theta}\:=\:+1$.  Solid lines correspond to $b\:=\:1$ and
  dashed lines, $b\:=\:-1$ in (\ref{eq:w1}).  The dotted lines mark
  the upper limit of the kinematically disallowed region.}
\label{fig:nonmark}
\end{figure}

In terms  of the evanescent  action $u$ given in  (\ref{eq:eact}), the
high-temperature limit of the one-loop $\Phi$ width reads:
\begin{align}
  \label{eq:hightlim}
  &\Gamma_{\Phi}^{(1),\, T\:\gg\:M}(q,t)\ =\ \frac{g^2}{32\pi^2Mt}
  \sum_{\{\alpha\},\,b\:=\:\pm 1}\alpha_1\alpha_2\!\int_{u_-(t)}^{u_+(t)}\!\D{}{u}
  \nonumber\\&\qquad \times\: \Bigg\{\frac{q_u(t)}{\lambda^{1/2}\big(q_u^2(t)
  \:-\:|\mathbf{q}|^2,m_1^2,m_2^2\big)}\Bigg[\frac{m_1^2\:+\:m_2^2}
  {q_u^2(t)\:-\:|\mathbf{q}|^2}\:-\:\Bigg(\frac{m_1^2\:-\:m_2^2}
  {q_u^2(t)\:-\:|\mathbf{q}|^2}\Bigg)^{\!2}\,\Bigg]\nonumber\\&\qquad\qquad +\:
  \frac{b\alpha_{\theta}}{2|\mathbf{q}|}\Bigg(1\:-\:\frac{\big(q_u^2(t)+|\mathbf{q}|^2\big)\big(m_1^2\:-\:m_2^2\big)}
  {\big(q_u^2(t)\:-\:|\mathbf{q}|^2\big)^2}\Bigg)\!\Bigg\}
  \Bigg[\Bigg(1\:+\:\frac{T}{\omega^{(b)}_1(q,u,t)}\:+\:\frac{T}{q_u(t)\:-\:
  \omega^{(b)}_2(q,u,t)}\Bigg)\,\mathrm{Si}(u)\nonumber\\&\qquad \qquad -\:
  \frac{T}{\omega^{(b)}_1(q,u,t)}\Bigg(\!\sin\!\big(\omega^{(b)}_1(q,u,t)t\big)\,
  \mathrm{Ci}\big(\omega^{(b)}_2(q,u,t)t\big)\:-\:
  \cos\!\big(\omega^{(b)}_1(q,u,t)t\big)\,\mathrm{Si}\big(\omega^{(b)}_2(q,u,t)t\big)
  \!\Bigg)\Bigg]\;,
\end{align}
where $\omega^{(b)}_1(q,u,t)$  and $\omega^{(b)}_2(q,u,t)$ are time-dependent
non-Markovian frequencies defined as
\begin{subequations}
  \begin{align}
    \label{eq:w1}
    \omega^{(b)}_1(q,u,t)&\ =\ q_0\:-\:\frac{\big(q_u^2(t)\,-\,
    |\mathbf{q}|^2\,+\,m_1^2\,-\,m_2^2\big)q_u(t)\:+\:b\alpha_{\theta}
    |\mathbf{q}|\lambda^{1/2}\big(q_u^2(t)-
    |\mathbf{q}|^2,m_1^2,m_2^2\big)}{2\big(q_u^2(t)\:-\:
    |\mathbf{q}|^2\big)}\;,\\
    \omega^{(b)}_2(q,u,t)&\ =\ \frac{\big(q_u^2(t)\,-\,
    |\mathbf{q}|^2\,-\,m_1^2\,+\,m_2^2\big)q_u(t)\:-\:b\alpha_{\theta}
    |\mathbf{q}|\lambda^{1/2}\big(q_u^2(t)-
    |\mathbf{q}|^2,m_1^2,m_2^2\big)}
    {2\big(q_u^2(t)\:-\:|\mathbf{q}|^2\big)}\;.
\end{align}
\end{subequations}
The time dependence of  the frequency $\omega^{(b)}_1(q,u,t)$ is shown
in  Figure~\ref{fig:nonmark}  for  $\alpha_{\theta}\:=\:+1$, with  the
on-shell   constraint  $q^2\:=\:M^2$.    The   $b\:=\:1$  contribution
persists  for  late  times,  at  which  point  the  frequency  of  the
modulations  have decayed  to  zero.  The  $b\:=\:-1$ contribution  is
disallowed  for   late  times,  such   that  the  amplitude   of  this
constant-frequency    modulation    is    damped   to    zero.     For
$\alpha_{\theta}\:=\:-1$, the  $b\:=\:+1$ and $b\:=\:-1$ contributions
are  interchanged,   such  that   the  frequency  of   the  $b\:=\:-1$
contribution  reduces to  zero for  late times.   Thus, we  obtain the
expected kinematics and equilibrium  behavior in the late-time limit.
We   have  not   plotted   $\omega^{(b)}_2$,  as   its  behavior   is
indistinguishable  from  $\omega^{(b)}_1$, for $m_1\:=\:m_2\:=\:m$.

Looking   again  at   (\ref{eq:hightlim}),  we   observe   that  these
non-Markovian oscillations  occur only for $T\:\neq\:  0$. We conclude
therefore  that  this behavior  signifies  a genuine  non-equilibrium
statistical effect.

\subsection{Perturbative Time Evolution Equations}
\label{sec:toyeom}

In  this  section, we  carry  out a  loopwise  expansion  of the  time
evolution equations derived  in Section~\ref{sec:eom} to leading order
in  the coupling  $g$,  evaluating their  one-loop  structure for  our
simple  scalar  model.  For  perturbatively  small  couplings, such  a
loopwise expansion  is expected  to accurately capture  the early-time
dynamics of  non-equilibrium systems. This  is a regime, in  which the
applicability of  truncated gradient expansions  becomes questionable,
according to our  discussion in Section~\ref{sec:gradapp}.  Given
the  closed  analytic form  of  the  free  CTP propagators,  both  the
amplitudes  of contributing  processes and  the  resulting phase space
integrals  describing the kinematics  can be  analytically determined.
The systematic  treatment of these  kinematic effects is  essential to
understand   the  consequences   on  the   early-time   dynamics.   In
particular, we  illustrate the significance of  contributions from the
energy-non-conserving evanescent regime to the dynamics of the system.

Let   us   first   return   to   the   Boltzmann--like   equation   in
(\ref{eq:boltz}),   in   which    we   artificially   imposed   energy
conservation. The collision terms for the real scalar $\Phi$ are
\begin{equation}
  \label{eq:phicol}
  \Pi_{\Phi,\,>}(p,t)\Delta_{\Phi,\,<}(p,t)\:-\:
  \Pi_{\Phi,\,<}(p,t)\Delta_{\Phi,\,>}(p,t)\;.
\end{equation}
Since we have assumed that the two subsystems $\mathscr{S}_{\Phi}$ and
$\mathscr{S}_{\chi}$  are separately  in thermodynamic  equilibrium at
the  same temperature  $T$  at the  initial  time $t\:=\:0$,  the
$\Phi$ propagators  and self-energies  will initially satisfy  the KMS
relation, i.e.
\begin{equation}
  \Delta_{\Phi,\,>}(p,0)\ =\ e^{\beta p_0}\Delta_{\Phi,\,<}(p,0)\;, \qquad
  \Pi_{\Phi,\,>}(p,0)\ =\ e^{\beta p_0}\Pi_{\Phi,\,<}(p,0)\;.
\end{equation}
As a consequence of the KMS relation, the collision terms in
(\ref{eq:phicol}) are identical to zero.  This is true also in the
collision terms of the truncated gradient expansion of the
Kadanoff--Baym kinetic equation in (\ref{eq:kinetic}).  With no
external sources present to perturb the combined system $\mathscr{S}$
from this non-interacting equilibrium, we are faced with the problem
that the statistical distribution functions will \emph{not} evolve
from their initial forms.  In reality, after the quantum quench, we
anticipate that the system should evolve to some new interacting
thermodynamic equilibrium for late times.  We conclude therefore that
the energy-non-conserving evanescent regime described by our approach
is entirely necessary to account correctly for the evolution of this
system.

We  now turn  our  attention  to the  master  time evolution  equation
(\ref{eq:evo}).  Truncating to leading  order in the coupling $g$, the
RHS  of this  time evolution  equation contains  free  propagators and
one-loop  self-energies.  For instance,  the one-loop  collision terms
in~(\ref{eq:evo})~are obtained from
\begin{align}
  \mathscr{C}^{(1)}(p+\tfrac{P}{2},p-\tfrac{P}{2},t;0)\ & \equiv\ 
  \frac{1}{2}\!\int\!\!\frac{\D{4}{q}}{(2\pi)^4}\;
  \Big[\,i\Pi^{(1)}_>(p+\tfrac{P}{2},q,t;0)i\Delta^0_<(q,p-\tfrac{P}{2},t;0)\nonumber\\
  &\qquad -\:i\Pi^{(1)}_<(p+\tfrac{P}{2},q,t)\,\Big(\,i\Delta^0_>(q,p-\tfrac{P}{2},t;0)
  \:-\:2i\Delta^0_{\mathcal{P}}(q,p-\tfrac{P}{2},t;0)\,\Big)\Big]\;.
\end{align}
We    insert   the    one-loop   integrals    (\ref{eq:phiself})   and
(\ref{eq:chiself}),  containing  the  homogeneous and  spectrally-free
$\Phi$  and $\chi$ distribution  functions, $f_{\Phi}(|\mathbf{p}|,t)$
and  $f_{\chi}^{(C)}(|\mathbf{p}|,t)$, into  the  full time  evolution
equation (\ref{eq:evo})  for the $\Phi$ and $\chi$  fields.  Note that
the  $f$ distribution  functions  are real in  this  spatially homogeneous  limit.  We  tacitly
assume  that a system  initially prepared  in a  spatially homogeneous
state  remains spatially homogeneous  throughout its  evolution.  This
assumption is reasonable if the system has~infinite~spatial~extent.

We may perform the loop and convolution integrals using the techniques
and results of  Appendix~\ref{app:loops}.  After carrying out the
$p_0$ and  $P_0$ integrals in (\ref{eq:evo}), we  obtain the following
one-loop  time  evolution   equation  for  the  spatially  homogeneous
statistical  distribution function  $f_{\Phi}(|\mathbf{p}|,t)$  of the
real scalar field:
\begin{align}
  \label{eq:phievo}
  \partial_tf_{\Phi}(|\mathbf{p}|,t)\ &=\ -\frac{g^2}{2}\sum_{\{\alpha\}}
  \int\!\!\frac{\D{3}{\mathbf{k}}}{(2\pi)^3}\,\frac{1}{2E_{\Phi}(\mathbf{p})}
  \,\frac{1}{2E_{\chi}(\mathbf{k})}\,\frac{1}{2E_{\chi}(\mathbf{p}-\mathbf{k})}
  \nonumber\\&\qquad \times\: \frac{t}{2\pi}\,\mathrm{sinc}
  \big[\big(\alpha E_{\Phi}(\mathbf{p})\:-\:\alpha_1E_{\chi}(\mathbf{k})
  \:-\:\alpha_2E_{\chi}(\mathbf{p}-\mathbf{k})\big)t/2\big]\nonumber\\&
  \qquad \times\: \Big\{\,\pi\:+\:2\mathrm{Si}\big[\big(\alpha E_{\Phi}(\mathbf{p})
  \:+\:\alpha_1 E_{\chi}(\mathbf{k})\:+\:\alpha_2
  E_{\chi}(\mathbf{p}-\mathbf{k})\big)t/2\big]\,\Big\}\nonumber\\&\qquad \times \:
  \Big\{\Big(\,\theta(-\alpha)\:+\:f_{\Phi}(|\mathbf{p}|,t)\,\Big)
  \Big[\,\theta(\alpha_1)\Big(\,1\:+\:f_{\chi}(|\mathbf{k}|,t)\,\Big)\:+\:
  \theta(-\alpha_1)f_{\chi}^C(|\mathbf{k}|,t)\,\Big]\nonumber\\&
  \qquad \quad \quad \quad \quad
  \times\: \Big[\,\theta(\alpha_2)\Big(\,1\:+\:
  f^C_{\chi}(|\mathbf{p}-\mathbf{k}|,t)\,\Big)\:+\:\theta(-\alpha_2)
  f_{\chi}(|\mathbf{p}-\mathbf{k}|,t)\,\Big]\nonumber\\&\qquad -\:
  \Big(\,\theta(\alpha)\:+\:f_{\Phi}(|\mathbf{p}|,t)\,\Big)
  \Big[\,\theta(\alpha_1)f_{\chi}(|\mathbf{k}|,t)\:+\:\theta(-\alpha_1)
  \Big(\,1\:+\:f_{\chi}^C(|\mathbf{k}|,t)\,\Big)\Big]\nonumber\\&
  \qquad \quad \quad \quad \quad \times\: \Big[\,\theta(\alpha_2)
  f^C_{\chi}(|\mathbf{p}-\mathbf{k}|,t)\:+\:\theta(-\alpha_2)
  \Big(\,1\:+\:f_{\chi}(|\mathbf{p}-\mathbf{k}|,t)\,\Big)\Big]\Big\}\;.
\end{align}
where $\mathrm{Si}(x)$ is the  sine integral function and $\{\alpha\}$
is    the    short-hand    notation    for    the    summation    over
$\alpha,\ \alpha_1,\ \alpha_2\:=\:\pm 1$. With the summation over
$\{\alpha\}$, the statistical  factors in the braces of  the last four
lines  of (\ref{eq:phievo})  contain the  difference  of contributions
from  the  four processes  shown  in  Figure~\ref{fig:kins} and  their
inverse processes. For early  times, all of these evanescent processes
and   inverse   processes   contribute,    as   can   be   seen   from
Figure~\ref{fig:seps}.  The  presence of~$\mathrm{Si}(x)$ ensures
that  the  correct late-time  limit  and  kinematics  are obtained  on
restoration  of energy conservation.   This factor  is missing  in the
existing      descriptions,      see     e.g.~\cite{Boyanovsky:1998pg,
  DeSimone:2007rw}.   The  dispersive  force term  and  off-shell
collision term vanish in the spatially homogeneous case, thanks to the
symmetry of the self-energy under $P\:\to\:-P$.

In the large-time limit $t\:\to\:\infty$, we have
\begin{align}
  &\frac{t}{2\pi}\,\mathrm{sinc}\big[\big(\alpha E_{\Phi}(\mathbf{p})
  \:-\:\alpha_1E_{\chi}(\mathbf{k})\:-\:\alpha_2
  E_{\chi}(\mathbf{p}-\mathbf{k})\big)t/2\big]\Big\{\,\pi\:+\:2\mathrm{Si}\big[\big(\alpha
  E_{\Phi}(\mathbf{p})\:+\:\alpha_1E_{\chi}(\mathbf{k})\:+\:\alpha_2
  E_{\chi}(\mathbf{p}-\mathbf{k})\big)t/2\big]\,\Big\}\nonumber\\&\qquad 
  \underset{t\:\to\:\infty}{\longrightarrow} \
  2\pi\theta(\alpha)\delta\big(E_{\Phi}(\mathbf{p})\:-\:
  \alpha_1E_{\chi}(\mathbf{k})\:-\:\alpha_2E_{\chi}(\mathbf{p}-\mathbf{k})
  \big)\;.
\end{align}
The          kinematic         constraints          then         force
$\alpha\:=\:\alpha_1\:=\:\alpha_2\:=\:+1$ and we obtain
\begin{align}
  \label{eq:BElim}
  &\partial_tf_{\Phi}(|\mathbf{p}|,t)\ =\ -\frac{g^2}{2}\!\int\!\!
  \frac{\D{3}{\mathbf{k}}}{(2\pi)^3}\,\frac{1}{2E_{\Phi}(\mathbf{p})}\,
  \frac{1}{2E_{\chi}(\mathbf{k})}\,
  \frac{1}{2E_{\chi}(\mathbf{p}-\mathbf{k})}\;2\pi
  \delta\big(E_{\Phi}(\mathbf{p})\:-\:E_{\chi}(\mathbf{k})\:-\:
  E_{\chi}(\mathbf{p}-\mathbf{k})\big)\nonumber\\&
  \qquad \times\: \Big[\,f_{\Phi}(|\mathbf{p}|,t)\Big(\,1\:+\:
  f_{\chi}(|\mathbf{k}|,t)\,\Big)\Big(\,1\:+\:
  f^C_{\chi}(|\mathbf{p}-\mathbf{k}|,t)\,\Big)\:-\: 
  \Big(\,1\:+\:f_{\Phi}(|\mathbf{p}|,t)\,\Big)f_{\chi}(|\mathbf{k}|,t)
  f^C_{\chi}(|\mathbf{p}-\mathbf{k}|,t)\,\Big]\;.
\end{align}
Equation (\ref{eq:BElim})  corresponds  to  the  semi-classical
Boltzmann equation [cf. (\ref{eq:boltz})].

By   analogy,   the  one-loop   time   evolution   equation  for   the
spatially homogeneous      statistical      distribution      function
$f_{\chi}(|\mathbf{p}|,t)$ of the complex scalar field $\chi$ is given
by
\begin{align}
  \label{eq:chievo}
  \partial_tf_{\chi}(|\mathbf{p}|,t)\ &=\ -\frac{g^2}{2}\sum_{\{\alpha\}}
  \int\!\!\frac{\D{3}{\mathbf{k}}}{(2\pi)^3}\,
  \frac{1}{2E_{\Phi}(\mathbf{p})}\,\frac{1}{2E_{\chi}(\mathbf{k})}\,
  \frac{1}{2E_{\chi}(\mathbf{p}-\mathbf{k})}\nonumber\\&\qquad \times \:
  \frac{t}{2\pi}\,\mathrm{sinc}\big[\big(\alpha E_{\chi}(\mathbf{p})
  \:-\:\alpha_1E_{\Phi}(\mathbf{k})\:-\:
  \alpha_2E_{\chi}(\mathbf{p}-\mathbf{k})\big)t/2\big]\nonumber\\&
  \qquad \times\: \Big\{\,\pi\:+\:2\mathrm{Si}\big[
  \big(\alpha E_{\chi}(\mathbf{p})\:+\:\alpha_1 E_{\Phi}(\mathbf{k})\:+\:
  \alpha_2 E_{\chi}(\mathbf{p}-\mathbf{k})\big)t/2\big]\,\Big\}\nonumber\\
  &\qquad \times\: \Big\{\Big[\,\theta(\alpha)f_{\chi}(|\mathbf{p}|,t)\:+\:
  \theta(-\alpha)\Big(\,1\:+\:f_{\chi}^C(|\mathbf{p}|,t)\,\Big)\Big]
  \Big(\,\theta(\alpha_1)\:+\:f_{\Phi}(|\mathbf{k}|,t)\,\Big)\nonumber\\&
  \qquad \quad \quad \quad \quad
  \times\: \Big[\,\theta(\alpha_2)\Big(\,1\:+\:
  f_{\chi}(|\mathbf{p}-\mathbf{k}|,t)\,\Big)\:+\:
  \theta(-\alpha_2)f^C_{\chi}(|\mathbf{p}-\mathbf{k}|,t)\,\Big]
  \nonumber\\&\qquad -\: \Big[\,\theta(\alpha)\Big(\,1\:+\:
  f_{\chi}(|\mathbf{p}|,t)\,\Big)\:+\:\theta(-\alpha)
  f_{\chi}^C(|\mathbf{p}|,t)\,\Big]\Big(\,\theta(-\alpha_1)\:+\:
  f_{\Phi}(|\mathbf{p}|,t)\,\Big)\nonumber\\&\qquad \quad \quad \quad \quad
  \times\: \Big[\,\theta(\alpha_2)f_{\chi}(|\mathbf{p}-\mathbf{k}|,t)\:+\:
  \theta(-\alpha_2)\Big(\,1\:+\:f^C_{\chi}(|\mathbf{p}-\mathbf{k}|,t)\,\Big)
  \Big]\Big\}\;.
\end{align}
In  the large-time  limit $t\:\to\:\infty$,  the  kinematics restrict
$\alpha\:=\:\alpha_1\:=\:-\alpha_2\:=\:+1$, giving
\begin{align}
  &\partial_tf_{\chi}(|\mathbf{p}|,t)\ =\ -\frac{g^2}{2}\!\int\!\!
  \frac{\D{3}{\mathbf{k}}}{(2\pi)^3}\,\frac{1}{2E_{\Phi}(\mathbf{p})}\,
  \frac{1}{2E_{\chi}(\mathbf{k})}\,
  \frac{1}{2E_{\chi}(\mathbf{p}-\mathbf{k})}\;2\pi
  \delta\big(E_{\Phi}(\mathbf{k})\:-\:E_{\chi}(\mathbf{p})\:-\:
  E_{\chi}(\mathbf{p}-\mathbf{k})\big)\nonumber\\&\qquad \times\: 
  \Big[\,\Big(\,1\:+\:f_{\Phi}(|\mathbf{k}|,t)\,\Big)
  f_{\chi}(|\mathbf{p}|,t)f_{\chi}^C(|\mathbf{p}-\mathbf{k}|,t)\:
  -\: f_{\Phi}(|\mathbf{k}|,t)\Big(\,1\:+\:f_{\chi}(|\mathbf{p}|,t)\,\Big)
  \Big(\,1\:+\:f^C_{\chi}(|\mathbf{p}-\mathbf{k}|,t)\,\Big)\Big]\;.
\end{align}
This  result is  consistent with  (\ref{eq:BElim}), differing  by an
overall sign, as one should expect.

At  $t\:=\:0$,  the  semi-classical  Boltzmann transport  equation  in
(\ref{eq:BElim}) becomes
\begin{align}
  \label{eq:BElimRHS}
  &\partial_tf_{\Phi}(|\mathbf{p}|,t)\big|_{t\:=\:0}\ = \ 
   -\frac{g^2}{2}\!\int\!\!\frac{\D{3}{\mathbf{k}}}{(2\pi)^3}
  \,\frac{1}{2E_{\Phi}(\mathbf{p})}\,\frac{1}{2E_{\chi}(\mathbf{k})}\,
  \frac{1}{2E_{\chi}(\mathbf{p}-\mathbf{k})}\;2\pi
  \delta\big(E_{\Phi}(\mathbf{p})\:-\:E_{\chi}(\mathbf{k})\:-\:
  E_{\chi}(\mathbf{p}-\mathbf{k})\big)\nonumber\\&\qquad \times\: 
  \Big[\,f_{\mathrm{B}}\big(E_{\Phi}(\mathbf{p})\big)\Big(1\:+\:
  f_{\mathrm{B}}\big(E_{\chi}(\mathbf{k})\big)\:+\:
  f^C_{\mathrm{B}}\big(E_{\chi}(\mathbf{p}-\mathbf{k})\big)\Big)\:-\: 
  f_{\mathrm{B}}\big(E_{\chi}(\mathbf{k})\big)
  f^C_{\mathrm{B}}\big(E_{\chi}(\mathbf{p}-\mathbf{k})\big)\Big]\;,
\end{align}
where we  remind the reader  that energy conservation  is artificially
imposed. By virtue  of this energy conservation, the  first product of
statistical factors in (\ref{eq:BElimRHS}) satisfies the identity
\begin{equation}
  f_{\mathrm{B}}\big(E_{\Phi}(\mathbf{p})\big)\Big(1\:+\:
  f_{\mathrm{B}}\big(E_{\chi}(\mathbf{k})\big)\:+\:
  f^C_{\mathrm{B}}\big(E_{\chi}(\mathbf{p}-\mathbf{k})\big)\Big)\ =\ 
  f_{\mathrm{B}}\big(E_{\chi}(\mathbf{k})\big)
  f^C_{\mathrm{B}}\big(E_{\chi}(\mathbf{p}-\mathbf{k})\big)\;.
\end{equation}
As  a  result,  the  RHS   of  (\ref{eq:BElim})  is  exactly  zero  at
$t\:=\:0$.   Consequently,  the energy-non-conserving  evanescent
regime plays a fundamental role in the description of the evolution of
the system $\mathscr{S}$.

We now wish to show that the master time evolution equations
(\ref{eq:phievo}) and (\ref{eq:chievo}) of our perturbative approach
describe a non-trivial evolution of the system $\mathscr{S}$.  For
this purpose, we consider the time evolution equation of the heavy
scalar $\Phi$ in (\ref{eq:phievo}) and assume that the $\chi$
statistical distribution functions $f^{(C)}_{\chi}(|\mathbf{p}|,t)$ of
the RHS remain in their initial equilibrium forms for all times,
i.e.~$f^{(C)}_{\chi}(|\mathbf{p}|,t)\: =\:
f_{\mathrm{B}}\big(E_{\chi}(\mathbf{p})\big)$ ($\mu/T\:\ll\:1$).  With
this assumption, we see from Figure~\ref{fig:phitime} that the RHS of
(\ref{eq:phievo}) is non-zero for early times.  Thus, the system
$\mathscr{S}$ is indeed perturbed from its non-interacting equilibrium
by the evanescent processes described by our perturbative time
evolution equations (\ref{eq:phievo}) and (\ref{eq:chievo}).

\begin{figure}
\begin{center}
\includegraphics[scale=0.45]{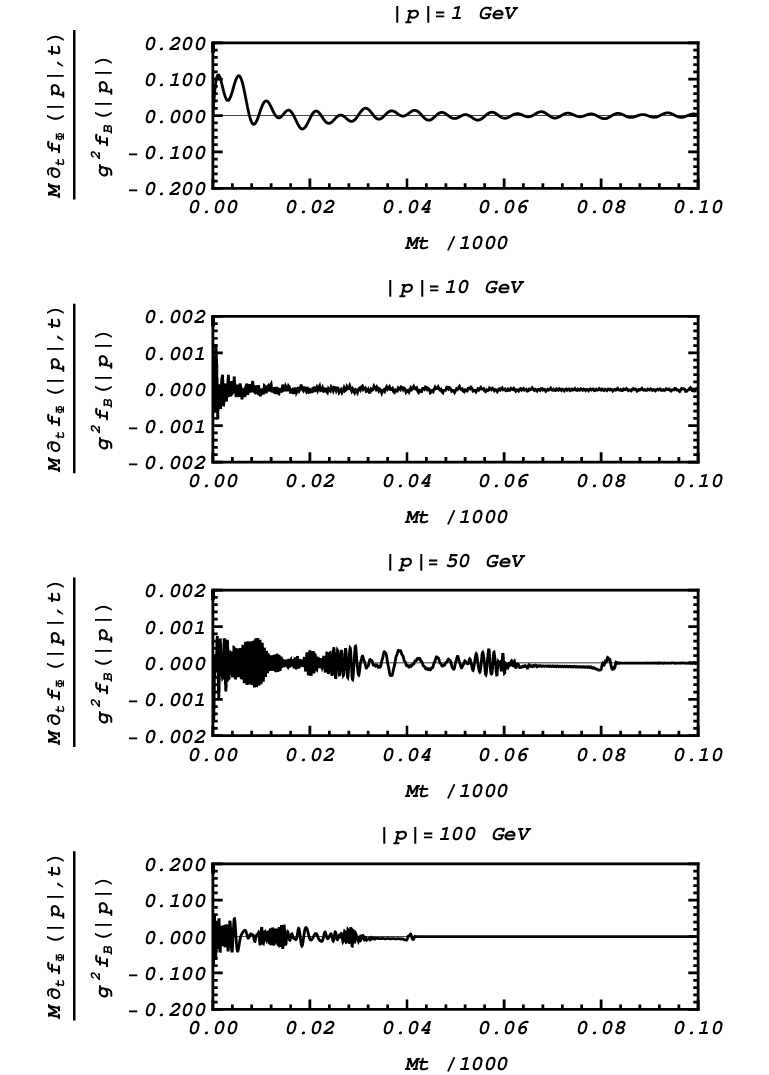}
\end{center}
\caption{Numerical estimates of $\partial_t f_{\Phi}(|\mathbf{p}|,t)$,
  as  a  function  of  $Mt$,  assuming  that  the  $\chi$  statistical
  distribution  functions $f^{(C)}_{\chi}(|\mathbf{p}|,t)$ on  the RHS
  of  the  time  evolution equation~(\ref{eq:phievo})  maintain  their
  equilibrium Bose--Einstein form for all times.}
\label{fig:phitime}
\end{figure}

Let us  finally have  a closer look  at the early-time  behavior that
immediately follows  after the switching  on of the  interactions.  We
expect  that this prompt  behavior is  dominated by  the ultra-violet
contribution   to   the   phase-space   integral   on   the   RHS   of
(\ref{eq:phievo}) due  to the Heisenberg  uncertainty principle.  From
Figure~\ref{fig:twidth},  we see  that this  shock  regime contributes
dominantly  to  the  prompt   evolution  of  the  infra-red  modes  of
$f_{\Phi}(|\mathbf{p}|,t)$.     In   this    regime,    we   take
$|\mathbf{p}|\:=\:0$ and $m\:=\:0$  in (\ref{eq:phievo}) and introduce
the ultra-violet  cut-off $\Lambda$ in  the limits of  the phase-space
integrals of the RHS.  Assuming that the $\chi$ distribution functions
are tempered, vanishing faster than  a power law for large momenta, we
may  ignore  their contribution  in  the  ultra-violet limit.   Hence,
$\partial_tf_{\Phi}(|\mathbf{p}|=0,t)$ can be approximated by
\begin{equation}
  \label{eq:cutoff}
  \partial_t f_{\Phi}(|\mathbf{p}|=0,t)\ \sim\ 
  -\frac{g^2}{32\pi^3M}\Big(\,1\:+\:2f_{\Phi}(|\mathbf{p}|=0,t)\,\Big)
  \lim_{\Lambda\:\to\:\infty}\mathrm{Si}^2(\Lambda t)\;.
\end{equation}
Clearly,  $  \partial_t  f_{\Phi}(|\mathbf{p}|=0,t)$ vanishes  in  the
limit       $t\:\ll\:1/\Lambda\:\to\:0$,      as       we      expect.
Expanding~$\mathrm{Si}\,(\Lambda t)$ about $\Lambda t\:=\:0$, i.e.~for
times infinitesimally close to zero, we obtain
\begin{equation}
  \partial_t f_{\Phi}(|\mathbf{p}|=0,t)\ \sim\ 
  -\frac{g^2}{32\pi^3 M}\Big(\,1\:+\:2f_{\Phi}(|\mathbf{p}|= 0,0)\,\Big)
  \lim_{\Lambda\:\to\:\infty}(\Lambda t)^2\;.
\end{equation}

For  small but  finite times,  ultra-violet contributions  are rapidly
varying.     It    is    apparent    from    this    discussion    and
Figures~\ref{fig:twidth} and \ref{fig:phitime}  that the effect of the
transient behavior of  the system  upon its  subsequent  dynamics is
significant, since the quantum memories  persist on long  time scales
$Mt\: \gg  \:1$.  We  may therefore conclude  that as  opposed to
other methods relying on  a truncated gradient expansion, our approach
consistently  describes this  highly-oscillatory  and rapidly-evolving
early-time behavior of the system.

\subsection{Inclusion of Thermal Masses}

In  this  section,   we  describe  how  local  thermal-mass
corrections may be incorporated consistently into our approach.

The  local part of  the one-loop  $\chi$ self-energy  shown in
Figure~\ref{fig:loc} has the explicit form
\begin{align}
  &\Pi_{\chi}^{\mathrm{loc}(1)}(p,p',\tilde{t}_f;\tilde{t}_i)\ =\ -\frac{\lambda}{2}(2\pi)^4
  \delta_t^{(4)}(p\:-\:p')(2\pi\mu)^{2\epsilon}e^{i(p_0\:-\:p_0')\tilde{t}_f}\nonumber\\
  & \qquad \qquad \times\: \!\iint\!\!\frac{\D{d}{k}}{(2\pi)^d}\,
  \frac{\D{4}{k'}}{(2\pi)^4}\;\bigg(\,\frac{i}{k^2\:-\:m^2\:+\:i\epsilon}
  (2\pi)^4\delta^{(4)}(k\:-\:k')\nonumber\\&\qquad \qquad +\: 2\pi\delta(k^2\:-\:m^2)
  |2k_0|^{1/2}\tilde{f}_{\chi}(k,k',t)e^{i(k_0\:-\:k_0')\tilde{t}_f}|2k_0'|^{1/2}2\pi
  \delta({k'}^2\:-\:m^2)\,\bigg)\;,
\end{align}
with  $d\:=\:4\:-\:2\epsilon$   (cf.~Appendix  \ref{app:loops}).   The
first   term  yields   the   standard  zero-temperature   UV
divergence,  which is  usually removed  by mass  renormalization.  The
second term yields
\begin{equation}
  \Pi_{\chi}^{\mathrm{loc}(1)}(p,p',\tilde{t}_f;\tilde{t}_i)\ =\ -(2\pi)^4\delta_t^{(4)}(p\:-\:p')
  e^{i(p_0\:-\:p_0')\tilde{t}_f}m_{\mathrm{th}}^2(\tilde{t}_f;\tilde{t}_i)\;,
\end{equation}
where         the          time-dependent         thermal         mass
$m_{\mathrm{th}}(\tilde{t}_f;\tilde{t}_i)$, given by
\begin{align}
  \label{eq:mth2}
  m_{\mathrm{th}}^2(\tilde{t}_f;\tilde{t}_i)\ &=\ \frac{\lambda}{2}\!\int\!\!
  \frac{\D{3}{\mathbf{k}}}{(2\pi)^3}\,\frac{1}{\sqrt{2E_{\chi}(\mathbf{k})}}
  \int\!\!\frac{\D{3}{\mathbf{k}'}}{(2\pi)^3}\,\frac{1}
  {\sqrt{2E_{\chi}(\mathbf{k}')}}\nonumber\\&\qquad \times\:
  \Big(\,f_{\chi}(\mathbf{k},\mathbf{k}',t)e^{i[E(\mathbf{k})\:-\:E(\mathbf{k}')]\tilde{t}_f}
  \:+\:f^{C*}_{\chi}(-\mathbf{k},-\mathbf{k}',t)
  e^{-i[E(\mathbf{k})\:-\:E(\mathbf{k}')]\tilde{t}_f}\,\Big)\;,
\end{align}
is UV finite. Notice that a  unique thermal mass may not be defined in
the  spatially inhomogeneous case  due to  the explicit  dependence on
$\tilde{t}_f$.

From the  Schwinger--Dyson equation in  (\ref{eq:sddmom}), the inverse
quasi-particle                      CTP                     propagator
$\Delta^{-1}_{\chi,\,ab}(p,p',\tilde{t}_f;\tilde{t}_i)$ of the complex
scalar field $\chi$ takes the form
\begin{equation}
  \label{eq:invchi}
  \Delta^{-1}_{\chi,\,ab}(p,p',\tilde{t}_f;\tilde{t}_i)\ =\ (2\pi)^4\delta^{(4)}_t(p\:-\:p')
  e^{i(p_0\:-\:p_0')\tilde{t}_f}\Big[\Big(\,p'^2\:-\:m_{\mathrm{th}}^2(\tilde{t}_f;\tilde{t}_i)\,\Big)
  \eta_{ab}\:+\:i\epsilon\mathbb{I}_{ab}\,\Big]\;,
\end{equation}
in which we have  assumed $m_{\mathrm{th}}(t)\:\gg\:m$. If the quartic
$\chi$ self-interaction  is switched on sufficiently  long before $t\:
=\: 0$, then for $t\:\geq\: 0$, we may replace the $\delta_t$ function
in the inverse quasi-particle propagator (\ref{eq:invchi}) by an exact
energy-conserving   delta   function.   This  imposition   of   energy
conservation constitutes  a quasi-particle approximation,  allowing us
to   invert  (\ref{eq:invchi})   exactly,  using   the   arguments  of
Sections~\ref{sec:CTPprop}  and  \ref{sec:nonhomprop},  to obtain  the
following quasi-particle $\chi$ propagators:
\begin{align}
  \label{eq:qpprops}
  \Delta^{0,\,ab}_{\chi}(p,p',\tilde{t}_f;\tilde{t}_i) \ &= \ 
  \!\begin{bmatrix}
    \big(p^2\,-\,m_{\mathrm{th}}^2(\tilde{t}_f;\tilde{t}_i)\,+\,i\epsilon)^{-1} & -i2\pi
    \theta(-p_0)\delta\big(p^2\,-\,m_{\mathrm{th}}^2(\tilde{t}_f;\tilde{t}_i)\big) \\
    -i2\pi\theta(p_0)\delta\big(p^2\,-\,m_{\mathrm{th}}^2(\tilde{t}_f;\tilde{t}_i)\big) &
    -\big(p^2\,-\,m_{\mathrm{th}}^2(\tilde{t}_f;\tilde{t}_i)\,-\,i\epsilon\big)^{-1}
  \end{bmatrix}\!(2\pi)^4\delta^{(4)}(p\,-\,p')\nonumber\\
  & \qquad -\:i2\pi|2p_0|^{1/2}\delta\big(p^2\,-\,m_{\mathrm{th}}^2(\tilde{t}_f;\tilde{t}_i)\big)
  \tilde{f}_{\chi}(p,p',t)e^{i(p_0\,-\,p_0')\tilde{t}_f}2\pi|2p_0'|^{1/2}
  \delta\big(p'^2\,-\,m_{\mathrm{th}}^2(\tilde{t}_f;\tilde{t}_i)\big)
  \!\begin{bmatrix}
    1 & 1 \\
    1 & 1
  \end{bmatrix}\!\;.
\end{align}

If the  subsystem $\mathscr{S}_{\chi}$ is in a  state of thermodynamic
equilibrium at the initial time $t\:=\:0$, the thermal mass reduces to
the known result
\begin{equation}
  m_{\mathrm{th}}^2(t\:=\:0)\ =\ \frac{\lambda T^2}{24}\;,
\end{equation}
for $m\:=\:0$ and $\mu\:\ll\: T$. In order to describe completely
the   dynamics    of   the   combined    system   $\mathscr{S}\:   =\:
\mathscr{S}_{\Phi}\:\cup\:\mathscr{S}_{\chi}$,  we couple  the
evolution  of  the  thermal mass  $m_{\mathrm{th}}(t)$  to  the
perturbative   time    evolution   equations   (\ref{eq:phievo})   and
(\ref{eq:chievo}) for  the $\Phi$ and  $\chi$ statistical distribution
functions.  This  is achieved by  differentiating (\ref{eq:mth2}) with
respect to $t$, such that
\begin{equation}
  \partial_t m_{\mathrm{th}}(t)\ =\ \frac{\lambda}{2m_{\mathrm{th}}(t)}\!\int
  \!\!\frac{\D{3}{\mathbf{k}}}{(2\pi)^3}\,\frac{1}{2E_{\chi}(\mathbf{k})}
  \;\frac{1}{2}\Big(\,\partial_tf_{\chi}(|\mathbf{k}|,t)\:+\:
  \partial_tf^C_{\chi}(|\mathbf{k}|,t)\,\Big)\;.
\end{equation}
Here,    $\partial_t    f_{\chi}(|\mathbf{k}|,t)$   and    $\partial_t
f_{\chi}^C(|\mathbf{k}|,t)$        depend        implicitly       upon
$m_{\mathrm{th}}(t)$   via   the   coupled  first-order   differential
equations (\ref{eq:phievo})  and (\ref{eq:chievo}), derived  using the
quasi-particle  $\chi$   propagators  in~(\ref{eq:qpprops}).   On
dimensional grounds, we may  estimate that the leading contribution to
$\partial_t m_{\mathrm{th}}(t)$  is of order  $\lambda^{1/2}g T$.
The  latter estimate  provides firm  support that  our  time evolution
equations   may  consistently  incorporate   thermal  masses   and  so
describe the main non-perturbative dynamics of the system.

\section{Conclusions}
\label{sec:con}

We  have  developed a  new  perturbative  approach to  non-equilibrium
thermal quantum field theory.  Our perturbative approach is based upon
non-homogeneous  free propagators  and time-dependent  vertices, which
explicitly  break  space-time  translational invariance  and  properly
encode the dependence  of the system on the  time of observation.  The
forms   of   these   propagators   are  constrained   by   fundamental
field-theoretic  requirements, such  as \emph{CPT}  invariance  of the
action, Hermiticity properties of correlation functions, causality and
unitarity.  We have shown that our perturbative approach gives rise to
time-dependent  diagrammatic perturbation  series, which  are  free of
pinch singularities.  The absence of these pinch singularities results
from the  systematic inclusion of  finite-time effects and  the proper
consideration of the  time of observation.  We emphasize  that this is
achieved  without invoking  \emph{ad hoc}  prescriptions  or effective
resummations of finite widths.  In  our formalism, we have derived the
{\em   new}  master  time   evolution  equations   (\ref{eq:kbn})  and
(\ref{eq:evo})   for  particle   number   densities  and   statistical
distribution  functions,  which  are  valid \emph{to  all  orders}  in
perturbation theory  and \emph{to  all orders} in  gradient expansion.
Furthermore,  the master time  evolution equations  (\ref{eq:kbn}) and
(\ref{eq:evo}) respect  $CPT$ invariance and are  also invariant under
time translations of the CTP contour.  As opposed to other methods, in
our  perturbative approach, we  do not  need to  employ quasi-particle
approximation and  no assumption is necessary about  the separation of
time scales.

We have shown how the effect of a finite time interval since the start
of evolution of the system leads to violation of energy conservation,
as dictated by the Heisenberg uncertainty principle. Our approach
permits the systematic treatment of the pertinent generalized
kinematics in this evanescent regime.  We have found that the
available phase space of $1\:\to\: 2$ decays increases and would-be
kinematically disallowed processes can still take place at early
times, contributing significantly to our time evolution equations.
Within a simple scalar model, we have illustrated that these
kinematically forbidden evanescent processes are the $2\:\to\: 1$
processes of Landau damping, as well as $3 \to 0$ processes of {\em
  total annihilation} into the thermal bath. The processes of Landau
damping and total annihilation have been shown to contribute promptly
to the particle width, to a level as high as $20\%$.  The switching on
of the interactions leads to a quantum quench in the system, which
manifests itself as a rapid change in both the particle width and the
collision terms of the time evolution equations.  These early-time
effects give rise to an oscillating pattern, which persists even at
later times. We have demonstrated that these late-time memory effects
exhibit a non-Markovian evolution characterized by oscillations with
time-dependent frequencies.  The latter constitutes a distinctive
feature of proper non-equilibrium dynamics, which is consistently
predicted by our perturbative approach.

We note that the rapid transient behavior of the system makes the
method of gradient expansion unsuitable for early times. We emphasize
that in our approach no assumption was made as to the relative rate of
thermalization of either the statistical or spectral behavior of the
system.  For the considered initial conditions of thermal equilibrium,
we have found that the spectral evolution resulting from evanescent
contributions is critical to the early-time statistical dynamics of
the system. Consequently, it would have been inappropriate to assume a
separation of time scales at early times. A more accurate numerical
solution to our time evolution equations turns out to be
computationally intensive and may be presented elsewhere.

Finite-time  effects,  which are  systematically  incorporated in  our
perturbative approach, are also relevant to many-body systems whenever
a natural  characteristic time scale for perturbations  arises in such
systems.  Thus, in addition  to possible applications to reheating and
preheating,   of    particular   interest   are    first-order   phase
transitions. For instance, such characteristic time scales result from
the  bubble  wall velocity  of  nucleation  in  the electroweak  phase
transition,  e.g.~see~\cite{Huber:2000mg, Huber:2011aa}.   In the
vicinity of  the bubble wall, evanescent contributions  may weaken the
constraints of  decay and inverse decay  thresholds and so  affect the
washout  phenomena  and  the  generation  of  relic  densities.   This
evanescent      regime      is      particularly      relevant      to
prethermalization~\cite{Berges:2004ce}                              and
isotropization~\cite{Berges:2005ai} time scales, which are known to be
shorter than the time scale of thermalization.

It  is  straightforward to  generalize  our  perturbative approach  to
theories that  include fermions and  gauge fields.  Thus, it  would be
very interesting to extend the classical approaches \cite{Sigl:1992fn,
  Boyanovsky:2006yg}  to  kinetic  equations  of  particle  mixing  to
non-homogeneous  backgrounds,   by  including  finite-time  evanescent
effects in  line with the non-equilibrium formalism  presented in this
paper.

\begin{acknowledgments}
The authors thank Stephan Huber and Daniele Teresi for useful comments
and discussions.  The work of  PM and AP  is supported in part  by the
Lancaster--Manchester--Sheffield  Consortium  for Fundamental  Physics
under STFC  grant ST/J000418/1.  AP also  acknowledges partial support
by an IPPP associateship from Durham University.
\end{acknowledgments}

\appendix

\section{Propagator Properties and Identities}
\label{app:rel}

In this  appendix, we  give a detailed  summary of  the transformation
properties and identities that  relate the various propagators defined
in Section~\ref{sec:canon}.   In detail, the  pertinent two-point
correlation functions for the complex scalar field $\chi$ are given by
\begin{subequations}
  \label{eq:pdefs}
  \begin{align}
    i\Delta(x,y)&\ \equiv\ \braket{[\,\chi(x),\ \chi^{\dag}(y)\,]}\;,\\
    i\Delta_1(x,y)&\ \equiv\ \braket{\{\,\chi(x),\ \chi^{\dag}(y)\,\}}\;,\\
    i\Delta_{\mathrm{R}}(x,y)&\ \equiv\ \theta(x_0\:-\:y_0)i\Delta(x,y)\;,\\
    i\Delta_{\mathrm{A}}(x,y)&\ \equiv\ -\theta(y_0\:-\:x_0)i\Delta(x,y)\;,\\
    i\Delta_{\mathcal{P}}(x,y)&\ \equiv\ \tfrac{1}{2}\varepsilon(x_0\:-\:y_0)
    \braket{[\,\chi(x),\ \chi^{\dag}(y)\,]}\;,\\
    i\Delta_>(x,y)&\ \equiv\ \braket{\chi(x)\chi^{\dag}(y)}\;,\\
    i\Delta_<(x,y)&\ \equiv\ \braket{\chi^{\dag}(y)\chi(x)}\;,\\
    i\Delta_{\mathrm{F}}(x,y)&\ \equiv\ \braket{\,\mathrm{T}\,
      [\,\chi(x)\chi^{\dag}(y)\,]}\;,\\
    i\Delta_{\mathrm{D}}(x,y)&\ \equiv\ \braket{\bar{\,\mathrm{T}}\,
      [\,\chi(x)\chi^{\dag}(y)\,]}\;.
  \end{align}
\end{subequations}
The definitions  of the  charge-conjugate propagators follow  from the
unitary transformation
\begin{subequations}
  \begin{align}
    \label{eq:ccon}
    U_C^{\dag}\chi(x)U_C&\ =\ \chi^C(x)\ =\ \eta\chi^{\dag}(x)\;,\\
    U_C^{\dag}\chi^{\dag}(x)U_C&\ =\ \chi^{C\dag}(x)\ =\ \eta^*\chi(x)\;,
  \end{align}
\end{subequations}
where the complex phase $\eta$ satisfies $|\eta|^2\:=\:1$.

It  follows from  the  definitions that  the  propagators satisfy  the
transformations listed below under charge- and Hermitian-conjugation:
\begin{subequations}
  \label{eq:hermit}
  \begin{align}
    \Delta(x,y)&\ =\ -\Delta^{*}(y,x)\ =\ -\Delta^{C}(y,x)\;,\\
    \Delta_1(x,y)&\ =\ -\Delta_1^{*}(y,x)\ =\ \Delta_1^{C}(y,x)\;,\\
    \Delta_{\mathcal{P}}(x,y)&\ =\ \Delta_{\mathcal{P}}^{*}(y,x)\ =\
    \Delta_{\mathcal{P}}^{C}(y,x)\;,\\
    \Delta_{\mathrm{R}}(x,y)&\ =\ \Delta^{C*}_{\mathrm{R}}(x,y)\ =\
    \Delta_{\mathrm{A}}^{*}(y,x)\ =\ \Delta^{C}_{\mathrm{A}}(y,x)\;,\\
    \Delta_>(x,y)&\ =\ -\Delta_>^{*}(y,x)\ =\ \Delta^{C}_<(y,x)\ =\ 
    -\Delta_<^{C*}(x,y)\;,\\
    \Delta_{\mathrm{F}}(x,y)&\ =\ \Delta^{C}_{\mathrm{F}}(y,x)\ =\
    -\Delta^{*}_{\mathrm{D}}(y,x)\ =\ -\Delta^{C*}_{\mathrm{D}}(x,y)\;,
\end{align}
\end{subequations}
where the  action of  charge-conjugation is trivial  in the case  of a
real scalar  field.   In  the  double momentum  representation,  these
identities take the form:
\begin{subequations}
  \label{eq:hermitdm}
  \begin{align}
    \Delta(p,p')&\ =\ -\Delta^{*}(p',p)\ =\ -\Delta^{C}(-p',-p)\;,\\
    \Delta_1(p,p')&\ =\ -\Delta_1^{*}(p',p)\ =\ \Delta_1^{C}(-p',-p)\;,\\
    \Delta_{\mathcal{P}}(p,p')&\ =\ \Delta_{\mathcal{P}}^{*}(p',p)\ =\ 
    \Delta_{\mathcal{P}}^{C}(-p',-p)\;,\\
    \Delta_{\mathrm{R}}(p,p')&\ =\ \Delta_{\mathrm{R}}^{C*}(-p,-p')\ =\ 
    \Delta_{\mathrm{A}}^{*}(p',p)\ =\ \Delta^{C}_{\mathrm{A}}(-p',-p)\;,\\
    \Delta_>(p,p')&\ =\ -\Delta_>^{*}(p',p)\ =\ \Delta_<^{C}(-p',-p)\ =\
    -\Delta_<^{C*}(-p,-p')\;,\\
    \label{eq:hermitdmF}
    \Delta_{\mathrm{F}}(p,p')&\ =\ \Delta_{\mathrm{F}}^{C}(-p',-p)\ =\ 
    -\Delta_{\mathrm{D}}^{*}(p',p)\ =\ -\Delta_{\mathrm{D}}^{C*}(-p,-p')\;.
  \end{align}
\end{subequations}
Finally,  in the  Wigner representation,  the propagators  satisfy the
following properties:
\begin{subequations}
  \label{eq:hermitwig}
  \begin{align}
    \Delta(q,X)&\ =\ -\Delta^{*}(q,X)\ =\ -\Delta^{C}(-q,X)\;,\\
    \Delta_1(q,X)&\ =\ -\Delta_1^{*}(q,X)\ =\ \Delta_1^{C}(-q,X)\;,\\
    \Delta_{\mathcal{P}}(q,X)&\ =\ \Delta_{\mathcal{P}}^*(q,X)\ =\ \Delta_{\mathcal{P}}^{C}(-q,X)\;,\\
    \Delta_{\mathrm{R}}(q,X)&\ =\ \Delta^{C*}_{\mathrm{R}}(-q,X)\ =\
    \Delta_{\mathrm{A}}^{*}(q,X)\ =\ \Delta^{C}_{\mathrm{A}}(-q,X)\;,\\
    \Delta_>(q,X)&\ =\ -\Delta_>^{*}(q,X)\ =\ \Delta_<^{C}(-q,X)\ =\
    -\Delta_<^{C*}(-q,X)\;,\\
    \Delta_{\mathrm{F}}(q,X)&\ =\ \Delta_{\mathrm{F}}^{C}(-q,X)\ =\
    -\Delta^{*}_{\mathrm{D}}(q,X)\ =\ -\Delta^{C*}_{\mathrm{D}}(-q,X)\;.
  \end{align}
\end{subequations}

We also list the following set of useful identities:
\begin{subequations}
  \label{eq:propidens}
  \begin{align}
    \Delta(x,y)&\ =\ \Delta_>(x,y)\:-\:\Delta_<(x,y)\ = \
    \Delta_{\mathrm{R}}(x,y)\:-\:\Delta_{\mathrm{A}}(x,y)\;,\\
    \label{eq:PJrealF}
    & \ = \ \varepsilon(x_0\:-\:y_0)\Big(\,\Delta_{\mathrm{F}}(x,y)\:-\:
    \Delta_{\mathrm{D}}(x,y)\,\Big)\;,\\
    \Delta_1(x,y)&\ = \ \Delta_>(x,y)\:+\:\Delta_<(x,y)\ = \
    \Delta_{\mathrm{F}}(x,y)\:+\:\Delta_{\mathrm{D}}(x,y)\;,\\
    \label{eq:useiden}
    \Delta_{\mathrm{R}(\mathrm{A})}(x,y)&\ =\ \Delta_{\mathrm{F}}(x,y)\:-\:
    \Delta_{<(>)}(x,y)\ =\ -\Delta_{\mathrm{D}}(x,y)\:+\:\Delta_{>(<)}(x,y)\;,\\
    \label{eq:decomF}
    \Delta_{\mathrm{F}(\mathrm{D})}(x,y)&\ =\ \frac{1}{2}
    \Big(\,(-)2\Delta_{\mathcal{P}}(x,y)\:+
    \:\Delta_>(x,y)\:+\:\Delta_<(x,y)\,\Big)\;,\\
    \Delta_{\mathcal{P}}(x,y)&\ =\ \frac{1}{2}
    \Big(\,\Delta_{\mathrm{R}}(x,y)\:+\:\Delta_{\mathrm{A}}(x,y)\,\Big)\ = \
    \frac{1}{2}\Big(\,\Delta_{\mathrm{F}}(x,y)\:-\:\Delta_{\mathrm{D}}(x,y)\,\Big).
\end{align}
\end{subequations}
We  note   that  analogous   relations  hold  for   the  corresponding
self-energies  and that these  identities and  relations are  true for
free and resummed propagators.

\section{Correspondence between Imaginary and Real Time Formalisms}
\label{app:ITF}

In this appendix,  we briefly outline a number  of relevant details of
the  Imaginary Time  Formalism (ITF)  of thermal  field  theory. These
details are discussed  in the context of the  real scalar field theory
introduced in  Section~\ref{sec:canon}. Subsequently, we  identify the
correspondence of  the ITF  $\Phi$-scalar propagator and  its one-loop
non-local self-energy with results calculated explicitly in real time,
using the  CTP formalism  of Section~\ref{sec:CTP} in  the equilibrium
limit discussed in Section~\ref{sec:eq}.

The    equilibrium    density    operator   $\rho_{\mathrm{eq}}$    in
(\ref{eq:eqdens}) permits  a path-integral representation  in negative
imaginary time.  The ITF generating functional is
\begin{equation}
  \label{eq:imgen}
  \mathcal{Z}[J]\ =\ \!\int\![\D{}{\Phi}]\,\exp\!\bigg(-\bar{S}[\Phi]\:+\:
    \!\int_0^{\beta}\!\!\D{}{\tau_x}\int\!\D{3}{\mathbf{x}}\;
    J(\bar{x})\Phi(\bar{x})\bigg)\;,
\end{equation}
with action
\begin{equation}
  \bar{S}[\Phi]\ =\ \!\int_0^{\beta}\!\!\D{}{\tau_x}\int\!\D{3}{\mathbf{x}}
  \;\Big(\,\tfrac{1}{2}\partial_{\mu}\Phi(\bar{x})\partial_{\mu}\Phi(\bar{x})
  \:+\:\tfrac{1}{2}M^2\Phi^2(\bar{x})\:+\:\tfrac{1}{3!}g\Phi^3(\bar{x})
  \:+\:\tfrac{1}{4!}\lambda\Phi^4(\bar{x})\,\Big)\;.
\end{equation}
We emphasize  the restricted domain of integration  over the imaginary
time    $\tau_x\:\in\:[0,\   \beta]$.     Four-dimensional   Euclidean
space-time   coordinates   are    denoted   by   a   horizontal   bar,
i.e.  {$\bar{x}_{\mu}\:\equiv\:(\tau_x,\mathbf{x})$}.   In  the  limit
$\beta\:\to\: \infty$, (\ref{eq:imgen}) is precisely the Wick rotation
to Euclidean  space-time of the  Minkowski-space generating functional
via the analytic continuation $x_0\:\to\:-i\tau_x$.

The free imaginary-time propagator  $\bar{\Delta}^0$ may be written as
follows:
\begin{equation}
  \bar{\Delta}^0(\bar{x}-\bar{y})\ =\ 
\frac{1}{\beta}\sum_{\ell\:=\:-\infty}^{+\infty} 
  \int\!\!\frac{\D{3}{\mathbf{p}}}{(2\pi)^3}\;
e^{i[\omega_{\ell}(\tau_x\:-\:\tau_y)\:+\:\mathbf{p}\cdot(\mathbf{x}\:-\:\mathbf{y})]}
  \bar{\Delta}_0(i\omega_{\ell},\mathbf{p})\;,
\end{equation}
where
\begin{equation}
  \label{eq:matsprop}
  \bar{\Delta}^0(i\omega_{\ell},\mathbf{p})\ =\
  \frac{1}{\omega_{\ell}^2\:+\:\mathbf{p}^2\:+\:M^2}
\end{equation}
is  the  so-called  Matsubara  propagator.   The  discrete  Matsubara
frequencies                         $\omega_{\ell}\:=\:2\pi\ell/\beta$,
$\ell\:\in\:\mathbb{Z}$,   arise   from   the   periodicity   of   the
imaginary time  direction  in  order   to  satisfy  the  KMS  relation
[cf.~(\ref{eq:KMSgl})]
\begin{equation}
  \bar{\Delta}^0(\bar{x}-\bar{y})\ =\
  \bar{\Delta}^0(\bar{x}-\bar{y}+\beta)\;.
\end{equation}
The resummed Matsubara  propagator is  given by
the imaginary-time Schwinger--Dyson equation
\begin{equation}
  \label{eq:sdim}
  \bar{\Delta}^{-1}(i\omega_{\ell},\mathbf{p})\ =\
  \bar{\Delta}^{0,\,-1}(i\omega_{\ell},\mathbf{p})\:+\:
  \bar{\Pi}(i\omega_{\ell},\mathbf{p})\;,
\end{equation}
where  $\bar{\Pi}(\omega_{\ell},\mathbf{p})$   is  the  imaginary-time
self-energy. Equation (\ref{eq:sdim}) may be inverted directly to obtain
\begin{equation}
  \bar{\Delta}(i\omega_{\ell},\mathbf{p})\ =\
  \frac{1}{\omega_{\ell}^2\:+\:\mathbf{p}^2\:+\:M^2\:+\:
  \bar{\Pi}(i\omega_{\ell},\mathbf{p})}\;.
\end{equation}

The  free  Matsubara propagator  in  (\ref{eq:matsprop})  may also  be
written in the following spectral representation:
\begin{equation}
  \bar{\Delta}^0(i\omega_{\ell},\mathbf{p})\ =\ 
  -i\!\int\!\frac{\D{}{k_0}}{2\pi}\;
  \frac{\Delta^0(k_0,\mathbf{p})}{i\omega_{\ell}\:-\:k_0}\;,
\end{equation}
where
\begin{equation}
  i\Delta^0(k_0,\mathbf{p})\ =\ 2\pi\varepsilon(k_0)
  \delta(k_0^2\:-\:\mathbf{p}^2\:-\:M^2)
\end{equation}
is  the  single-momentum  representation  of  the  free  Pauli--Jordan
propagator,  consistent  with   (\ref{eq:pj}).   Making  the  analytic
continuation   $i\omega_{\ell}\:\to\:   p_0\:+\:i\epsilon$   to   real
frequencies  and comparing  with  the spectral  representation of  the
retarded  propagator $\Delta_{\mathrm{R}}$  in  (\ref{eq:specrep}), we
may establish the correspondence
\begin{equation}
  \label{eq:delcorres}
  \bar{\Delta}^0(i\omega_{\ell}\to p_0+i\epsilon,\mathbf{p})
  \ =\ -\Delta^0_{\mathrm{R}}(p)\;.
\end{equation}
This  correspondence  must  also   hold  for  the  resummed  Matsubara
propagator  $\bar{\Delta}$  via  (\ref{eq:sdim}).  As  such,  the
analytic continuation of the ITF self-energy $\bar{\Pi}$,
\begin{equation}
  \label{eq:picorres}
  \bar{\Pi}(i\omega_{\ell} \to 
  p_0+i\epsilon,\mathbf{p})\ =\ -\Pi_{\mathrm{R}}(p)\;,
\end{equation}
yields the equilibrium retarded self-energy $\Pi_{\mathrm{R}}$.

Thus,  in thermodynamic  equilibrium, an  exact correspondence  can be
established  between the  ITF and  the real-time  approach of  the CTP
formalism,    by   means   of    {\em   retarded}    propagators   and
self-energies.  The  full   complement  of  propagators  exhibited  in
Table~\ref{tab:props}  and  the  corresponding  self-energies  may  be
obtained using  the constraints of  causality (\ref{eq:causality}) and
unitarity  (\ref{eq:unitarity}) in combination  with the  KMS relation
(\ref{eq:KMS})    and    the    condition    of    detailed    balance
(\ref{eq:db}).   The relationships between  the retarded  and CTP
propagators    and   self-energies    are    listed   explicitly    in
(\ref{eq:retident}) and (\ref{eq:pirels}).

To illustrate  the correspondence in  (\ref{eq:picorres}), we consider
the   one-loop  bubble   diagram  of   the  real   scalar   theory  in
(\ref{eq:scallag}).  The real and imaginary parts of the retarded
self-energy $\Pi_{\mathrm{R}}$ may be  calculated in the CTP formalism
from  the time-ordered $\Pi$  and positive-frequency  Wightman $\Pi_>$
self-energies, respectively, using the relations in (\ref{eq:pirels}).

The    real    part   of    the    one-loop   retarded    self-energy
$\Pi_{\mathrm{R}}^{(1)}$ is given by
\begin{equation}
  \mathrm{Re}\,\Pi_{\mathrm{R}}^{(1)}(p)\ =\ \mathrm{Re}
  \bigg[-i\frac{(-ig)^2}{2}\!\int\!\!\frac{\D{4}{k}}{(2\pi)^4}\;
  i\Delta_{\mathrm{F}}^0(k)i\Delta_{\mathrm{F}}^0(p-k)\bigg]\;,
\end{equation}
where   $i\Delta^0_{\mathrm{F}}(k)$  is   the  equilibrium   CTP  Feynman
propagator in (\ref{eq:eqF}). Explicitly, we have
\begin{align}
  \mathrm{Re}\,\Pi_{\mathrm{R}}^{(1)}(p)\ &=\ -\frac{g^2}{2}
  \int\!\!\frac{\D{3}{\mathbf{k}}}{(2\pi)^3}\;\frac{1}{4E_1E_2}\bigg[\,
  \sum_{\alpha_1\:=\:\pm1}\;\int\limits_{(p-k)^2\neq M^2}\!\!\!\!\!\!\!\!\!\!\!\D{}{k_0}\;
  \frac{2E_2}{(p-k)^2-M^2}\,\delta\big(k_0\:-\:\alpha_1 E_1\big)
  \Big(\,\tfrac{1}{2}\:+\:f_{\mathrm{B}}(k_0)\,\Big)\nonumber\\
  &\qquad + \: \sum_{\alpha_2\:=\:\pm 1}\;\int\limits_{k^2\neq M^2}\!\!\!\!\!\!\D{}{k_0}
  \;\frac{2E_1}{k^2-M^2}\,
  \delta\big(p_0\:-\:k_0\:-\:\alpha_2 E_2\big)
  \Big(\,\tfrac{1}{2}\:+\:f_{\mathrm{B}}(p_0-k_0)\,\Big)\bigg]\;,
\end{align}
where $E_1\:\equiv\: E(\mathbf{k})$ and $E_2\:\equiv
\:E(\mathbf{p}-\mathbf{k})$.  The integral subscripts remind us that
the integration around the on-shell poles is understood in the Cauchy
principal value sense. Integration over $k_0$ yields the result
\begin{equation}
  \label{eq:reret}
  \mathrm{Re}\,\Pi_{\mathrm{R}}^{(1)}(p)\ =\ -\frac{g^2}{2}\sum_{\{\alpha\}}
  \!\int\!\!\frac{\D{3}{\mathbf{k}}}{(2\pi)^3}\,
  \frac{\alpha_1\alpha_2}{4E_1E_2}\,
  \frac{1\:+\:f_{\mathrm{B}}(\alpha_1E_1)\:+\:
  f_{\mathrm{B}}(\alpha_2E_2)}
  {p_0\:-\:\alpha_1E_1\:-\:\alpha_2E_2}\;,
\end{equation}
where  we have  used the  short-hand notation  $\{\alpha\}$  to denote
summation over $\alpha_1,\ \alpha_2\:=\:\pm 1$.

The one-loop positive-frequency Wightman self-energy is given by
\begin{equation}
  i\Pi^{(1)}_>(p)\ =\ \frac{(-ig)^2}{2}\!\int\!\!\frac{\D{4}{k}}{(2\pi)^4}
  \;i\Delta^0_>(k)i\Delta^0_>(p-k)\;.
\end{equation}
Using the signum form  of the equilibrium positive-frequency Wightman
propagator $i\Delta^0_>(k)$ in (\ref{eq:equilpropg}), we obtain
\begin{equation}
  \label{eq:PIgrsub}
  \Pi^{(1)}_>(p)\ =\ i\pi g^2\sum_{\{\alpha\}}
  \int\!\!\frac{\D{3}{\mathbf{k}}}{(2\pi)^3} \int\!\D{}{k_0}\;
  \frac{\alpha_1\alpha_2}{4E_1E_2}\,\delta(k_0\:-\:\alpha_1E_1)
  \delta(p_0\:-\:k_0\:-\:\alpha_2E_2)
  \Big(\,1\:+\:f_{\mathrm{B}}(k_0)\,\Big)
  \Big(\,1\:+\:f_{\mathrm{B}}(p_0-k_0)\,\Big)\;.
\end{equation}
The  product  of  Bose--Einstein distributions  in  (\ref{eq:PIgrsub})
satisfies the following relation:
\begin{equation}
  f_{\mathrm{B}}(k_0)f_{\mathrm{B}}(p_0-k_0)\ = \ f_{\mathrm{B}}(p_0)\,
  \Big(\,1\:+\:f_{\mathrm{B}}(k_0)\:+\:f_{\mathrm{B}}(p_0-k_0)\,\Big)\;.
\end{equation}
Thus, upon integration over $k_0$, we find
\begin{equation}
  \Pi^{(1)}_>(p)\ =\ i\pi g^2\Big(\,1\:+\:f_{\mathrm{B}}(p_0)\,\Big)
  \sum_{\{\alpha\}}\int\!\!
  \frac{\D{3}{\mathbf{k}}}{(2\pi)^3}\;
  \frac{\alpha_1\alpha_2}{4E_1E_2}\,\delta(p_0\:-\:\alpha_1E_1\:-\:
  \alpha_2E_2)
  \Big(\,1\:+\:f_{\mathrm{B}}(\alpha_1E_1)\:+\:
  f_{\mathrm{B}}(\alpha_2E_2)\,\Big)\;.
\end{equation}
Using the  relation in (\ref{eq:pirelsg}),  the imaginary part  of the
one-loop retarded self-energy may then be written down as
\begin{equation}
  \label{eq:imret}
  \mathrm{Im}\,\Pi_{\mathrm{R}}^{(1)}(p)\ =\ \frac{\pi g^2}{2}
  \sum_{\{\alpha\}}\int\!\!\frac{\D{3}{\mathbf{k}}}{(2\pi)^3}
  \;\frac{\alpha_1\alpha_2}{4E_1E_2}\,\delta(p_0\:-\:\alpha_1E_1\:-\:\alpha_2E_2)
  \Big(\,1\:+\:f_{\mathrm{B}}(\alpha_1E_1)\:+\:
  f_{\mathrm{B}}(\alpha_2E_2)\,\Big)\;.
\end{equation}

In the ITF, the one-loop self-energy is given by
\begin{equation}
  -\bar{\Pi}^{(1)}(i\omega_{\ell},\mathbf{p})\ =\ \frac{(-g)^2}{2\beta}
  \sum_{n\:=\:-\infty}^{+\infty}\int\!\!\frac{\D{3}{\mathbf{k}}}{(2\pi)^3}\;
  \bar{\Delta}(i\omega_n,\mathbf{k})
  \bar{\Delta}(i(\omega_{\ell}-\omega_n),\mathbf{p}-\mathbf{k})\;.
\end{equation}
After   performing  the   summation   over  $n$   (see  for   instance
\cite{LeBellac2000}), we obtain
\begin{align}
  \label{eq:itfloop}
  \bar{\Pi}^{(1)}(i\omega_{\ell},\mathbf{p})\ &=\ \frac{g^2}{2}\sum_{\{\alpha\}}
  \int\!\!\frac{\D{3}{\mathbf{k}}}{(2\pi)^3}\;
  \frac{\alpha_1\alpha_2}{4E_1E_2}\,
  \frac{1\:+\:f_{\mathrm{B}}(\alpha_1E_1)\:+\:
  f_{\mathrm{B}}(\alpha_2E_2)}
  {i\omega_{\ell}\:-\:\alpha_1E_1\:-\:
  \alpha_2E_2}\;.
\end{align}
Making     the     analytic    continuation     $i\omega_{\ell}\:\to\:
p_0\:+\:i\epsilon$ and subsequently  extracting the real and imaginary
parts, the  result in (\ref{eq:itfloop})  agrees with (\ref{eq:reret})
and (\ref{eq:imret}) via the correspondence in~(\ref{eq:picorres}).

\section{The Complex Scalar Field}
\label{app:complex}

This  appendix   describes  the  generalization  of   the  results  in
Sections~\ref{sec:nonhomprop}  and  \ref{sec:eq}  to  the  case  of  a
complex  scalar  field $\chi$.   In  particular,  we  derive the  full
complement of non-homogeneous free propagators and expand upon the ITF
correspondence identified in Appendix~\ref{app:ITF}.

Our starting point is the complex-scalar Lagrangian
\begin{equation}
  \label{eq:cscal}
  \mathcal{L}(x)\ =\ \partial_{\mu}\chi^{\dag}(x)\partial^{\mu}\chi(x)
  \:-\:m^2\chi^{\dag}(x)\chi(x)\:-\:
  \tfrac{1}{4}\lambda\big[\chi^{\dag}(x)\chi(x)\big]^2\;.
\end{equation}
In analogy to (\ref{eq:PhiI0}), the complex scalar field $\chi(x)$ may
be written in the interaction picture as
\begin{equation}
  \label{eq:ChiI}
  \chi(x;\tilde{t}_i)\ =\ \!\int\!\D{}{\Pi_{\mathbf{p}}}\;
  \Big(\,a(\mathbf{p},0;\tilde{t}_i)e^{-iE(\mathbf{p})x_0}e^{i\mathbf{p}\cdot \mathbf{x}}
  \:+\:b^{\dag}(\mathbf{p},0;\tilde{t}_i)e^{iE(\mathbf{p})x_0}e^{-i\mathbf{p}\cdot \mathbf{x}}\,\Big)\;,
\end{equation}
where             $a^{\dag}(\mathbf{p},0;\tilde{t}_i)$             and
$a(\mathbf{p},0;\tilde{t}_i)$     ($b^{\dag}(\mathbf{p},0;\tilde{t}_i)$
and   $b(\mathbf{p},0;\tilde{t}_i)$)   are   the   interaction-picture
particle   (anti-particle)   creation   and  annihilation   operators,
respectively.    Under  $C$-conjugation   [cf.~(\ref{eq:ccon})]  these
creation and annihilation operators satisfy the transformations
\begin{equation}
    U_C^{\dag}\,a(\mathbf{p},\tilde{t};\tilde{t}_i)U_C\ =\
    \eta b(\mathbf{p},\tilde{t};\tilde{t}_i)\;, \qquad
    U_C^{\dag}\,b^{\dag}(\mathbf{p},\tilde{t};\tilde{t}_i)U_C\ =\ \eta
    a^{\dag}(\mathbf{p},\tilde{t};\tilde{t}_i)\;.
\end{equation}
Introducing the four-dimensional LIPS measure from (\ref{eq:lips}),
the field operator (\ref{eq:ChiI}) and its Hermitian conjugate may be
recast in the form
\begin{equation}
    \chi(x;\tilde{t}_i)\ =\ \!\int\!\!\frac{\D{4}{p}}{(2\pi)^4}\;
    e^{-ip\cdot x}\,\chi(p;\tilde{t}_i)\;, \qquad
    \chi^{\dag}(x;\tilde{t}_i)\ =\ \!\int\!\!\frac{\D{4}{p}}{(2\pi)^4}\;
    e^{-ip\cdot x}\,\chi^{\dag}(-p;\tilde{t}_i)\;,
\end{equation}
where the Fourier amplitudes are given by
\begin{equation}
  \chi(p;\tilde{t}_i)\ =\ 2\pi\delta(p^2\:-\:m^2)
  \Big(\,\theta(p_0)a(\mathbf{p},0;\tilde{t}_i)
  \:+\:\theta(-p_0)b^{\dag}(-\mathbf{p},0;\tilde{t}_i)\,\Big)\;.
\end{equation}

For the real scalar field,  the quantization scheme was only dependent
on the  restriction placed upon the  form of the  field commutator. In
the case of  the complex scalar field, we have  two degrees of freedom
to  fix.   Thus,  we  begin  with the  following  two  commutators  of
interaction-picture fields:
\begin{equation}
    \big[\,\chi(x),\ \chi(y)\,\big]\ =\ 0\;,\qquad
    \label{eq:cmplxpj}
    \big[\,\chi(x),\ \chi^{\dag}(y)\,\big]\ =\ i\Delta^0(x,y;m^2)\;,
  \end{equation}
where  the   Pauli--Jordan  propagator  has  precisely   the  form  in
(\ref{eq:pjfunc}). In  analogy to the  real scalar field,  we may
derive from (\ref{eq:cmplxpj}) the equal-time commutation relations
\begin{subequations}
  \begin{align}
    i\Delta^0(x,y;m^2)\big|_{x^0\:=\:y^0\:=\:\tilde{t}}&\ =\
    \big[\,\chi(\tilde{t},\mathbf{x}),\
      \chi^{\dag}(\tilde{t},\mathbf{y})\,\big]\ =\ 0\;,\\
    \partial_{x_0}i\Delta^0(x,y;m^2)\big|_{x^0\:=\:y^0\:=\:\tilde{t}}&\ =\
    \big[\,\pi^{\dag}(\tilde{t},\mathbf{x}),\
      \chi^{\dag}(\tilde{t},\mathbf{y})\,\big]\ =\
    -i\delta^{(3)}(\mathbf{x}\:-\:\mathbf{y})\;,\\
    \partial_{y_0}i\Delta^0(x,y;m^2)\big|_{x^0\:=\:y^0\:=\:\tilde{t}}&\ =\
    \big[\,\chi(\tilde{t},\mathbf{x}),\ \pi(\tilde{t},\mathbf{y})\,\big]
    \ =\ i\delta^{(3)}(\mathbf{x}\:-\:\mathbf{y})\;,\\
    \partial_{x_0}\partial_{y_0}i\Delta^0(x,y;m^2)\big|_{x^0\:=\:y^0\:=\:\tilde{t}}
    &\ =\ \big[\,\pi^{\dag}(\tilde{t},\mathbf{x}),\ \pi(\tilde{t},\mathbf{y})\,\big]
    \ =\ 0\;,
\end{align}
\end{subequations}
where $\pi(x)\:=\:\partial_{x_0}\chi^{\dag}(x)$  is the conjugate momentum
operator.  The particle  and anti-particle  creation  and annihilation
operators necessarily satisfy the algebra
\begin{equation}
  \label{eq:abquant}
  \big[\,a(\mathbf{p},\tilde{t}),\ a^{\dag}(\mathbf{p}',\tilde{t}')\,\big]
  \ =\ \big[\,b(\mathbf{p},\tilde{t}),\ b^{\dag}(\mathbf{p}',\tilde{t}')\,\big]
  \ =\ (2\pi)^32E(\mathbf{p})\delta^{(3)}(\mathbf{p}\:-\:\mathbf{p}')
  e^{-iE(\mathbf{p})(\tilde{t}\:-\:\tilde{t}')}\;,
\end{equation}
with all other commutators vanishing.   We stress again that the phase
factor  $e^{-iE(\mathbf{p})(\tilde{t}\:-\:\tilde{t}')}$   on  the  RHS
of~(\ref{eq:abquant})   has  appeared   due  to   the   difference  in
microscopic times of the interaction-picture creation and annihilation
operators.

In   addition   to    the   $C$-conserving   propagators   listed   in
Appendix~\ref{app:rel}, we may  also define $C$-violating propagators.
As    an    example,    the    $C$-violating    Hadamard    propagator
$\Delta_{1\slashed{C}}(x,y)$ would read:
\begin{equation}
  \label{eq:cvprop}
  i\Delta_{1\slashed{C}}(x,y)\ =\
  \braket{\,\{\,\chi(x),\ \chi(y)\,\}\,}\;,
\end{equation}
which satisfies
\begin{equation}
  \label{eq:cvproprel}
  \Delta_{1\slashed{C}}(x,y)\ =\ \Delta_{1\slashed{C}}(y,x)\ =\
  -\eta^2\Delta^{C*}_{1\slashed{C}}(x,y)\;.
\end{equation}
This Hadamard  correlation function may,  in general, be  non-zero for
early times, thus  permitting extra $C$-violating evanescent processes
in addition to those described in Section~\ref{sec:toy}.

In analogy to  (\ref{eq:fsgs}), we write the following  set of EEVs of
two-point  products   of  particle  and   anti-particle  creation  and
annihilation operators:
\begin{subequations}
  \begin{align}
    \braket{a^{\dag}(\mathbf{p}',\tilde{t}_f;\tilde{t}_i)
      a(\mathbf{p},\tilde{t}_f;\tilde{t}_i)}_t&\ =\
    2\mathscr{E}(\mathbf{p},\mathbf{p}')f(\mathbf{p},\mathbf{p}',t)\;,\\
    \braket{b(\mathbf{p}',\tilde{t}_f;\tilde{t}_i)
      a(\mathbf{p},\tilde{t}_f;\tilde{t}_i)}_t&\ =\
    2\mathscr{E}(\mathbf{p},\mathbf{p}')g(\mathbf{p},\mathbf{p}',t)\;,\\
    \braket{a(\mathbf{p}',\tilde{t}_f;\tilde{t}_i)
      a(\mathbf{p},\tilde{t}_f;\tilde{t}_i)}_t&\ =\
    2\mathscr{E}(\mathbf{p},\mathbf{p}')h(\mathbf{p},\mathbf{p}',t)\;,\\
    \braket{b^{\dag}(\mathbf{p}',\tilde{t}_f;\tilde{t}_i)
      a(\mathbf{p},\tilde{t}_f;\tilde{t}_i)}_t&\ =\
    2\mathscr{E}(\mathbf{p},\mathbf{p}')d(\mathbf{p},\mathbf{p}',t)\;,
  \end{align}
\end{subequations}
where  the  remaining  EEVs  are  obtained  by  Hermitian  and  charge
conjugation.     The   four    statistical    distribution   functions
$f,\ g,\ h,\ d$ satisfy the following identities:
\begin{subequations}
  \begin{align}
    f(\mathbf{p},\mathbf{p}',t)&\ =\ f^*(\mathbf{p}',\mathbf{p},t)\;,\\
    g(\mathbf{p},\mathbf{p}',t)&\ =\ g^C(\mathbf{p}',\mathbf{p},t)\;,\\
    h(\mathbf{p},\mathbf{p}',t)&\ =\ h(\mathbf{p}',\mathbf{p},t)\;,\\
    d(\mathbf{p},\mathbf{p}',t)&\ =\ \eta^2d^{C*}(\mathbf{p}',\mathbf{p},t)\;.
  \end{align}
\end{subequations}

The  non-homogeneous  free propagators  of  the  complex scalar  field
$\chi$  may be  written as  listed in  Table~\ref{tab:props}  with the
substitution of the following ensemble function:
\begin{align}
  \tilde{f}(p,p',t)\ &=\ \theta(p_0)\theta(p_0')f(\mathbf{p},\mathbf{p}',t)
  \:+\:\theta(-p_0)\theta(-p_0')f^{C*}(-\mathbf{p},-\mathbf{p}',t)\nonumber\\
  &\qquad +\: \theta(p_0)\theta(-p_0')g(\mathbf{p},-\mathbf{p}',t)\:+\:
  \theta(-p_0)\theta(p_0')g^{C*}(-\mathbf{p},\mathbf{p}',t)\;,
\end{align}
satisfying                        the                       relations:
$\tilde{f}(p,p',t)\:=\:\tilde{f}^C(-p',-p,t)\:=
\:\tilde{f}^{C*}(-p,-p',t)$  in  accordance with  (\ref{eq:hermitdm}).
The        free        $C$-violating        Hadamard        propagator
$\Delta^0_{1\slashed{C}}(p,p',\tilde{t}_f;\tilde{t}_i)$,   defined  in
(\ref{eq:cvprop}), may be written down as
\begin{equation}
  \Delta_{1\slashed{C}}(p,p',\tilde{t}_f;\tilde{t}_i)
  \ =\ -i2\pi\delta(p^2\:-\:m^2)|2p_0|^{1/2}
  2\tilde{d}(p,p',t)e^{i(p_0\:-\:p_0')\tilde{t}_f}
  |2p_0'|^{1/2}2\pi\delta(p'^2\:-\:m^2)\;,
\end{equation}
where   we   have   defined   the  $C$-violating   ensemble   function
$\tilde{d}(p,p',t)$. Evaluating the EEVs directly, we find
\begin{align}
  \tilde{d}(p,p',t)\ &=\
  \theta(p_0)\theta(p_0')d(\mathbf{p},\mathbf{p}',t)
  \:+\:\theta(-p_0)\theta(-p_0')\eta^2d^{C*}(-\mathbf{p},-\mathbf{p}',t)
  \nonumber\\
  &\qquad +\: \theta(p_0)\theta(-p_0')h(\mathbf{p},-\mathbf{p}',t)
  \:+\:\theta(-p_0)\theta(p_0')\eta^2h^{C*}(-\mathbf{p},\mathbf{p}',t)\;.
\end{align}
satisfying         the         relations:        $\tilde{d}(p,p',t)\:=
\:\tilde{d}^C(-p',-p,t)\:= \:\eta^2\tilde{d}^{C*}(-p,-p',t)$.

The inclusion of the $C$-violating distribution functions requires the
addition of the $C$-violating source  $l_{ab}$ to the expansion of the
kernel  of the  density  operator in  the  CTP generating  functional,
according to our  discussion in Section~\ref{sec:ctpsd}. Omitting
the $t$ dependence of the  sources for notational convenience, the CTP
generating functional  for the complex  scalar field $\chi$  takes the
form
\begin{align}
  \mathcal{W}[j_a,k_{ab},l_{ab}]&\ =\ -i\hbar
  \ln\!\iint\![\D{}{(\chi^{a\dag},\chi^a)}]\exp\!
  \bigg\{\frac{i}{\hbar}\bigg[S\big[\chi^{a\dag},\chi^a\big]
  \:+\:\!\int_{\Omega_t}\!\D{4}{x}\;
  \Big(\,j_a^{\dag}(x)\chi^a(x)\:+\:\chi_a^{\dag}(x)j^a(x)\,\Big)\nonumber\\&\qquad
  +\:\!\iint_{\Omega_t}\!\D{4}{x}\,\D{4}{x'}\;
  \Big(\,\chi^{a\dag}(x)k_{ab}(x,x')\chi^b(x')\nonumber\\&\qquad +\:
  \frac{1}{2}\,\chi^a(x)l_{ab}^{\dag}(x,x')\chi^b(x')\:+\:
  \frac{1}{2}\,\chi^{a\dag}(x)l_{ab}(x,x')\chi^{b\dag}(x')\,\Big)\:+\:\cdots\bigg]\bigg\}\;.
\end{align}
The  bi-local sources  $k_{ab}$ and  $l_{ab}$ necessarily  satisfy the
identities
\begin{equation}
    k_{ab}(x,x')\ =\ k_{ba}^{\dag}(x',x)\;,\qquad 
    l_{ab}(x,x')\ =\ l_{ba}(x',x)\;,
\end{equation}
 to  ensure  that  the   exponent  of  the  generating  functional  is
 $CPT$-invariant.  The subsequent derivation  of the  effective action
 then follows analogously to Section~\ref{sec:nonhom}.

The  Lagrangian  in (\ref{eq:cscal})  is  invariant  under the  global
$U(1)$ transformation
\begin{equation}
    \chi(x)\ \to\ \chi'(x)\ =\ e^{-i\alpha}\,\chi(x)\;, \qquad
    \chi^{\dag}(x)\ \to\ \chi'^{\dag}(x)\ =\ e^{i\alpha}\,\chi^{\dag}(x)\; ,
\end{equation}
entailing the conserved Noether current
\begin{equation}
  j_{\mu}(x)\ =\ i\Big(\,\chi^{\dag}(x)\partial_{\mu}\chi(x)\:-\:
  \big(\partial_{\mu}\chi^{\dag}(x)\big)\chi(x)\,\Big)\;,
\end{equation}
with corresponding conserved charge
\begin{equation}
  :\mathcal{Q}(x_0):\ =\ \!\int\!\D{3}{\mathbf{x}}\;:j_0(x):\ =\ 
  \!\int\!\D{}{\Pi_{\mathbf{p}}}\;\Big(\,
  a^{\dag}(\mathbf{p},0)a(\mathbf{p},0)\: -\:
  b^{\dag}(\mathbf{p},0)b(\mathbf{p},0)\,\Big)\;,
\end{equation}
where $:\ :$ denotes  normal ordering. The existence of this conserved
charge  necessitates the  introduction of  a chemical  potential $\mu$
and, as such, the equilibrium density operator is of the grand-canonical
form
\begin{equation}
  \label{eq:grancon}
  \rho(\beta,\mu)\ =\ e^{-\beta(H-\mu Q)}\;.
\end{equation}
In  the presence  of  this  chemical potential,  the  KMS relation  in
(\ref{eq:KMSgl}) generalizes to
\begin{equation}
  \Delta_{>}(x^0-y^0,\mathbf{x}-\mathbf{y})\ =\ e^{-\beta\mu}
  \Delta_{<}(x^0-y^0+i\beta,\mathbf{x}-\mathbf{y})
\end{equation}
or, in the momentum representation,
\begin{equation}
  \label{eq:KMSgen}
  \Delta_>(p)\ =\ e^{\beta (p_0-\mu)}\Delta_<(p)\;.
\end{equation}
Proceeding  as  in  Section~\ref{sec:CTP},  we find  that  the  final
constraint on $\tilde{f}(p)$, generalizing (\ref{eq:fconst}), is
\begin{equation}
  \tilde{f}(p)\ =\ 
  \theta(p_0)f_{\mathrm{B}}(p_0)\:+\:\theta(-p_0)f^{C}_{\mathrm{B}}(-p_0)\;,
\end{equation}
where   $f^{(C)}_{\mathrm{B}}(p_0)\:=\:(e^{\beta[p_0-(+)\mu]}-1)^{-1}$  is
the particle  (anti-particle) Bose--Einstein distribution  function. In
equilibrium, translational invariance is  restored and the elements of
the  free  CTP  propagator  may  be  written  in  the  single momentum
representations
\begin{subequations}
  \label{eq:eqchi}
  \begin{align}
    i\Delta^0_{\mathrm{F}}(p)&\ =\ i\Big(\,p^2\:-\:m^2\:+\:
        i\epsilon\,\Big)^{-1}\:+\:2\pi\,\Big(\,\theta(p_0)f_{\mathrm{B}}(p_0)\:+
    \:\theta(-p_0)f^{C}_{\mathrm{B}}(-p_0)\,\Big)\delta(p^2\:-\:m^2)\;,\\
i\Delta^0_>(p)&\ =\ 2\pi\Big[\,\theta(p_0)\Big(\,1\:+\:f_{\mathrm{B}}(p_0)\,\Big)
    \:+\:\theta(-p_0)f^{C}_{\mathrm{B}}(-p_0)\,\Big]\delta(p^2\:-\:m^2)\;,\\
    i\Delta^0_<(p)&\ =\ 2\pi\Big[\,\theta(p_0)f_{\mathrm{B}}(p_0)\:+\:\theta(-p_0)
    \Big(\,1\:+\:f^{C}_{\mathrm{B}}(-p_0)\,\Big)\Big]\delta(p^2\:-\:m^2)\;.
  \end{align}
\end{subequations}

In  order  to define  the  ITF  generating  functional for  the  grand
canonical partition  function in (\ref{eq:grancon}),  one may consider
the Hamiltonian form  of the path integral directly  (see for instance
\cite{LeBellac2000}). We may write
\begin{align}
    \mathcal{Z}[j]\ &=\ \!\iint\![\D{}{(\pi^{\dag},\pi)}]\,[\D{}{(\chi^{\dag},\chi)}]
    \;\exp\!\bigg\{-\!\int_0^{\beta}\!\!\D{}{\tau_x}\int\!\D{3}{\mathbf{x}}
    \;\Big[\,\mathcal{H}(\pi^{(\dag)},\chi^{(\dag)})\:-\:
    i\,\Big(\,\pi(\bar{x})\partial_{\tau_x}\chi(\bar{x})\nonumber\\ &\qquad +\:
    \pi^{\dag}(\bar{x})\partial_{\tau_x}\chi^{\dag}(\bar{x})\Big)\: -\:
    i\mu\,\Big(\,\pi^{\dag}(\bar{x})\chi^{\dag}(\bar{x})\:-\:
    \pi(\bar{x})\chi(\bar{x})\,\Big)\:-\:
    j^{\dag}(\bar{x})\chi(\bar{x})\:-\:\chi^{\dag}(\bar{x})j(\bar{x})\,\Big]\bigg\}\;,
\end{align}
where
\begin{equation}
  \mathcal{H}(\pi^{(\dag)},\chi^{(\dag)})\ =\ \pi^{\dag}(\bar{x})\pi(\bar{x})\:+\:
  \mathbf{\nabla}\chi^{\dag}(\bar{x})\cdot\mathbf{\nabla}\chi(\bar{x})\:+\:m^2\chi^{\dag}(\bar{x})\chi(\bar{x})
  \:+\:\mathcal{H}^{\mathrm{int}}(\chi^{\dag},\chi)
\end{equation}
is the Hamiltonian density and $\mathcal{H}^{\mathrm{int}}$ is the
interaction part.  Expanding the fields and conjugate momenta in terms
of two real degrees of freedom $\chi_{1,2}(\bar{x})$ and
$\pi_{1,2}(\bar{x})$ as
\begin{equation}
    \chi(\bar{x})\ =\ \frac{1}{\sqrt{2}}\,\Big(\,\chi_1(\bar{x})\:+\:
    i\chi_2(\bar{x})\,\Big)\;,\qquad
    \pi(\bar{x})\ =\ \frac{1}{\sqrt{2}}\,\Big(\,\pi_1(\bar{x})\:-\:
    i\pi_2(\bar{x})\,\Big)\;,
\end{equation}
we   may   analytically   calculate   the  Gaussian   integrals   over
$\pi_{1,2}(\bar{x})$, yielding
\begin{align}
  \label{eq:cmplximgen}
  \mathcal{Z}[j]\ &=\ \!\int\![\D{}{(\chi^{\dag},\chi)}]\;\exp\!
  \bigg[-\!\int_0^{\beta}\!\!\D{}{\tau_x}\int\!\D{3}{\mathbf{x}}\;
  \Big(\,(\partial_{\tau_x}+\mu)\chi^{\dag}(\bar{x})
  (\partial_{\tau_x}-\mu)\chi(\bar{x})\:+\:
  \mathbf{\nabla}\chi^{\dag}(\bar{x})\cdot\mathbf{\nabla}\chi(\bar{x})
  \nonumber\\&\qquad +\:m^2\chi^{\dag}(\bar{x})\chi(\bar{x})\:+\:\mathcal{H}^{\mathrm{int}}(\chi^{\dag},\chi)\:-\:
  j^{\dag}(\bar{x})\chi(\bar{x})\:-\:\chi^{\dag}(\bar{x})j(\bar{x})\,\Big)
\bigg]\;.
\end{align}
In order  to derive the form  of the ITF  $\chi$-propagator, we insert
into (\ref{eq:cmplximgen}) the Fourier transform
\begin{equation}
  \chi(\bar{x})\ =\ \frac{1}{\beta}\sum_{\ell\:=\:-\infty}^{+\infty}\int\!\!
\frac{\D{3}{\mathbf{p}}}{(2\pi)^3}\;
e^{i[\omega_{\ell}\tau_x\:+\:\mathbf{p}\cdot\mathbf{x}]} 
  \chi(i\omega_{\ell},\mathbf{p})\;,
\end{equation}
where $\omega_{\ell}$ are the discrete Matsubara frequencies described
in Appendix~\ref{app:ITF}. The effect of the chemical potential is to
shift the poles of the Matsubara propagator, which becomes
\begin{equation}
  \bar{\Delta}^0(i\omega_{\ell}-\mu,\mathbf{p})\ =\
  \frac{1}{(\omega_{\ell}\:+\:i\mu)^2\:+\:\mathbf{p}^2\:+\:m^2}\;.
\end{equation}
In generalization  of the correspondence  in (\ref{eq:delcorres}), the
equilibrium retarded  propagator $\Delta_{\mathrm{R}}$ is  obtained by
the      analytic      continuation     $i\omega_{\ell}\:-\:\mu\:\to\:
p_0\:+\:i\epsilon$ via
\begin{equation}
  \bar{\Delta}^0(i\omega_{\ell}-\mu\to
  p_0+i\epsilon,\mathbf{p})\ =\ -\Delta^0_{\mathrm{R}}(p)\;.
\end{equation}
Correspondingly,     the      equilibrium     retarded     self-energy
$\Pi_{\mathrm{R}}$ is given by
\begin{equation}
  \bar{\Pi}(i\omega_{\ell}-\mu\to p_0+i\epsilon)\ =\ -\Pi_{\mathrm{R}}(p)\;,
\end{equation}
thereby generalizing (\ref{eq:picorres}).

\newpage

\section{Non-Homogeneous Density Operator}
\label{app:dens}

In  (\ref{eq:densop}),  we  highlight  the  full form  of  the  series
expansion of  the general Gaussian-like density  operator described in
Section~\ref{sec:nonhomprop}.      Symmetric    and    asymmetric
multi-particle  states  are built  up  by  summing  over all  possible
convolutions of  the $W$  amplitudes.  To facilitate  our presentation
task, the time dependence and phase-space integrals have been omitted.
In fact,  all momenta  are integrated with  the LIPS measure  given in
(\ref{eq:lips}).
\begin{align}
  \label{eq:densop}
  \rho\ &=\ \Big(\,1\:+\:\frac{1}{2}W_{10}(\mathbf{p}_1:0)W_{01}(0:\mathbf{p}_1)
  \:+\:\frac{1}{4}W_{20}(\mathbf{p}_1,\mathbf{p}_2:0)
  W_{02}(0:\mathbf{p}_2,\mathbf{p}_1)\:+\:\cdots\,\Big)\Big\{\,\ket{0}\!\bra{0}\nonumber\\
  &\qquad +\: \Big(\,-W_{10}(\mathbf{k}_1:0)\:+\:
  \frac{1}{2}W_{11}(\mathbf{k}_1:\mathbf{q}_1)W_{10}(\mathbf{q}_1:0)
  \nonumber\\& \qquad \quad \quad \quad
  \:+\:\frac{1}{2}W_{20}(\mathbf{k}_1,\mathbf{q}_1:0)
  W_{01}(0:\mathbf{q}_1)\:+\:\cdots\,\Big)\ket{\mathbf{k}_1}\!\bra{0}\nonumber\\
  &\qquad +\: \Big(\,-W_{01}(0:\mathbf{k}_1')\:+\:
  \frac{1}{2}W_{01}(0:\mathbf{q}_1)W_{11}(\mathbf{q}_1:\mathbf{k}_1')
  \nonumber\\&\qquad \quad \quad \quad
  \:+\:\frac{1}{2}W_{02}(0:\mathbf{k}_1',\mathbf{q}_1)
  W_{10}(\mathbf{q}_1:0)\:+\:\cdots\,\Big)\ket{0}\!\bra{\mathbf{k}_1'}\nonumber\\
  &\qquad +\: \Big(\,(2\pi)^32E(\mathbf{k}_1)\delta^{(3)}(\mathbf{k}_1\:-\:\mathbf{k}_1')
  \:-\:W_{11}(\mathbf{k}_1:\mathbf{k}_1')\:+\:
  W_{10}(\mathbf{k}_1:0)W_{01}(0:\mathbf{k}_1')\nonumber\\&\qquad 
  \quad \quad \quad\:+\:\frac{1}{2}W_{11}(\mathbf{k}_1:\mathbf{q}_1)
  W_{11}(\mathbf{q}_1:\mathbf{k}_1')
  \:+\:\frac{1}{4}W_{20}(\mathbf{k}_1,\mathbf{q}_1:0)
  W_{02}(0:\mathbf{q}_1,\mathbf{k}_1')\:+\:\cdots\,\Big)
  \ket{\mathbf{k}_1}\!\bra{\mathbf{k}_1'}\nonumber\\
  &\qquad + \: \frac{1}{2}\Big(\,-W_{20}(\mathbf{k}_1,\mathbf{k}_2:0)
  \:+\:W_{10}(\mathbf{k}_1:0)W_{10}(\mathbf{k}_2:0)\:+\:
  \frac{1}{2}W_{11}(\mathbf{k}_1:\mathbf{q}_1)
  W_{20}(\mathbf{q}_1,\mathbf{k}_2:0)\nonumber\\
  &\qquad \quad \quad \quad
  \:+\:\frac{1}{2}W_{11}(\mathbf{k}_2:\mathbf{q}_1)
  W_{20}(\mathbf{q}_1,\mathbf{k}_1:0)\:+\:\cdots\,\Big)
  \ket{\mathbf{k}_1,\mathbf{k}_2}\!\bra{0}\nonumber\\
  &\qquad + \: \frac{1}{2}\Big(\,-W_{02}(0:\mathbf{k}_1',\mathbf{k}_2')
  \:+\:W_{01}(0:\mathbf{k}_1')W_{01}(0:\mathbf{k}_2')\:+\:
  \frac{1}{2}W_{02}(0:\mathbf{k}_1',\mathbf{q}_1)
  W_{11}(\mathbf{q}_1:\mathbf{k}_2')\nonumber\\&
  \qquad \quad \quad \quad
  \:+\:\frac{1}{2}W_{02}(0:\mathbf{k}_2',\mathbf{q}_1)
  W_{11}(\mathbf{q}_1:\mathbf{k}_1')\:+\:\cdots\,\Big)
  \ket{0}\!\bra{\mathbf{k}_1',\mathbf{k}_2'}\nonumber\\
  &\qquad +\: \frac{1}{2}\Big(W_{11}(\mathbf{k}_1:\mathbf{k}_1')
  W_{10}(\mathbf{k}_2:0)\:+\:W_{11}(\mathbf{k}_2:\mathbf{k}_1')
  W_{10}(\mathbf{k}_1:0)\nonumber\\&
  \qquad \quad \quad \quad
  \:+\:W_{20}(\mathbf{k}_1,\mathbf{k}_2:0)W_{01}(0:\mathbf{k}_1')\:+\:\cdots\,\Big)
  \ket{\mathbf{k}_1,\mathbf{k}_2}\!\bra{\mathbf{k}_1'}\nonumber\\
  &\qquad +\: \frac{1}{2}\Big(W_{11}(\mathbf{k}_1:\mathbf{k}_1')
  W_{01}(0:\mathbf{k}_2')\:+\:W_{11}(\mathbf{k}_1:\mathbf{k}_2')
  W_{01}(0:\mathbf{k}_1')\nonumber\\&
  \qquad \quad \quad \quad
  \:+\:W_{02}(0:\mathbf{k}_1',\mathbf{k}_2')W_{10}(\mathbf{k}_1:0)\:+\:\cdots\,\Big)
  \ket{\mathbf{k}_1}\!\bra{\mathbf{k}_1',\mathbf{k}_2'}\nonumber\\
  &\qquad +\: \frac{1}{2}\Big(W_{20}(\mathbf{k}_1,\mathbf{k}_2:0)W_{10}(\mathbf{k}_3:0)\:+\:
  \cdots\,\Big)\ket{\mathbf{k}_1,\mathbf{k}_2,\mathbf{k}_3}\!\bra{0}\nonumber\\
  &\qquad +\: \frac{1}{2}\Big(W_{02}(0:\mathbf{k}_1',\mathbf{k}_2')W_{01}(0:\mathbf{k}_3')
  \:+\:\cdots\,\Big)\ket{0}\!\bra{\mathbf{k}_1',\mathbf{k}_2',\mathbf{k}_3'}\nonumber\\
  &\qquad +\: \Big[\,\frac{1}{2}\Big(\,(2\pi)^32E(\mathbf{k}_1)
  \delta^{(3)}(\mathbf{k}_1\:-\:\mathbf{k}_1')\:-\:
  W_{11}(\mathbf{k}_1:\mathbf{k}_1')\:+\:W_{10}(\mathbf{k}_1:0)W_{01}(0:\mathbf{k}_1')
  \nonumber\\& \qquad \quad \quad \quad \quad \quad \quad
  \:+\:\frac{1}{2}W_{11}(\mathbf{k}_1:\mathbf{q}_1)W_{11}(\mathbf{q}_1:\mathbf{k}_1')
  \:+\:\frac{1}{4}W_{20}(\mathbf{k}_1,\mathbf{q}_1:0)
  W_{02}(0:\mathbf{q}_1,\mathbf{k}_1')\:+\:\cdots\,\Big)\nonumber\\
  & \qquad \quad \quad \quad \:\times\:\Big(\,(2\pi)^32E(\mathbf{k}_2)
  \delta^{(3)}(\mathbf{k}_2\:-\:\mathbf{k}_2')\:-\:
  W_{11}(\mathbf{k}_2:\mathbf{k}_2')\:+\:W_{10}(\mathbf{k}_2:0)
  W_{01}(0:\mathbf{k}_2')\nonumber\\&
  \qquad \quad \quad \quad \quad \quad \quad
  \:+\:\frac{1}{2}W_{11}(\mathbf{k}_2:\mathbf{q}_2)
  W_{11}(\mathbf{q}_2:\mathbf{k}_2')\:+\:
  \frac{1}{4}W_{20}(\mathbf{k}_2,\mathbf{q}_2:0)W_{02}(0:\mathbf{q}_2,\mathbf{k}_2')
  \:+\:\cdots\,\Big)\nonumber\\&\quad \quad
  \:+\:\frac{1}{4}\Big(\,-W_{20}(\mathbf{k}_1,\mathbf{k}_2:0)\:+\:
  W_{10}(\mathbf{k}_1:0)W_{10}(\mathbf{k}_2:0)\:+\:
  \frac{1}{2}W_{11}(\mathbf{k}_1:\mathbf{q}_1)W_{20}(\mathbf{q}_1,\mathbf{k}_2:0)
  \nonumber\\&\qquad \quad \quad \quad \quad \quad \quad
  \:+\:\frac{1}{2}W_{11}(\mathbf{k}_2:\mathbf{q}_1)W_{20}(\mathbf{q}_1,\mathbf{k}_1:0)
  \:+\:\cdots\,\Big)\nonumber\\&\quad \quad \quad \quad
  \:\times\:\Big(\,-W_{02}(0:\mathbf{k}_1',\mathbf{k}_2')\:+\:
  W_{01}(0:\mathbf{k}_1')W_{01}(0:\mathbf{k}_2')\:+\:
  \frac{1}{2}W_{02}(0:\mathbf{k}_1',\mathbf{q}_1)W_{11}(\mathbf{q}_1:\mathbf{k}_2')
  \nonumber\\&\qquad \quad \quad \quad \quad \quad \quad
  \:+\:\frac{1}{2}W_{02}(0:\mathbf{k}_2',\mathbf{q}_1)
  W_{11}(\mathbf{q}_1:\mathbf{k}_1')\:+\:\cdots\,\Big)\Big]
  \ket{\mathbf{k}_1,\mathbf{k}_2}\!\bra{\mathbf{k}_1',\mathbf{k}_2'}
  \: +\: \cdots\,\Big\}\;.
\end{align}

\section{Gradient Expansion of Kadanoff--Baym Equations}
\label{app:kb}

In this appendix, we outline  the derivation and gradient expansion of
the  Kadanoff--Baym equations  \cite{Baym:1961zz,  Kadanoff1989}.  The
resulting time  evolution equations  are included here  for comparison
with  the   new  results  described  in   Sections  \ref{sec:eom}  and
\ref{sec:toy}.   The  Kadanoff--Baym  equations  are obtained  by  the
partial  inversion   of  the  Schwinger--Dyson   equation  derived  in
Section~\ref{sec:ctpsd}.   Herein,  we  omit  the  $\tilde{t}_f$  and
$\tilde{t}_i$ dependence  of  all  propagators and  self-energies  for
notational convenience.

Inverse   Fourier  transforming   (\ref{eq:doubKG}),  we   obtain  the
coordinate-space    representation     of    the    partially inverted
Schwinger--Dyson equation
\begin{multline}
  \label{eq:kbcoord}
  -\Big(\,\Box_x^2\:+\:M^2\,\Big)\Delta_{\gtrless}(x,y)\:+\:
  \!\int_{\Omega_t}\!\!\D{4}{z}\;\Big(\,
  \Pi_{\gtrless}(x,z)\Delta_{\mathcal{P}}(z,y)\:+\:
  \Pi_{\mathcal{P}}(x,z)\Delta_{\gtrless}(z,y)\,\Big)\\
  =\ \frac{1}{2}\!\int_{\Omega_t}\!\!\D{4}{z}\;
  \big(\,\Pi_<(x,z)\Delta_>(z,y)\:-\:
  \Pi_>(x,z)\Delta_<(z,y)\,\Big)\;,
\end{multline}
where we  have also  included the positive-frequency  contribution for
completeness.  Introducing   the  relative  and   central  coordinates
$R^{\mu}$ and  $X^{\mu}$ via (\ref{eq:relcen}),  we may show  that the
d'Alembertian operator may be rewritten as
\begin{equation}
  \Box_x^2\ =\ \Box_R^2\:+\:\partial_{R,\mu}\partial_X^{\mu}\:+\:
  \frac{1}{4}\Box_X^2\;.
\end{equation}
Performing   a  gradient   expansion  of   the  Wigner   transform  of
(\ref{eq:kbcoord}),  as  described  in Section~\ref{sec:gradapp},  we
obtain
\begin{align}
  \label{eq:gradkb}
  &\Big(\,q^2\:-\:M^2\:+\:iq\!\cdot\!\partial_X\:-\:
  \frac{1}{4}\Box_X^2\,\Big)
  \Delta_{\gtrless}(q,X)\:+\: \!\int\!\!\frac{\D{4}{Q}}{(2\pi)^4}\;
  (2\pi)^4\delta_t^{(4)}(Q)\,\exp\!\Big[-i\,\Big(\,Q\cdot
  X\:+\:\diamondsuit^{-}_{q,X}\:+\:2\diamondsuit^{+}_{Q,X}\,\Big)\Big]
  \nonumber\\& \qquad
  \Big(\,\{\Pi_{\gtrless}(q+\tfrac{Q}{2},X)\}\{\Delta_{\mathcal{P}}(q-\tfrac{Q}{2},X)\}
  \:+\:\{\Pi_{\mathcal{P}}(q+\tfrac{Q}{2},X)\}\{\Delta_{\gtrless}(q-\tfrac{Q}{2},X)\}\,\Big)
  \nonumber\\&
  \ =\ \frac{1}{2}\!\int\!\!\frac{\D{4}{Q}}{(2\pi)^4}\;
  (2\pi)^4\delta_t^{(4)}(Q)\,\exp\!\Big[ -i\,\Big(\,Q\cdot X\:+\:
  \diamondsuit^{-}_{q,X}\:+\:2\diamondsuit^{+}_{Q,X}\,\Big)\Big]
  \nonumber\\&
  \qquad \Big(\,\{\Pi_<(q+\tfrac{Q}{2},X)\}\{\Delta_>(q-\tfrac{Q}{2},X)\}\:-\:
  \{\Pi_>(q+\tfrac{Q}{2},X)\}\{\Delta_<(q-\tfrac{Q}{2},X)\}\,\Big)\;.
\end{align}
The  diamond operators  $\diamondsuit^{\pm}\{\bullet\}\{\bullet\}$ are
defined     in      (\ref{eq:diam})     and     (\ref{eq:pois})     of
Section~\ref{sec:gradapp}.  Subsequently separating the Hermitian
and anti-Hermitian  parts of (\ref{eq:gradkb}), we  find the so-called
constraint and kinetic equations
\begin{subequations}
  \begin{align}
    &\Big(\,q^2\:-\:M^2\:-\:\frac{1}{4}\Box_X^2\,\Big)
    \Delta_{\gtrless}(q,X)\:+\:\!\int\!\!\frac{\D{4}{Q}}{(2\pi)^4}\;
    (2\pi)^4\delta_t^{(4)}(Q)\,\cos\Big(\,Q\cdot X\:+\:
    \diamondsuit^{-}_{q,X}\:+\:2\diamondsuit^{+}_{Q,X}\,\Big)\nonumber\\
    &\qquad \Big(\,\{\Pi_{\gtrless}(q+\tfrac{Q}{2},X)\}\{\Delta_{\mathcal{P}}(q-\tfrac{Q}{2},X)\}
    \:+\:\{\Pi_{\mathcal{P}}(q+\tfrac{Q}{2},X)\}\{\Delta_{\gtrless}(q-\tfrac{Q}{2},X)\}\,
    \Big)\nonumber\\&\
     =\ \frac{i}{2}\!\int\!\!\frac{\D{4}{Q}}{(2\pi)^4}\;
    (2\pi)^4\delta_t^{(4)}(Q)\,\sin\Big(\,Q\cdot X\:+\:
    \diamondsuit^{-}_{q,X}\:+\:2\diamondsuit^{+}_{Q,X}\,\Big)\nonumber\\&
    \qquad \Big(\,\{\Pi_>(q+\tfrac{Q}{2},X)\}\{\Delta_<(q-\tfrac{Q}{2},X)\}\:+\:
    \{\Pi_<(q+\tfrac{Q}{2},X)\}\{\Delta_>(q-\tfrac{Q}{2},X)\}\,\Big)\;,\\
    \label{eq:kinfull}
    &q\cdot\partial_X\Delta_{\gtrless}(q,X)\:-\:\!\int\!\!\frac{\D{4}{Q}}{(2\pi)^4}
    \;(2\pi)^4\delta_t^{(4)}(Q)\,\sin\Big(\,Q\cdot X\:+\:\diamondsuit^{-}_{q,X}\:+\:
    2\diamondsuit^{+}_{Q,X}\,\Big)\nonumber\\&
    \qquad \Big(\,\{\Pi_{\gtrless}(q+\tfrac{Q}{2},X)\}\{\Delta_{\mathcal{P}}(q-\tfrac{Q}{2},X)\}
    \:+\:\{\Pi_{\mathcal{P}}(q+\tfrac{Q}{2},X)\}\{\Delta_{\gtrless}(q-\tfrac{Q}{2},X)\}\,
    \Big)\nonumber\\& \ =\ \frac{i}{2}\!\int\!\!\frac{\D{4}{Q}}{(2\pi)^4}\;
    (2\pi)^4\delta_t^{(4)}(Q)\,\cos\Big(\,Q\cdot X\:+\:\diamondsuit^{-}_{q,X}
    \:+\:2\diamondsuit^{+}_{Q,X}\,\Big)\nonumber\\&
    \qquad \Big(\,\{\Pi_>(q+\tfrac{Q}{2},X)\}\{\Delta_<(q-\tfrac{Q}{2},X)\}
    \:-\:\{\Pi_<(q+\tfrac{Q}{2},X)\}\{\Delta_>(q-\tfrac{Q}{2},X)\}\,\Big)\;,
  \end{align}
\end{subequations}
respectively.

In  the  late-time limit  $t\:\to\:  \infty$,  the  $Q$ dependence  is
removed and the microscopic violation of energy conservation resulting
from the  uncertainty principle is  neglected.  As was  established in
Section~\ref{sec:toy},   this    microscopic   violation   of   energy
conservation is significant to  the early-time dynamics of the system
Subsequently, keeping terms to zeroth order in the gradient expansion,
the   above   expressions  reduce   to   the  following   differential
equations~\cite{Baym:1961zz,       Kadanoff1989,       Blaizot:2001nr,
  Prokopec:2003pj, Prokopec:2004ic}:
\begin{subequations}
  \begin{align}
    \Big(\,q^2\:-\:M^2\,\Big)\Delta_{\gtrless}(q,X)&\ =\ 
    -\,\Big(\,\Pi_{\gtrless}(q,X)\Delta_{\mathcal{P}}(q,X)\:+\:
    \Pi_{\mathcal{P}}(q,X)\Delta_{\gtrless}(q,X)\,\Big)\;,\\
    \label{eq:kinetic}
    q\cdot\partial_X\Delta_{\gtrless}(q,X)&\ =\ \frac{i}{2}\,
    \Big(\,\Pi_>(q,X)\Delta_<(q,X)\:-\:\Pi_<(q,X)\Delta_>(q,X)\Big)\;.
  \end{align}
\end{subequations}
The kinetic equation (\ref{eq:kinetic}) is to be compared with the
semi-classical Boltzmann transport equation and the
energy-conserving limit of our time evolution equations in
(\ref{eq:boltz}).

\section{Non-Homogeneous Loop Integrals}
\label{app:loops}

In this final  appendix, we outline the techniques  we have been using
to  perform   the  loop  integrals   that  occur  with   the  modified
time-dependent  Feynman rules  of this  new perturbative  approach. In
particular,   we    develop   a   method   for    dealing   with   the
energy-non-conserving  vertices that  utilizes the  Laplace transform.
Lastly,  we  summarize  the  time-\emph{independent}  equilibrium  and
time-\emph{dependent} spatially homogeneous limits of these integrals.

We begin  by defining a  generalization of the  zero-temperature $B_0$
function~\cite{'tHooft:1978xw}
\begin{equation}
  \label{eq:B0zero}
  B_0^{T\:=\:0}(q,m_1,m_2)\ =\ (2\pi\mu)^{4-d}\!\int\!\!
  \frac{\D{d}{k}}{i\pi^2}\;\frac{1}{k^2\:-\:m_1^2\:+\:i\epsilon}\,
  \frac{1}{(k-q)^2\:-\:m_1^2\:+\:i\epsilon}\;.
\end{equation}
To this end,  we define the non-homogeneous CTP  $B_0$ function, shown in Figure~\ref{fig:B0}, which
may be written in the $2 \times 2$ matrix form as
\begin{equation}
  \label{eq:B0matrix}
  B_0^{ab}\ =\
  \begin{bmatrix}
    B_0 & B^<_0 \\
    B^>_0 & -B^*_0
  \end{bmatrix}\!\;,
\end{equation}
consistent with the propagators and self-energies of the CTP formalism
in  Sections~\ref{sec:CTP}~and~\ref{sec:nonhom}. The  elements of
$B_0^{ab}$ are given by the following integral:
\begin{align}
  \label{eq:nonhomb0}
  &B^{ab}_0(q_1,q_2,m_1,m_2,\tilde{t}_f;\tilde{t}_i)\ \equiv\ (2\pi\mu)^{4-d}e^{i(q_1^0\:-\:q_2^0)\tilde{t}_f}
  \!\idotsint\!\frac{\D{d}{k_1}}{i\pi^2}\,\frac{\D{4}{k_1'}}{(2\pi)^4}
  \,\frac{\D{4}{k_2}}{(2\pi)^4}\,\frac{\D{4}{k_2'}}{(2\pi)^4}\nonumber\\
  &\qquad \times\: (2\pi)^4\delta_t^{(4)}(q_1\:-\:k_1\:-\:k_2)
  (2\pi)^4\delta_t^{(4)}(q_2\:-\:k_1'\:-\:k_2')
  \eta^{acd}\Delta^0_{ce}(k_1,k_1',m_1,\tilde{t}_f;\tilde{t}_i)
  \Delta^{C,\, 0}_{df}(k_2,k_2',m_2,\tilde{t}_f;\tilde{t}_i)\eta^{efb},
\end{align}
where   $\delta^{(4)}_t$   is   defined  in   (\ref{eq:deltat4})   and
(\ref{eq:deltat}),   $\Delta^{0,\,ab}(p,p',t)$   is   the   free   CTP
propagator derived in  Section~\ref{sec:nonhomprop} and the upper-case
Roman $C$  denotes charge-conjugate free CTP  propagator. We reiterate
that  lower-case   Roman  CTP  indices  are  raised   and  lowered  by
contraction     with    the     $\mathbb{SO}\,(1,\     1)$    `metric'
$\eta_{ab}\:=\:\mathrm{diag}\,(1,\  -1)$ and  that $\eta_{abc}\:=\:+1$
is     for      $a\:=\:b\:=\:c\:=\:1$,     $\eta_{abc}\:=\:-1$     for
$a\:=\:b\:=\:c\:=\:2$ and $\eta_{abc}\:=\:0$ otherwise.

\begin{figure}
\begin{center}
\includegraphics{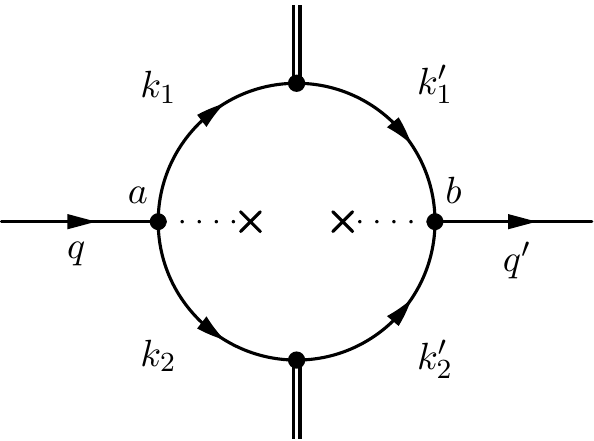}
\end{center}
\caption{The non-homogeneous $B^{ab}_0$ function.}
\label{fig:B0}
\end{figure}

In the  definition (\ref{eq:nonhomb0}),  we  have assumed
that  the  statistical  distribution  functions  are  cut off  in  the
ultra-violet.  Thus, any ultra-violet divergences anticipated from the
superficial degree  of divergence of  the integral will be  those that
result  from  the  homogeneous  zero-temperature contribution.   As  a
result,  the  dimensional  regularization  of the  integral  has  been
restricted to the $k_1$ dependence only.

In   order    to   deal   with   the    product   of   time-dependent
energy-non-conserving vertices, we make the following replacement:
\begin{equation}
  \label{eq:laprepl}
  \delta_t(x)\delta_t(y)\ =\ \!\int_{\sigma-i\infty}^{\sigma+i\infty}\!\!
  \frac{\D{}{s}}{2\pi i}\;e^{st}\,\frac{2}{\pi^2}\,
  \frac{4s}{(x\:-\:y)^2\:+\:4s^2}\,\frac{1}{(x\:+\:y)^2\:+\:4s^2}\;,
\end{equation}
where    the   RHS    is   the    inverse   Laplace    transform   and
$s\:\in\:\mathbb{C}$ is  a complex variable.  The  Bromwich contour is
chosen so that $\sigma\:\in\:\mathbb{R}$  is larger than the real part
of  the  right-most  pole  in   the  integrand,  in  order  to  ensure
convergence.      We     then     introduce     the     representation
$B^{ab}_0(q_1,q_2,m_1,m_2,s)$   of  the   non-homogeneous   CTP  $B_0$
function through
\begin{equation}
  B^{ab}(q_1,q_2,m_1,m_2,\tilde{t}_f;\tilde{t}_i)\ =\ \int_{\sigma-i\infty}^{\sigma+i\infty}\!\!
  \frac{\D{}{s}}{2\pi i}\;e^{st}\,B^{ab}_0(q_1,q_2,m_1,m_2,s)\;.
\end{equation}
Note  that  $B^{ab}_0(q_1,q_2,m_1,m_2,s)$   is  \emph{not}  the  exact
Laplace transform of  $B^{ab}_0(q_1,q_2,m_1,m_2,t)$, since we have not
transformed  the  $t$ dependence  of  the distribution  functions.  We
suppress   the    dependence   of   $B^{ab}_0(q_1,q_2,m_1,m_2,s)$   on
$\tilde{t}_f$ and $\tilde{t}_i$ for notational convenience.

After     making     the     replacement     (\ref{eq:laprepl})     in
(\ref{eq:nonhomb0}), we obtain the integral
\begin{align}
  &B^{ab}_0(q_1,q_2,m_1,m_2,s)\ =\ 8(2\pi\mu)^{4-d}e^{i(q_1^0\:-\:q_2^0)\tilde{t}_f}
  \!\idotsint\!\frac{\D{d}{k_1}}{i\pi^2}\,\frac{\D{4}{k_1'}}{(2\pi)^4}
  \,\frac{\D{4}{k_2}}{(2\pi)^4}\,\frac{\D{4}{k_2'}}{(2\pi)^4}\nonumber\\
  &\qquad \times\: (2\pi)^3\delta^{(3)}(\mathbf{q}_1\:-\:\mathbf{k}_1\:-\:\mathbf{k}_2)
  (2\pi)^3\delta^{(3)}(\mathbf{q}_2\:-\:\mathbf{k}_1'\:-\:\mathbf{k}_2')
  \nonumber\\&\qquad \times\:
  4s\Big[\Big(\,q_1^0\:-\:q_2^0\:-\:k_1^0\:+\:{k_1'}^0\:-\:k_2^0\:+\:{k_2'}^0\,\Big)^2
  \:+\:4s^2\,\Big]^{-1}
  \Big[\Big(\,q_1^0\:+\:q_2^0\:-\:k_1^0\:-\:{k_1'}^0\:-\:k_2^0\:-\:{k_2'}^0\,\Big)^2
  \:+\:4s^2\,\Big]^{-1}\nonumber\\&\qquad \times\:
  \eta^{acd}\Delta^{0}_{ce}(k_1,k_1',m_1,\tilde{t}_f;\tilde{t}_i)
  \Delta^{0,\,C}_{df}(k_2,k_2',m_2,\tilde{t}_f;\tilde{t}_i)\eta^{efb}\;,
\end{align}
in  which   the  analytic  structure   of  the  product   of  formerly
$t$-dependent vertex functions is now manifest.

For conciseness, throughout this appendix we use the short-hand notation:
\begin{equation}
  \sum_{\{\alpha\}}
\end{equation}
for    summation   over    only   the    $\alpha_i\:=\:\pm    1$   and
$\alpha_i'\:=\:\pm  1$  that   appear  explicitly  in  an  expression.
On-shell energies are denoted by
\begin{equation}
  E_i\:\equiv\:E_i(\mathbf{k}_i)\:=\:\sqrt{\mathbf{k}_i^2+m_i^2}\;, \qquad
  E_i'\:\equiv\:E_i'(\mathbf{k}_i')\:=\:\sqrt{{\mathbf{k}_i'}^2+m_i^2}\;,
\end{equation}
and on-shell four-momenta by
\begin{equation}
  \hat{k}_i\ \equiv\ \hat{k}^{\mu}_i\ =\ (\alpha_iE_i(\mathbf{k}_i),\mathbf{k}_i)\;, \qquad
  \hat{k}_i'\ \equiv\ {\hat{k}_i'}{}^{\mu}\ =\ (\alpha_i'E_i'(\mathbf{k}_i'),\mathbf{k}_i')\;.
\end{equation}
For the energy factor in (\ref{eq:efact}), we use the notation
\begin{equation}
  \mathscr{E}_i\ \equiv \ \mathscr{E}_i(\mathbf{k}_i,\mathbf{k}_i')
  \ =\ \sqrt{E_i(\mathbf{k}_i)E_i(\mathbf{k}_i')}\;.
\end{equation}
Products of unit-step functions are denoted by
\begin{equation}
  \label{eq:thetashort}
  \theta(x,y,\cdots,z)\ \equiv\ \theta(x)\theta(y)\cdots\theta(z)\;.
\end{equation}

\subsection*{Time-Ordered Functions}

The  $(a,b)\:=\:(1,1)$ element  of (\ref{eq:nonhomb0})  coincides with
the  time-ordered function  $B_0$,  which we  may  separate into  four
contributions (suppressing all arguments):
\begin{equation}
  \label{eq:B04terms}
  B_0\ =\ I^{(\mathrm{i})}\: +\: I^{(\mathrm{ii})}\: +\:
  I^{(\mathrm{iiia})}\:+\:I^{(\mathrm{iiib})}\;.
\end{equation}
These four  contributions are the zero  temperature limit ($I^{(i)}$),
the purely thermal term ($I^{(ii)}$) and the cross terms ($I^{(iiia)}$
and $I^{(iiib)}$).

\noindent (i)  {\bfseries The zero-temperature part}  may be extracted
from  the product  of  terms that  remain  in the  limit of  vanishing
statistical distribution functions. This part is
\begin{align}
  &I^{(\mathrm{i})}(q_1,q_2,m_1,m_2,s)\ =\ 8(2\pi\mu)^{4-d}e^{i(q_1^0\:-\:q_2^0)\tilde{t}_f}
  \!\idotsint\!\frac{\D{d}{k_1}}{i\pi^2}\,\frac{\D{4}{k_1'}}{(2\pi)^4}\,
  \frac{\D{4}{k_2}}{(2\pi)^4}\,\frac{\D{4}{k_2'}}{(2\pi)^4}\nonumber\\
  &\qquad \times\: (2\pi)^3\delta^{(3)}(\mathbf{q}_1\:-\:\mathbf{k}_1\:-\:\mathbf{k}_2)
  (2\pi)^3\delta^{(3)}(\mathbf{q}_2\:-\:\mathbf{k}_1'\:-\:\mathbf{k}_2')\nonumber\\
  &\qquad \times \
  4s\Big[\Big(\,q_1^0\:-\:q_2^0\:-\:k_1^0\:+\:{k_1'}^0\:-\:k_2^0\:+\:{k_2'}^0\,\Big)^2
  \:+\:4s^2\,\Big]^{-1}
  \Big[\Big(\,q_1^0\:+\:q_2^0\:-\:k_1^0\:-\:{k_1'}^0\:-\:k_2^0\:-\:{k_2'}^0\,\Big)^2
  \:+\:4s^2\,\Big]^{-1}\nonumber\\&\qquad \times\
  \frac{1}{k_1^2\:-\:m_1^2
\:+\:i\epsilon}(2\pi)^4\delta^{(4)}(k_1\:-\:k_1')\frac{1}{k_2^2\:-\:m_2^2\:+\:i\epsilon}(2\pi)^4\delta^{(4)}(k_2\:-\:k_2')\;.
\end{align}
The ${k_1'}^0$ and ${k_2'}^0$ integrations  are performed by means of the
delta functions, giving
\begin{align}
  \label{eq:I1ret}
  &I^{(\mathrm{i})}(q_1,q_2,m_1,m_2,s)\ =\ (2\pi\mu)^{4-d}e^{i(q_1^0\:-\:q_2^0)\tilde{t}_f}
  \!\idotsint\!\frac{\D{d-1}{k_1}}{i\pi^2}\,\frac{\D{3}{\mathbf{k}_1'}}{(2\pi)^3}
  \,\frac{\D{3}{\mathbf{k}_2}}{(2\pi)^3}\,\frac{\D{3}{\mathbf{k}_2'}}{(2\pi)^3}
  \,\D{}{k_1^0}\,\D{}{k_2^0}\nonumber\\&\qquad \times\: 
  (2\pi)^3\delta^{(3)}(\mathbf{q}_1\:-\:\mathbf{k}_1\:-\:\mathbf{k}_2)
  (2\pi)^3\delta^{(3)}(\mathbf{q}_2\:-\:\mathbf{k}_1'\:-\:\mathbf{k}_2')
  \nonumber\\&\qquad \times\: \frac{4s}{\pi}\Big[\Big(\,q_1^0\:-\:q_2^0\Big)^2\:+\:4s^2\,\Big]^{-1}
  \Big[\Big(\tfrac{q_1^0\:+\:q_2^0}{2}\:-\:k_1^0\:-\:k_2^0\,\Big)^2\:+\:s^2\,\Big]^{-1}
  \nonumber\\&\qquad \times\: \frac{1}{(k_1^0)^2\:-\:E_1^2\:+\:i\epsilon}
  \,\frac{1}{(k_2^0)^2\:-\:E_2^2\:+\:i\epsilon}
  (2\pi)^3\delta^{(3)}(\mathbf{k}_1\:-\:\mathbf{k}_1')
  (2\pi)^3\delta^{(3)}(\mathbf{k}_2\:-\:\mathbf{k}_2')\;.
\end{align}

By  virtue of  the residue  theorem, we  may perform  the  $k_1^0$ and
$k_2^0$ integrations  by closing contours  in the lower halves  of the
$k_1^0$  and $k_2^0$  complex  planes. After  collecting together  the
resulting terms, we find
\begin{align}
  &I^{(\mathrm{i})}(q_1,q_2,m_1,m_2,s)\ =\ -2(2\pi)^3\mu^{4-d}e^{i(q_1^0\:-\:q_2^0)\tilde{t}_f}\sum_{\{\alpha\}}
  \idotsint\!\frac{\D{d-1}{\mathbf{k}_1}}{(2\pi)^{d-1}}\,\frac{\D{3}{\mathbf{k}_1'}}{(2\pi)^3}
  \,\frac{\D{3}{\mathbf{k}_2}}{(2\pi)^3}\,\frac{\D{3}{\mathbf{k}_2'}}{(2\pi)^3}
  \nonumber\\&\qquad \times\: \frac{1}{E_1E_2}\,(2\pi)^3\delta^{(3)}(\mathbf{q}_1\:-\:\mathbf{k}_1\:-\:\mathbf{k}_2)
  (2\pi)^3\delta^{(3)}(\mathbf{q}_2\:-\:\mathbf{k}_1'\:-\:\mathbf{k}_2')
  \nonumber\\&\qquad \times\: \frac{1}{\pi}\Big[\Big(\,q_1^0\:-\:q_2^0\,\Big)^2\:+\:4s^2\,\Big]^{-1}
  \,\alpha_1\,\Big[\,\tfrac{q_1^0\:+\:q_2^0}{2}\:-\:
  \alpha_1\Big(\,E_1\:+\:E_2\:-\:is\,\Big)\Big]^{-1}
  \nonumber\\&\qquad \times\: (2\pi)^3\delta^{(3)}(\mathbf{k}_1\:-\:\mathbf{k}_1')
  (2\pi)^3\delta^{(3)}(\mathbf{k}_2\:-\:\mathbf{k}_2')\;,
\end{align}
or,  in  terms  of   the  Lorentz-invariant  phase-space  measures  in
(\ref{eq:lips}),
\begin{align}
  \label{eq:itosep}
  &I^{(\mathrm{i})}(q_1,q_2,m_1,m_2,s)\ =\ -8(2\pi)^3\mu^{4-d}e^{i(q_1^0\:-\:q_2^0)\tilde{t}_f}\sum_{\{\alpha\}}
  \idotsint\D{}{\Pi^{d-1}_{\mathbf{k}_1}}\,\D{}{\Pi_{\mathbf{k}_1'}}
  \,\D{}{\Pi_{\mathbf{k}_2}}\,\D{}{\Pi_{\mathbf{k}_2'}}\,\nonumber\\
  &\qquad \times\: (2\pi)^3\delta^{(3)}(\mathbf{q}_1\:-\:\mathbf{k}_1\:-\:\mathbf{k}_2)
  (2\pi)^3\delta^{(3)}(\mathbf{q}_2\:-\:\mathbf{k}_1'\:-\:\mathbf{k}_2')
  \nonumber\\&\qquad \times\: \frac{1}{\pi}\Big[\Big(\,q_1^0\:-\:q_2^0\,\Big)^2\:+\:4s^2\,\Big]^{-1}
  \,\alpha_1\,\Big[\,\tfrac{q_1^0\:+\:q_2^0}{2}\:-\:\alpha_1
  \Big(\,E_1\:+\:E_2\:-\:is\,\Big)\Big]^{-1}
  \nonumber\\&\qquad \times\: (2\pi)^32E_1'
  \delta^{(3)}(\mathbf{k}_1\:-\:\mathbf{k}_1')
  (2\pi)^32E_2'\delta^{(3)}(\mathbf{k}_2\:-\:\mathbf{k}_2')\;,
\end{align}
where  the   superscript  $d-1$  indicates   that  the  $\mathbf{k}_1$
integration is to be taken over a {$(d-1)$-dimensional} phase space.

The  Laplace  transform  $\mathcal{F}(s)$ of  a
complex-valued function $F(t)$ has the following properties:
\begin{subequations}
  \begin{align}
    \mathcal{L}_t[\mathrm{Re}\,F](s)&\ =\ \frac{1}{2}
    \Big[\,\mathcal{F}(s)\:+\:\Big(\mathcal{F}(s^*)\Big)^*\,\Big]\;,\\
    \mathcal{L}_t[\mathrm{Im}\,F](s)&\ =\ \frac{1}{2i}
    \Big[\,\mathcal{F}(s)\:-\:\Big(\mathcal{F}(s^*)\Big)^*\,\Big]\;.
  \end{align}
\end{subequations}
We  may then  identify  the  dispersive and  absorptive  parts of  the
{$t$-dependent}   $B_0$  function   with  the   parts   symmetric  and
anti-symmetric   in   $s$,   respectively.   With   this   separation,
(\ref{eq:itosep}) becomes
\begin{align}
  \label{eq:i}
  &I^{(\mathrm{i})}(q_1,q_2,m_1,m_2,s)\ =\ 8(2\pi)^3\mu^{4-d}
  e^{i(q_1^0\:-\:q_2^0)\tilde{t}_f}\sum_{\{\alpha\}}\idotsint\!\D{}{\Pi^{d-1}_{k_1}}\,\D{}{\Pi_{\mathbf{k}_1'}}
  \,\D{}{\Pi_{\mathbf{k}_2}}\,\D{}{\Pi_{\mathbf{k}_2'}}\nonumber\\&\qquad \times \:
  (2\pi)^3\delta^{(3)}(\mathbf{q}_1\:-\:\mathbf{k}_1\:-\:\mathbf{k}_2)
  (2\pi)^3\delta^{(3)}(\mathbf{q}_2\:-\:\mathbf{k}_1'\:-\:\mathbf{k}_2')
  \nonumber\\&\qquad \times\: \frac{1}{\pi}\Big[\Big(\,q_1^0\:-\:q_2^0\,\Big)^2\:+\:4s^2\,\Big]^{-1}
  \,\Big\{\Big[\,\tfrac{q_1^0\:+\:q_2^0}{2}\:-\:\alpha_1
  \Big(\,E_1\:+\:E_2\,\Big)\Big]^2\:+\:s^2\,\Big\}^{-1}
  \nonumber\\&\qquad \times\:
  \Big\{\,-\alpha_1\Big[\,\tfrac{q_1^0\:+\:q_2^0}{2}\:-\:\alpha_1
  \Big(\,E_1\:+\:E_2\,\Big)\Big]\:+\:is\,\Big\}
  (2\pi)^32E_1'
  \delta^{(3)}(\mathbf{k}_1\:-\:\mathbf{k}_1')
  (2\pi)^32E_2'\delta^{(3)}(\mathbf{k}_2\:-\:\mathbf{k}_2')\;,
\end{align}
where  the  delineation between  dispersive  and  absorptive parts  is
identified in the second set of curly brackets.

Isolating the  dispersive part of  this result and performing  all but
one of the remaining phase-space integrals, we find
\begin{align}
  &\mathrm{Disp}\,I^{(\mathrm{i})}(q_1,q_2,m_1,m_2,s)\ =\ (2\pi)^4\frac{1}{\pi}\,
  \frac{1}{\big(\,q_1^0\:-\:q_2^0\,\big)^2\:+\:4s^2}
  \delta^{(3)}(\mathbf{q}_1\:-\:\mathbf{q}_2) 
  e^{i(q_1^0\:-\:q_2^0)\tilde{t}_f}\nonumber\\& \qquad \times \sum_{\{\alpha\}}
  \Big\{-(2\pi\mu)^{4-d}\!\int\!\D{d-1}{\mathbf{k}_1}\,
  \frac{E_1\:+\:E_2}{E_1
  E_2}\,\frac{1}{\pi}
  \Big[\Big(\,\tfrac{q_1^0+q_2^0}{2}\:-\:i\alpha s\,\Big)^2\:-\:
  \Big(\,E_1\:+\:E_2\,\Big)^2\,\Big]^{-1}\Big\}\;,
\end{align}
where $\mathrm{Disp}$ stands for  the dispersive part. We compare this
result   to  the   form   of  the   zero-temperature  $B_0$   function
(\ref{eq:B0zero}) after the $k_0$ integration has been performed:
\begin{align}
  B_0^{T\:=\:0}(q,m_1,m_2)\ &=\ -(2\pi\mu)^{4-d}
  \!\int\!\D{d-1}{\mathbf{k}_1}\;\frac{E_1\:+\:E_2}
  {E_1E_2}\,\frac{1}{\pi}\,\Big[\,q^2_0\:-\:\Big(\,E_1\:+\:
  E_2\,\Big)^2\,\Big]^{-1}\;.
\end{align}
Hence, we may write
\begin{align}
  \mathrm{Disp}\,I^{(\mathrm{i})}(q_1,q_2,m_1,m_2,s)\ &=\ (2\pi)^4\frac{1}{\pi}\,
  \frac{1}{\big(\,q_1^0\:-\:q_2^0\,\big)^2\:+\:4s^2}
  \delta^{(3)}(\mathbf{q}_1\:-\:\mathbf{q}_2)e^{i(q_1^0\:-\:q_2^0)\tilde{t}_f}
  \nonumber\\&\qquad\times\:\sum_{\{\alpha\}}B^{T\:=\:0}_0\Big(\tfrac{q_1^0+q_2^0}{2}-i \alpha
  s, \tfrac{\mathbf{q}_1+\mathbf{q}_2}{2},m_1,m_2\Big)\;,
\end{align}
where, for $d\:=\:4\:-\:2\epsilon$ dimensions,
\begin{align}
  \label{eq:BT0}
  B^{T=0}_0(q^0-i\alpha s,\mathbf{q},m_1,m_2)\ &=\
  \frac{1}{\epsilon}\:-\:\gamma_{\mathrm{E}}\:+\:\ln\frac{4\pi\mu^2}
  {m_1m_2}\ +\ \frac{1}{\big(q^0\:-\:i\alpha
  s\big)^2\:-\:\mathbf{q}^2}
  \bigg[\,\big(m_2^2\:-\:m_1^2\big)\ln\frac{m_1^2}{m_2^2}\nonumber\\&\qquad +\:
    \lambda^{1/2}\big((q^0-i\alpha s)^2-\mathbf{q}^2,m_1^2,m_2^2\big)
    \cosh^{-1}\bigg(\,\frac{m_1^2\:+\:m_2^2\:-\:(q^0-i\alpha
      s)^2\:+\:\mathbf{q}^2} 
         {2m_1m_2}\,\bigg)\bigg]\;,
\end{align}
containing             the             zero-temperature             UV
divergence~\cite{'tHooft:1978xw}.        In       (\ref{eq:BT0}),
$\gamma_{\mathrm{E}}$ is the Euler constant  and $\mu$ is the 't Hooft
mass scale.

\noindent (ii)  {\bfseries The purely thermal part} is
more straightforward to analyze:
\begin{align}
  \label{eq:iistart}
  &I^{(\mathrm{ii})}(q_1,q_2,m_1,m_2,s)\ =\ -8(2\pi\mu)^{4-d}e^{i(q_1^0\:-\:q_2^0)\tilde{t}_f}
  \!\idotsint\!\frac{\D{d}{k_1}}{i\pi^2}\,\frac{\D{4}{k_1'}}{(2\pi)^4}\,
  \frac{\D{4}{k_2}}{(2\pi)^4}\,\frac{\D{4}{k_2'}}{(2\pi)^4}\nonumber\\
  & \qquad \times\: (2\pi)^3\delta^{(3)}(\mathbf{q}_1\:-\:\mathbf{k}_1\:-\:\mathbf{k}_2)
  (2\pi)^3\delta^{(3)}(\mathbf{q}_2\:-\:\mathbf{k}_1'\:-\:\mathbf{k}_2')\nonumber\\
  & \qquad \times\: 4s\Big[\Big(\,q_1^0\:-\:q_2^0\:-\:k_1^0\:+\:{k_1'}^0\:-\:k_2^0\:+\:{k_2'}^0\,\Big)^2\:+\:4s^2\,\Big]^{-1}
  \Big[\Big(\,q_1^0\:+\:q_2^0\:-\:k_1^0\:-\:{k_1'}^0\:-\:k_2^0\:-\:{k_2'}^0\,\Big)^2\:+\:4s^2\,\Big]^{-1}
  \nonumber\\& \qquad \times\ 
  2\pi\delta(k_1^2\:-\:m_1^2)|2k_1^0|^{1/2}\tilde{f}_1(k_1,k_1',t)e^{i(k_1^0\:-\:{k_1'}^0)\tilde{t}_f}
  |2{k_1'}^0|^{1/2}2\pi\delta({k_1'}^2\:-\:m_1^2)\nonumber\\& \qquad \times \
  2\pi\delta(k_2^2\:-\:m_2^2)|2k_2^0|^{1/2}\tilde{f}^C_2(k_2,k_2',t)e^{i(k_2^0\:-\:{k_2'}^0)\tilde{t}_f}
  |2{k_2'}^0|^{1/2}2\pi\delta({k_2'}^2\:-\:m_2^2)\;.
\end{align}
Performing the  four zeroth-component momentum  integrations by virtue
of   the  on-shell   delta  functions   $\delta(k_i^2\:-\:m_i^2)$  and
$\delta({k_i'}^2\:-\:m_i^2)$ in (\ref{eq:iistart}), we find
\begin{align}
  \label{eq:ii}
  &I^{(\mathrm{ii})}(q_1,q_2,m_1,m_2,s)\ =\ 8(2\pi)^3\mu^{4-d}e^{i(q_1^0\:-\:q_2^0)\tilde{t}_f}
  \sum_{\{\alpha\}}\idotsint\!\D{}{\Pi^{d-1}_{k_1}}\,\D{}{\Pi_{\mathbf{k}_1'}}
  \,\D{}{\Pi_{\mathbf{k}_2}}\,\D{}{\Pi_{\mathbf{k}_2'}}\nonumber\\
  & \qquad \times\: (2\pi)^3\delta^{(3)}(\mathbf{q}_1\:-\:\mathbf{k}_1\:-\:\mathbf{k}_2)
  (2\pi)^3\delta^{(3)}(\mathbf{q}_2\:-\:\mathbf{k}_1'\:-\:\mathbf{k}_2')\nonumber\\
  & \qquad \times\: \Big[\Big(\,q_1^0\:-\:q_2^0\:-\:\alpha_1E_1
    \:+\:\alpha_1'E_1'\:-\:\alpha_2E_2\:+\:
    \alpha_2'E_2'\,\Big)^2\:+\:4s^2\,\Big]^{-1}\nonumber\\
  & \qquad \times\: \Big[\Big(\,\tfrac{q_1^0\:+\:q_2^0}{2}\:-\:
    \tfrac{\alpha_1E_1\:+\:\alpha_1'E_1'}{2}\:-\:
    \tfrac{\alpha_2E_2\:+\:\alpha_2'E_2'}{2}\,\Big)^2\:+\:s^2\,\Big]^{-1}\:2is
  \nonumber\\& \qquad \times \
  2\mathscr{E}_1\tilde{f}_1(\hat{k}_1,\hat{k}_1',t)
  e^{i(\alpha_1E_1\:-\:\alpha_1'E_1')\tilde{t}_f}2\mathscr{E}_2
  \tilde{f}_2^C(\hat{k}_2,\hat{k}_2',t)
  e^{i(\alpha_2E_2\:-\:\alpha_2'E_2')\tilde{t}_f}\;.
\end{align}
Note that the purely thermal part in~(\ref{eq:ii}) contains only absorptive contributions.

\noindent (iii) {\bfseries The  cross-terms} yield both dispersive and
absorptive contributions. The first of the cross terms yields
\begin{align}
  \label{eq:iiiastart}
  &I^{(\mathrm{iiia})}(q_1,q_2,m_1,m_2,s)\ =\ -8i(2\pi \mu)^{4-d}e^{i(q_1^0\:-\:q_2^0)\tilde{t}_f}
  \!\idotsint\!\frac{\D{d}{k_1}}{i\pi^2}\,\frac{\D{4}{k_1'}}{(2\pi)^4}\,
  \frac{\D{4}{k_2}}{(2\pi)^4}\,\frac{\D{4}{k_2'}}{(2\pi)^4}\nonumber\\
  &\qquad \times\: (2\pi)^3
  \delta^{(3)}(\mathbf{q}_1\:-\:\mathbf{k}_1\:-\:\mathbf{k}_2)
  (2\pi)^3\delta^{(3)}(\mathbf{q}_2\:-\:\mathbf{k}_1'\:-\:\mathbf{k}_2')\nonumber\\
  &\qquad \times \
  4s\Big[\Big(\,q_1^0\:-\:q_2^0\:-\:k_1^0\:+\:{k_1'}^0\:-\:k_2^0\:+\:{k_2'}^0\,\Big)^2\:+\:4s^2\,\Big]^{-1}
  \Big[\Big(\,q_1^0\:+\:q_2^0\:-\:k_1^0\:-\:{k_1'}^0\:-\:k_2^0\:-\:{k_2'}^0\,\Big)^2\:+\:4s^2\,\Big]^{-1}
  \nonumber\\&\qquad \times\: \frac{1}{k_1^2\:-\:m_1^2\:+\:i\epsilon}
  (2\pi)^4\delta^{(4)}(k_1\:-\:k_1')\nonumber\\&\qquad \times\
  2\pi\delta(k_2^2\:-\:m_2^2)|2k_2^0|^{1/2}\tilde{f}_2^C(k_2,k_2',t)e^{i(k_2^0\:-\:{k_2'}^0)\tilde{t}_f}
  |2{k_2'}^0|^{1/2}2\pi\delta({k_2'}^2\:-\:m_2^2)\;.
\end{align}
After evaluating  the ${k_1'}^0$, $k_2^0$ and  ${k_2'}^0$ integrals by
virtue  of the  delta functions,  we perform  the $k_1^0$  integral by
closing  a contour in  the lower  half of  the $k_1^0$  complex plane.
Equation~(\ref{eq:iiiastart}) may then be  written in the more compact
form
\begin{align}
  &I^{(\mathrm{iiia})}(q_1,q_2,m_1,m_2,s)\ =\ -8(2\pi)^3\mu^{4-d}e^{i(q_1^0\:-\:q_2^0)\tilde{t}_f}\sum_{\{\alpha\}}
  \idotsint\!\D{}{\Pi^{d-1}_{\mathbf{k}_1}}\,\D{}{\Pi_{\mathbf{k}_1'}}\,
  \D{}{\Pi_{\mathbf{k}_2}}\,\D{}{\Pi_{\mathbf{k}_2'}}\nonumber\\
  &\qquad \times\: (2\pi)^3\delta^{(3)}(\mathbf{q}_1\:-\:\mathbf{k}_1\:-\:\mathbf{k}_2)
  (2\pi)^3\delta^{(3)}(\mathbf{q}_2\:-\:\mathbf{k}_1'\:-\:\mathbf{k}_2')\nonumber\\
  &\qquad \times\: \frac{1}{\pi}
  \Big[\Big(\,q_1^0\:-\:q_2^0\:-\:\alpha_2E_2\:+\:\alpha_2'E_2'\,\Big)^2
  \:+\: 4s^2\,\Big]^{-1}
  \alpha_1\,\Big[\,\tfrac{q_1^0\:+\:q_2^0}{2}\:-\:\tfrac{\alpha_2E_2\:+\:\alpha_2'E_2'}{2}
    \:-\:\alpha_1\Big(\,E_1\:-\:is\,\Big)\Big]^{-1}\nonumber\\& \qquad \times (2\pi)^32E_1'
  \delta^{(3)}(\mathbf{k}_1\:-\:\mathbf{k}_1')2\mathscr{E}_2
  \tilde{f}_2^C(\hat{k}_2,\hat{k}'_2,t)e^{i(\alpha_2E_2\:-\:\alpha_2'E_2')\tilde{t}_f}\;.
\end{align}
Separating  again   into  dispersive   and  absorptive  parts   as  in
(\ref{eq:i}), $I^{(\mathrm{iiia})}$ may be written as
\begin{align}
  \label{eq:iiia}
  &I^{(\mathrm{iiia})}(q_1,q_2,m_1,m_2,s)\ =\ 8(2\pi)^3\mu^{4-d}e^{i(q_1^0\:-\:q_2^0)\tilde{t}_f}
  \sum_{\{\alpha\}}\idotsint\!\D{}{\Pi^{d-1}_{\mathbf{k}_1}}\,\D{}{\Pi_{\mathbf{k}_1'}}\,
  \D{}{\Pi_{\mathbf{k}_2}}\,\D{}{\Pi_{\mathbf{k}_2'}}\nonumber\\
  &\qquad \times\: (2\pi)^3\delta^{(3)}(\mathbf{q}_1\:-\:\mathbf{k}_1\:-\:\mathbf{k}_2)
  (2\pi)^3\delta^{(3)}(\mathbf{q}_2\:-\:\mathbf{k}_1'\:-\:\mathbf{k}_2')\nonumber\\
  &\qquad \times\: \frac{1}{\pi}
  \Big[\Big(\,q_1^0\:-\:q_2^0\:-\:\alpha_2E_2\:+\:\alpha_2'E_2'\,\Big)^2
  \:+\:4s^2\,\Big]^{-1}
  \Big[\Big(\,\tfrac{q_1^0\:+\:q_2^0}{2}\:-\:\alpha_1E_1\:-\:
  \tfrac{\alpha_2E_2\:+\:\alpha_2'E_2'}{2}\,\Big)^2\:+\:s^2\,\Big]^{-1}
  \nonumber\\&\qquad \times\
  \Big\{\,-\alpha_1\Big(\,\tfrac{q_1^0\:+\:q_2^0}{2}\:-\:\alpha_1E_1\:-\:
  \tfrac{\alpha_2E_2\:+\:\alpha_2'E_2'}{2}\,\Big)\:+\:is\,\Big\}
  \nonumber\\&\qquad \times\
  (2\pi)^32E_1'\delta^{(3)}(\mathbf{k}_1\:-\:\mathbf{k}_1')
  2\mathscr{E}_1\tilde{f}_2^C(\hat{k}_2,\hat{k}_2',t)
  e^{i(\alpha_2E_2\:-\:\alpha_2'E_2')\tilde{t}_f}\;.
\end{align}

Similarly,  for the  second   cross term  $I^{(\mathrm{iiib})}$
in~(\ref{eq:B04terms}) , we obtain
\begin{align}
  \label{eq:iiib}
  &I^{(\mathrm{iiib})}(q_1,q_2,m_1,m_2,s)\ =\ 8(2\pi)^3\mu^{4-d}e^{i(q_1^0\:-\:q_2^0)\tilde{t}_f}\sum_{\{\alpha\}}
  \idotsint\!\D{}{\Pi^{d-1}_{\mathbf{k}_1}}\,\D{}{\Pi_{\mathbf{k}_1'}}
  \,\D{}{\Pi_{\mathbf{k}_2}}\,\D{}{\Pi_{\mathbf{k}_2'}}\nonumber\\
  &\qquad  \times\: (2\pi)^3\delta^{(3)}(\mathbf{q}_1\:-\:\mathbf{k}_1\:-\:\mathbf{k}_2)
  (2\pi)^3\delta^{(3)}(\mathbf{q}_2\:-\:\mathbf{k}_1'\:-\:\mathbf{k}_2')
  \nonumber\\&\qquad \times\: \frac{1}{\pi}
  \Big[\Big(\,q_1^0\:-\:q_2^0\:-\:\alpha_1E_1\:+\:\alpha_1'E_1'\,\Big)^2
  \:+\:4s^2\,\Big]^{-1}
  \Big[\Big(\,\tfrac{q_1^0\:+\:q_2^0}{2}\:-\:\tfrac{\alpha_1E_1\:+\:\alpha_1'E_1'}{2}
  \:-\:\alpha_2E_2\,\Big)^2\:+\:s^2\,\Big]^{-1}\nonumber\\
  &\qquad \times\: \Big\{\,-\alpha_2\Big(\,\tfrac{q_1^0\:+\:q_2^0}{2}\:-\:
  \tfrac{\alpha_1E_1\:+\:\alpha_1'E_1'}{2}\:-\:
  \alpha_2E_2\,\Big)\:+\:is\,\Big\}\nonumber\\&\qquad \times \
  (2\pi)^32E_2'\delta^{(3)}(\mathbf{k}_2\:-\:\mathbf{k}_2')
  2\mathscr{E}_1\tilde{f}_1(\hat{k}_1,\hat{k}_1',t)
  e^{i(\alpha_1E_1\:-\:\alpha_1'E_1')\tilde{t}_f}\;.
\end{align}

Collecting  the four  contributions from  (\ref{eq:i}), (\ref{eq:ii}),
(\ref{eq:iiia})  and  (\ref{eq:iiib})   together,  we  find  the  full
time-ordered function
\begin{align}
  \label{eq:B0tord}
  &B_0(q_1,q_2,m_1,m_2,s)\ =\ 8(2\pi)^3\mu^{4-d}e^{i(q_1^0\:-\:q_2^0)\tilde{t}_f}
  \sum_{\{\alpha\}}\idotsint\!\D{}{\Pi^{d-1}_{\mathbf{k}_1}}\,\D{}{\Pi_{\mathbf{k}_1'}}
  \,\D{}{\Pi_{\mathbf{k}_2}}\,\D{}{\Pi_{\mathbf{k}_2'}}\nonumber\\
  &\qquad \times\: (2\pi)^3\delta^{(3)}(\mathbf{q}_1\:-\:\mathbf{k}_1\:-\:\mathbf{k}_2)
  (2\pi)^3\delta^{(3)}(\mathbf{q}_2\:-\:\mathbf{k}_1'\:-\:\mathbf{k}_2')
  2\mathscr{E}_1e^{i(\alpha_1E_1\:-\:\alpha_1'E_1')\tilde{t}_f}2\mathscr{E}_2
  e^{i(\alpha_2E_2\:-\:\alpha_2'E_2')\tilde{t}_f}
  \nonumber\\&\qquad \times\: \frac{1}{\pi}
  \Big[\Big(\,q_1^0\:-\:q_2^0\:-\:\alpha_1E_1\:+\:\alpha_1'E_1'
  \:-\:\alpha_2E_2\:+\:\alpha_2'E_2'\,\Big)^2\:+\:4s^2\,\Big]^{-1}
  \nonumber\\&\qquad \times\
  \Big[\Big(\,\tfrac{q_1^0\:+\:q_2^0}{2}\:-\:\tfrac{\alpha_1E_1\:+\:\alpha_1'E_1'}{2}
  \:-\:\tfrac{\alpha_2E_2\:+\:\alpha_2'E_2'}{2}\,\Big)^2\:+\:s^2\,\Big]^{-1}
  \nonumber\\&\qquad \times\: \Big\{\,-\Big(\,\tfrac{q_1^0\:+\:q_2^0}{2}
  \:-\:\tfrac{\alpha_1E_1\:+\:\alpha_1'E_1'}{2}
  \:-\:\tfrac{\alpha_2E_2\:+\:\alpha_2'E_2'}{2}\,\Big)
  F^{\mathrm{R}}(\{\hat{k}\},t)\:+\:isF^1(\{\hat{k}\},t)\,\Big\}\;,
\end{align}
where  $\{\hat{k}\} \: \equiv  \: \{\hat{k}_1,  \hat{k}_1', \hat{k}_2,
\hat{k}_2'\}$ is  the set of  on-shell four momenta.   The statistical
structure      is     contained      within      the     distributions
$F^{\mathrm{R}}(\{\hat{k}\},t)$  and  $F^1(\{\hat{k}\},t)$ defined  as
follows:
\begin{subequations}
  \begin{align}
    F^{\mathrm{R}}(\{\hat{k}\},t)\ &=\ (2\pi)^3
    \delta^{(3)}(\mathbf{k}_1\:-\:\mathbf{k}_1')
    (2\pi)^3\delta^{(3)}(\mathbf{k}_2\:-\:\mathbf{k}_2')
    \Big(\,\theta(\alpha_1,\alpha_1',\alpha_2,\alpha_2')\:-\:
    \theta(-\alpha_1,-\alpha_1',-\alpha_2,-\alpha_2')\,\Big)
    \nonumber\\&\qquad + \: (2\pi)^3\delta^{(3)}(\mathbf{k}_1\:-\:\mathbf{k}_1')
    \Big(\,\theta(\alpha_1,\alpha_1')\:-\:\theta(-\alpha_1,-\alpha_1')\,\Big)
    \tilde{f}_2^C(\hat{k}_2,\hat{k}_2',t)\nonumber\\
    &\qquad +\: \tilde{f}_1(\hat{k}_1,\hat{k}_1',t)
    (2\pi)^3\delta^{(3)}(\mathbf{k}_2\:-\:\mathbf{k}_2')
    \Big(\,\theta(\alpha_2,\alpha_2')\:-\:\theta(-\alpha_2,-\alpha_2')\,\Big)\;,\\
    F^1(\{\hat{k}\},t)\ &=\ (2\pi)^3\delta^{(3)}(\mathbf{k}_1\:-\:\mathbf{k}_1')
    (2\pi)^3\delta^{(3)}(\mathbf{k}_2\:-\:\mathbf{k}_2')
    \Big(\,\theta(\alpha_1,\alpha_1',\alpha_2,\alpha_2')\:+\:
    \theta(-\alpha_1,-\alpha_1',-\alpha_2,-\alpha_2')\,\Big)\nonumber\\&
    \qquad +\: (2\pi)^3\delta^{(3)}(\mathbf{k}_1\:-\:\mathbf{k}_1')
    \Big(\,\theta(\alpha_1,\alpha_1')\:+\:\theta(-\alpha_1,-\alpha_1')\,\Big)
    \tilde{f}_2^C(\hat{k}_2,\hat{k}_2',t)\nonumber\\&
    \qquad +\: \tilde{f}_1(\hat{k}_1,\hat{k}_1',t)
    (2\pi)^3\delta^{(3)}(\mathbf{k}_2\:-\:\mathbf{k}_2')
    \Big(\,\theta(\alpha_2,\alpha_2')\:+\:\theta(-\alpha_2,-\alpha_2')\,\Big)
    \:+\: 2\tilde{f}_1(\hat{k}_1,\hat{k}_1',t)
    \tilde{f}^C_2(\hat{k}_2,\hat{k}_2',t)\;.
\end{align}
\end{subequations}

For  completeness,  the   anti-time-ordered  function,  obtained  from
the RHS of (\ref{eq:B0tord}) by  taking $s\to -s$  and multiplying by  an overall
minus sign, is given by
\begin{align}
&-B_0^*(q_1,q_2,m_1,m_2,s)\ =\ 8(2\pi)^3\mu^{4-d}e^{i(q_1^0\:-\:q_2^0)\tilde{t}_f}
  \sum_{\{\alpha\}}\idotsint\!\D{}{\Pi^{d-1}_{\mathbf{k}_1}}\,\D{}{\Pi_{\mathbf{k}_1'}}
  \,\D{}{\Pi_{\mathbf{k}_2}}\,\D{}{\Pi_{\mathbf{k}_2'}}\nonumber\\
  &\qquad \times\: (2\pi)^3\delta^{(3)}(\mathbf{q}_1\:-\:\mathbf{k}_1\:-\:\mathbf{k}_2)
  (2\pi)^3\delta^{(3)}(\mathbf{q}_2\:-\:\mathbf{k}_1'\:-\:\mathbf{k}_2')
  2\mathscr{E}_1e^{i(\alpha_1E_1\:-\:\alpha_1'E_1')\tilde{t}_f}2\mathscr{E}_2
  e^{i(\alpha_2E_2\:-\:\alpha_2'E_2')\tilde{t}_f}
  \nonumber\\&\qquad \times\ \frac{1}{\pi}
  \Big[\Big(\,q_1^0\:-\:q_2^0\:-\:\alpha_1E_1\:+\:\alpha_1'E_1'
  \:-\:\alpha_2E_2\:+\:\alpha_2'E_2'\,\Big)^2\:+\:4s^2\,\Big]^{-1}
  \nonumber\\&\qquad \times\
  \Big[\Big(\,\tfrac{q_1^0\:+\:q_2^0}{2}\:-\:\tfrac{\alpha_1E_1\:+\:\alpha_1'E_1'}{2}
  \:-\:\tfrac{\alpha_2E_2\:+\:\alpha_2'E_2'}{2}\,\Big)^2\:+\:s^2\,\Big]^{-1}
  \nonumber\\&\qquad \times\ \Big\{\,\Big(\,\tfrac{q_1^0\:+\:q_2^0}{2}
  \:-\:\tfrac{\alpha_1E_1\:+\:\alpha_1'E_1'}{2}
  \:-\:\tfrac{\alpha_2E_2\:+\:\alpha_2'E_2'}{2}\,\Big)
  F^{\mathrm{R}}(\{\hat{k}\},t)\:+\:isF^1(\{\hat{k}\},t)\,\Big\}\;,
\end{align}
Finally, from (\ref{eq:hadself}), the `Hadamard' function $B_0^1$ is given by
\begin{align}
  &B^{1}_0(q_1,q_2,m_1,m_2,s)\ =\ 16(2\pi)^3\mu^{4-d}e^{i(q_1^0\:-\:q_2^0)\tilde{t}_f}
  \sum_{\{\alpha\}}\idotsint\!\D{}{\Pi^{d-1}_{\mathbf{k}_1}}\,\D{}{\Pi_{\mathbf{k}_1'}}
  \,\D{}{\Pi_{\mathbf{k}_2}}\,\D{}{\Pi_{\mathbf{k}_2'}}\nonumber\\
  &\qquad \times\: (2\pi)^3\delta^{(3)}(\mathbf{q}_1\:-\:\mathbf{k}_1\:-\:\mathbf{k}_2)
  (2\pi)^3\delta^{(3)}(\mathbf{q}_2\:-\:\mathbf{k}_1'\:-\:\mathbf{k}_2')
  2\mathscr{E}_1e^{i(\alpha_1E_1\:-\:\alpha_1'E_1')\tilde{t}_f}2\mathscr{E}_2
  e^{i(\alpha_2E_2\:-\:\alpha_2'E_2')\tilde{t}_f}
  \nonumber\\&\qquad \times\ \frac{1}{\pi}
  \Big[\Big(\,q_1^0\:-\:q_2^0\:-\:\alpha_1E_1\:+\:\alpha_1'E_1'
  \:-\:\alpha_2E_2\:+\:\alpha_2'E_2'\,\Big)^2\:+\:4s^2\,\Big]^{-1}
  \nonumber\\&\qquad \times\
  \Big[\Big(\,\tfrac{q_1^0\:+\:q_2^0}{2}\:-\:\tfrac{\alpha_1E_1\:+\:\alpha_1'E_1'}{2}
  \:-\:\tfrac{\alpha_2E_2\:+\:\alpha_2'E_2'}{2}\,\Big)^2\:+\:s^2\,\Big]^{-1}isF^1(\{\hat{k}\},t)\;, 
\end{align}
which depends only on $F^1(\{\hat{k}\},t)$, as one should expect.

\subsection*{Absolutely-Ordered Functions}

We   turn   our   attention   now   to   the   $(a,b)\:=\:(2,1)$   and
$(a,b)\:=\:(1,2)$ elements of (\ref{eq:B0matrix}). These are
the  positive-  and  negative-frequency absolutely-ordered  functions,
respectively:
\begin{align}
  &B^{>(<)}_0(q_1,q_2,m_1,m_2,s)\ =\ -8(2\pi\mu)^{4-d}e^{i(q_1^0\:-\:q_2^0)\tilde{t}_f}
  \!\idotsint\!\frac{\D{d}{k_1}}{i\pi^2}\,\frac{\D{4}{k_1'}}{(2\pi)^4}\,
  \frac{\D{4}{k_2}}{(2\pi)^4}\,\frac{\D{4}{k_2'}}{(2\pi)^4}\nonumber\\
  &\qquad \times\: (2\pi)^3\delta^{(3)}(\mathbf{q}_1\:-\:\mathbf{k}_1\:-\:\mathbf{k}_2)
  (2\pi)^3\delta^{(3)}(\mathbf{q}_2\:-\:\mathbf{k}_1'\:-\:\mathbf{k}_2')\nonumber\\
  &\qquad \times\
  4s\Big[\Big(\,q_1^0\:-\:q_2^0\:-\:k_1^0\:+\:{k_1'}^0\:-\:k_2^0\:+\:{k_2'}^0\,\Big)^2\:+\:4s^2\,\Big]^{-1}
  \Big[\Big(\,q_1^0\:+\:q_2^0\:-\:k_1^0\:-\:{k_1'}^0\:-\:k_2^0\:-\:{k_2'}^0\,\Big)^2\:+\:4s^2\,\Big]^{-1}
  \nonumber\\&\qquad \times\:
  2\pi\delta(k_1^2\:-\:m_1^2)|2k_1^0|^{1/2}
  \Big(\,\theta((-)k_1^0,(-){k_1'}^0)(2\pi)^3\delta^{(3)}(\mathbf{k}_1\:-\:\mathbf{k}_1')\:+\:
  \tilde{f}_1(k_1,k_1',t)\,\Big)\nonumber\\&\qquad \qquad \times\:e^{i(k_1^0\:-\:{k_1'}^0)\tilde{t}_f}|2{k_1'}^0|^{1/2}
  2\pi\delta({k_1'}^2\:-\:m_1^2)\nonumber\\&\qquad \times\:
  2\pi\delta(k_2^2\:-\:m_2^2)|2k_2^0|^{1/2}
  \Big(\,\theta((-)k_2^0,(-){k_2'}^0)(2\pi)^3\delta^{(3)}(\mathbf{k}_2\:-\:\mathbf{k}_2')
  \:+\: \tilde{f}_2^C(k_2,k_2',t)\,\Big)\nonumber\\&\qquad \qquad \times\:
  e^{i(k_2^0\:-\:{k_2'}^0)\tilde{t}_f}|2{k_2'}^0|^{1/2}2\pi\delta({k_2'}^2\:-\:m_2^2)\;,
\end{align}
where  $\theta(x,y)$   is  defined  in   (\ref{eq:thetashort}).

After carrying out the zeroth-component momentum integrals, we obtain
\begin{align}
  &B^{>(<)}_0(q_1,q_2,m_1,m_2,s)\ =\ 16(2\pi)^3\mu^{4-d}e^{i(q_1^0\:-\:q_2^0)\tilde{t}_f}
  \sum_{\{\alpha\}}\idotsint\!\D{}{\Pi^{d-1}_{\mathbf{k}_1}}\,\D{}{\Pi_{\mathbf{k}_1'}}
  \,\D{}{\Pi_{\mathbf{k}_2}}\,\D{}{\Pi_{\mathbf{k}_2'}}\nonumber\\
  &\qquad \times\: (2\pi)^3\delta^{(3)}(\mathbf{q}_1\:-\:\mathbf{k}_1\:-\:\mathbf{k}_2)
  (2\pi)^3\delta^{(3)}(\mathbf{q}_2\:-\:\mathbf{k}_1'\:-\:\mathbf{k}_2')
  2\mathscr{E}_1e^{i(\alpha_1E_1\:-\:\alpha_1'E_1')\tilde{t}_f}2\mathscr{E}_2
  e^{i(\alpha_2E_2\:-\:\alpha_2'E_2')\tilde{t}_f}
  \nonumber\\&\qquad \times\: \frac{1}{\pi}
  \Big[\Big(\,q_1^0\:-\:q_2^0\:-\:\alpha_1E_1\:+\:\alpha_1'E_1'
  \:-\:\alpha_2E_2\:+\:\alpha_2'E_2'\,\Big)^2\:+\:4s^2\,\Big]^{-1}
  \nonumber\\&\qquad \times\
  \Big[\Big(\,\tfrac{q_1^0\:+\:q_2^0}{2}\:-\:\tfrac{\alpha_1E_1\:+\:\alpha_1'E_1'}{2}
  \:-\:\tfrac{\alpha_2E_2\:+\:\alpha_2'E_2'}{2}\,\Big)^2\:+\:s^2\,\Big]^{-1}
  i s F^{>(<)}(\{\hat{k}\},t)\;,
\end{align}
where
\begin{align}
  F^{>(<)}(\{\hat{k}\},t)\ &=\ \Big(\,\theta((-)\alpha_1,(-)\alpha_1')
  (2\pi)^3\delta^{(3)}(\mathbf{k}_1\:-\:\mathbf{k}_1')\:+\:
  \tilde{f}_1(\hat{k}_1,\hat{k}_1',t)\,\Big)\nonumber\\
  &\qquad \times\: \Big(\,\theta((-)\alpha_2,(-)\alpha_2')
  (2\pi)^3\delta^{(3)}(\mathbf{k}_2\:-\:\mathbf{k}_2')\:+\:
  \tilde{f}_2^C(\hat{k}_2,\hat{k}_2',t)\,\Big)\;.
\end{align}
We may confirm the following relations between the $F$'s:
\begin{subequations}
  \label{eq:Frels}
  \begin{align}
    F^{\mathrm{R}}(\{\hat{k}\},t)\ &=\ F^>(\{\hat{k}\},t)\:-\:F^<(\{\hat{k}\},t)\;,\\
    F^{1}(\{\hat{k}\},t)\ &=\ F^>(\{\hat{k}\},t)\:+\:F^<(\{\hat{k}\},t)\;,
  \end{align}
\end{subequations}
which  are  in  line  with   our  notation  for  the  propagators  and
the self-energies.

\subsection*{Causal Functions}

We are now in a position  to obtain the causal counterparts. Given the
relations in  (\ref{eq:useiden}), the retarded  and advanced functions
$B_0^{\mathrm{R}(\mathrm{A})}$ are given by
\begin{align}
  &B^{\mathrm{R}(\mathrm{A})}_0(q_1,q_2,m_1,m_2,s)\ =\ -8(2\pi)^3\mu^{4-d}e^{i(q_1^0\:-\:q_2^0)\tilde{t}_f}
  \sum_{\{\alpha\}}\idotsint\!\D{}{\Pi^{d-1}_{\mathbf{k}_1}}\,\D{}{\Pi_{\mathbf{k}_1'}}\,
  \D{}{\Pi_{\mathbf{k}_2}}\,\D{}{\Pi_{\mathbf{k}_2'}}\nonumber\\&\qquad 
  \times\: (2\pi)^3\delta^{(3)}(\mathbf{q}_1\:-\:\mathbf{k}_1\:-\:\mathbf{k}_2)
  (2\pi)^3\delta^{(3)}(\mathbf{q}_2\:-\:\mathbf{k}_1'\:-\:\mathbf{k}_2')
  2\mathscr{E}_1e^{i(\alpha_1E_1\:-\:\alpha_1'E_1')\tilde{t}_f}2\mathscr{E}_2
  e^{i(\alpha_2E_2\:-\:\alpha_2'E_2')\tilde{t}_f}\nonumber\\
  &\qquad \times\: \frac{1}{\pi}\Big[\Big(\,q_1^0\:-\:q_2^0\:-\:\alpha_1E_1
  \:+\:\alpha_1'E_1'\:-\:\alpha_2E_2\:+\:
  \alpha_2'E_2'\,\Big)^2\:+\:4s^2\,\Big]^{-1}\nonumber\\
  &\qquad \times\: \Big(\,\tfrac{q_1^0\:+\:q_2^0}{2}\:-\:\tfrac{\alpha_1E_1
  \:+\:\alpha_1'E_1'}{2}\:-\:\tfrac{\alpha_2E_2
  \:+\:\alpha_2'E_2'}{2}\:+(-)\:is\,\Big)^{-1}F^{\mathrm{R}(\mathrm{A})}(\{\hat{k}\},t)\;,
\end{align}
where $F^{\mathrm{R}}(\{\hat{k}\},t)\:=\:F^{\mathrm{A}}(\{\hat{k}\},t)$.

\subsection*{The Thermodynamic Equilibrium Limit}
\label{sec:eqself}

In the thermodynamic equilibrium limit, we  expect to be  able to
recover  the results  from  the discussions  of Section~\ref{sec:eq},
using  the correspondence identified  in (\ref{eq:eqlim}).  It follows
that  the   various  distribution  functions   satisfy  the  following
factorization:
\begin{align}
  F(\{\hat{k}\},t)\ &\rightarrow \ (2\pi)^3\delta^{(3)}(\mathbf{k}_1\:-\:\mathbf{k}_1')
  \Big(\,\theta(\alpha_1,\alpha_1')\:+\:\theta(-\alpha_1,-\alpha_1')\,\Big)\nonumber\\
  &\qquad \times\: (2\pi)^3\delta^{(3)}(\mathbf{k}_2\:-\:\mathbf{k}_2')
  \Big(\,\theta(\alpha_2,\alpha_2')\:+\:\theta(-\alpha_2,-\alpha_2')\,\Big)
  \alpha_1\alpha_2F_{\mathrm{eq}}(\alpha_1E_1,\alpha_2E_2)\;,
\end{align}
where
\begin{subequations}
  \begin{align}
    F^{>}_{\mathrm{eq}}(\alpha_1E_1,\alpha_2E_2)&\ =\
    \Big(\,1\:+\:f_{\mathrm{B}}(\alpha_1E_1)\,\Big)
    \Big(\,1\:+\:f^{C}_{\mathrm{B}}(\alpha_2E_2\,)\Big)\;,\\
    F^{<}_{\mathrm{eq}}(\alpha_1E_1,\alpha_2E_2)&\ =\
    f_{\mathrm{B}}(\alpha_1E_1)
    f^{C}_{\mathrm{B}}(\alpha_2E_2)\;.
\end{align}
\end{subequations}
Using  the  above expressions,  we  may perform  all  but  one of  the
three-dimensional  phase-space integrations and  both of  the `primed'
summations  in  the  various  $B_0$  functions.  We  then  obtain  the
following set of `equilibrium' results:
\begin{subequations}
\begin{align}
  \label{eq:eqtc}
  &B_0(-B_0^*)(q_1,q_2,m_1,m_2,s)\ =\ (2\pi)^4\frac{1}{\pi}\,\frac{1}{\big(q_1^0\:-\:q_2^0\big)^2\:+\:4s^2}
  e^{i(q_1^0\:-\:q_2^0)\tilde{t}_f}\delta^{(3)}(\mathbf{q}_1\:-\:\mathbf{q}_2)\nonumber\\
  &\qquad \times\: (2\pi\mu)^{4-d}\sum_{\{\alpha\}}
  \int\!\D{d-1}{\mathbf{k}_1}\,\frac{1}{\pi}\,
  \frac{\alpha_1\alpha_2}{E_1E_2}\,\Big[\Big(\,\tfrac{q_1^0\:+\:q_2^0}{2}\:-\:\alpha_1E_1\:-\:
  \alpha_2E_2\,\Big)^2\:+\:s^2\,\Big]^{-1}\nonumber\\
  &\qquad \times\: \Big[\,-(+)\Big(\,\tfrac{q_1^0\:+\:q_2^0}{2}\:-\:
  \alpha_1E_1\:-\:\alpha_2E_2\,\Big)\,
  F^{\mathrm{R}}_{\mathrm{eq}}(\alpha_1E_1,\alpha_2E_2)\:+\: isF^1_{\mathrm{eq}}(\alpha_1E_1,\alpha_2E_2)\,\Big]\;,\\
  &B^{>,<,1}_0(q_1,q_2,m_1,m_2,s)\ =\ (2\pi)^4\frac{1}{\pi}\,\frac{1}{\big(q_1^0\:-\:q_2^0\big)^2\:+\:4s^2}
  e^{i(q_1^0\:-\:q_2^0)\tilde{t}_f}\delta^{(3)}(\mathbf{q}_1\:-\:\mathbf{q}_2)\nonumber\\
  &\qquad \times\: (2\pi\mu)^{4-d}\sum_{\{\alpha\}}
  \int\!\D{d-1}{\mathbf{k}_1}\,\frac{1}{\pi}\,
  \frac{\alpha_1\alpha_2}{E_1E_2}\,\Big[\Big(\,\tfrac{q_1^0\:+\:q_2^0}{2}\:-\:\alpha_1E_1
  \:-\:\alpha_2E_2\,\Big)^2\:+\:s^2\,\Big]^{-1}
  2i s F^{>,<,1}_{\mathrm{eq}}(\alpha_1E_1,\alpha_2E_2)\;,\\
  \label{eq:eqtc2}
  &B^{\mathrm{R}(\mathrm{A})}_0(q_1,q_2,m_1,m_2,s)\ =\ -(2\pi)^4\frac{1}{\pi}\,
  \frac{1}{\big(q_1^0\:-\:q_2^0\big)^2\:+\:4s^2}e^{i(q_1^0\:-\:q_2^0)\tilde{t}_f}\delta^{(3)}(\mathbf{q}_1\:-\:\mathbf{q}_2)
  \nonumber\\&\qquad \times\ (2\pi\mu)^{4-d}\sum_{\{\alpha\}}
  \int\!\D{d-1}{\mathbf{k}_1}\,\frac{1}{\pi}\,
  \frac{\alpha_1\alpha_2}{E_1E_2}\,\Big(\,\tfrac{q_1^0\:+\:q_2^0}{2}\:-\:\alpha_1E_1
  \:-\:\alpha_2E_2\:+(-)\:is\,\Big)^{-1}
  F^{\mathrm{R\,(A)}}_{\mathrm{eq}}(\alpha_1E_1,\alpha_2E_2)\;,
\end{align}
\end{subequations}
with      $F^{\mathrm{R}}_{\mathrm{eq}}(\alpha_1E_1,\alpha_2E_2)     =
F^{\mathrm{A}}_{\mathrm{eq}}(\alpha_1E_1,\alpha_2E_2)$.

At     this     point,      we     caution     the     reader     that
(\ref{eq:eqtc})--(\ref{eq:eqtc2}) are not  the exact equilibrium $B_0$
functions. For late times, we may use the final value theorem
\begin{equation}
  \lim_{t\:\to\:\infty}\,F(t)\ = \ \lim_{s\:\to\: 0}\,s\mathcal{L}_{t}\big[F\big](s)
\end{equation}
to obtain the true time-invariant equilibrium functions:
\begin{subequations}
\label{eq:tinveqb0}
\begin{align}
  &B_0(-B_0^*)(q_1,q_2,m_1,m_2)\ = \ (2\pi)^4\delta^{(4)}(q_1\:-\:q_2)(2\pi\mu)^{4-d}
  \sum_{\{\alpha\}}\int\!\D{d-1}{\mathbf{k}_1}\,
  \frac{\alpha_1\alpha_2}{E_1E_2}\nonumber\\&\qquad \times \
  \bigg(\,-(+)\frac{1}{2\pi}\,\frac{1}{q_1^0\:-\:\alpha_1E_1
  \:-\:\alpha_2E_2}\,F^{\mathrm{R}}_{\mathrm{eq}}(\alpha_1E_1,\alpha_2E_2)
  \:+\:\frac{i}{2}\,\delta(q_1^0\:-\:
  \alpha_1E_1\:-\:\alpha_2E_2)
  F^1_{\mathrm{eq}}(\alpha_1E_1,\alpha_2E_2)\,\bigg)\;,\\
  &B^{>,<,1}_0(q_1,q_2,m_1,m_2)\ =\ i(2\pi)^4\delta^{(4)}(q_1\:-\:q_2)(2\pi\mu)^{4-d}
  \sum_{\{\alpha\}}\int\!\D{d-1}{\mathbf{k}_1}\,
  \frac{\alpha_1\alpha_2}{E_1E_2}\nonumber\\
  &\qquad \times \: \delta(q_1^0\:-\:\alpha_1E_1\:-\:
  \alpha_2E_2)F^{>,<,1}_{\mathrm{eq}}(\alpha_1E_1,\alpha_2E_2)\;,\\
  &B^{\mathrm{R}(\mathrm{A})}_0(q_1,q_2,m_1,m_2)\ =\ -(2\pi)^4\delta^{(4)}(q_1\:-\:q_2)
  (2\pi\mu)^{4-d}\sum_{\{\alpha\}}\int\!\D{d-1}{\mathbf{k}_1}\;
  \frac{\alpha_1\alpha_2}{E_1E_2}\nonumber\\
  &\qquad \times\: \frac{1}{2\pi}\,\frac{1}{q_1^0\:-\:\alpha_1E_1
  \:-\:\alpha_2E_2\:+(-)\:i\epsilon}\,
  F^{\mathrm{R}}_{\mathrm{eq}}(\alpha_1E_1,\alpha_2E_2)\;.
\end{align}
\end{subequations}
The above expressions are  consistent with known results calculated in
the ITF or equilibrium CTP formalism (see Appendix~\ref{app:ITF}).

\subsection*{The Homogeneous Limit}

Lastly,  we summarize  the  time-dependent homogeneous limit  of
the $B_0$ functions.  In the  spatially homogeneous case,  the $F$'s
satisfy the following factorization
\begin{align}
  F(\{\hat{k}\},t)\ &\rightarrow \  (2\pi)^3
  \delta^{(3)}(\mathbf{k}_1\:-\:\mathbf{k}_1')
  \Big(\,\theta(\alpha_1,\alpha_1')\:+\:\theta(-\alpha_1,-\alpha_1')\,\Big)
  \nonumber\\&\qquad \times \:(2\pi)^3\delta^{(3)}(\mathbf{k}_2\:-\:\mathbf{k}_2')
  \Big(\,\theta(\alpha_2,\alpha_2')\:+\:\theta(-\alpha_2,-\alpha_2')\,\Big)
  F_{\mathrm{hom}}(\alpha_1,\alpha_2,\mathbf{k}_1,\mathbf{k}_2)\;.
\end{align}
The set of homogeneous distributions $F_{\mathrm{hom}}$ can be obtained from
\begin{align}
    F^{>(<)}_{\mathrm{hom}}(\alpha_1,\alpha_2,\mathbf{k}_1,\mathbf{k}_2,t)\ &=\ 
    \Big(\,\theta((-)\alpha_1)\:+\:\theta(\alpha_1)f\big(|\mathbf{k}_1|,t\big)\:+\:
    \theta(-\alpha_1)f^C\big(|\mathbf{k}_1|,t\big)\,\Big)\nonumber\\
    &\qquad \times\: \Big(\,\theta((-)\alpha_2)\:+\:\theta(\alpha_2)f^C\big(|\mathbf{k}_2|,t\big)
    \:+\:\theta(-\alpha_2)f\big(|\mathbf{k}_2|,t\big)\,\Big)\;,
\end{align}
using  the  relations  in (\ref{eq:Frels}).   Substituting  these
distributions into the non-homo\-geneous $B_0$ functions, we
perform all but one of the three-dimensional phase-space integrals and
both `primed'  summations.  Subsequently, with  the aid of  an inverse
Laplace  transformation with  respect  to $s$,  the  following set  of
$t$-dependent homogeneous $B_0$ functions are obtained:
\begin{subequations}
\begin{align}
  &B_0(-B_0^*)(q_1,q_2,m_1,m_2,\tilde{t}_f;\tilde{t}_i)\ =\ (2\pi)^4\delta^{(3)}(\mathbf{q}_1\:-\:\mathbf{q}_2)
  (2\pi\mu)^{4-d}e^{i(q_1^0\:-\:q_2^0)\tilde{t}_f}\nonumber\\&\qquad \times \sum_{\{\alpha\}}\int\!\D{d-1}{\mathbf{k}_1}\,
  \frac{1}{E_1E_2}\bigg[\,-(+)\frac{1}{2\pi}\,
  \frac{\tfrac{q_1^0\:+\:q_2^0}{2}\:-\:\alpha_1E_1\:-\:\alpha_2 E_2}
  {\big(q_1^0\:-\:\alpha_1E_1\:-\:\alpha_2E_2\big)\big(q_2^0\:-\:\alpha_1E_1\:-\:\alpha_2E_2\big)}
  \nonumber\\& \qquad \times\: \Big(\,\delta_t(q_1^0\:-\:q_2^0)\:-\:
  \delta_t\big(q_1^0\:+\:q_2^0\:-\:2\alpha_1E_1\:-\:2\alpha_2E_2\big)\,\Big)
  F^{\mathrm{R}}_{\mathrm{hom}}(\alpha_1,\alpha_2,\mathbf{k}_1,\mathbf{q}_1-\mathbf{k}_1,t)\nonumber\\&
  \qquad \qquad +\: \frac{i}{2}\,\delta_t\big(q_1^0\:-\:\alpha_1E_1\:-\:\alpha_2E_2\big)
  \delta_t\big(q_2^0\:-\:\alpha_1E_1\:-\:\alpha_2E_2\big)
  F^1_{\mathrm{hom}}(\alpha_1,\alpha_2,\mathbf{k}_1,\mathbf{q}_1-\mathbf{k}_1,t)\,\bigg]\;,\\
  &B^{>,<,1}_0(q_1,q_2,m_1,m_2,\tilde{t}_f;\tilde{t}_i)\ =\ i(2\pi)^4\delta^{(3)}(\mathbf{q}_1\:-\:\mathbf{q}_2)
  (2\pi\mu)^{4-d}e^{i(q_1^0\:-\:q_2^0)\tilde{t}_f}\nonumber\\&\qquad \times \sum_{\{\alpha\}}\int\!\D{d-1}{\mathbf{k}_1}\,
  \frac{1}{E_1E_2}\,\delta_t\big(q_1^0\:-\:\alpha_1E_1\:-\:\alpha_2E_2\big)
  \delta_t\big(q_2^0\:-\:\alpha_1E_1\:-\:\alpha_2E_2\big)
  F^{>,<,1}_{\mathrm{hom}}(\alpha_1,\alpha_2,\mathbf{k}_1,\mathbf{q}_1-\mathbf{k}_1,t)\;,\\
  &B^{\mathrm{R}(\mathrm{A})}_0(q_1,q_2,m_1,m_2,\tilde{t}_f;\tilde{t}_i)\ =\ (2\pi)^4\delta^{(3)}(\mathbf{q}_1\:-\:\mathbf{q}_2)
  (2\pi\mu)^{4-d}e^{i(q_1^0\:-\:q_2^0)\tilde{t}_f}\nonumber\\&\qquad \times \sum_{\{\alpha\}}\int\!\D{d-1}{\mathbf{k}_1}\,
  \frac{1}{E_1E_2}\bigg[\,-\frac{1}{2\pi}\,
  \frac{\tfrac{q_1^0\:+\:q_2^0}{2}\:-\:\alpha_1E_1\:-\:\alpha_2E_2}
  {\big(q_1^0\:-\:\alpha_1E_1\:-\:\alpha_2E_2\big)\big(q_2^0\:-\:\alpha_1E_1\:-\:\alpha_2E_2\big)}
  \nonumber\\& \qquad \times\: \Big(\,\delta_t(q_1^0\:-\:q_2^0)\:-\:
  \delta_t\big(q_1^0\:+\:q_2^0\:-\:2\alpha_1E_1\:-\:2\alpha_2E_2\big)\,\Big)\nonumber\\
  &\qquad \qquad +(-)\: \frac{i}{2}\, \delta_t\big(q_1^0\:-\:\alpha_1E_1\:-\:\alpha_2E_2\big)
  \delta_t\big(q_2^0\:-\:\alpha_1E_1\:-\:\alpha_2E_2\big)\,\bigg]
  F^{\mathrm{R}(\mathrm{A})}_{\mathrm{hom}}(\alpha_1,\alpha_2,\mathbf{k}_1,\mathbf{q}_1-\mathbf{k}_1,t)\;.
\end{align}
\end{subequations}
In the limit $t\:\to\:  \infty$, using (\ref{eq:deltatlim}) and making the
replacement
\begin{equation}
  F_{\mathrm{hom}}(\alpha_1,\alpha_2,\mathbf{k}_1,\mathbf{q}_1-\mathbf{k}_1,t)
  \ \rightarrow\ \alpha_1\alpha_2
  F_{\mathrm{eq}}(\alpha_1,\alpha_2,\mathbf{k}_1,\mathbf{q}_1-\mathbf{k}_1)\;,
\end{equation}
we    recover    the    time-independent   equilibrium    results    in
(\ref{eq:tinveqb0}).

\newpage

\bibliographystyle{apsrev4-1}
\bibliography{pTFT}

\end{document}